\pdfoutput=1
\documentclass[cernpreprint, atlasdraft=false, UKenglish, texmf, orcidlogo]{atlasdoc}

\usepackage{atlaspackage}
\usepackage{atlasbiblatex}

\usepackage{atlasphysics}
\usepackage{multirow}
\usepackage{comment}
\usepackage{amsmath}

\graphicspath{{logos/}{figures/}}

\usepackage{ANA-TOPQ-2018-24-PAPER-defs}

\AtlasTitle{Measurements of inclusive and differential cross-sections of $t\bar{t}\gamma$ production in $pp$ collisions at $\sqrt{s}=13$~TeV with the ATLAS detector}

\AtlasAbstract{%
Inclusive and differential cross-sections are measured at particle level for the associated production of a top quark pair and a photon ($t\bar{t}\gamma$). The analysis is performed using an integrated luminosity of 140\,fb$^{-1}$ of proton--proton collisions at a centre-of-mass energy of 13~TeV collected by the ATLAS detector. The measurements are performed in the single-lepton and dilepton top quark pair decay channels focusing on $t\bar{t}\gamma$ topologies where the photon is radiated from an initial-state parton or one of the top quarks. The absolute and normalised differential cross-sections are measured for several variables characterising the photon, lepton and jet  kinematics as well as the angular separation between those objects. The observables are found to be in good agreement with the Monte Carlo predictions. The photon transverse momentum differential distribution is used to set limits on effective field theory parameters related to the electroweak dipole moments of the top quark. The combined limits using the photon and the $Z$ boson transverse momentum measured in $t\bar{t}$ production in associations with a $Z$ boson are also set. }

\AtlasRefCode{TOPQ-2018-24}

\PreprintIdNumber{CERN-EP-2024-052}

\AtlasJournal{JHEP}
\AtlasJournalRef{JHEP 10 (2024) 191}
\AtlasDOI{10.1007/JHEP10(2024)191}

\hypersetup{pdftitle={ATLAS document},pdfauthor={The ATLAS Collaboration}}

\begin{document}

\maketitle

\section{Introduction}

Precise measurements of the associated production of a top quark pair (\ttbar) with a high-energy photon provide crucial information about the predictions of the Standard Model (SM) in the top quark sector, specifically of the top-photon electroweak coupling. Therefore, measurements of the inclusive and differential cross-sections of the \tty process are a probe for possible extensions of the SM, being sensitive to new physics through anomalous dipole moments of the top quark~\cite{Baur:2004uw, Bouzas:2012av, Schulze:2016qas}, studied in the context of effective field theories (EFT)~\cite{Bylund:2016phk}.

The first evidence for the production of \tty was reported by the CDF Collaboration~\cite{Aaltonen:2011sp}, while the observation of the \tty process was established by the ATLAS Collaboration with the data collected at $\sqrt{s}=7~\TeV$~\cite{TOPQ-2012-07}. Since those measurements were published, efforts have concentrated on improving the precision and extending the scope of the inclusive and differential cross-section measurements~\cite{TOPQ-2015-21, CMS-TOP-14-008, TOPQ-2017-14, TOPQ-2020-03, CMS-TOP-18-010, CMS-TOP-21-004} and the study of the production properties~\cite{TOPQ-2021-29}.

In \tty final states, the photon can be emitted in the top quark production stage or in the decay stage (including the $W$ boson and $W$ boson decay products). Theoretical studies~\cite{Bevilacqua:2018woc,Bevilacqua:2019quz} show that in the narrow-width approximation, which is valid in the considered kinematic regime as shown in
Ref.~\cite{Bevilacqua:2019quz},
the cross-section of the \tty process with a high \pt photon in the final state can be factorised into two contributions:
the first one describes the photon emission in the production part of the process, while the second corresponds to the emission from the top quark decay products. The interference among the production and decay contributions is negligible, both in the SM case and when considering the EFT effects. Of the two contributions the production one is the most sensitive to the top-photon coupling.

\begin{figure}[ht!]
\centering
\subfloat[]{\includegraphics[width=0.45\textwidth]{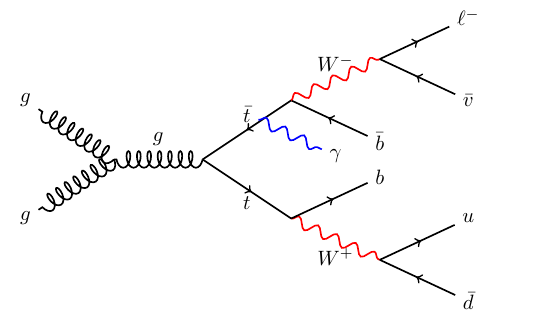}}%
\subfloat[]{\includegraphics[width=0.45\textwidth]{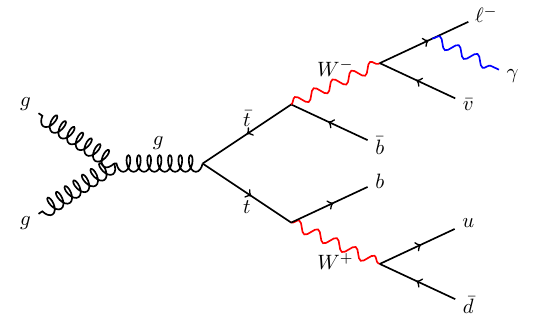}}
\caption{Examples of leading-order Feynman diagrams for (a) \tty production and (b) \tty decay in the single-lepton final state. The diagrams correspond to the cases where the photon is radiated by an off-shell top quark and by a charged lepton, respectively.}
\label{fig:Feynman}
\end{figure}

The main focus of this \paperornote is the measurement of the inclusive and differential cross-sections of top quark pair production with an additional photon where the photon is radiated in the production part, i.e., from an initial-state parton or from an off-shell top quark, referred to as \tty production, following the notation of Ref.~\cite{TOPQ-2021-29}.
Thus, the analysis strategy is designed to improve the separation of the \tty production events from the \tty events where the photons are radiated from any of the charged decay products of the top quark (including the $W$ boson), referred to as \tty decay. Examples of Feynman diagrams at leading order for \tty production and \tty decay are illustrated in Figure~\ref{fig:Feynman}. In the simulation of the processes, all possible diagrams are considered. This approach enhances the sensitivity of the measurement to the top--photon coupling. The cross-sections of the \tty process, regardless of the origin of the photon, are also measured to facilitate the comparison with a previous measurement using a partial data set at 13~\TeV~\cite{TOPQ-2017-14}.

The measurements are performed using the data sample collected during Run 2 with the ATLAS detector at the Large Hadron Collider (LHC), between 2015 and 2018, corresponding to an integrated luminosity of 140~fb$^{-1}$, exploiting both the single-lepton and dilepton \ttbar channels at stable particle level in a fiducial phase space. The differential cross-sections are measured in the same fiducial region as functions of photon kinematic variables, angular variables related to the photon and the leptons or the jets, and, in the dilepton channel, as functions of
angular separations between the two leptons in the event.

In the single-lepton channel, events with exactly one photon, one lepton and at least four jets, with at least one jet identified as
coming from the hadronisation of a $b$-quark ($b$-tagged), are selected. In the dilepton channel, the events are required to contain exactly one photon, two oppositely charged leptons (electrons or muons), and at least two jets with at least one of them being $b$-tagged.
To separate the \tty signal from the main background sources, multivariate discriminants are built using neural networks (NNs). They are used to define signal and control regions, enriched in signal or in background events, respectively. In the single-lepton channel, a multi-class approach is used, considering three different background categories, while in the dilepton channel a binary classifier is trained due to the smaller contribution of background processes.

The inclusive cross-section is obtained via a profile-likelihood fit to the discriminating variables in the signal and control regions. The differential cross-sections are corrected to stable particle level using a
profile-likelihood unfolding approach. The results are interpreted in the context of EFT, testing the sensitivity of the \tty production process to possible modifications of the coupling between the top quark and the photon. The limits on the relevant Wilson coefficients in the SM effective field theory (SMEFT) framework~\cite{Bylund:2016phk} are obtained from the differential cross-section measurement of the photon \pt. The production of \ttbar in association with a $Z$ boson (\ttZ) is also sensitive to modifications of the electroweak couplings between the $Z$ boson and the top quark. The combined limits are extracted from the simultaneous measurement of the photon \pt and $Z$ boson \pt, presented in Ref.~\cite{ATLAS:2023eld}.

This \paperornote is organised as follows. The ATLAS detector is described in Section~\ref{sec:detector}. The simulation of signal and background processes and the event reconstruction and selection are discussed in Sections~\ref{sec:modelling} and ~\ref{sec:selection}, respectively. The estimate of the background processes is summarised in Section~\ref{sec:background}. The sources of systematic uncertainties considered in the measurements are described in Section~\ref{sec:systematics}. Details of the analysis strategy and the NNs employed to discriminate between signal and background events are given in Section~\ref{sec:NN}.  The definition of the fiducial regions and the results of the inclusive and differential cross-section measurements are presented in Sections~\ref{sec:fiducial}, ~\ref{sec:inclusive} and~\ref{sec:differential}, respectively. The combination with the \ttZ process and the EFT interpretation are discussed in Section~\ref{sec:EFT}. Finally, a summary of the results is given in Section~\ref{sec:conclusion}.

\section{ATLAS detector}
\label{sec:detector}
\newcommand{\AtlasCoordFootnote}{%
ATLAS uses a right-handed coordinate system with its origin at the nominal interaction point (IP)
in the centre of the detector and the \(z\)-axis along the beam pipe.
The \(x\)-axis points from the IP to the centre of the LHC ring,
and the \(y\)-axis points upwards.
Polar coordinates \((r,\phi)\) are used in the transverse plane,
\(\phi\) being the azimuthal angle around the \(z\)-axis.
The pseudorapidity is defined in terms of the polar angle \(\theta\) as \(\eta = -\ln \tan(\theta/2)\) and is equal to the rapidity
$ y = \frac{1}{2} \ln \left( \frac{E + p_z c}{E - p_z c} \right) $ in the relativistic limit.
Angular distance is measured in units of \(\Delta R \equiv \sqrt{(\Delta y)^{2} + (\Delta\phi)^{2}}\).}

The ATLAS detector~\cite{PERF-2007-01} at the LHC covers nearly the entire solid angle around the collision point.\footnote{\AtlasCoordFootnote}
It consists of an inner tracking detector surrounded by a thin superconducting solenoid, electromagnetic and hadronic calorimeters,
and a muon spectrometer incorporating three large superconducting air-core toroidal magnets.

The inner-detector system (ID) is immersed in a \qty{2}{\tesla} axial magnetic field
and provides charged-particle tracking in the range \(|\eta| < 2.5\).
The high-granularity silicon pixel detector covers the vertex region and typically provides four measurements per track,
the first hit generally being in the insertable B-layer (IBL) installed before Run~2~\cite{ATLAS-TDR-19,PIX-2018-001}.
It is followed by the SemiConductor Tracker (SCT), which usually provides eight measurements per track.
These silicon detectors are complemented by the transition radiation tracker (TRT),
which enables radially extended track reconstruction up to \(|\eta| = 2.0\).
The TRT also provides electron identification information
based on the fraction of hits (typically 30 in total) above a higher energy-deposit threshold corresponding to transition radiation.

The calorimeter system covers the pseudorapidity range \(|\eta| < 4.9\).
Within the region \(|\eta|< 3.2\), electromagnetic calorimetry is provided by barrel and
endcap high-granularity lead/liquid-argon (LAr) calorimeters,
with an additional thin LAr presampler covering \(|\eta| < 1.8\)
to correct for energy loss in material upstream of the calorimeters.
Hadronic calorimetry is provided by the steel/scintillator-tile calorimeter,
segmented into three barrel structures within \(|\eta| < 1.7\), and two copper/LAr hadronic endcap calorimeters.
The solid angle coverage is completed with forward copper/LAr and tungsten/LAr calorimeter modules
optimised for electromagnetic and hadronic energy measurements respectively.

The muon spectrometer (MS) comprises separate trigger and
high-precision tracking chambers measuring the deflection of muons in a magnetic field generated by the superconducting air-core toroidal magnets.
The field integral of the toroids ranges between \num{2.0} and \qty{6.0}{\tesla\metre}
across most of the detector.
Three layers of precision chambers, each consisting of layers of monitored drift tubes, cover the region \(|\eta| < 2.7\),
complemented by cathode-strip chambers in the forward region, where the background is highest.
The muon trigger system covers the range \(|\eta| < 2.4\) with resistive-plate chambers in the barrel, and thin-gap chambers in the endcap regions.

The luminosity is measured mainly by the LUCID--2~\cite{LUCID2} detector that records Cherenkov light produced in the quartz windows of photomultipliers located close to the beampipe.

Events are selected by the first-level trigger system implemented in custom hardware,
followed by selections made by algorithms implemented in software in the high-level trigger~\cite{TRIG-2016-01}.
The first-level trigger accepts events from the \qty{40}{\MHz} bunch crossings at a rate below \qty{100}{\kHz},
which the high-level trigger further reduces in order to record complete events to disk at about \qty{1}{\kHz}.

A software suite~\cite{ATL-SOFT-PUB-2021-001} is used in data simulation, in the reconstruction
and analysis of real and simulated data, in detector operations, and in the trigger and data acquisition
systems of the experiment.

\section{Simulation of signal and background processes}
\label{sec:modelling}
A set of Monte Carlo (MC) simulated samples described in detail in Ref.~\cite{TOPQ-2021-29} is used in this paper to model the different signal and background processes.
The response of the ATLAS detector was simulated~\cite{SOFT-2010-01} with \GEANT~\cite{Agostinelli:2002hh}. Some of the alternative samples used to evaluate systematic uncertainties and in the EFT interpretation were processed using a fast simulation (\textsc{AtlFast-II}), which relies on a parameterisation of the calorimeter response~\cite{ATL-PHYS-PUB-2010-013}.
All samples were reconstructed using the same software as for the data.

Two samples to model the \tty process were simulated with the \MGNLO~\cite{Alwall:2014hca} generator. The first MC sample of \tty events with the photon produced from the top quark or from initial-state radiation was simulated as a $2\to 3$ process at next-to-leading (NLO) accuracy in quantum chromodynamics (QCD) with the on-shell top quarks in the final state being decayed at leading order (LO) using \MADSPIN~\cite{Frixione:2007zp,Artoisenet:2012st} to preserve spin correlations. The interference effects between initial-state photon radiation (from the incoming partons) and the final-state photon radiation (from the off-shell top quarks) were taken into account in the simulation. This sample is referred to as the `\tty production sample’ in the following.
The second \tty sample, where the photon arises from any of the decay products of the top quarks or from one of the on-shell top quarks, was simulated with the same version of \MGNLO but at LO precision as a $2 \to 2$ process followed by the decay of the top quarks, also simulated with \MGNLO at LO precision. This sample is referred to in the following as the `\tty decay sample’. Both samples were generated by using the \NNPDF[3.0nlo]~\cite{Ball:2014uwa} parton distribution function (PDF) set. The renormalisation and factorisation scales were set to $0.5\times \sum_i \sqrt{m^2_i+p^2_{\text{T},i}}$, where $m_i$ and $p_{\text{T},i}$ are the masses and transverse momenta of the particles generated from the matrix element (ME) calculation. In both samples, photons were required to have $\pT > 15~\GeV$ and to be isolated according to a smooth-cone hadronic isolation criterion %
~\cite{Frixione:1998jh}, which avoids infrared divergences.
The \tty production sample is normalised to the NLO cross-section given by the MC simulation, while the normalisation of the \tty decay sample is corrected by a NLO/LO inclusive $K$-factor of 1.5 obtained in Ref.~\cite{TOPQ-2021-29}.

The \tWy events were generated at LO accuracy with the \MGNLO generator in the five-flavour scheme (5FS). To simulate the full process, two complementary samples were generated: one as a $2\rightarrow3$ process assuming a stable top quark and the other as a $2 \to 2$ process, where the photon is radiated from any other final-state charged particle. The decay of the top quarks is also simulated with \MGNLO. To avoid infrared divergences, the photon was required to have $\pT > 15~\GeV$ and $|\eta| < 5.0$ and to be separated by $\Delta R>0.2$ from any parton. Both samples make use of the \NNPDF[2.3lo] PDF set~\cite{Pumplin:2002vw} and are normalised to the cross-section provided by the MC simulation.

The production of \ttbar and single-top-quark events ($tW$, $t$- and $s$-channels) was modelled at NLO in QCD using \powhegbox~\cite{Nason:2004rx,Frixione:2007nw,Frixione:2007vw,Alioli:2010xd}. For the \ttbar sample, the \hdamp parameter that controls the \pt of the first additional emission,
was set to 1.5 times the top quark mass~\cite{ATL-PHYS-PUB-2016-020}. The \ttbar and single-top-quark simulated samples are normalised to the cross-sections calculated at next-to-next-to-leading order (NNLO) in QCD including the
resummation of next-to-next-to-leading-logarithmic (NNLL) soft-gluon terms~\cite{Czakon:2011xx} at NNLO or approximated NNLO~\cite{Kidonakis:2010ux,Kidonakis:2010tc,Kidonakis:2011wy}, respectively. The overlap between the \ttbar and the single top $tW$ final
states is removed using the diagram-removal scheme~\cite{Frixione:2008yi}.

The production of $V\gamma$, diboson processes ($VV$) with leptonic final states and $V$+jets processes with $V=W,Z$ were simulated using \SHERPA~\cite{Bothmann:2019yzt,Hoeche:2009rj} at NLO in QCD with the \NNPDF[3.0nnlo] PDF set.
The $V\gamma$ production is normalised to the cross-section provided by the MC simulation. The diboson processes are normalised to NLO QCD cross-sections~\cite{Campbell:1999ah}, and the $W$ and $Z$ boson samples are normalised to the cross-sections calculated at NNLO in QCD~\cite{STDM-2015-03}.
Events with a \ttbar pair produced in association with a $W$~or $Z$~boson ($t\bar{t}V$) were simulated at NLO in QCD with \MGNLO using the \NNPDF[3.0nlo] PDF set.

Two \tty MC samples used in previous publications~\cite{TOPQ-2020-03, TOPQ-2017-14} as nominal and to determine the parton shower uncertainty are used in this paper for comparison with the unfolded data in Section~\ref{sec:total_tty_diff}. These samples were simulated at LO using the \madgraph v2.3.3 generator~\cite{Alwall:2014hca} and the \NNPDF[2.3lo] PDF set and were interfaced either to \PYTHIA[8] or \Herwig[7]. The \tty events were generated as a doubly resonant $2\to7$~process. Diagrams where the photon is radiated from the initial state, from the intermediate top quarks, \bquarks, the intermediate $W$~bosons, and from the decay products of the $W$ bosons, were included. The LO MC samples are re-scaled to the NLO prediction using the $K$-factors derived in Ref.~\cite{TOPQ-2017-14}.

Additional MC samples were simulated at LO in QCD for the \tty and \ttZ production processes to perform the EFT interpretation. They were generated by using the \MGNLO generator with the same set-up as those discussed in Ref.~\cite{ATLAS:2023eld}, including as universal FeynRules output the \texttt{SMEFTsim 3.0} model~\cite{Brivio:2020onw}, in the $M_W$ electroweak input scheme~\cite{Brivio:2021yjb} with the \texttt{top} flavour restrictions (5FS). The renormalisation and factorisation scales were set to $\mu=\sum_i m_i$, where $i$ indicates the massive final state resonances. The events were generated as the SM process, and the \MGNLO reweighting module was used to obtain alternative event weights to model the dimension-6 EFT vertices and propagators. Further details can be found in Ref.~\cite{ATLAS:2023eld}.

All samples generated with \powhegbox and \MGNLO  were interfaced to \PYTHIA[8]~\cite{Sjostrand:2007gs} to simulate the parton shower, fragmentation, and underlying event. The A14 set of tuned parameters (tune)~\cite{ATL-PHYS-PUB-2014-021} and the \NNPDF[2.3lo] PDF set were used in \Pythia. The heavy-flavour hadron decays were modelled by EvtGen~\cite{Lange:2001uf}.
All samples generated with \Sherpa used the \Sherpa parton shower based on the Catani--Seymour dipole factorisation with a dedicated tune provided by the authors~\cite{Schumann:2007mg}.
The top quark mass was set to \SI{172.5}{\GeV}.
The overlap between the samples in which events were generated without explicitly including a photon in the ME in the final state and the samples where photons were included, such as \tty, \Wty and $V\gamma$, was removed following the same procedure as described in Ref.~\cite{TOPQ-2021-29}. In particular, events with parton-level photons with $\pT > 15~\GeV$ and separated by $\Delta R>0.2$ from any charge lepton are discarded. The number of the events from the inclusive samples subject to the overlap removal amounts to a few per cent of the total number of events.

Additional \pp~collisions in the same or neighbouring bunch crossings (pile-up) were modelled by overlaying minimum bias events, simulated
using \PYTHIA[8] with the A3 tune~\cite{ATL-PHYS-PUB-2016-017} and the \NNPDF[2.3lo] PDF set, on events from hard-scatter processes.
The MC events were reweighted to reproduce the distribution of the average number of interactions per bunch crossing observed in the data.

Corrections to the trigger, reconstruction and selection efficiencies, and the energy scales and resolutions, are applied to the MC simulated events to match the performance in the data.

\section{Event reconstruction and selection}
\label{sec:selection}

This analysis is performed using the $pp$ collision data collected with the ATLAS detector between 2015 and 2018 at $\sqrt{s} = \SI{13}{\TeV}$.
After the application of data-quality requirements~\cite{DAPR-2018-01},
the data sample corresponds to an integrated luminosity of 140~\ifb~\cite{DAPR-2021-01}.
Events were selected using single-lepton triggers with variable electron and
muon transverse momentum (\pT) thresholds, and various identification and isolation criteria depending on the lepton flavour and the data-taking period~\cite{TRIG-2018-01,TRIG-2018-05,TRIG-2019-04_sp,TRIG-2019-03}.
The events are also required to have at least one reconstructed collision vertex with two or more associated tracks with  $\pt>0.5$~\GeV. The primary vertex is chosen as the vertex with the highest $\sum \pt^2$ of the associated tracks and consistent with the average beam-spot position.

Photon candidates are reconstructed from energy deposits (clusters) in the central region of the electromagnetic calorimeter. Two types of photon candidates are considered: the  unconverted, with the cluster not matched to any reconstructed track in the ID system, and the converted, with the cluster matched to reconstructed tracks consistent with originating from a photon conversion and with a reconstructed conversion vertex. Both types of photons are required to satisfy a \emph{tight} identification criteria~\cite{EGAM-2018-01} and to be isolated~\cite{EGAM-2018-01} fulfilling the criteria $\left. E_{\mathrm{T}}^{\mathrm{iso}} \right|_{\Delta R = 0.4} < 0.022 \cdot E_{\mathrm{T}}(\gamma) + \SI{2.45}{\GeV}$ and $\left. \pT^{\mathrm{iso}} \right|_{\Delta R = 0.2} < 0.05 \cdot E_{\mathrm{T}}(\gamma)$, where $\left. E_{\mathrm{T}}^{\mathrm{iso}} \right|_{\Delta R = 0.4}$ is the calorimeter isolation within a cone of radius $\Delta R = 0.4$ in the direction of the photon candidate, $\left. \pT^{\mathrm{iso}} \right|_{\Delta R = 0.2}$ is the track isolation within $\Delta R = 0.2$ and $E_{\mathrm{T}}(\gamma)$ is the transverse energy of a photon.\footnote{The calorimeter isolation $\left. E_{\mathrm{T}}^{\mathrm{iso}} \right|_{\Delta R = 0.4}$ is calculated as the sum of the transverse energy of the topological clusters in the calorimeter in a cone of $\Delta R = 0.4$ around the barycentre of the photon candidate, with the core energy of the photon candidate subtracted. The track isolation $\left. \pT^{\mathrm{iso}} \right|_{\Delta R = 0.2}$ is obtained by adding the \pt of the tracks with $\pt>1$~GeV within $\Delta R = 0.2$ around the photon cluster. The tracks associated with photon conversions are excluded from the calculation.}. Photon candidates are selected if they satisfy $E_{\mathrm{T}}(\gamma)>$~20~\GeV, $|\eta_{\text{cluster}}|<2.37$ and ${|\eta_{\text{cluster}}|\not\in[1.37,1.52]}$, the transition region between the barrel and the endcap calorimeters.

Electron candidates are reconstructed from energy deposits in the central region of the electromagnetic calorimeter associated with reconstructed tracks from the ID system, and are required to satisfy the \emph{MediumLH} identification criteria~\cite{EGAM-2018-01} and have a pseudorapidity of  $|\eta_{\text{cluster}}|<2.47$, excluding candidates in the transition region %
(${|\eta_{\text{cluster}}|\not\in[1.37,1.52]}$).

Muon candidates are reconstructed with a combined algorithm, using the track segments in the various layers of the muon spectrometer and the tracks in the ID system. Muons are required to satisfy the \emph{Medium} identification quality criteria and to have
$|\eta| < 2.5$~\cite{MUON-2018-03}.

The electron and muon candidates are required to have $\pT>$~25~\GeV\ ($\pT>$~20~\GeV\ for the subleading lepton in the dilepton channel) and meet the
loose working point (WP) of the prompt-lepton isolation discriminant~\cite{HIGG-2017-02}, trained to separate prompt and non-prompt leptons. The transverse impact parameter divided by its estimated
uncertainty, ${|\dzero|/\sigma(\dzero)}$, is required to be less
than five (three) for electron (muon) candidates. The longitudinal impact
parameter must satisfy ${|\zzsth|<0.5}$~mm for both lepton flavours.

Jets are reconstructed using the anti-$k_{t}$ algorithm~\cite{Cacciari:2008gp} in the \textsc{FastJet} implementation~\cite{Fastjet} with a radius parameter $R=0.4$ using particle flow objects~\cite{PERF-2015-09}. The jet energy scale (JES) and resolution are calibrated using simulations with in situ corrections obtained from data~\cite{JETM-2018-05}. They are required to have $\pT > \SI{25}{\GeV}$ and $|\eta| < 2.5$. To reject jets from pile-up or other primary vertices, a \emph{jet vertex tagger} (JVT) discriminant~\cite{PERF-2014-03} is required to be larger than 0.59 for jets with $\pT < \SI{60}{\GeV}$ and $|\eta| < 2.4$. Jets arising from \bquark hadronisation, referred to as \bjets, are identified using the DL1r \btag algorithm~\cite{ATLAS:2022qxm}. The $b$-tagged jets are required to satisfy the WP corresponding to 70\% or 85\% efficiency for identifying $b$-quark initiated jets in \ttbar simulated events in the single-lepton or dilepton channels, respectively. To fully exploit the $b$-tagging information of an event, each jet is assigned a pseudo-continuous $b$-tagging score that defines if a jet satisfies a given WP but fails to satisfy the adjacent tighter one. A score of two, three, four or five is assigned to a jet satisfying the 85\%, 77\%, 70\% or 60\% WP, respectively. If a jet does not satisfy any WP, a score of one is assigned.

The missing transverse momentum vector, with magnitude \met, is defined as the negative sum of the transverse momenta of the reconstructed and calibrated physical objects, plus a `soft term’ built from all other tracks not matched to a reconstructed object that are associated with the primary vertex~\cite{ATLAS-CONF-2018-023}.

An overlap removal procedure is applied to avoid the double counting of detector signatures. Electron candidates sharing a track with a muon candidate are first removed. The closest jet found within a $\Delta R$ = 0.2 cone of an electron is removed and electrons within a $\Delta R$ = 0.4 cone of a remaining jet are rejected. Jets with less than three associated tracks and within $\Delta R = 0.2$ of the muon and muons within $\Delta R= 0.4$ of a jet with more than two associated tracks are rejected. Photons within a $\Delta R$ = 0.4 cone of a remaining lepton are removed. Finally, jets found in a $\Delta R$ = 0.4 cone around a photon are removed.

Events in the single-lepton channel are selected if they contain exactly one photon and one isolated electron or muon, fulfilling the requirements described above and matched to the corresponding trigger-level object. The \pt thresholds for the matched leptons are $25~\GeV$ in 2015 data, $27~\GeV$ in 2016 data, and $28~\GeV$ in 2017 and 2018 data, which are at least 1~\GeV\ above the \pt thresholds of the single-lepton triggers. Events are rejected if there are additional lepton candidates with $\pt > 7$~\GeV. Additionally, only events where the invariant mass of the electron and the photon is outside a $Z$ boson mass window of $\pm$5~\GeV\ are kept to reduce the background contribution from $Z \to ee$ events, where an electron is misidentified as a photon.
Events are required to contain at least four reconstructed jets, and at least one of these jets is required to be tagged
as a $b$-jet using a 70\% WP\@. In the dilepton channel, events are selected if they contain exactly two isolated leptons ($ee$, $e\mu$, $\mu\mu$) of opposite electric charges and at least one is matched to the corresponding trigger-level object, and at least two jets out of which at least one is $b$-tagged using a WP with 85\% efficiency. In the same lepton flavour channels, events with an invariant mass of the lepton pair smaller than 15~\GeV are rejected to suppress contribution from the decays of heavy-flavour resonances and low-mass Drell–Yan processes. Events in the $ee$ and $\mu\mu$ channels are also required to have an invariant mass outside a $Z$ boson mass window of $\pm$~5~\GeV, and to have $\met >$~30~\GeV to further reject backgrounds from $Z\gamma$ processes.
\section{Background estimation}
\label{sec:background}
Background processes are divided into categories based on the photon origin into the prompt and fake or misidentified photon background, where the photon is mimicked by another object, such as a misreconstructed hadron or electron. The contribution of events with prompt photons is estimated by using MC simulation while the fake backgrounds are estimated by using data-driven methods. In the following, the background categories and the methods used to estimate them are briefly summarised. Further details can be found in Ref.~\cite{TOPQ-2021-29}.

The largest prompt photon background contribution arises from \tty decay events, about 30\% (45\%) of the total number of events in the single-lepton (dilepton) channel.
All other background processes with a prompt photon ($tW\gamma$, single-top quark, $V\gamma$, $VV$, $t\bar{t}V$ and \ttbar), referred to as the `\catprompty' background category, constitute about 15\% of the selected events in both channels.

The contribution from processes with an electron mimicking a photon signature, `e-fake', is estimated in data from the numbers of electron--positron candidates from
$Z\rightarrow e^+e^-$ decays that are reconstructed as $ee$ or $e\gamma$ pairs~\cite{PERF-2017-02}. Two control regions are defined: one containing events with an electron--positron pair and the other, enriched in e-fake events, selecting events with an electron and a photon satisfying the selection criteria of the signal region.
The e-fake rate is determined by comparing the observed $ee$ and $e\gamma$ invariant mass spectra around the $Z$ boson peak in these two regions, after having subtracted the respective background contributions, estimated from the sidebands around the $Z$ boson peak. It is measured as a function of photon $\eta$ and \pt, separately for converted and unconverted photons.
The systematic uncertainty in the e-fake background estimate arises from the choice of the function describing the $Z$ boson mass peak and the mass range considered as well as from the assumed function for the background description. The ratio of the e-fake rates in data and simulation measured in Ref.~\cite{TOPQ-2021-29} is used to correct the e-fake background background predicted by the simulation in the signal regions. The e-fake background originates mainly from dilepton \ttbar events where an electron from the top quark decay chain is identified as a photon resulting in the same signature as the
single-lepton \tty channel. It represents about 15\% of the total number of selected events in this channel,  while the e-fake contribution is below 1\% in the dilepton channel.

The background contribution from events where the photon signature arises from hadronic energy depositions in the electromagnetic calorimeter or from hadron decays, referred to as `h-fake', is estimated from data using the `ABCD' data-driven method~\cite{CDF:1990kbl}. In this approach, three orthogonal regions enriched with h-fake photon events are defined either by inverting the photon isolation or by loosening some of the identification criteria. Counting the number of selected data events in these regions, after subtracting the prompt photon and e-fake background contributions, allows the number of h-fake events in the signal region to be evaluated and the correction factors relative to the MC prediction to be determined. These correction factors are obtained as a function of photon \pt and $\eta$, separately for converted and unconverted photons, and applied to the simulation to determine the h-fake background contribution. Since around 90\% of h-fake events originate from \ttbar production, all modelling uncertainties considered for the \ttbar process and described in Section~\ref{sec:systematics} are propagated to the correction factors and constitute the dominant source of the uncertainty. Additional contributions arise from the normalisation uncertainties in the prompt photon backgrounds and the uncertainties in the data-driven e-fake background entering the h-fake enriched regions.
This category corresponds to about 7\% (3\%) of the events in the single-lepton (dilepton) channel.

An additional small background contribution arising from events with a non-prompt or misidentified lepton, referred to as `lepton fake', is also obtained from data with the `matrix method'~\cite{EGAM-2019-01}. The contribution of this process in the single-lepton channel is estimated from the data events, satisfying lepton selection requirements with looser identification and isolation, corrected by a weight that depends on the efficiency of a real prompt and a fake or misidentified loose lepton to satisfy tight lepton selection criteria. %
Systematic uncertainties in the fake lepton background estimate arise from the statistical uncertainties and the choice of parameterisation of real and fake lepton efficiencies.
This contribution is negligible in the dilepton channel. It is taken from simulation and is not treated as a separate category.

The observed data and the expected event yields for the signal and background processes in both channels after applying all data-driven corrections are given in Table~\ref{tab:yield_prefit}. The `Other prompt $\gamma$' category corresponds to \catprompty background excluding the $tW\gamma$ and $W\gamma$ contribution in the single-lepton channel.

\begin{table}
\centering
\caption{The observed data and the expected event yields for the signal and backgrounds in the single-lepton and dilepton channels. All data-driven corrections are included. The \tty decay sample is scaled by the NLO $K$-factor. The $W\gamma$ contribution is included in the Other prompt $\gamma$ category in the dilepton channel. Lepton fakes in the dilepton channel are negligible and are included in the corresponding MC yields. The uncertainties correspond to the sum of the statistical and systematic uncertainties (see Section~\ref{sec:systematics}). The uncertainty of the \tty decay events includes the uncertainty in the $K$-factor.}
\label{tab:yield_prefit}
\begin{tabular}{l c c }
\hline
Category& Single-lepton channel & Dilepton channel \\
\hline
\tty production  &$12450 \pm \phantom{0}740$ & $2400 \pm \phantom{0}99$\\
\tty decay       &$13400 \pm 3100$ & $3100 \pm 640$\\
h-fake          &$\phantom{0}3600 \pm 1200$ & $\phantom{0}220 \pm \phantom{0}82$\\
e-fake          &$\phantom{0}6900 \pm \phantom{0}980$ & $\phantom{0}57.9 \pm 7.0$\\
$W\gamma$        &$\phantom{0}2700 \pm 1400$ & -- \\
$tW\gamma$       &$\phantom{0}1180 \pm \phantom{0}580$ & $\phantom{0}290 \pm 150$ \\
Other prompt $\gamma$   &$\phantom{0}2500 \pm \phantom{0}600$ & $ \phantom{0}820 \pm 170$\\
Lepton fake      &$\phantom{0}\phantom{0}640 \pm \phantom{0}110$ & --\\
\hline
Total            &$43900 \pm 4600$ & $6900 \pm 710$  \\
\hline
Data             & 47767  & 7379 \\
\hline
\end{tabular}
\end{table}
\section{Systematic uncertainties}
\label{sec:systematics}
Several sources of systematic uncertainty, arising from detector effects or theoretical assumptions, and uncertainties due to the limited number of events in the MC simulations, are considered in the inclusive and differential measurements in both channels. They affect the event yields and the shapes of the distributions of the observables of interest.

The correction factors applied to the simulated samples to improve the description of the photon and lepton identification and isolation efficiencies, momentum scale and resolution~\cite{EGAM-2018-01,MUON-2018-03}, and lepton trigger efficiencies are varied within their uncertainties to estimate the corresponding systematic uncertainty. The JES uncertainty~\cite{JETM-2018-05} accounts for contributions from jet-flavour composition, $\eta$-intercalibration, punch-through, single-particle response, calorimeter response to different jet flavours, and pile-up, resulting in 30 uncorrelated JES uncertainty subcomponents. The jet energy resolution has been measured separately for data and MC using two in situ techniques~\cite{JETM-2018-05}. The  systematic uncertainty is defined as the quadratic difference between the jet energy resolutions for data and simulation and split into 13 uncorrelated uncertainty components.
Additionally, the uncertainty associated with the JVT discriminant for pile-up jet rejection is obtained by varying the efficiency correction factors~\cite{PERF-2014-03}. The uncertainties in the $b$-jet tagging calibration are determined separately for $b$-jets, $c$-jets and light-flavour jets~\cite{PERF-2016-05,FTAG-2020-08,FTAG-2019-02} and are decomposed into several uncorrelated components for each category. Additionally, a 50\% normalisation uncertainty is applied to events with high-\pt jets above the \pt limit where the $b$-jet tagging algorithm is calibrated. The fractions of such events with high-\pt $b$-jets, $c$-jets or light jets are about 0.3\%, 0.8\% and 3.5\%, respectively. The uncertainty in \MET results from the propagation of the uncertainties in the energy scales and resolutions of photons, leptons and jets, and from the modelling of its soft term~\cite{ATLAS-CONF-2018-023}.

The uncertainty in the integrated luminosity is 0.83\%~\cite{DAPR-2021-01}, obtained using the LUCID-2 detector~\cite{LUCID2} for the primary luminosity measurements. The uncertainty in the pile-up modelling is obtained by varying the pile-up reweighting applied to match the distribution in the simulation to the data within its uncertainties.

The signal and background modelling uncertainties account for effects from the choice of QCD scales, parton shower (PS), amount of QCD initial-state radiation (ISR), and PDF set. They are treated as uncorrelated sources of uncertainty between different processes. The effect of the QCD scale uncertainty for each of the \tty, \Wty and \ttbar processes is obtained by separately varying the renormalisation and factorisation scales relative to their nominal values by a factor of two up and down. The uncertainty from the PS and hadronisation is estimated by comparing the nominal simulated samples interfaced to \PYTHIA[8] with samples interfaced to \HERWIG[7.2.1]~\cite{Bahr:2008pv,Bellm:2015jjp}. The uncertainty in the modelling of the ISR is estimated by varying the amount of radiation corresponding to the \textit{var3c} eigentune uncertainty~\cite{ATL-PHYS-PUB-2014-021} in the A14 tune. The impact of the ME corrections applied to all emissions is evaluated by comparing the nominal sample to a dedicated \powheg + \PYTHIA[8] sample with the corrections turned off.
An additional uncertainty is applied to the \ttbar process by comparing the nominal sample to the alternative one simulated with
\hdamp parameter in \powheg, increased by a factor of two~\cite{ATL-PHYS-PUB-2018-009}.
The uncertainty in the PDFs for the signal and background \tty processes is estimated by using the 30 PDF variations of the PDF4LHC15 prescription~\cite{Butterworth:2015oua}. Each PDF variation is considered as a separate nuisance parameter in the fit.

The total uncertainties associated with the data-driven estimates of the e-fake, h-fake, and lepton-fake backgrounds are propagated to the final result. A normalisation uncertainty of 20\% is assigned to the \tty decay process, based on the estimated uncertainty in the NLO $K$-factor~\cite{TOPQ-2017-14}, in the case of the measurement of the total, production and decay, \tty cross-section.
The uncertainty in the normalisation of the background processes estimated from MC is 6\% for \ttbar~\cite{Czakon:2011xx},
50\% for $W\gamma$~\cite{TOPQ-2021-29} and for all other minor background processes contributing to the \catprompty  category, i.e.,  single top quark, \ttV, diboson, and $Z\gamma$.
\section{Signal discrimination}
\label{sec:NN}
To separate the signal from the background processes multivariate analysis techniques are employed, in particular fully connected feed forward NNs. In the single-lepton channel, a multi-class NN is used to separate the \tty production signal from the background, which is divided into three categories: \tty decay, fake photon processes (\efakes and \hfakes) and other prompt photon backgrounds.
In the dilepton channel, where the background contribution from sources other than \tty decay is much smaller, a binary classification NN is employed to separate the \tty production signal from all background sources.

A total of 40 and 16 variables are used as input to the NN training in the single-lepton and the dilepton channels, respectively. These variables describe the kinematic properties of the photon and the leptons, the kinematic and flavour properties of the jets, and the invariant masses and angular distances between different combinations of objects.
Additionally, in the single-lepton channel, to improve the discrimination between the \tty production and \tty decay processes,  the \tty event is reconstructed. The two jets with the highest $b$-tagging score in the event are considered to be the two $b$-jets from a \ttbar decay. If two $b$-tagged jets have the same score, the one with the largest \pt is chosen. The rest of the jets are used to find a pair with an invariant mass closest to the $W$ boson mass to reconstruct the hadronically decaying $W$ boson (`hadronic $W$ boson'). The leptonically decaying $W$ boson (`leptonic $W$~boson') is reconstructed using the $W$ boson mass constraint, lepton and \met information, and solving the quadratic equation to determine the $z$-component of the neutrino momentum. For real solutions, the smallest value is chosen, while in the case of complex solutions, the real part of the solution is considered. The reconstructed $W$ bosons, $b$-jets and a photon are used to reconstruct the top quark and antiquark. Out of the four possible permutations corresponding to an assignment of $b$-jets and a photon to the hadronic or leptonic $W$ boson decay, the one that minimises the value of
$\Delta m^2=(m_1-m_{t})^2+(m_2-m_{t})^2$,
where $m_{1}$ and $m_{2}$ are the invariant masses of the two reconstructed top quarks and $m_t=172.5$~\GeV, the top quark mass in the simulation, is selected as the best solution. Several variables related to the reconstructed $W$ bosons, such as the invariant masses of the leptonic and hadronic $W$ bosons and $\Delta m^2$, are used in the NN training.
Since the selected photon is included in the event reconstruction as one of the \ttbar decay products,
these variables are expected to behave differently and provide separation between \tty decay and \tty production events, where the photon is not emitted by the decay products of the \ttbar system. The list of variables used in the NN training in each channel and examples of the variables with the largest separation power can be found in Appendix~\ref{sec:appendix_NN}.

The training of the NNs is performed using Keras~\cite{chollet2015keras} with the TensorFlow backend~\cite{tensorflow2015-whitepaper} and categorical or binary cross-entropy as a loss function in the single-lepton and dilepton channels, respectively. All MC simulated samples described in Section~\ref{sec:modelling} with the corresponding data-driven scale factors, if applicable, are used in the training. It is verified that the shapes of the input variables in the inclusive region and in each of the categories are well described by the simulations. The resulting output distributions are shown in Figure~\ref{fig:NNoutput_prefit} for the four NN nodes (\tty production, \tty decay, photon fakes and other prompt photon) in the single-lepton channel and for the binary NN in the dilepton channel. The shape of the distributions is well described by the simulation in all the categories. A slight underestimate of the data by the SM prediction in the regions where the fraction of \tty production events is significant is likely due to a smaller \tty production cross-section in MC simulation than in data that was observed in the previous measurements~\cite{TOPQ-2017-14, CMS-TOP-21-004}.

\begin{figure}[ht!]
\centering
\subfloat[]{\includegraphics[width=0.35\textwidth]{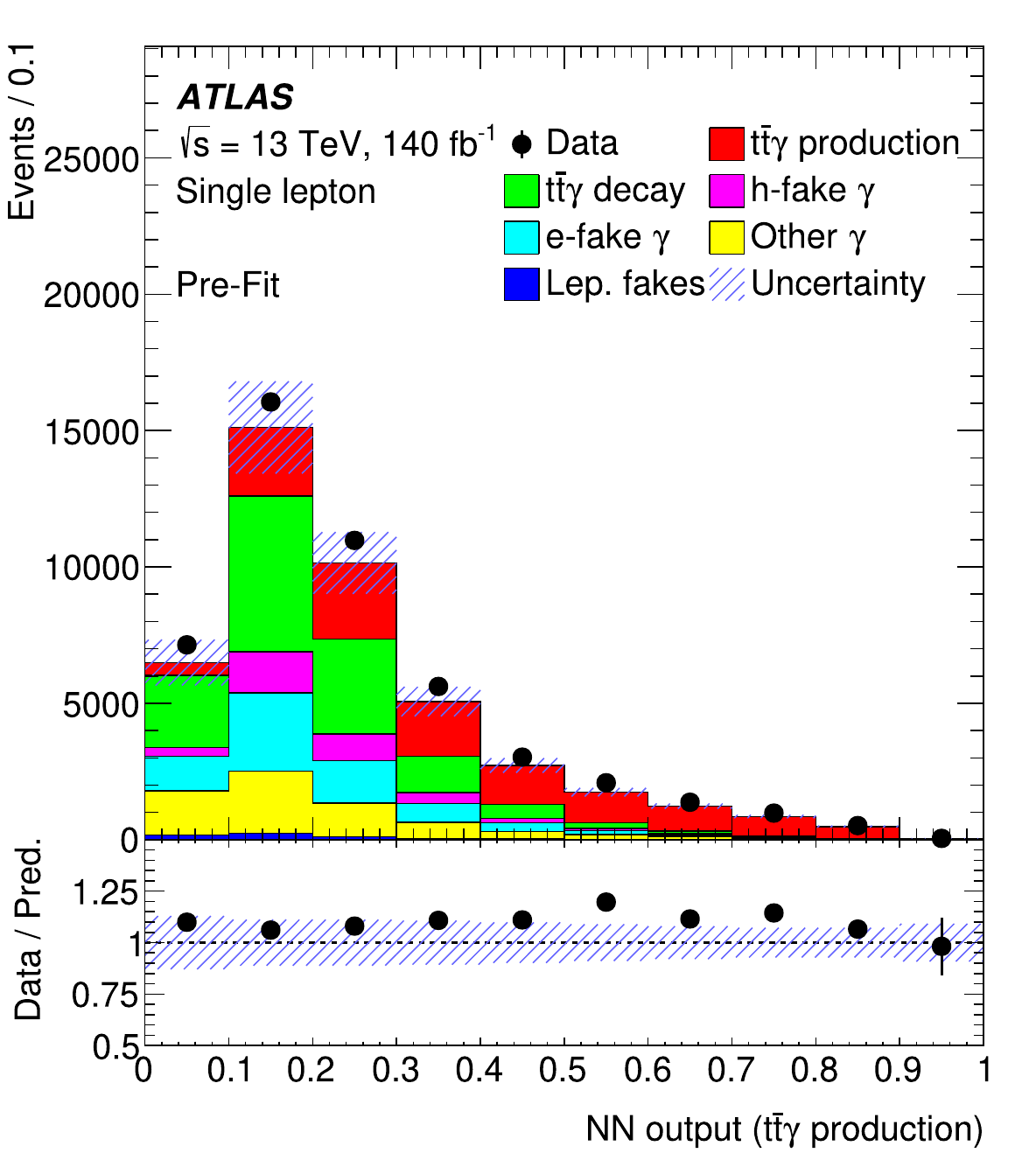}}%
\subfloat[]{\includegraphics[width=0.35\textwidth]{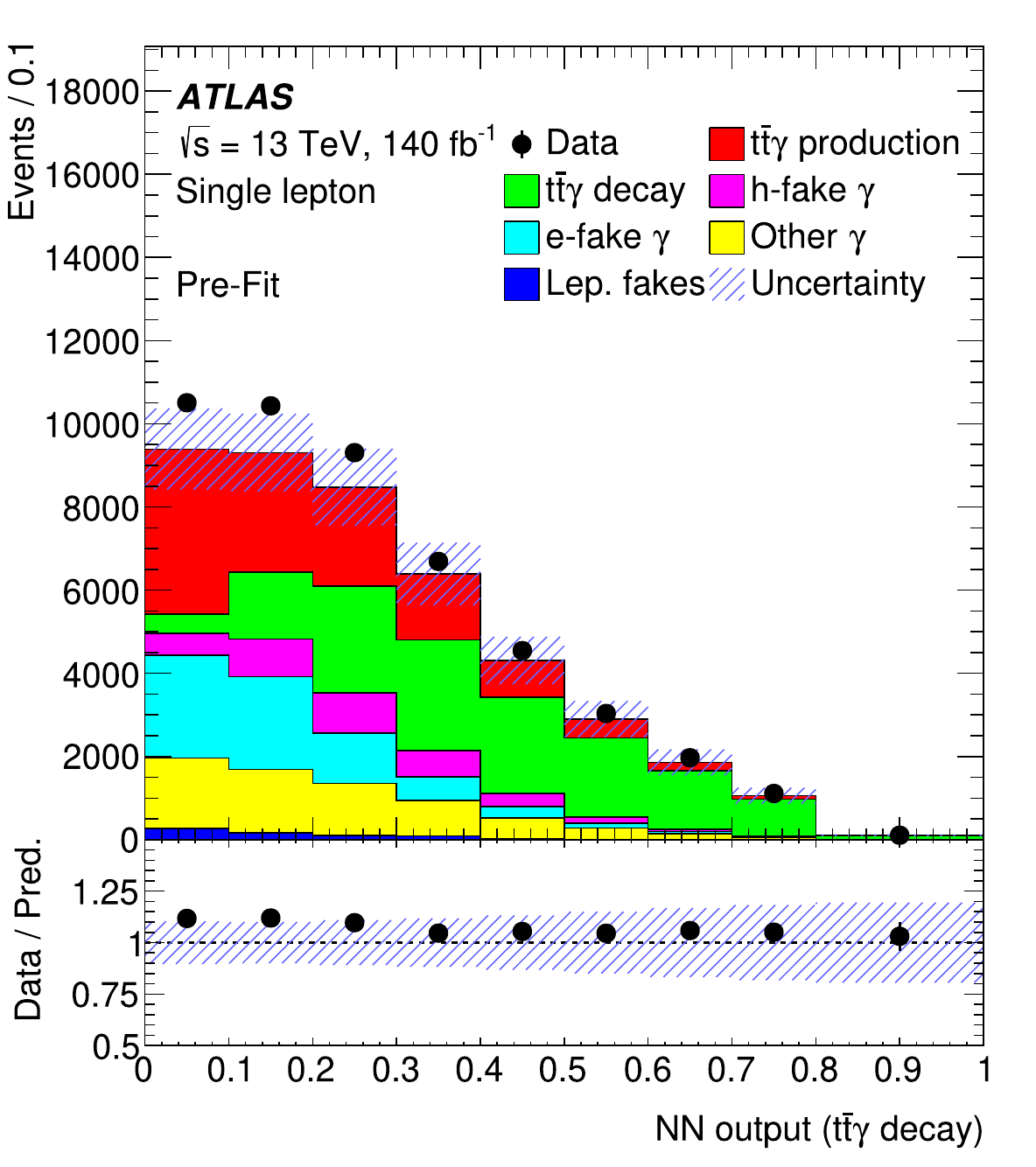}}\\
\subfloat[]{\includegraphics[width=0.35\textwidth]{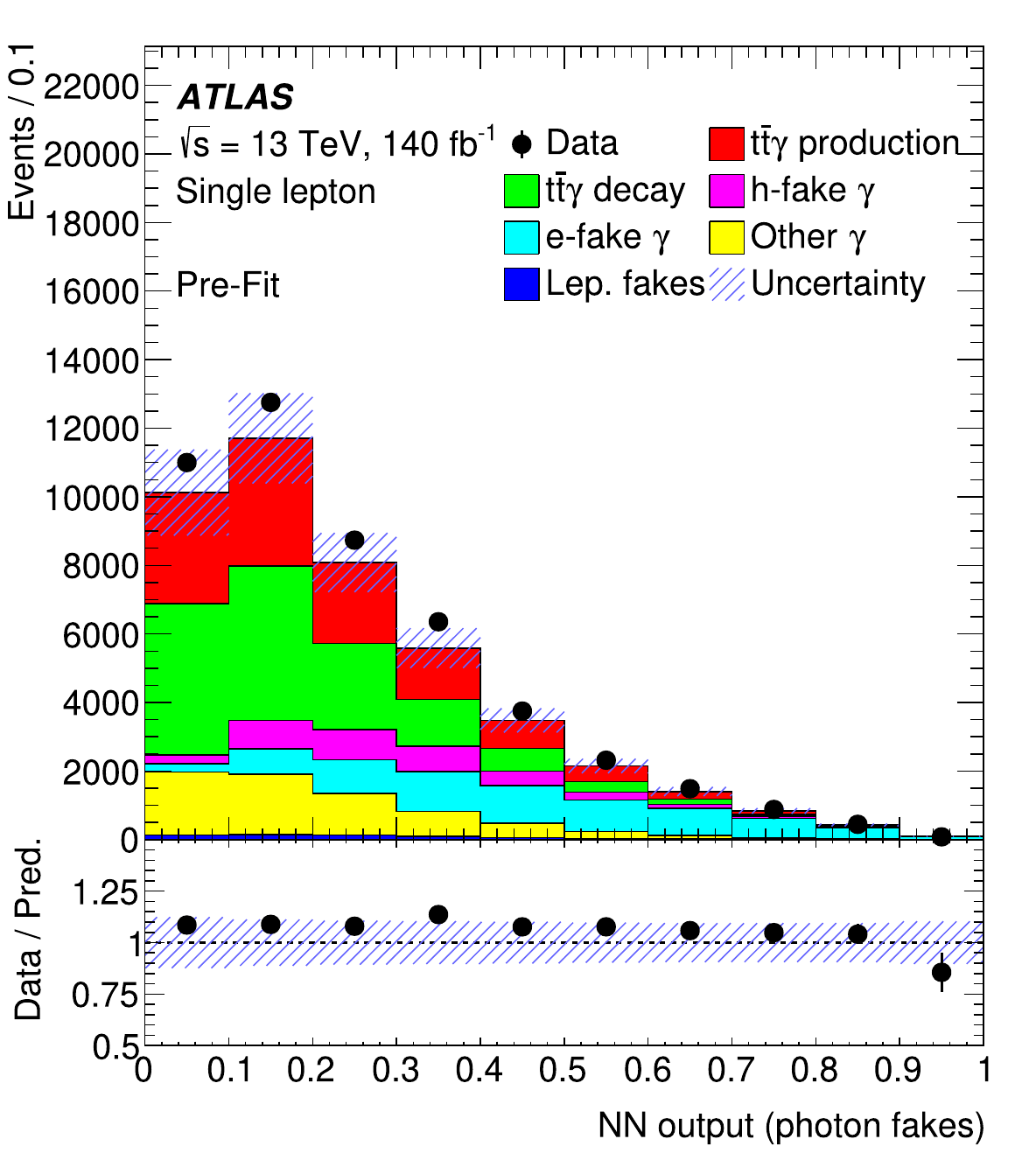}}%
\subfloat[]{\includegraphics[width=0.35\textwidth]{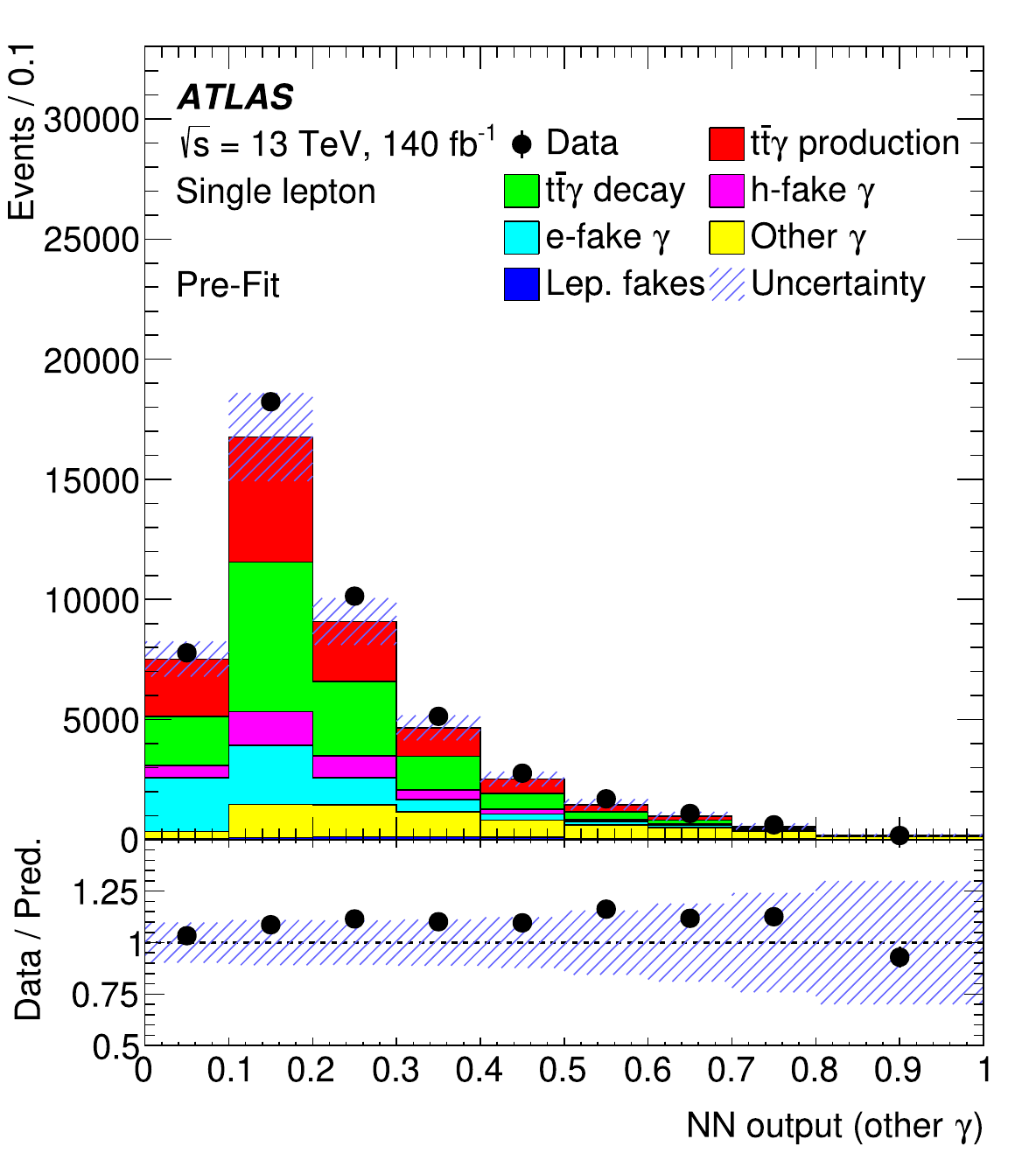}}\\
\subfloat[]{\includegraphics[width=0.35\textwidth]{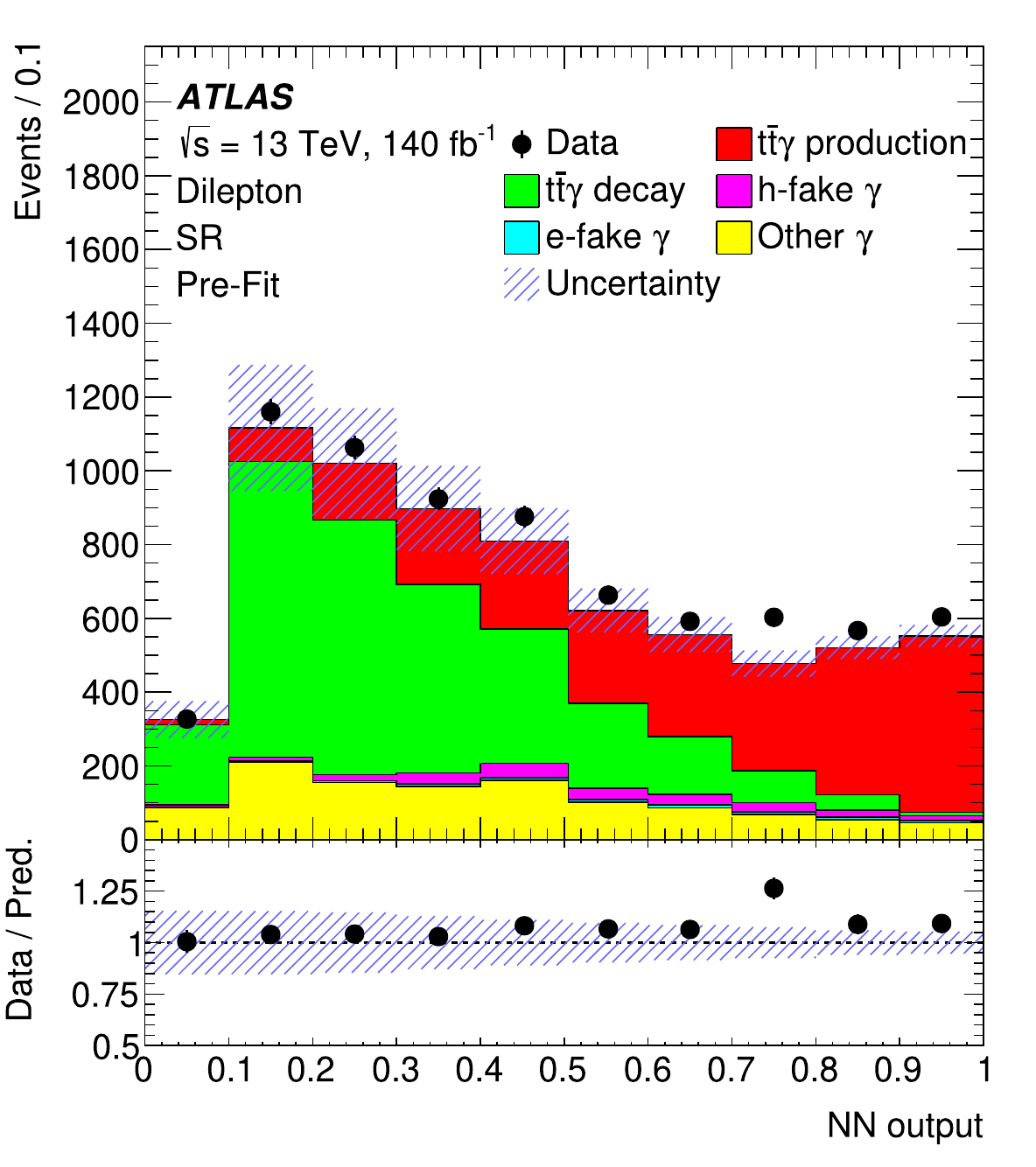}}
\caption{Distributions of the four NN output classifiers in the single-lepton channel: (a) \tty production node, (b) \tty decay node, (c) photon fakes and (d) other prompt $\gamma$, and (e) the NN output in the dilepton channel. The lower panels show the ratios of the data to the predictions. The uncertainty band represents the total uncertainties before the fit to data.}
\label{fig:NNoutput_prefit}
\end{figure}

\FloatBarrier

To enhance the separation of the \tty production process from  the different background categories and to better constrain the uncertainties in the different processes and increase the overall sensitivity, a signal and three control regions are defined based on the four NN output classifiers in the single-lepton channel. %
The \tty production signal region (SR) is defined first, followed by the control region (CR) enriched in \tty decay events, and the CR enriched in \hfakes and \efakes processes. The remaining events form the other prompt photon CR. The detailed criteria to define the signal and the control regions based on the requirements on the NN output classifiers are summarised in Table~\ref{tab:NN_regions}.

\begin{table}%
\centering
\caption{Summary of the criteria to define the signal and the control regions in the single-lepton channel, based on the requirements on the NN output classifiers. The last column of the table corresponds to the purity of the target processes (signal or specific background processes) in the particular region.}
\label{tab:NN_regions}
\begin{tabular}{l c c c c}
\hline
Category& \tty decay classifier & fake $\gamma$ classifier &  other prompt $\gamma$ classifier & purity\\
\hline
SR \tty production  & $<0.15$ & $<0.2$ & $<0.5$ & 73\%\\ %
CR \tty decay       & $>0.25$ & -- & $<0.4$ & 71\%\\ %
CR fake $\gamma$            & $<0.15$ &  $>0.2$  & $<0.5$ & 50\%\\ %
CR Other $\gamma$   &         & remaining events & & 26\%\\ %
\hline
\end{tabular}
\end{table}

\section{Cross-section definition}
\label{sec:fiducial}
Although the focus of this \paperornote is to consider only the \tty production process as a signal, the cross-sections are also measured including the \tty decay process as part of the signal for completeness and for a comparison with the previous measurements.
The inclusive and differential cross-sections are measured in all scenarios at particle level in a fiducial region, defined by the kinematic properties of the signal process, in which all selected final-state objects are produced within the detector acceptance.

Particle level refers to a collection of objects that are considered stable in MC simulation~(lifetime $\tau \geq 30$ ps) but without any simulation of the interaction of these particles with the detector components or additional $pp$ interactions. The objects at particle level are defined in the following.

Photons are required to not originate from a hadron decay, satisfy $\et > \SI{20}{\GeV}$ and $|\eta| < \SI{2.37}{}$, and that the sum of transverse momenta of all charged particles surrounding the photon within $\Delta R = 0.2$ is less than 5\% of its own \pt. Muons and electrons are required to have $\pt>25$~\GeV\ and $|\eta| < 2.5$, and must not originate from hadron decays. The momenta of nearby photons, within a $\Delta R = 0.1$ cone, are added to the lepton before applying the selection. Particle-level jets are clustered with the anti-$k_{t}$ algorithm with a radius parameter of $R=0.4$. All stable particles are considered in the clustering, except for the selected electrons, muons and photons, and the neutrinos originating from the top quarks. Jets are required to satisfy $\pt > 25~\GeV$ and $|\eta| < 2.5$. A jet is identified as a $b$-jet if a hadron with $\pT > \SI{5}{\GeV}$ containing a $b$-quark is matched to the jet through a ghost-matching method~\cite{Cacciari:2008gn}. Muons and electrons with separation $\Delta R < 0.4$ from a jet are excluded. Jets are removed if they are within $\Delta R = 0.4$ of an isolated photon candidate. Events are additionally required to satisfy $\Delta R (\gamma,\ell)>0.4$, where $\ell$ is the electron or muon.

The fiducial phase space in the single-lepton channel is defined by requiring exactly one photon, exactly one electron or muon, and at least four jets, and in the dilepton channel by requiring exactly one photon and exactly two leptons, and at least two jets. In both cases, at least one jet must be a $b$-tagged jet. Events are rejected if there are additional leptons with $\pt > 7$~\GeV. For the combination of the channels, a union of the single-lepton and dilepton fiducial phase spaces is used.

In the case of the measurement of \tty production, the efficiency of selecting and reconstructing events that are generated within the fiducial phase space is about 31\% in the single-lepton channel and about 35\% in the dilepton channel, while the fractions of events that satisfy the selection at the reconstruction level but fail to satisfy the particle-level requirements are about 17\% and 19\%, respectively.
\section{Inclusive cross-section}
\label{sec:inclusive}

The fiducial inclusive \tty production cross-sections are obtained using a binned profile-likelihood fit.
The expected signal and background distributions are modelled by template histograms from the MC simulated samples corrected by the data-driven estimates where applicable. The sum of signal and background contributions is fitted to the data. The likelihood function consists of Poisson terms for the event yields and Gaussian functions to constrain the nuisance parameters which represent the systematic uncertainties described in Section~\ref{sec:systematics}. For the uncertainties related to the finite number of simulated MC events, the Gaussian terms in the likelihood are replaced by Poisson terms, following the Barlow–Beeston `light' approach~\cite{BARLOW1993219}.

In the single-lepton channel, the fit is performed in the four regions defined in Section~\ref{sec:NN}. In the \tty production and the \tty decay regions, the output of the corresponding NN classifiers are used as fitted distributions while in the photon fakes CR and in the other prompt photon CR the distributions of the second largest pseudo-continuous $b$-tagging score are used.
The choice of the fitted variable in the photon fakes CR and in the other prompt photon CR is motivated by a requirement to have bins that are sufficiently well populated by the targeted processes that are statistically limited. Additionally, since most of the other prompt photon backgrounds have fewer $b$-jets than the signal, the second largest $b$-tagging score distribution provides a reasonably good discrimination. In the dilepton channel the NN output is used in the fit. The binning of the NN output distributions in both channels
is chosen to avoid bins with large statistical uncertainties and to ensure that distributions of the small backgrounds are smooth, while keeping a good separation between the process of interest and the backgrounds.

\subsection{Results for the \tty production measurement}

For the \tty production cross-section measurement, the normalisation of the \tty decay process is a free parameter in the profile-likelihood fit. Figures~\ref{fig:NNoutput_postfit_lj} and ~\ref{fig:NNoutput_postfit_dil} show the distributions of the fitted variables in the single-lepton and dilepton channel after the fit.
Good agreement is observed between data and the prediction in both channels. The uncertainty bands are calculated taking into account the full post-fit correlation matrix.

The measured fiducial \tty production cross-section in the single-lepton channel is
\begin{equation}\nonumber
\sigma^{\text{Single lepton}}_{t\bar{t}\gamma\; {\text{production}}} = 288^{+21}_{-19}\, \text{fb}= 288 \pm 5 ({ \text{stat}})^{+20}_{-19} ({\mathrm{syst}})\, \text{fb},
\end{equation}\nonumber
and, in the dilepton channel, it is
\begin{equation}\nonumber
\sigma^{\text{Dilepton}}_{t\bar{t}\gamma\; {\text{{production}}}} = 45.7 ^{+3.3}_{-3.1}\, \text{fb}= 45.7^{+1.4}_ {-1.3}  ({ \text{stat}})^{+3.0}_{-2.8} ({\text{syst}})\, \text{fb}.
\end{equation}\nonumber
The value of the normalisation factor of the \tty decay contribution determined by the fit is $0.99 \pm 0.07$ in both channels.
The total relative uncertainties in the measurements are 7.6\% (7.1\%) in the single-lepton (dilepton) channel. The systematic uncertainty is determined by subtracting the statistical uncertainty in quadrature from the total uncertainty. The statistical uncertainty is obtained from a fit where all nuisance parameters are fixed to their post-fit values.

Most of the nuisance parameters are not constrained by the fit and their best-fit values are well within one standard deviation of the prior uncertainties.
The fit slightly constrains the \tty production parton-shower uncertainty in the single-lepton channel, resulting in an uncertainty 20\% smaller than its initial value.

The measured cross-sections are in agreement within the uncertainties with the NLO predictions from \MGNLO interfaced to \Pythia8, which correspond to $255^{+25}_{-26} ({ \text{scale}})^{+6}_{-4} ({\text{PDF}})\, \text{fb}$ and $40.9^{+3.9}_{-4.0} ({ \text{scale}})^{+0.9}_{-0.5} ({\text{PDF}})\, \text{fb}$ in the single-lepton and dilepton channel, respectively, with a central value about 10\% larger than the corresponding predictions.

\begin{figure}[ht!]
\centering
\subfloat[]{\includegraphics[width=0.45\textwidth]{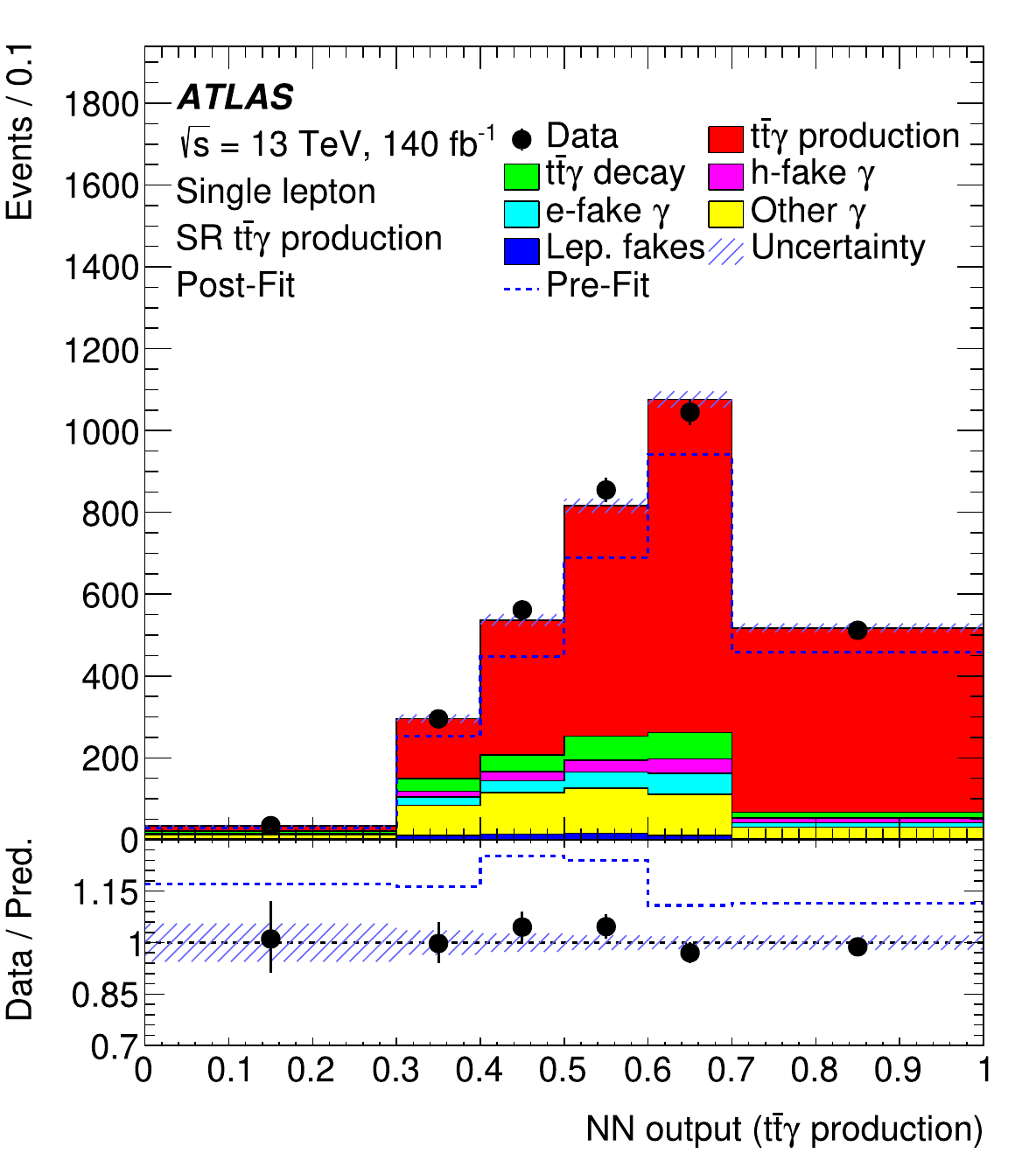}}%
\subfloat[]{\includegraphics[width=0.45\textwidth]{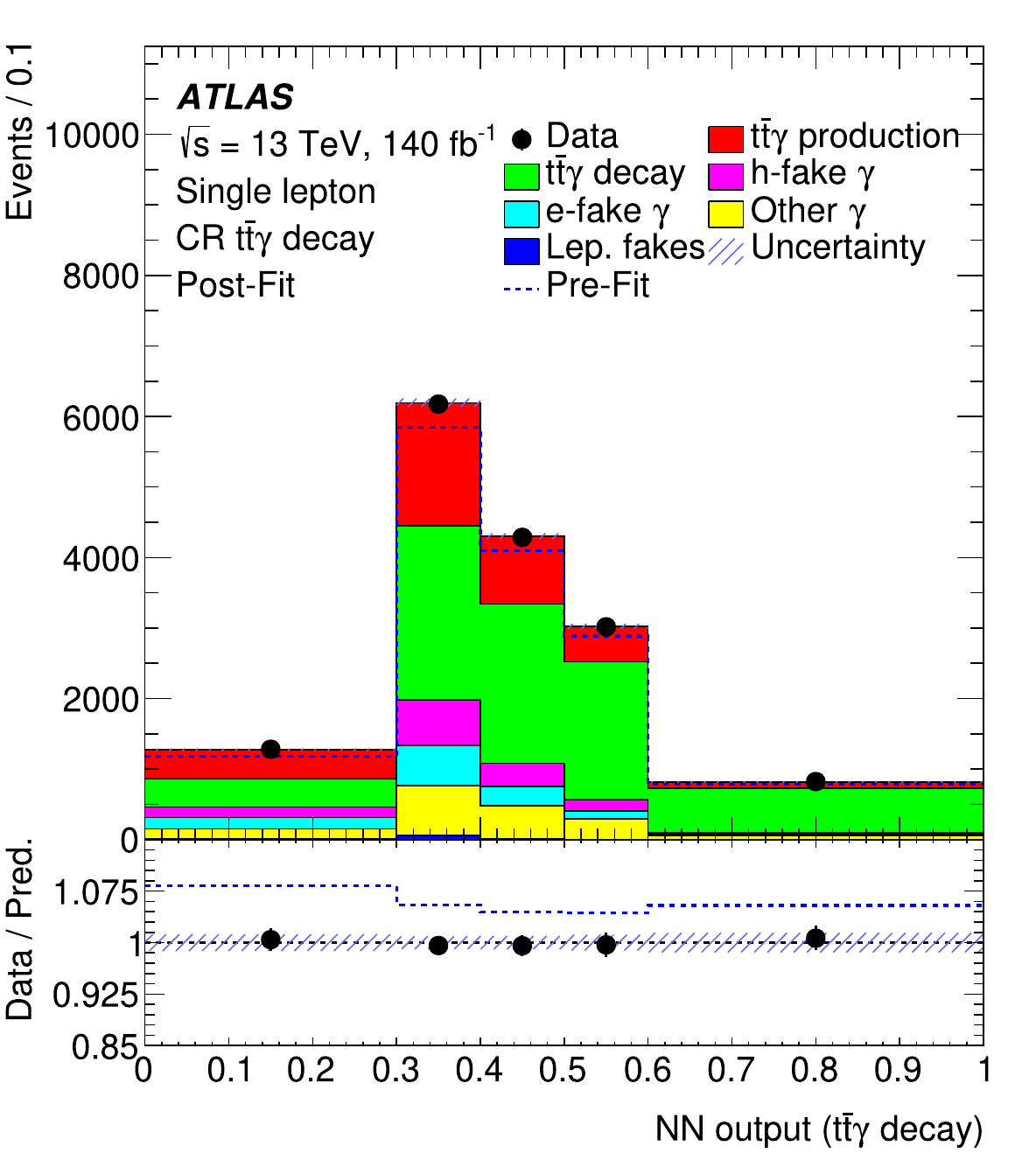}}\\
\subfloat[]{\includegraphics[width=0.45\textwidth]{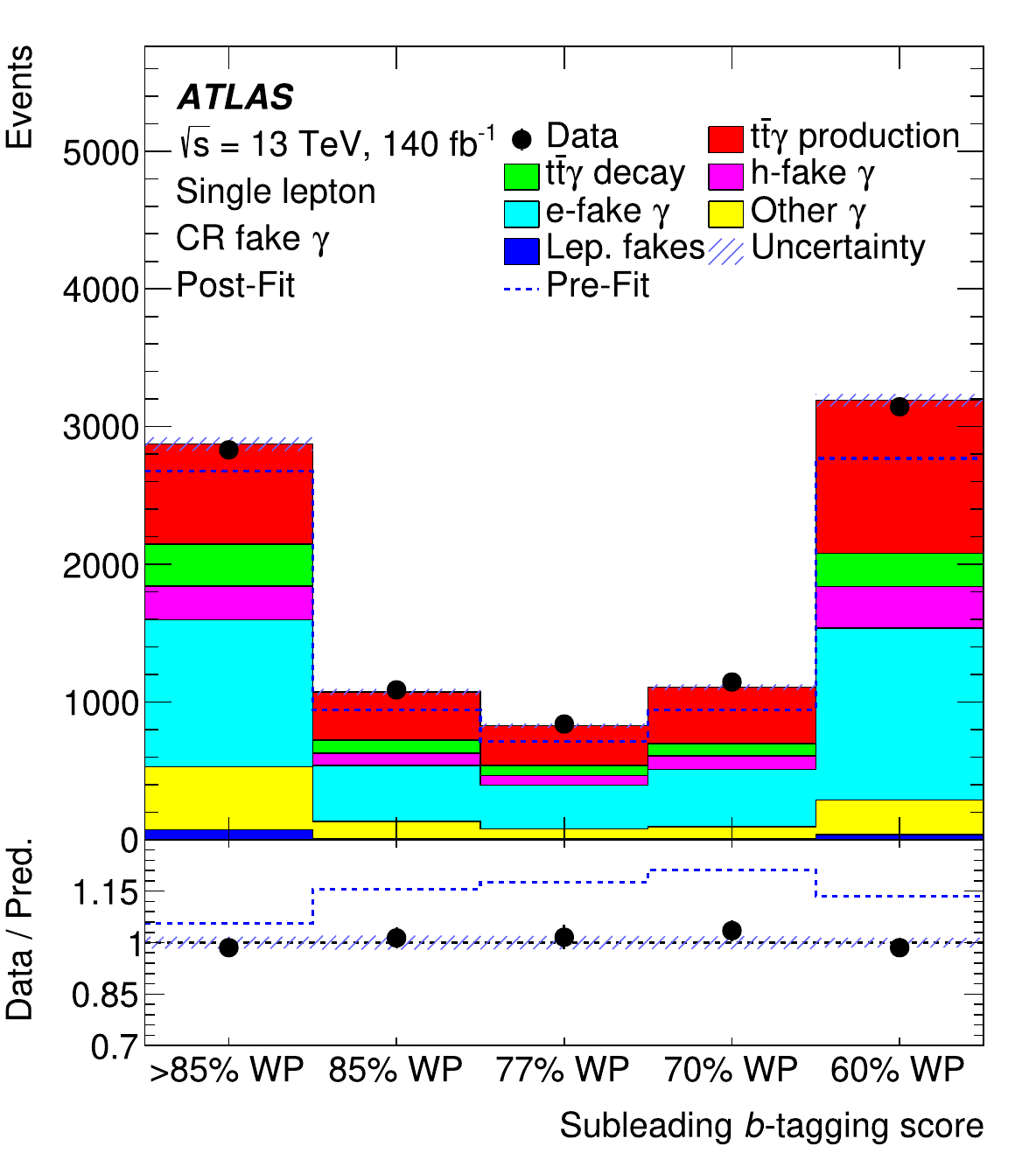}}%
\subfloat[]{\includegraphics[width=0.45\textwidth]{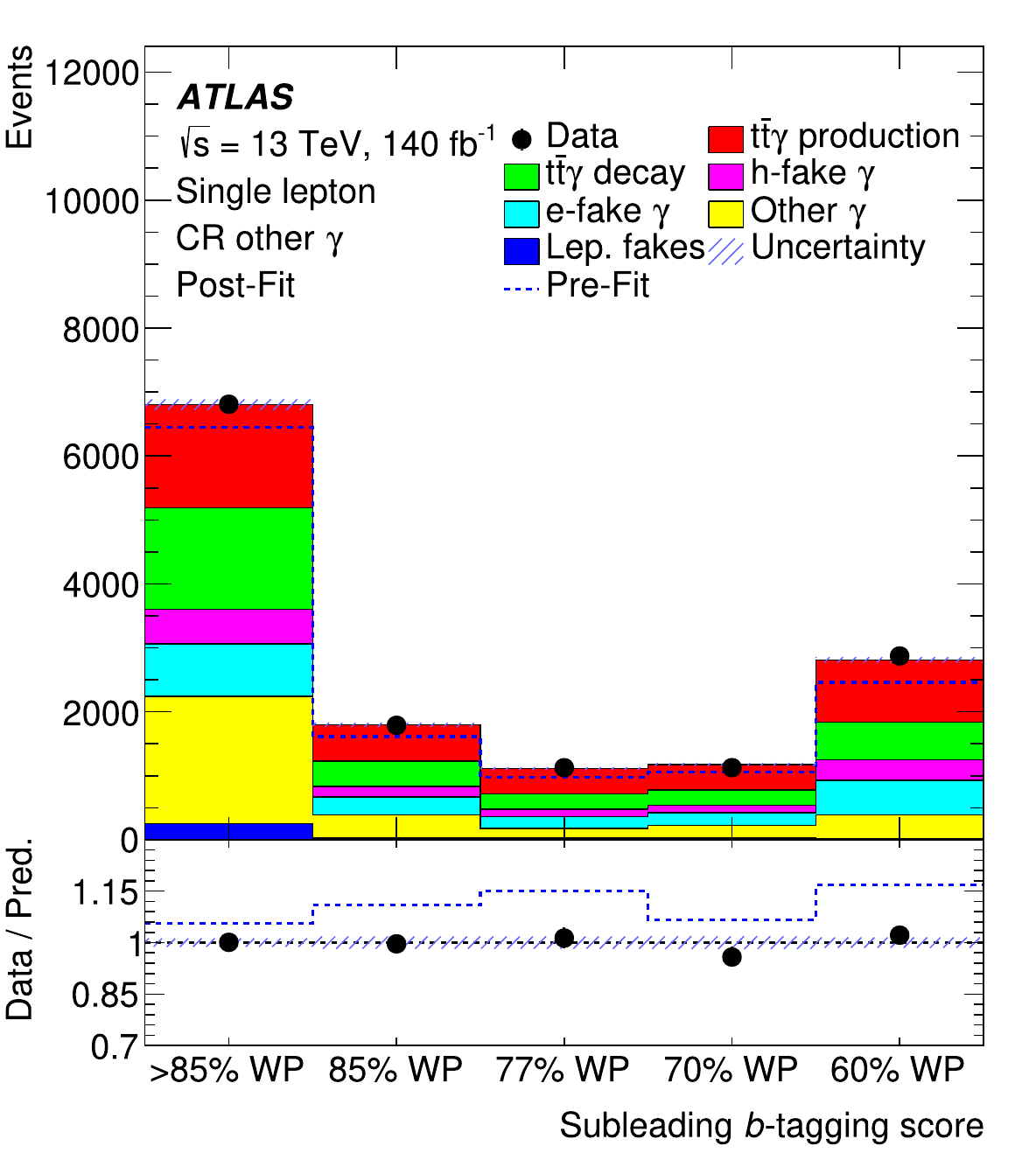}}\\
\caption{Distributions of (a) the \tty production classifier in the SR, (b) the \tty decay classifier in the \tty decay CR and   the second largest pseudo-continuous $b$-tagging score in (c) the photon fakes CR and (d) the other prompt photon CR
after the fit to data for the measurement of the \tty production cross-section in the single-lepton channel. The uncertainty band represents the total post-fit uncertainty in the prediction.
The lower panels show the ratios of the data to the total post-fit predictions. The dashed lines correspond to the pre-fit prediction (upper panel) and the ratio of the data to the total pre-fit prediction (lower panel).}
\label{fig:NNoutput_postfit_lj}
\end{figure}

\begin{figure}[t]
\centering
\includegraphics[width=0.45\textwidth]{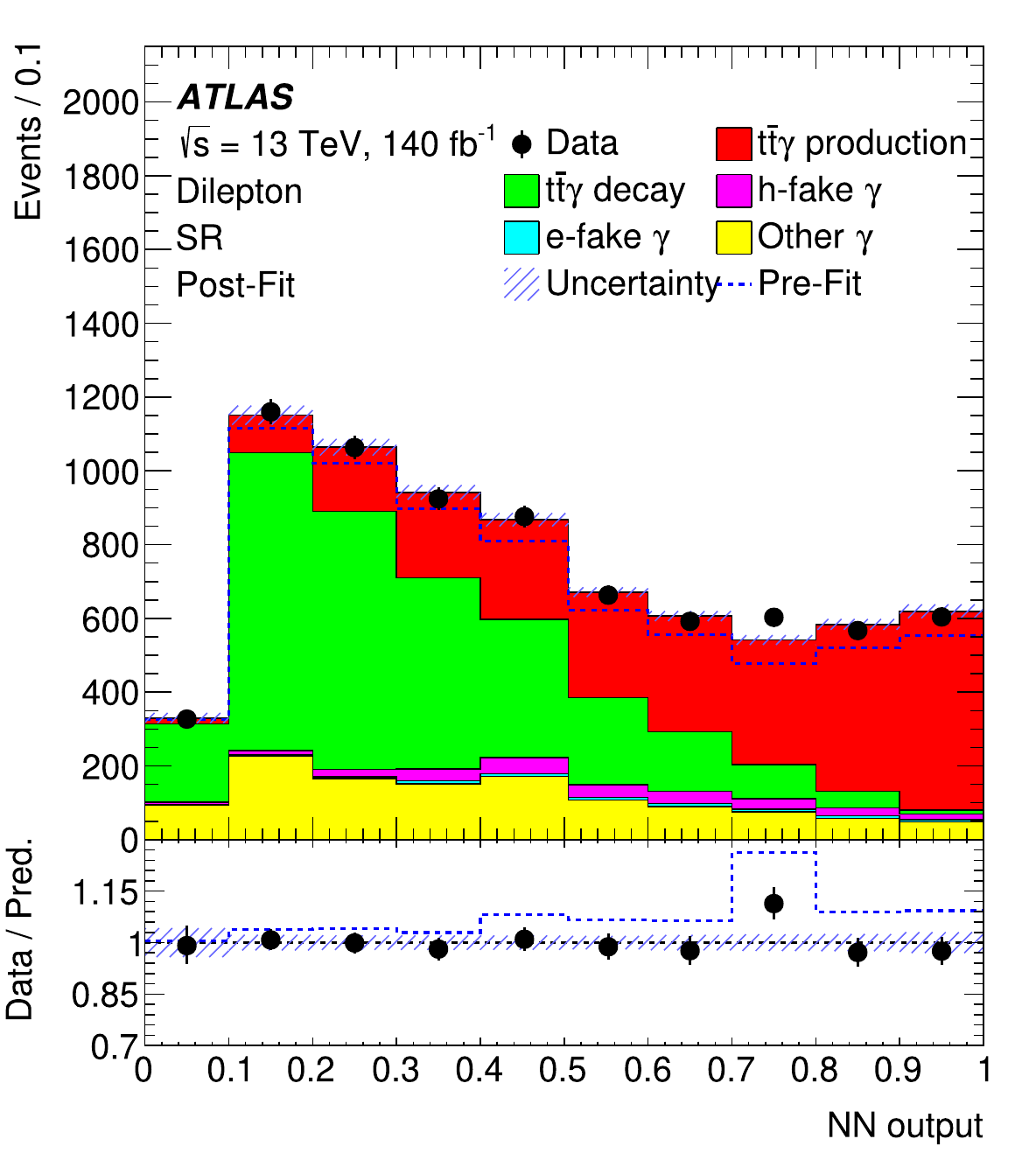}
\caption{Distribution of the NN output distribution after the fit to data for the measurement of the \tty production cross-section in the dilepton channel. The uncertainty band represents the total post-fit uncertainty in the prediction.
The lower panel shows the ratio of the data to the total post-fit prediction. The dashed lines correspond to the pre-fit prediction (upper panel) and the ratio of the data to the total pre-fit prediction (lower panel).}
\label{fig:NNoutput_postfit_dil}
\end{figure}

The \tty production cross-section is also obtained from a simultaneous profile-likelihood fit to the data in all the signal and control regions in the single-lepton and the dilepton channels treating all systematic uncertainties as correlated. The result yields:
\begin{equation}\nonumber
\sigma_{t\bar{t}\gamma\; {\text{production}}} = 319 \pm 15\, \text{fb} = 319 \pm 4 \, ({ \text{stat}})\, ^{+15}_{-14} ({\mathrm{syst}})\, \text{fb}.
\end{equation}\nonumber
The measured value is slightly lower than the sum of the two channels owing to the correlations among systematic uncertainties. In particular, the fits performed in the dilepton and single-lepton channels separately show different correlations after the fit between the parton-shower uncertainty in the \tty production and the cross-section. The difference between the correlations also leads to a reduction of this uncertainty in the combined fit. This observed behaviour of the parton-shower uncertainty is related to the jet-related variables having a different impact on the fitted distributions, also resulting in a reduction of some JES and flavour tagging uncertainties in the combination.
The relative uncertainty of the combined \tty production cross-section is 5.2\%, and the normalisation of the \tty decay component determined from the combined fit is 0.98 with an uncertainty of 6\%. The expected cross-section given by the NLO \MGNLO simulation is $296^{+29}_{-30} ({ \text{scale}})^{+6}_{-4} ({\text{PDF}})\, \text{fb}$.

The impact of the systematic uncertainties on the inclusive cross-section in the single-lepton and dilepton channels grouped into different categories is given in Table~\ref{tab:syst_impact}. The measurements are limited by the systematic uncertainties. The dominant sources are those related to the modelling of the \tty production in both channels and the normalisation of the \tty decay. Other significant contributions are spread over many different sources of experimental and modelling uncertainties, the largest contributors being the uncertainties associated with jets, $b$-tagging and the normalisation of the prompt backgrounds.

\begin{table}[htbp]
\centering
\caption{%
Summary of the impact of the systematic uncertainties on the \tty production fiducial inclusive cross-section in the single-lepton and dilepton channels and their combination grouped into different categories. The category `Jets’ corresponds to the combined effect of JES, jet resolution and JVT uncertainties, while the categories `Photon’ and `Lepton’ correspond to all experimental uncertainties related to photons and leptons (including trigger uncertainties), respectively. The relative uncertainties quoted  are obtained by repeating the fit, fixing the set of nuisance parameters of the sources corresponding to each category to their post-fit values, and subtracting in quadrature the resulting uncertainty from the total uncertainty of the nominal fit. The total uncertainty is different from the sum in quadrature of the components due to correlations among nuisance parameters. }
\label{tab:syst_impact}
\footnotesize
\begin{tabular}{l  c  c  c }
\hline
& \multicolumn{3}{c}{$\Delta\sigma_{t\bar{t}\gamma \, {\text{production}}}/\sigma_{t\bar{t}\gamma \, {\text{production}}}$ (\%)} \\
Source & Single lepton & Dilepton & Combination \\
\midrule
\textbf{Statistical uncertainty} & 1.8 & 3.3 & 1.5 \\
\hline
\textbf{MC statistical uncertainties} & 1.5 & 1.5 & 1.0 \\
\midrule
\textbf{Modelling uncertainties} & & &\\
\midrule
\tty production PS uncertainty & 2.4 & 3.7 & 0.9 \\
Other \tty production modelling & 5.1 & 1.6 & 3.0 \\
\tty decay modelling & 0.3 & 1.3 & 0.8 \\
\tty decay normalisation & 2.4 & 3.1 & 2.1\\
Prompt photon background normalisation & 1.5 & 2.0 & 2.0\\
Fake photon background estimate& 0.8& 1.5 & 1.6 \\
Fake lepton background estimate& 0.4& -- & 0.1 \\
Other Background modelling & 0.7 & 0.2 & 0.5\\
\midrule
\textbf{Experimental uncertainties} & &  & \\
\midrule
Jet uncertainties & 3.5 & 3.0 & 1.7\\
B-tagging uncertainties& 2.6 & 2.1 & 1.0 \\
Photon & 0.5 & 1.5 & 0.8 \\
Lepton & 1.3 & 1.4 & 1.3 \\
\MET & 0.3 & 0.4 & 0.4 \\
Pile-up & 0.3 & 0.7 & 0.5 \\
Luminosity &0.8 & 1.0 & 0.8\\
\midrule
\textbf{Total systematic uncertainty} & 7.6 & 7.1 & 5.0 \\
\midrule
\textbf{Total uncertainty} & 7.8 & 7.7 & 5.2 \\
\hline
\end{tabular}
\end{table}

\subsection{Results for the total \tty production and decay  measurement}
For the measurement of the \tty process, where all \tty events are considered to be a signal regardless of the origin of the photon, the \tty production and \tty decay templates are added together, and the normalisation of this combined template is treated as a free parameter of the fit. A 20\% uncertainty is assigned to the normalisation of the \tty decay template~\cite{TOPQ-2017-14,TOPQ-2021-29}.

The fiducial cross-sections yield
\begin{equation}\nonumber
\sigma^{\text{Single lepton}}_{t\bar{t}\gamma} = 704^{+49}_{-46} \, \text{fb}= 704 \pm 5 \, ({ \text{stat}})\, ^{+49}_{-46} \,({\mathrm{syst}})\, \text{fb}
\end{equation}\nonumber
in the single-lepton channel and
\begin{equation}\nonumber
\sigma^{\text{Dilepton}}_{t\bar{t}\gamma} = 116.1^{+8.2}_{-7.7}\, \text{fb} = 116.1 \pm 1.7\, ({ \text{stat}})\, ^{+8.0}_{-7.6} \,({\mathrm{syst}})\, \text{fb}
\end{equation}\nonumber
in the dilepton channel. The combined fiducial cross-section is measured to be
\begin{equation}\nonumber
\sigma_{t\bar{t}\gamma} = 788 ^{+38}_{-37} \, \text{fb}= 788 \pm 5\, ({ \text{stat}})\, ^{+38}_{-37} \,({\mathrm{syst}})\, \text{fb}.
\end{equation}\nonumber
As expected, the statistical uncertainty is reduced in this measurement while the total systematic uncertainty is slightly larger compared with the measurement of the \tty production cross-section. This difference is related to the \tty decay normalisation uncertainty having a larger impact, while there are no significant differences in the impact of the experimental and modelling uncertainties relative to the measurement of \tty production.

\FloatBarrier
\section{Differential cross-sections}
\label{sec:differential}

The measurements of the differential cross-sections of the \tty production and the \tty process including photons from production and decay, are performed as functions of photon, lepton and jet kinematic properties and angular separations of the photon, leptons and jets. Both absolute and normalised differential cross-sections are measured. The list of all the variables considered is given in Table~\ref{tab:listvariables}. The kinematic properties of the photon are sensitive to the top-photon coupling, in particular the photon \pt. Angular distances, such as $\Delta R (\gamma,\ell)_{\mathrm{min}}$, are related to the angle between the top quark and the radiated photon that gives insight into the structure of this coupling and are expected to be less sensitive to the top quark off-shell effects~\cite{Bevilacqua:2019quz}. Since these variables have significant shape differences between the \tty production and the \tty decay, they are measured for both processes. Following Refs.~\cite{CMS-TOP-21-004, TOPQ-2020-03}, additional differential distributions of leptonic observables are measured in the dilepton channel for the \tty total process, in particular, $\Delta \phi(\ell,\ell)$ and $|\Delta\eta(\ell,\ell)|$ that are sensitive to the \ttbar spin correlation.

In the single-lepton channel, the same four regions considered in the inclusive measurement are used. In the dilepton channel, all events are split into a signal-enriched region with the value of the NN discriminant above 0.6 and a background-enriched control region with NN below 0.6. This threshold is optimised to minimise the expected total uncertainty in the photon \pt differential distribution.
The measurements are performed by using the profile-likelihood unfolding~\cite{ATLAS:2023eld} in the signal and control regions.
The parameters of interest $\vec{\mu}$, the signal strength in each bin of the distribution at particle level, are free parameters of the fit.
The same likelihood but with only one parameter of interest is used for the inclusive cross-section measurement.
No regularisation is applied due to small bin-to-bin migrations for all the  variables considered.
In the case of the normalised distribution, the $\vec{\mu}$ is reparametrised such that the last element of the vector is the overall signal normalisation. In this case, the content of the last bin of the unfolded distribution is dropped from the $\vec{\mu}$, since it is no longer a free parameter, and it can be calculated based on the values in other bins and the overall normalisation.

The performance of the unfolding procedure is tested for possible biases from the choice of the input model using pseudodata. The unfolding procedure is performed using the nominal response matrices and modified \tty templates where the distributions are linearly reweighted or the shape is modified by the observed differences between data and the MC simulation at reconstruction level. In both cases, the procedure reproduces the altered shapes of the input distributions within the statistical uncertainties.

\begin{table}[ht]
\centering
\caption{List of variables measured differentially
in the single-lepton and dilepton channels. }
\label{tab:listvariables}
\begin{footnotesize}
\begin{tabular}{l l }
Variable & Description \\
\midrule
\multicolumn{2}{c}{Both channels} \\
\midrule
$\pt(\gamma)$ & Transverse momentum of the photon \\
$|\eta(\gamma)|$ & Absolute value of the pseudorapidity of the photon \\
\DRlph & Angular separation between the photon and the closest lepton \\
$\Delta R(\gamma, b)_{\text{min}}$ &  Angular separation between the photon and the closest $b$-jet \\
$\Delta R(\ell, j)_{\text{min}}$ & Smallest angular separation between any of the selected leptons and jets \\
$\pt(j_1)$ & Transverse momentum of the leading jet in \pt\\
\midrule
\multicolumn{2}{c}{Additional variables: dilepton channel} \\
\midrule
$\Delta R(\gamma, \ell_1)$ & Angular separation between the photon and the leading lepton\\
$\Delta R(\gamma, \ell_2)$ & Angular separation between the photon and the subleading lepton\\
\Detall & Pseudorapidity difference between the two leptons\\
\Dphill & Azimuthal angle difference between the two leptons\\
$\pt(\ell,\ell)$ & Transverse momentum of the dilepton system\\
\end{tabular}
\end{footnotesize}
\end{table}

\subsection{Differential distributions for the \tty production}

The photon \pt distributions in the four signal and control regions in the single-lepton channel and the migration matrices are shown for illustration in Figures~\ref{fig:ttyprod_input_pt_ljets} and ~\ref{fig:ttyprod_input_matrix_ljets}, respectively. The corresponding distributions in the dilepton signal and control regions are given in Figure~\ref{fig:ttyprod_input_pt_dl}. The binning is  optimised in the dilepton channel to ensure that the bin width is larger than twice the resolution of the observable and that the expected statistical uncertainty in the distribution at particle level is below 10\%. The resulting binning is found to be similar to the one used in Ref.~\cite{CMS-TOP-21-004} for most variables. Therefore, the latter is adopted to simplify the comparisons. The fraction of events reconstructed in the same bin of the differential distribution as the one they were generated in is above 90\% in every bin of the \pt distributions in all the SRs and CRs. This fraction ranges from 70\% for variables involving $b$-jets up to 99\% for the leptonic variables.

\begin{figure}[ht]
\centering
\subfloat[]{\includegraphics[width=0.45\textwidth]{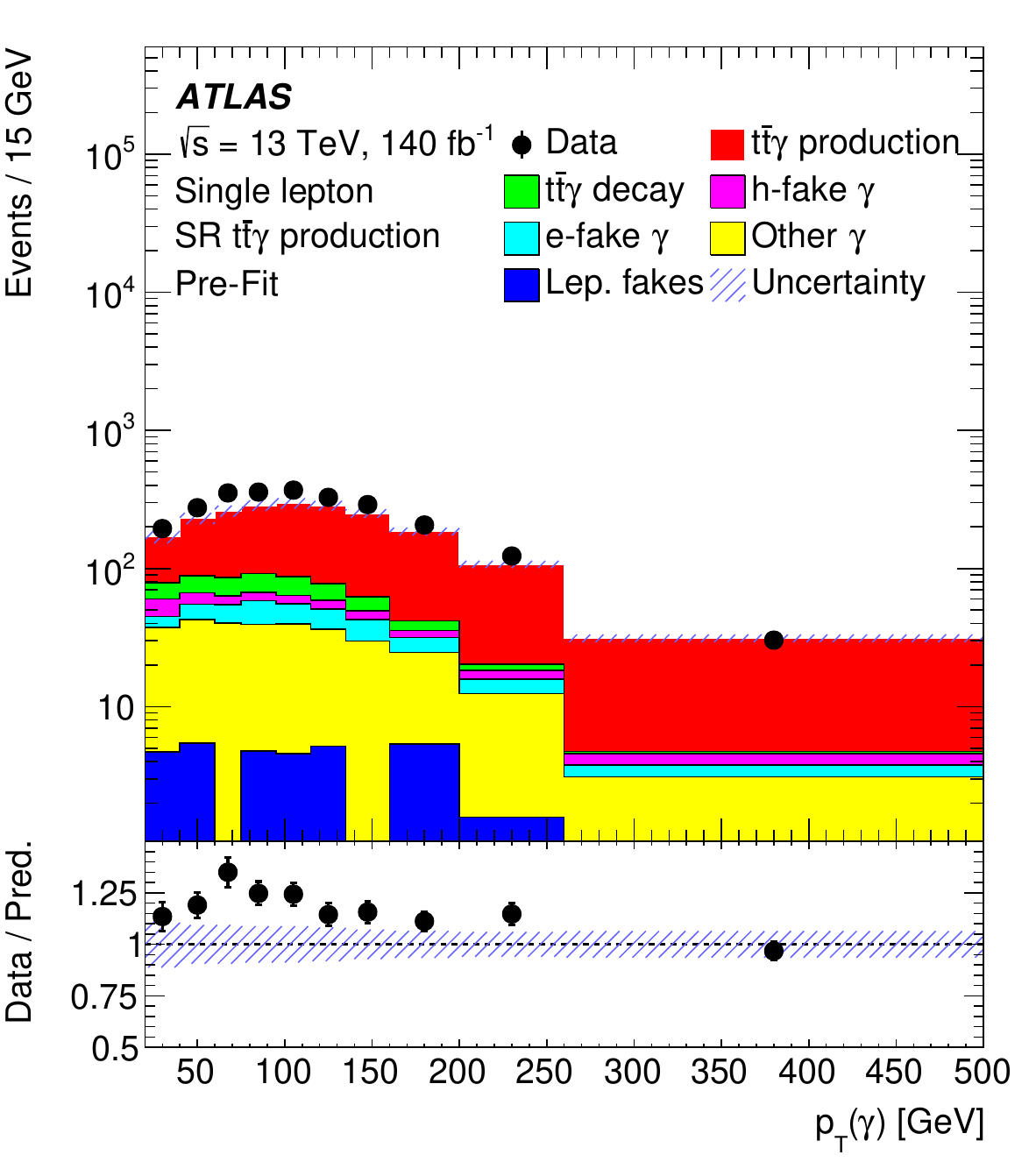}}
\subfloat[]{\includegraphics[width=0.45\textwidth]{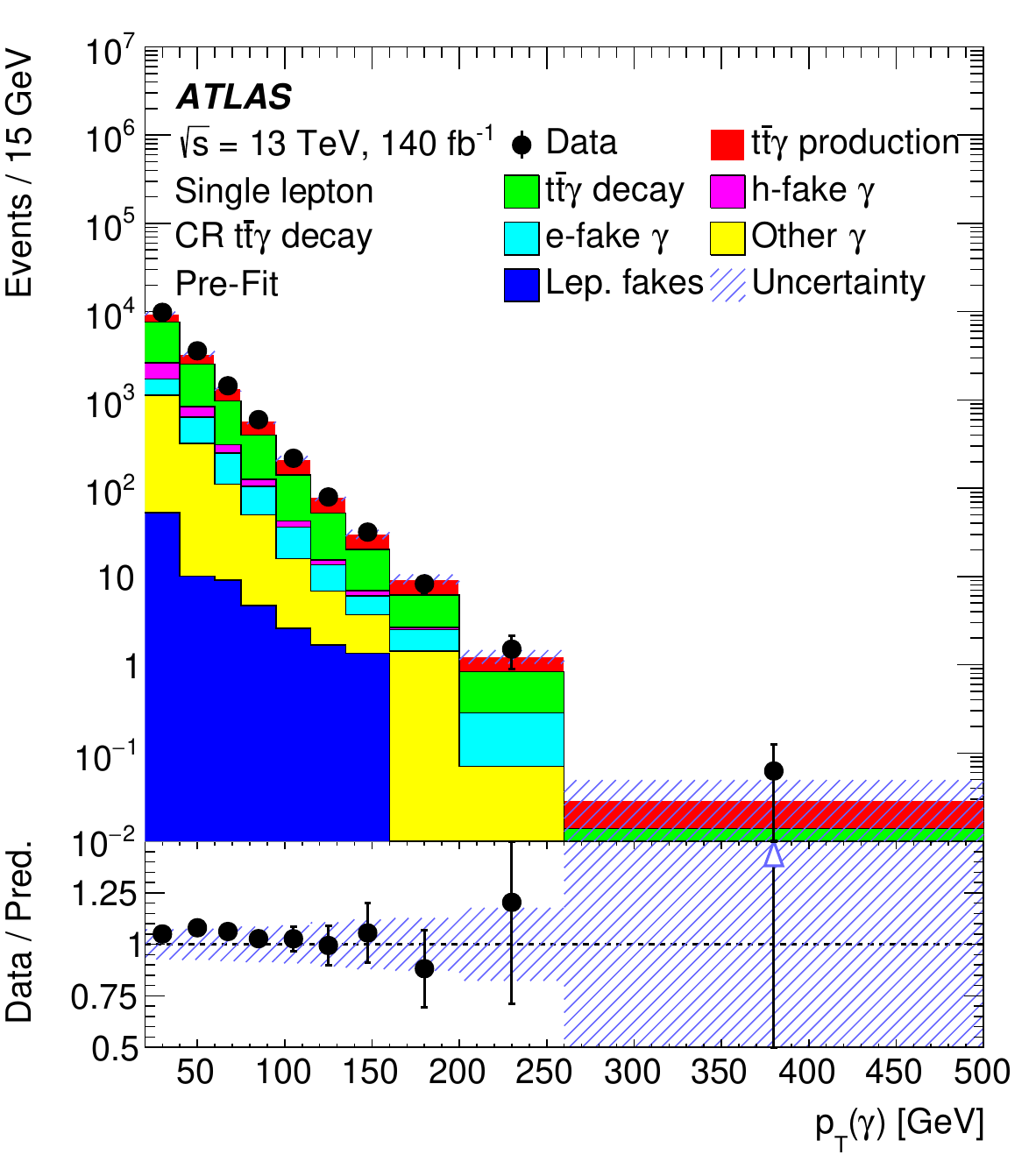}}\\
\subfloat[]{\includegraphics[width=0.45\textwidth]{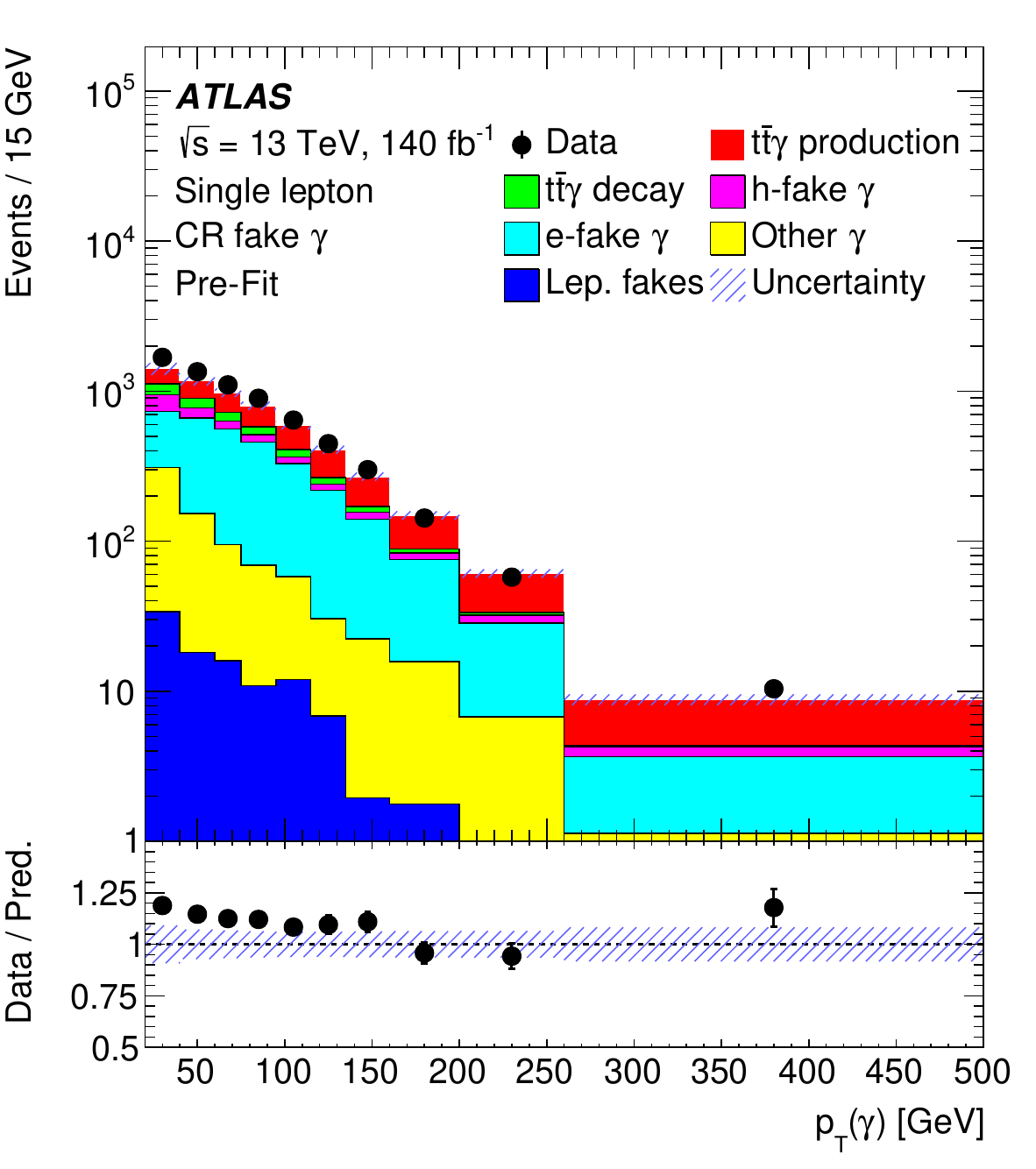}}
\subfloat[]{\includegraphics[width=0.45\textwidth]{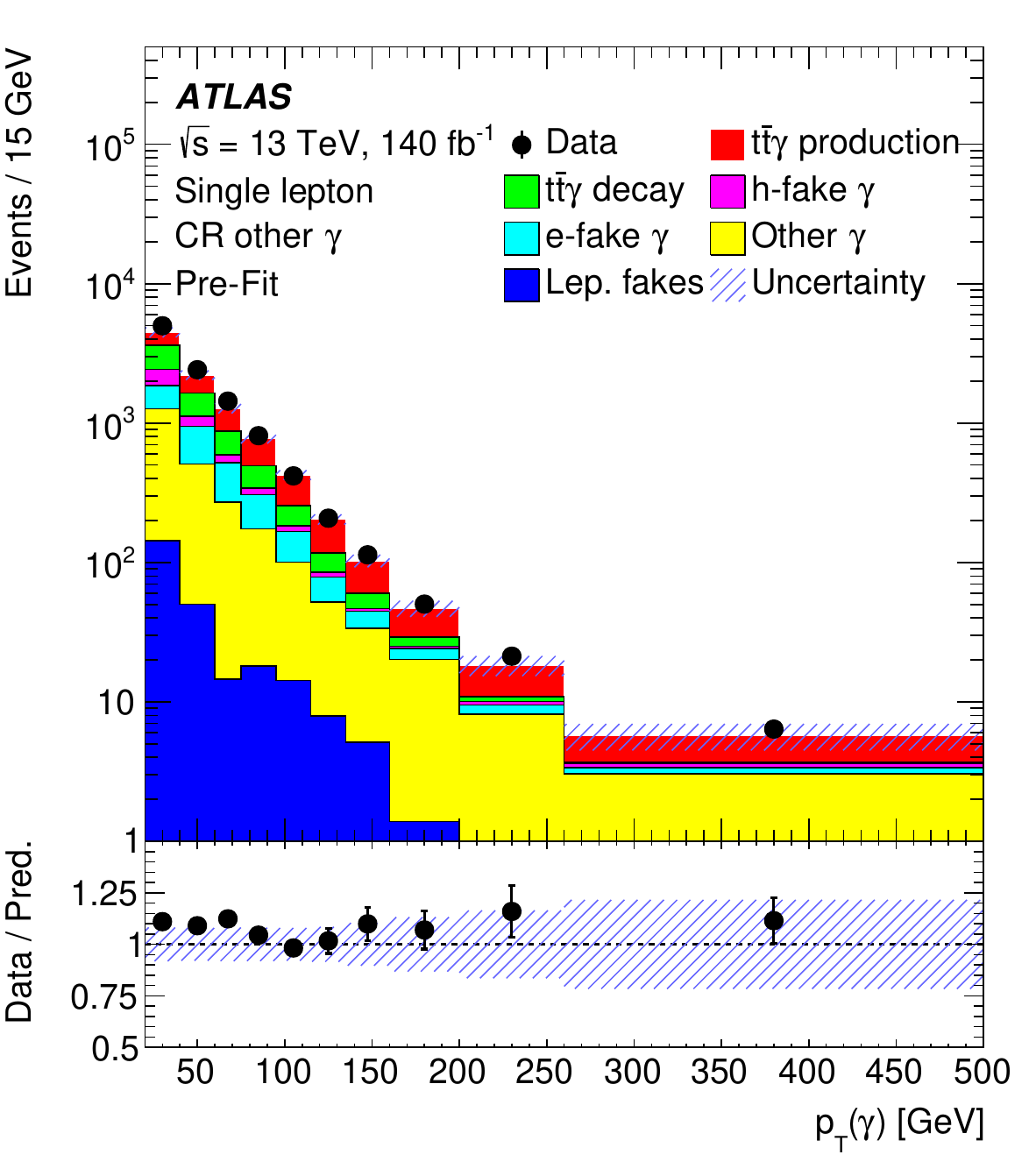}}\\
\caption{Photon \pt distributions at reconstruction level in the four regions of the single-lepton channel: (a) \tty production SR, (b) \tty decay CR, (c) photon fakes CR and (d) other prompt  $\gamma$ CR. The lower panels show the ratios of the data to the predictions. The uncertainty band represents the total uncertainties before the fit to data. The last bin of the distributions includes the overflow events. }
\label{fig:ttyprod_input_pt_ljets}
\end{figure}

\begin{figure}[ht]
\centering
\subfloat[]{\includegraphics[width=0.45\textwidth]{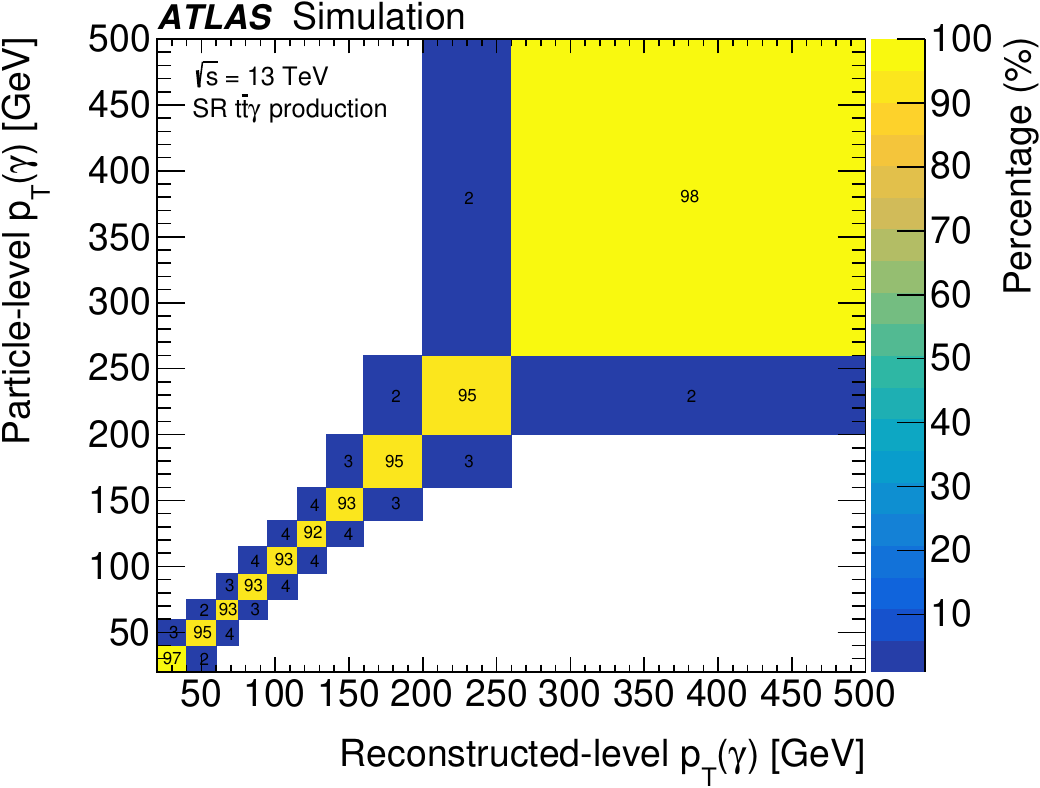}}
\hspace{0.02\textwidth}%
\subfloat[]{\includegraphics[width=0.45\textwidth]{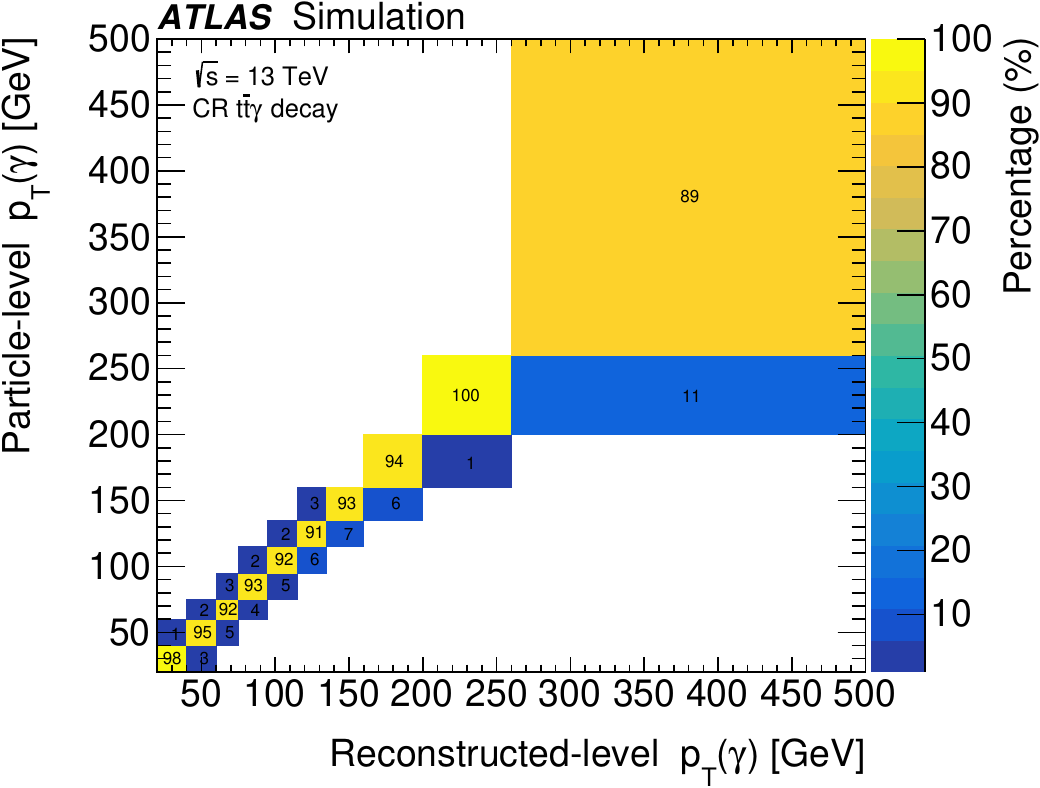}}\\
\subfloat[]{\includegraphics[width=0.45\textwidth]{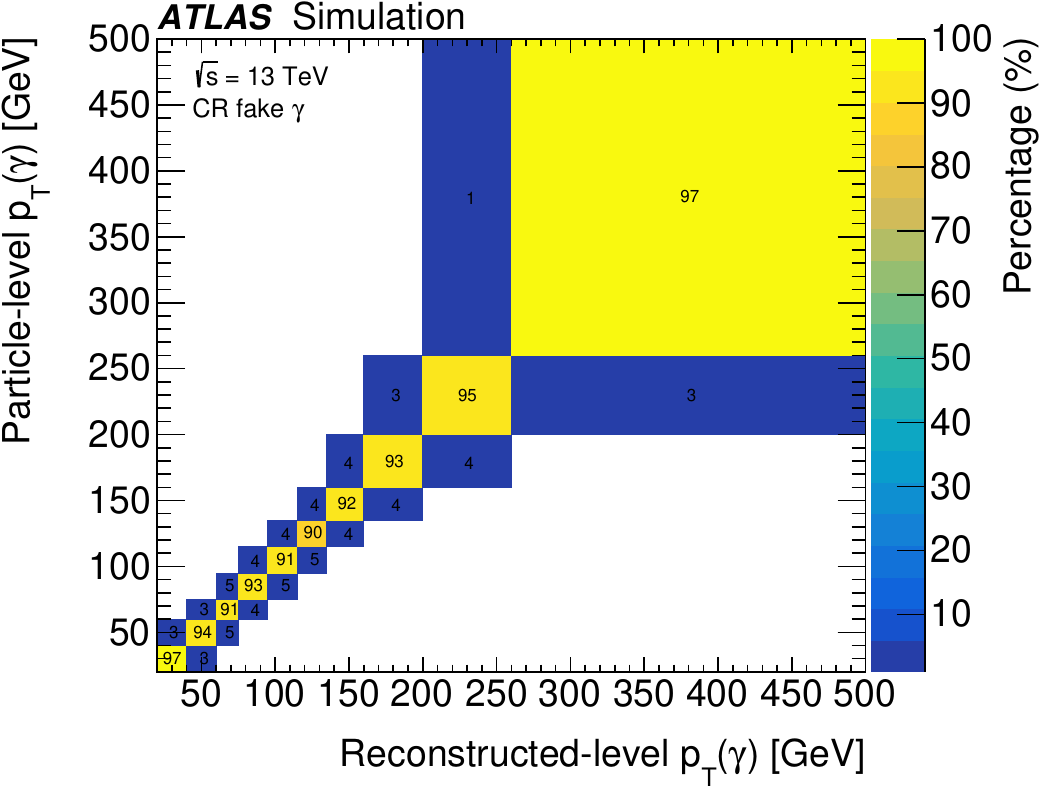}}%
\hspace{0.02\textwidth}%
\subfloat[]{\includegraphics[width=0.45\textwidth]{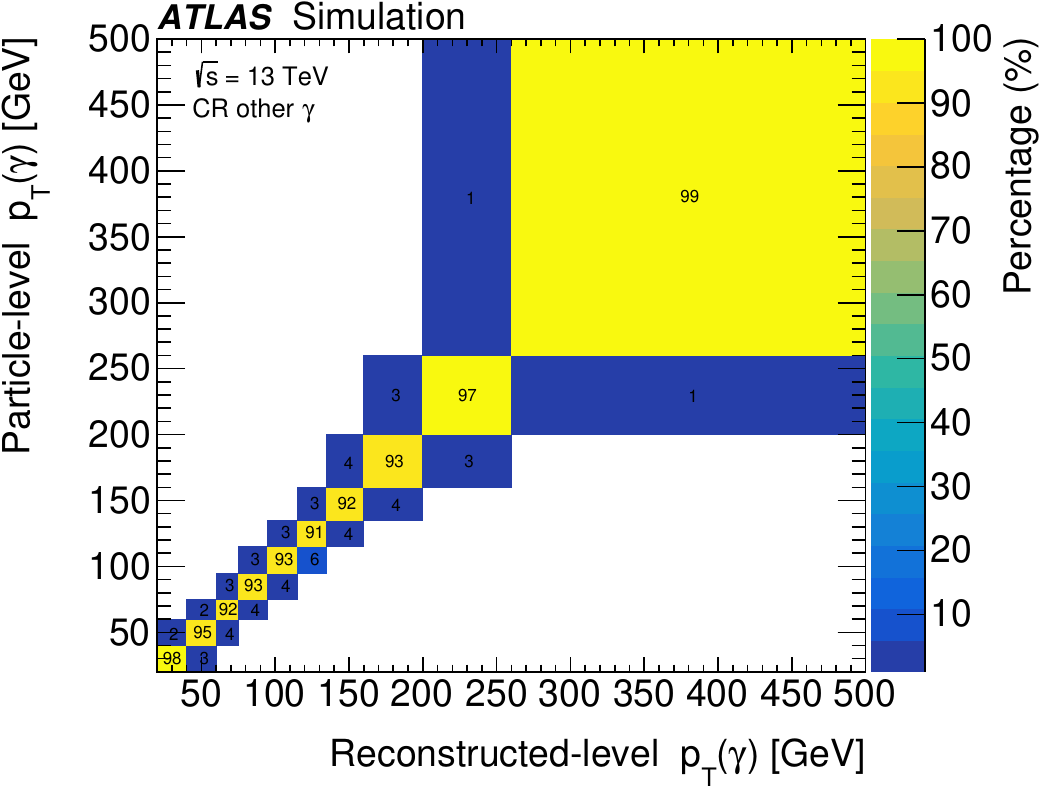}}\\
\caption{Migration matrices in the four regions of the single-lepton channel: (a) \tty production SR, (b) \tty decay CR, (c) photon fakes CR and (d) other prompt  $\gamma$ CR, relating the photon \pt at the reconstruction and particle levels in the fiducial phase space. The values correspond to the fraction of events in each bin normalised by column and shown as percentage. The last bin of the distributions includes the overflow events. }
\label{fig:ttyprod_input_matrix_ljets}
\end{figure}

\begin{figure}[ht]
\centering
\subfloat[]{\includegraphics[width=0.46\textwidth]{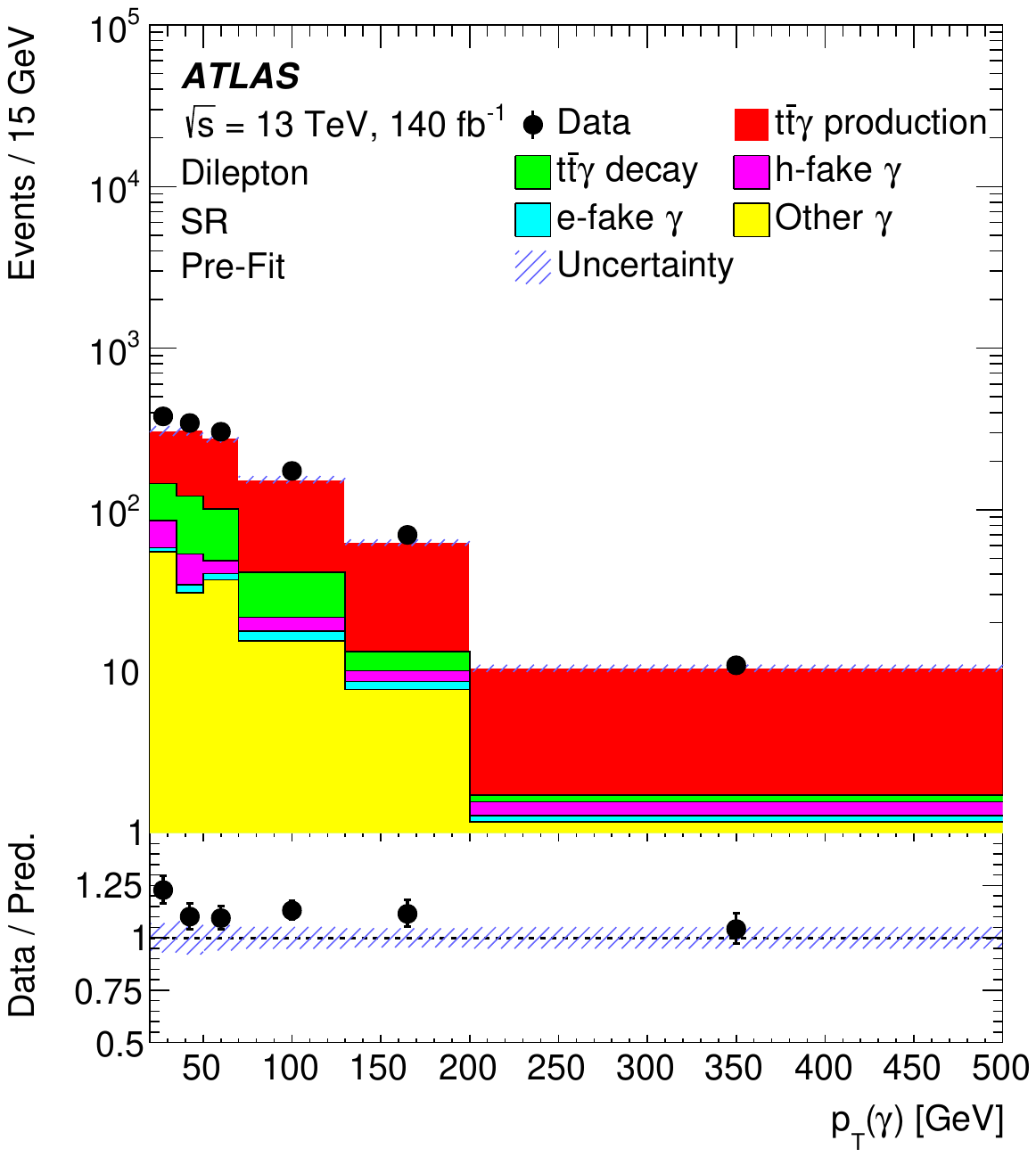}}%
\hspace{0.04\textwidth}%
\subfloat[]{\includegraphics[width=0.46\textwidth]{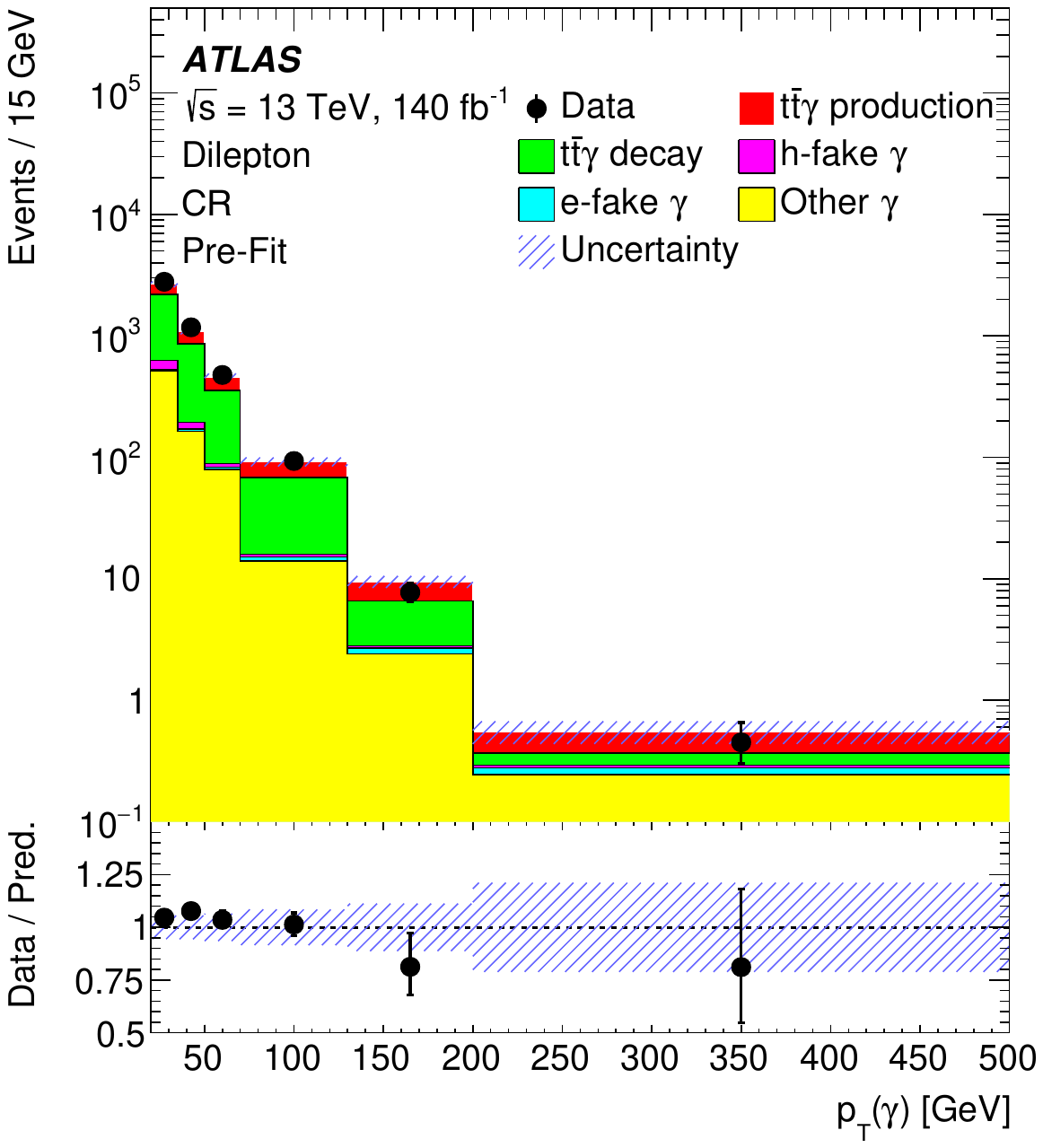}}\\
\subfloat[]{\includegraphics[width=0.48\textwidth]{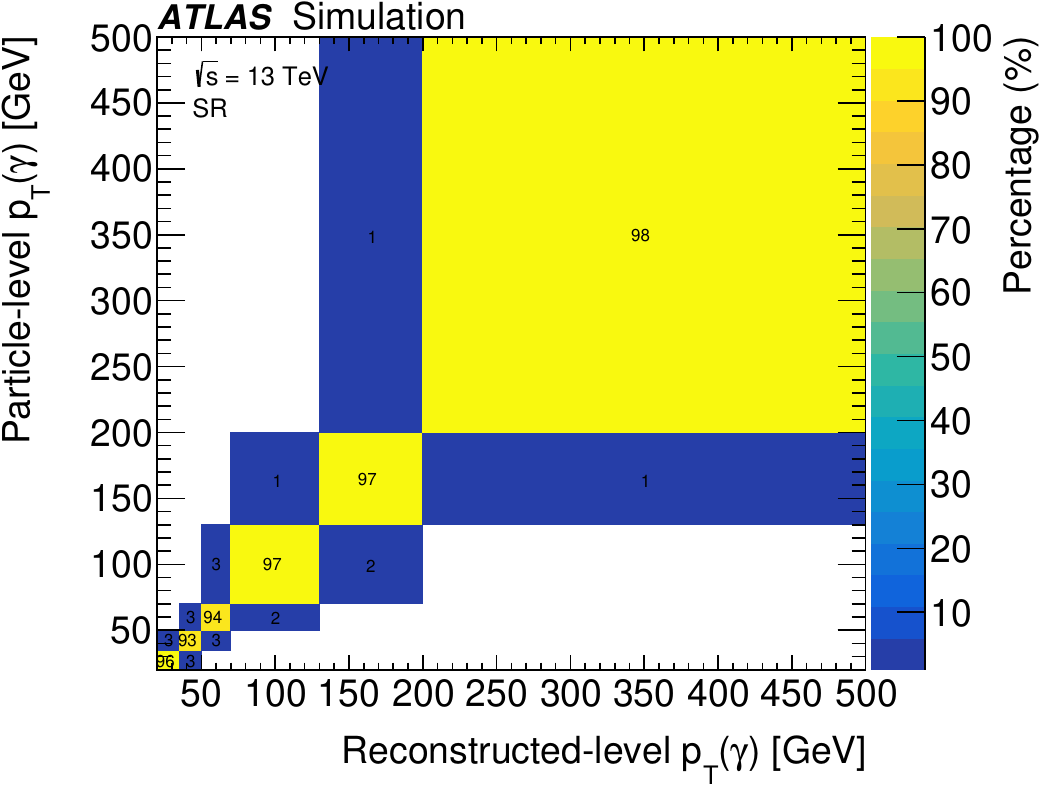}}%
\hspace{0.02\textwidth}%
\subfloat[]{\includegraphics[width=0.48\textwidth]{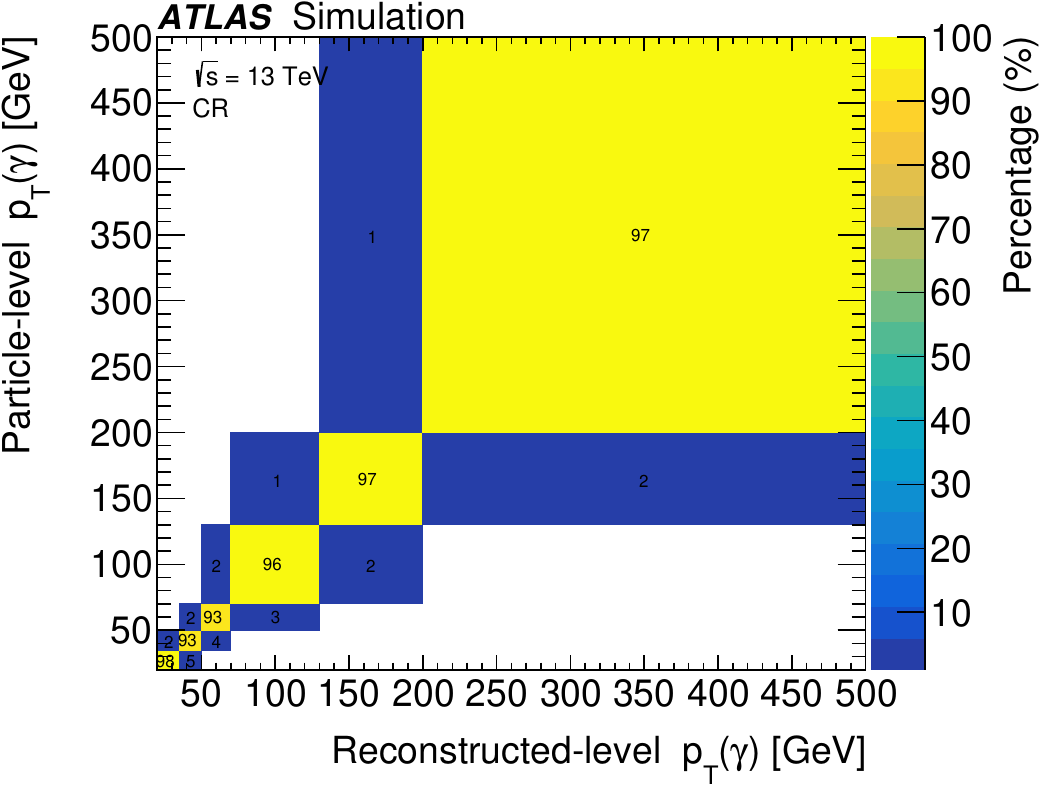}}\\
\caption{Top row: Photon \pt distributions at reconstruction level in the (a) signal and (b) control region defined by the NN output above or below the value of 0.6 in the dilepton channel. The lower panels show the ratios of the data to the predictions. The uncertainty band represents the total uncertainties before the fit to data. Bottom row: Migration matrices relating the photon \pT at the reconstruction and particle levels in the fiducial phase space in the (c) signal and (d) control region. The values correspond to the fraction of events in each bin normalised by column and shown as percentage. The last bin of the distributions includes the overflow events. }
\label{fig:ttyprod_input_pt_dl}
\end{figure}

The absolute and normalised \tty production cross-sections are presented in Figure~\ref{fig:ttyprod_diff_Ljets} for the single-lepton channel and in Figure~\ref{fig:ttyprod_diff_DL} for the dilepton channel as a function of angular distances between the photon and the closest lepton or $b$-jet and between the closest lepton--jet pair.
The results are compared with the NLO predictions from the \MGNLO simulation interfaced to \PYTHIA[8] and \HERWIG[7]. The MC predictions, normalised to the NLO cross-section given by the simulation, slightly underestimate the measured cross-section, in agreement with the measurements of the inclusive cross-section. There are no significant differences between the two MC simulations. Both describe the shape of the distributions well.
The total uncertainty in the absolute cross-section in the single-lepton channel ranges from 8\% to 20\%, depending on the variable and the bin of the distribution. The statistical uncertainty is about 3\%--10\% for most of the bins. The largest systematic uncertainties arise from the normalisation of the backgrounds with prompt photons, some components of the JES uncertainty and from the \tty modelling uncertainties. The total uncertainty in the dilepton channel varies from 8\% up to 20\%--30\% in the tails of the distributions, and it is dominated by the statistical uncertainty in the data.

The precision of the normalised cross-sections is below 10\% for most bins of the distributions, owing to a large cancellation of systematic uncertainties that affect the normalisation of the distributions, in particular, the normalisation uncertainties associated with the prompt background and the \tty decay contribution.

\begin{figure}[ht]
\centering
\subfloat[]{\includegraphics[width=0.40\textwidth]{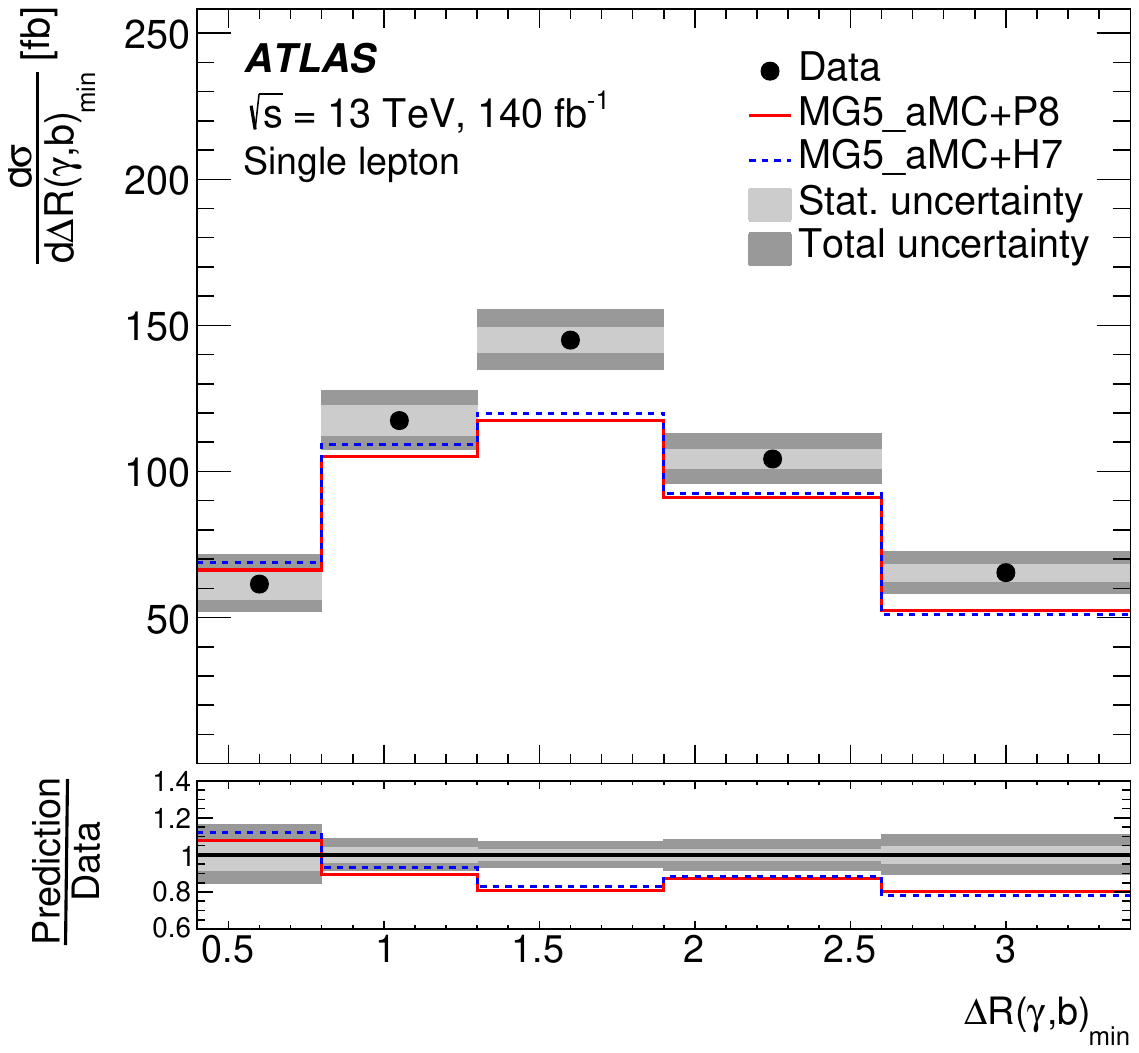}}%
\subfloat[]{\includegraphics[width=0.40\textwidth]{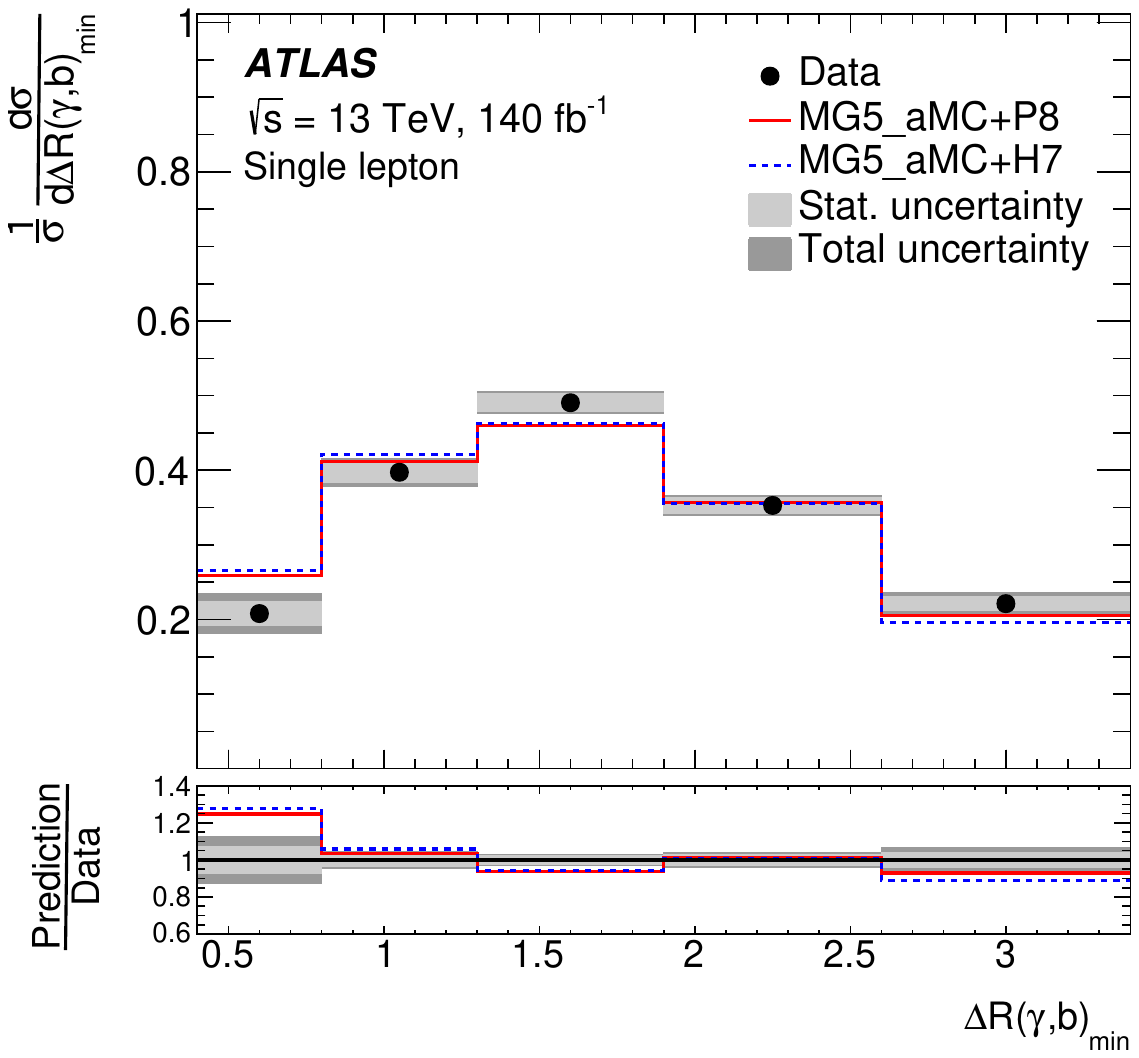}}\\
\subfloat[]{\includegraphics[width=0.40\textwidth]{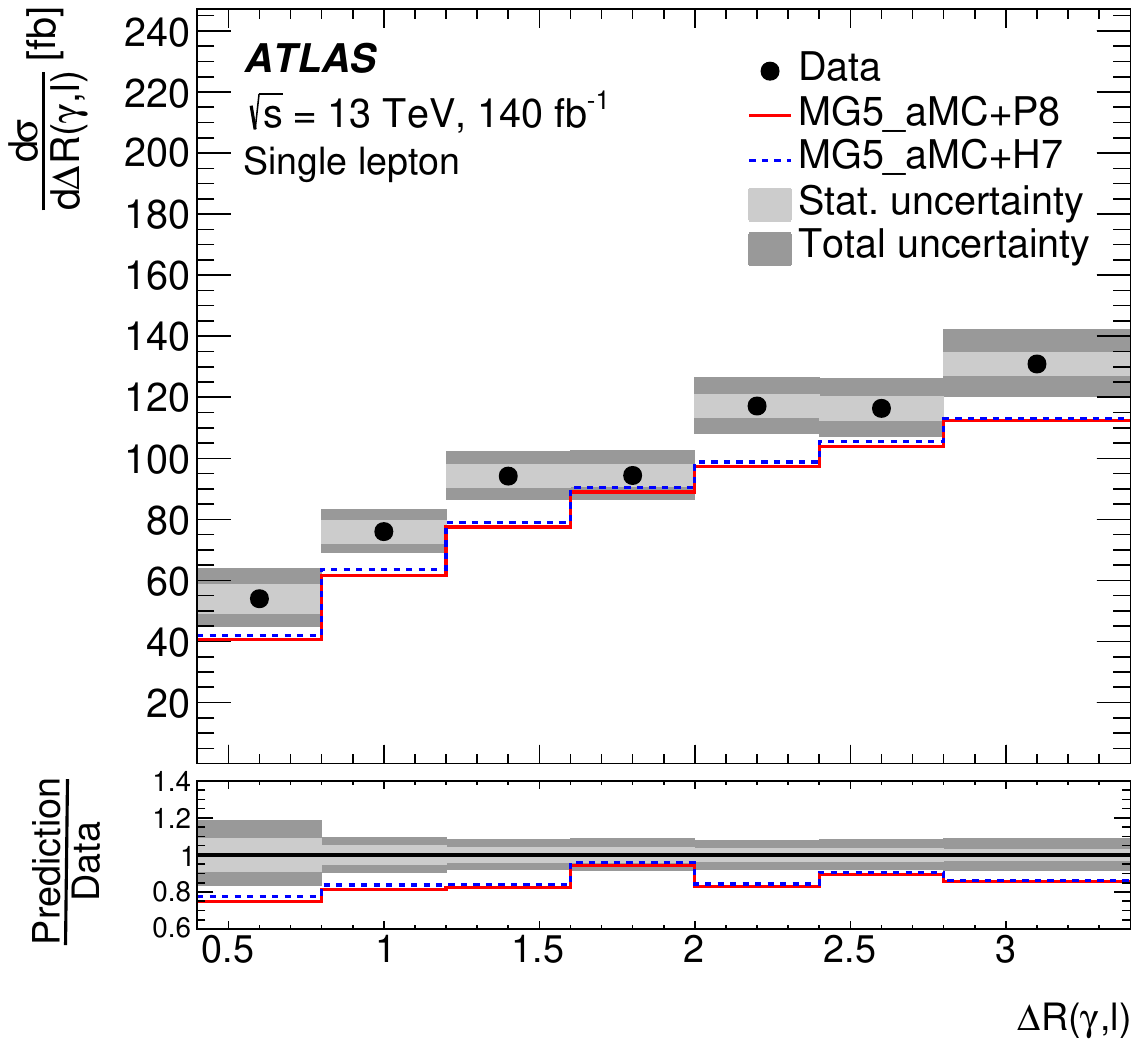}}%
\subfloat[]{\includegraphics[width=0.40\textwidth]{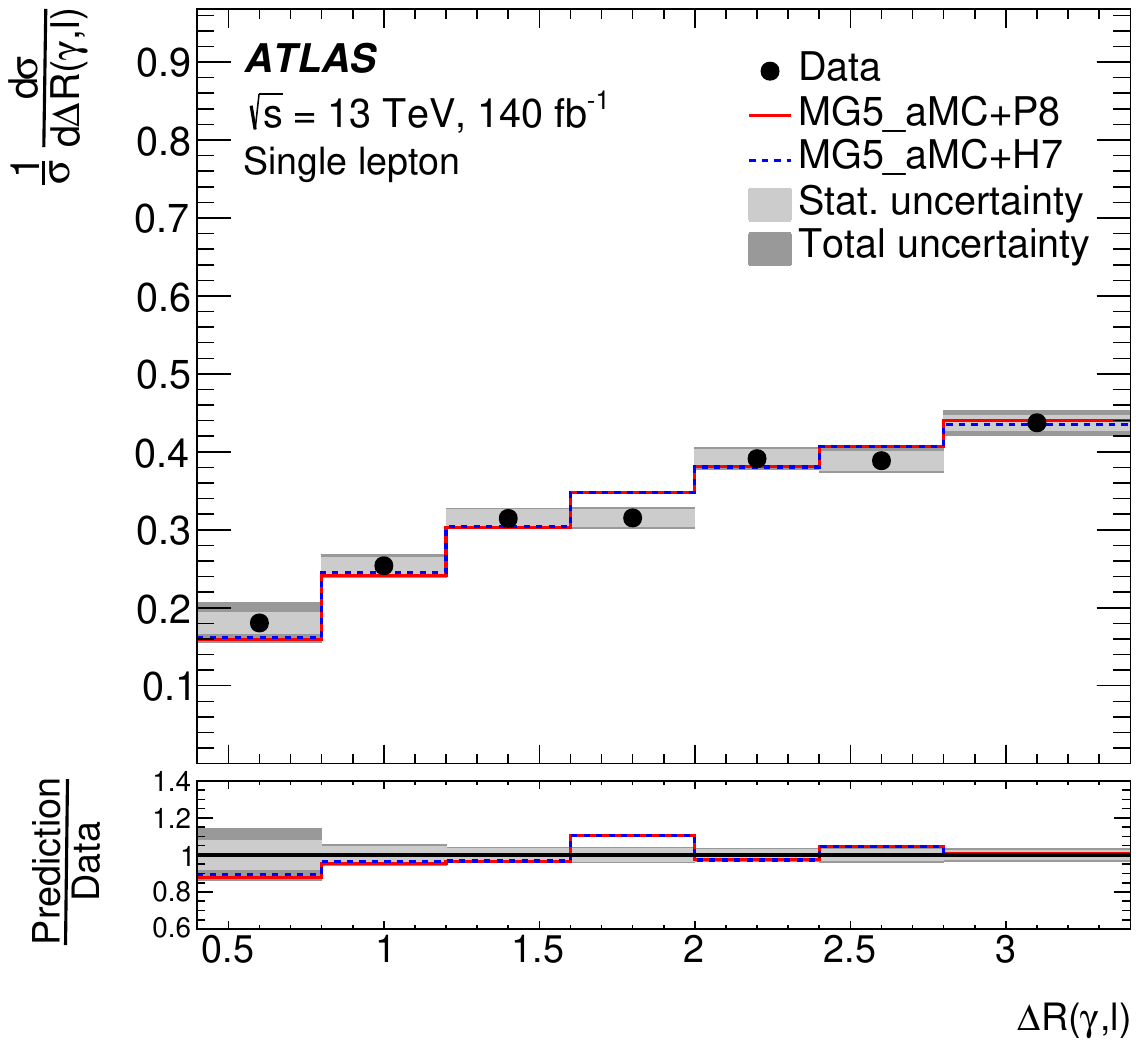}}\\
\subfloat[]{\includegraphics[width=0.40\textwidth]{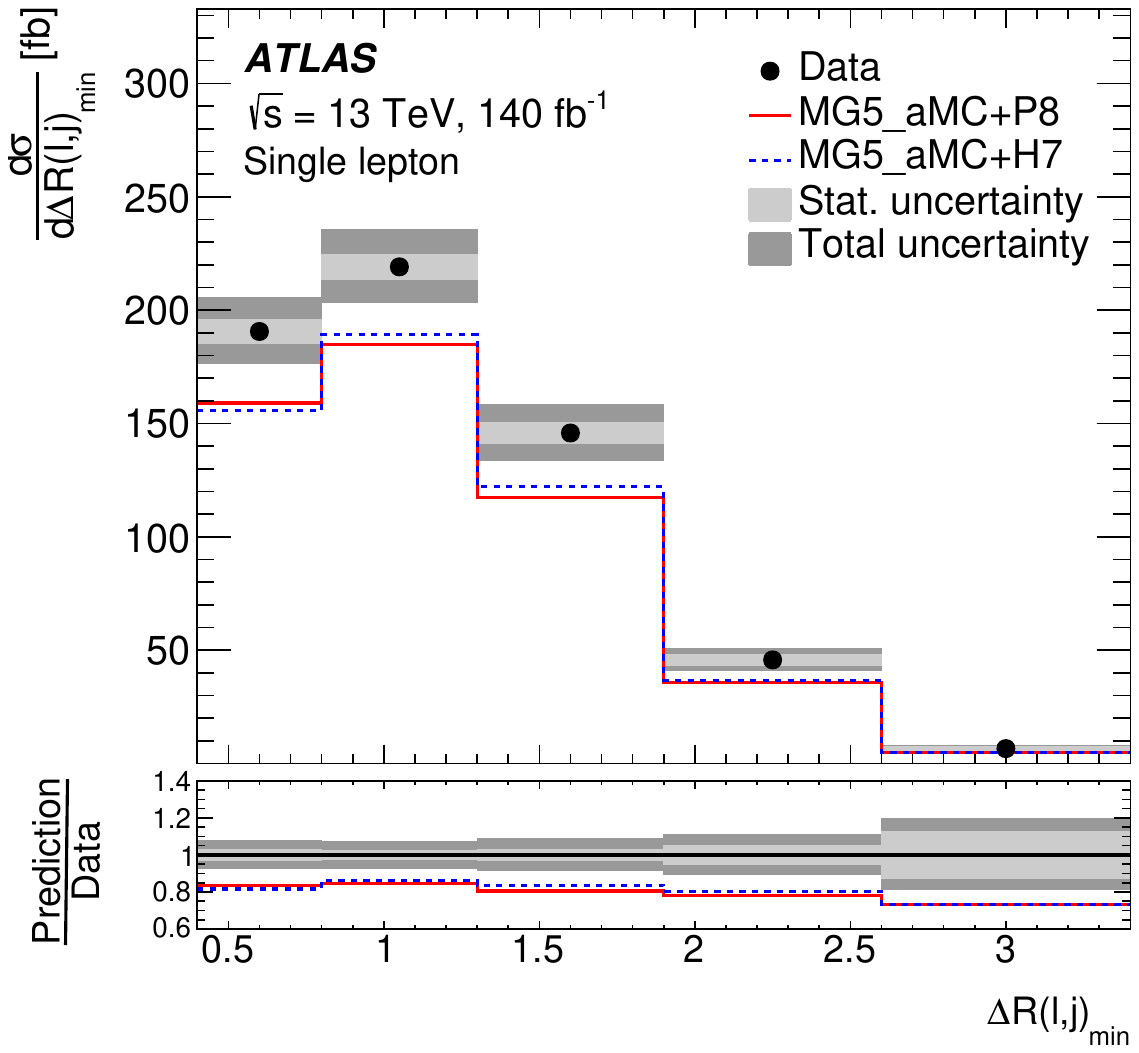}}%
\subfloat[]{\includegraphics[width=0.40\textwidth]{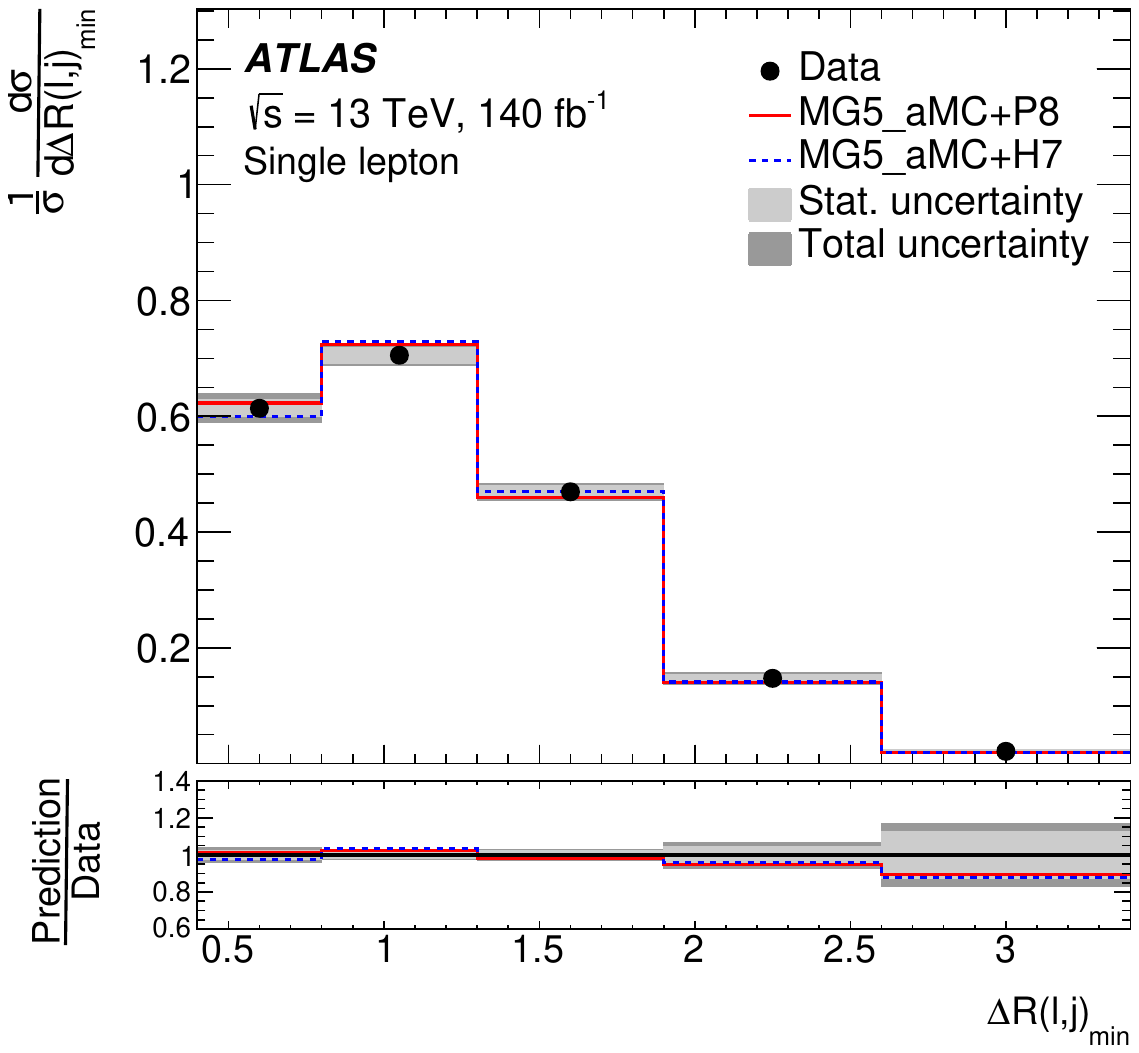}}\\
\caption{The (a, c, e) absolute and (b, d, f) normalised \tty production differential cross-sections measured in the fiducial phase space in the single-lepton channel as a function of (a, b) $\Delta R (\gamma, b)_{\text{min}}$, (c, d) $\Delta R (\gamma, \ell)$ and (e, f) $\Delta R (\ell,j)_{\text{min}}$. Data are compared with the NLO \MGNLO simulation interfaced to \PYTHIA[8] and \HERWIG[7]. The lower panels show the ratios of the predictions to the data. The last bin of the distributions includes the overflow events.}
\label{fig:ttyprod_diff_Ljets}
\end{figure}

\begin{figure}[ht]
\centering
\subfloat[]{\includegraphics[width=0.40\textwidth]{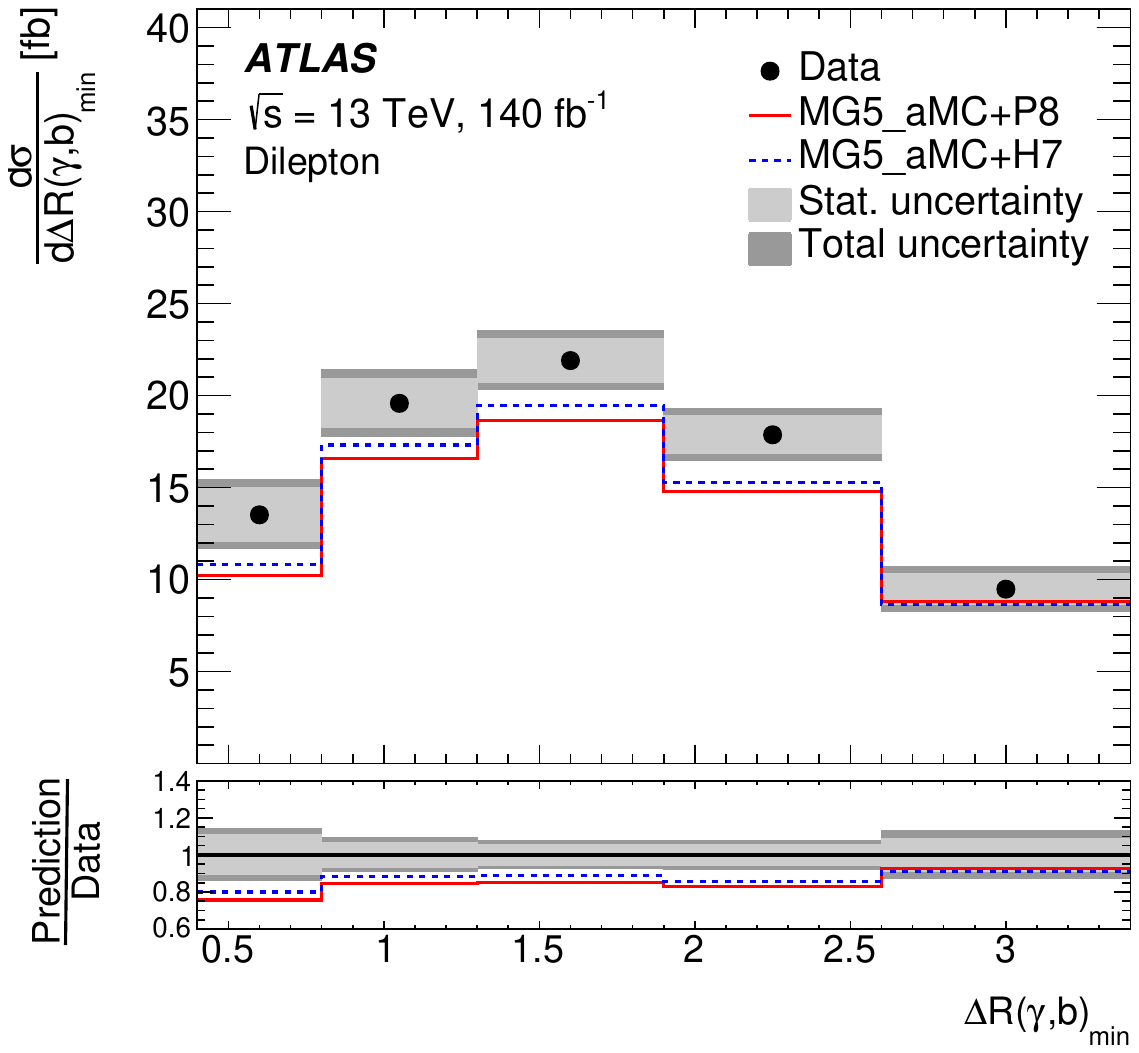}}%
\subfloat[]{\includegraphics[width=0.40\textwidth]{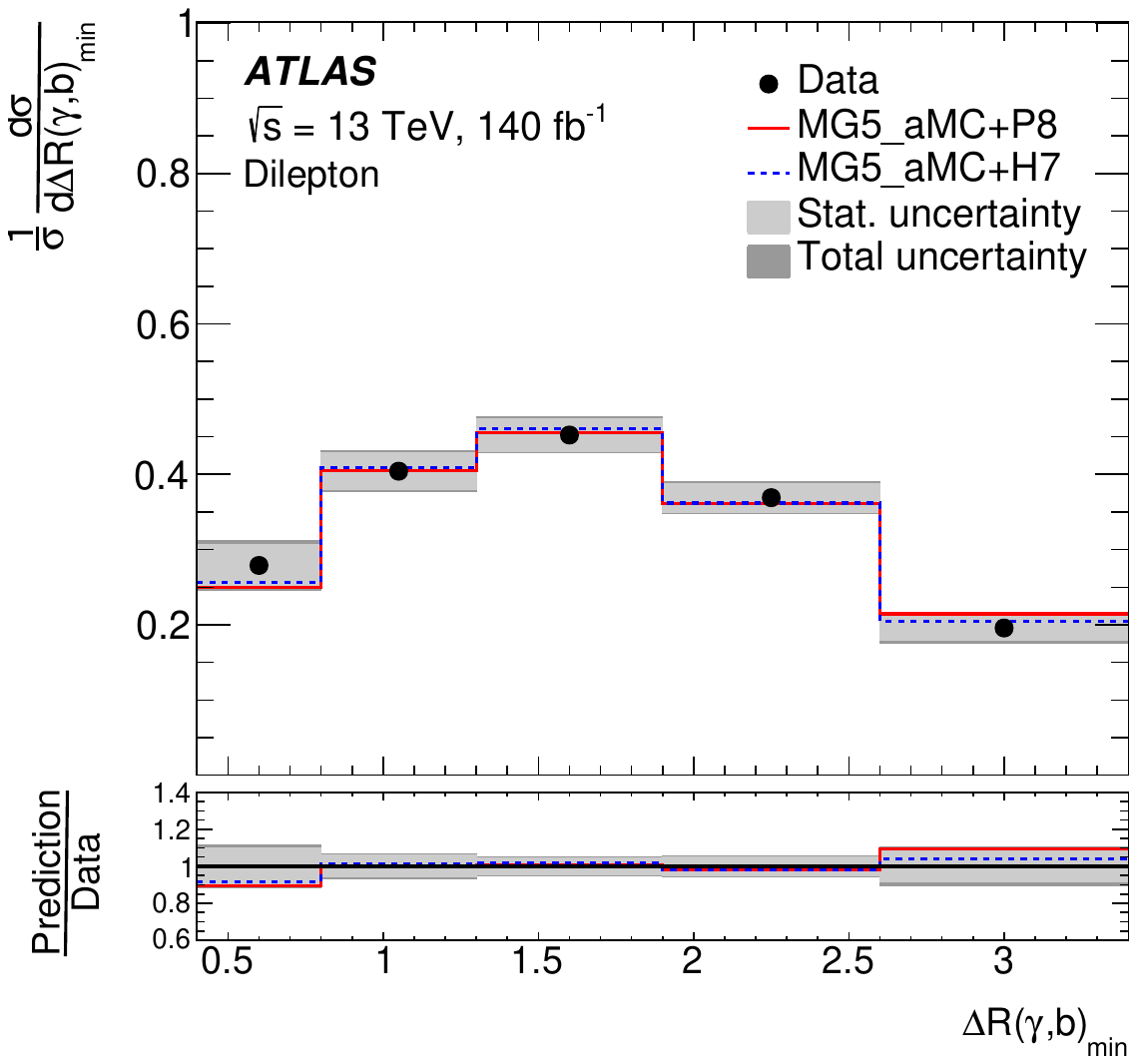}}\\
\subfloat[]{\includegraphics[width=0.40\textwidth]{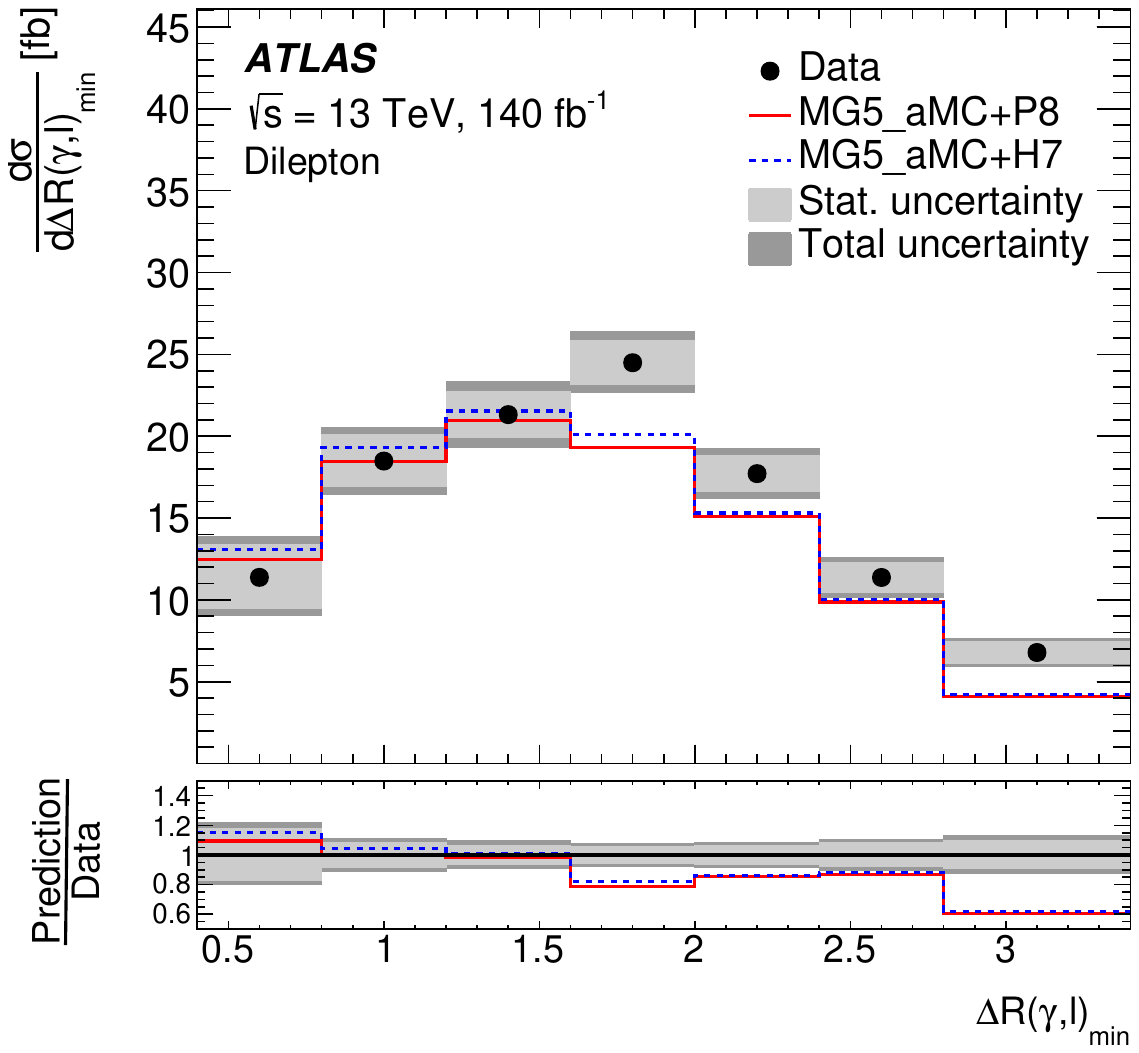}}%
\subfloat[]{\includegraphics[width=0.40\textwidth]{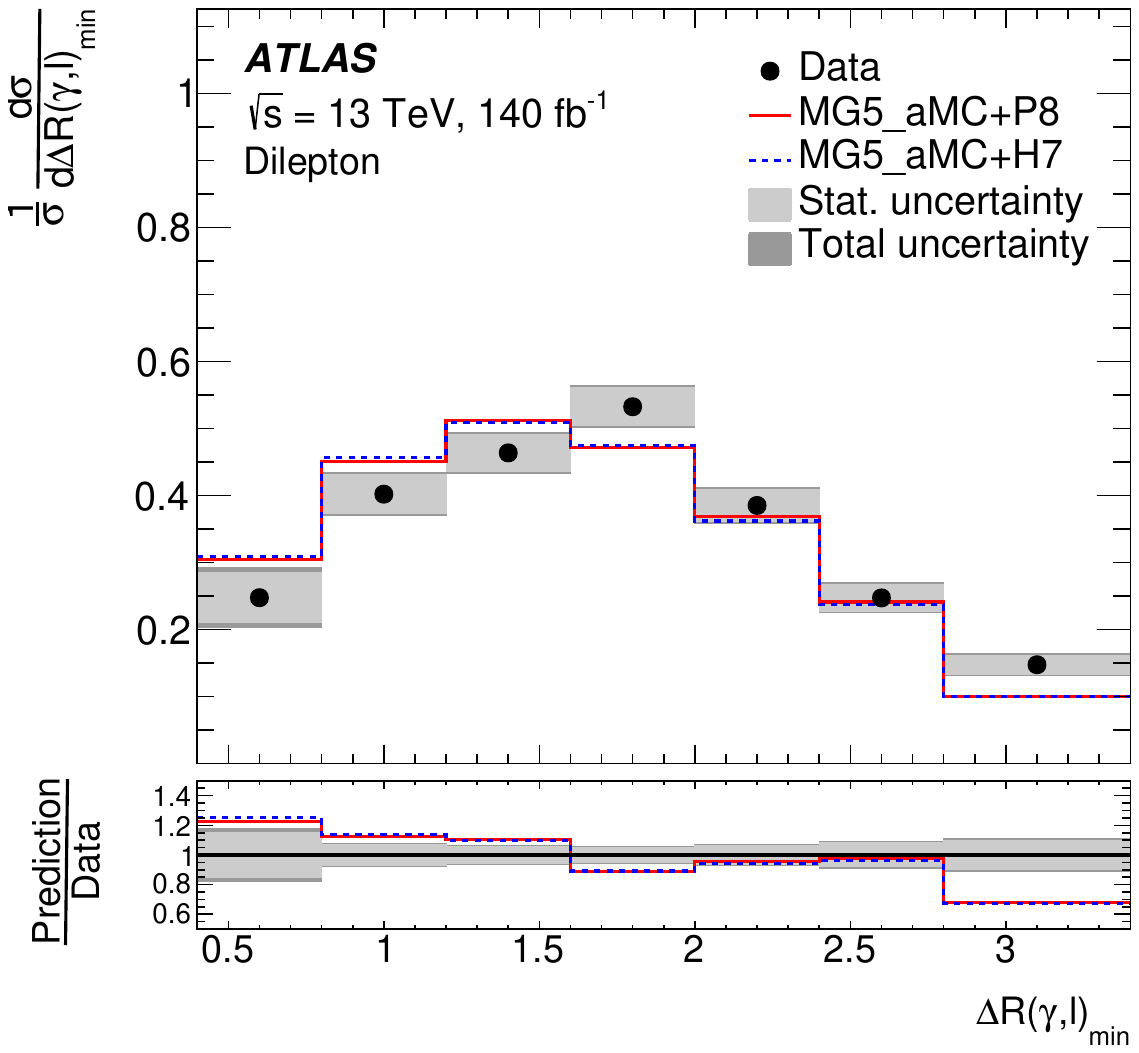}}\\
\subfloat[]{\includegraphics[width=0.40\textwidth]{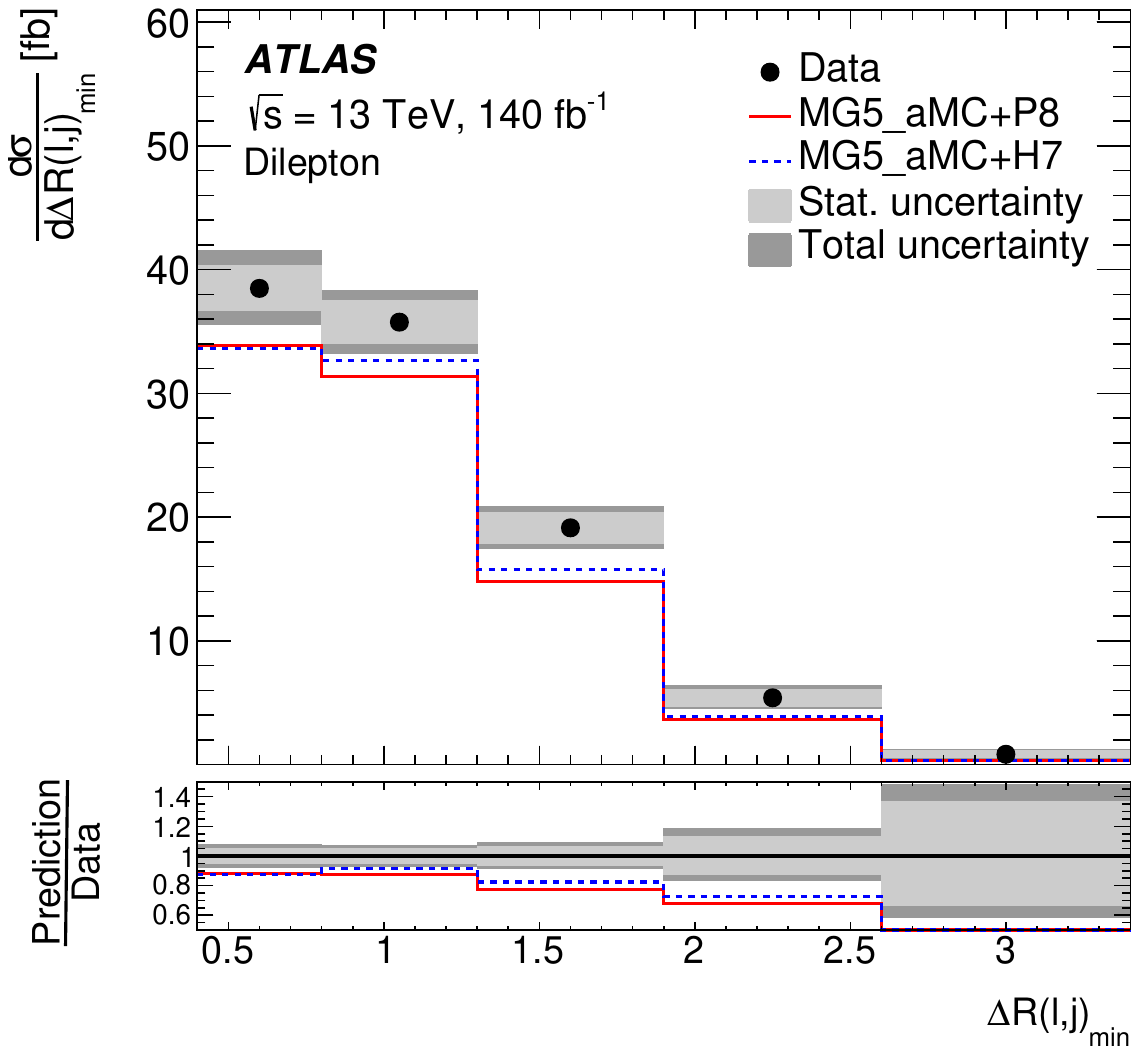}}%
\subfloat[]{\includegraphics[width=0.40\textwidth]{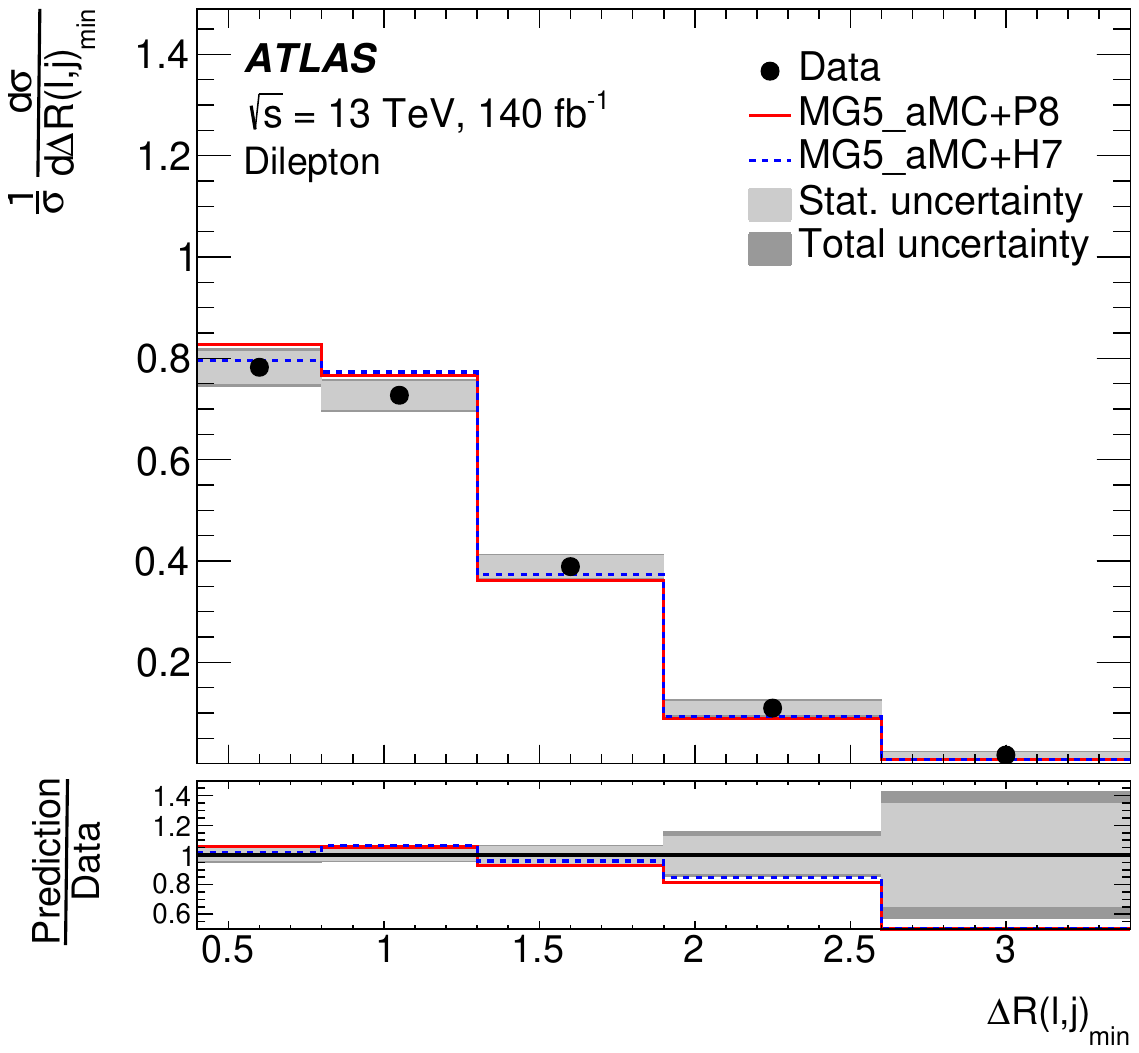}}
\caption{The (a, c, e) absolute and (b, d, f) normalised \tty production differential cross-sections measured in the fiducial phase space in the dilepton channel as a function of (a, b) $\Delta R (\gamma, b)_{\text{min}}$, (c, d) $\Delta R (\gamma, \ell)$ and (e, f) $\Delta R (\ell,j)_{\text{min}}$. Data are compared with the NLO \MGNLO simulation interfaced to \PYTHIA[8] and \HERWIG[7]. The lower panels show the ratios of the predictions to the data. The last bin of the distributions includes the overflow events.}
\label{fig:ttyprod_diff_DL}
\end{figure}

The photon \pt and $\eta$ distributions are measured in each channel and in the combined fiducial phase space. The combined measurement of photon \pt is used as input to obtain limits on the EFT parameters, as described in Section~\ref{sec:EFT}. The absolute and normalised cross-sections measured in the combined phase space are shown in Figure~\ref{fig:tty_prod_SLDL}, while the individual results and the measured leading jet \pt are included in Appendix~\ref{sec:appendix_diff_prod}. As expected, the measured absolute cross-sections are larger than the prediction. The total uncertainty per bin varies between 8\% and 12\%, the largest single contribution arises from the statistical uncertainty (about 5\%). The photon \pt distribution in data is somewhat softer than the MC prediction, while there are no significant differences between the shapes of the $|\eta|$ distributions.

\begin{figure}[ht]
\centering
\subfloat[]{\includegraphics[width=0.40\textwidth]{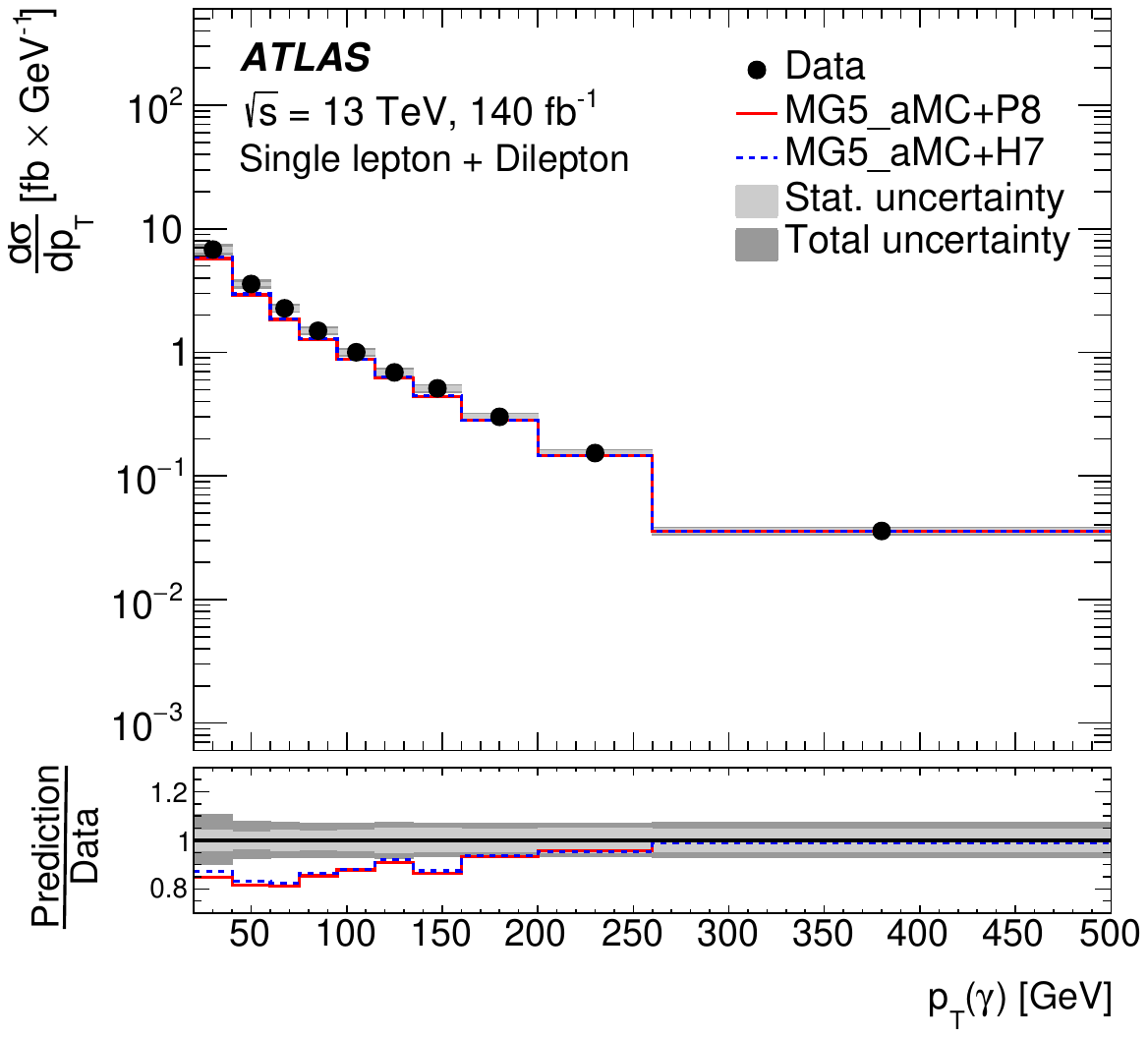}}%
\subfloat[]{\includegraphics[width=0.40\textwidth]{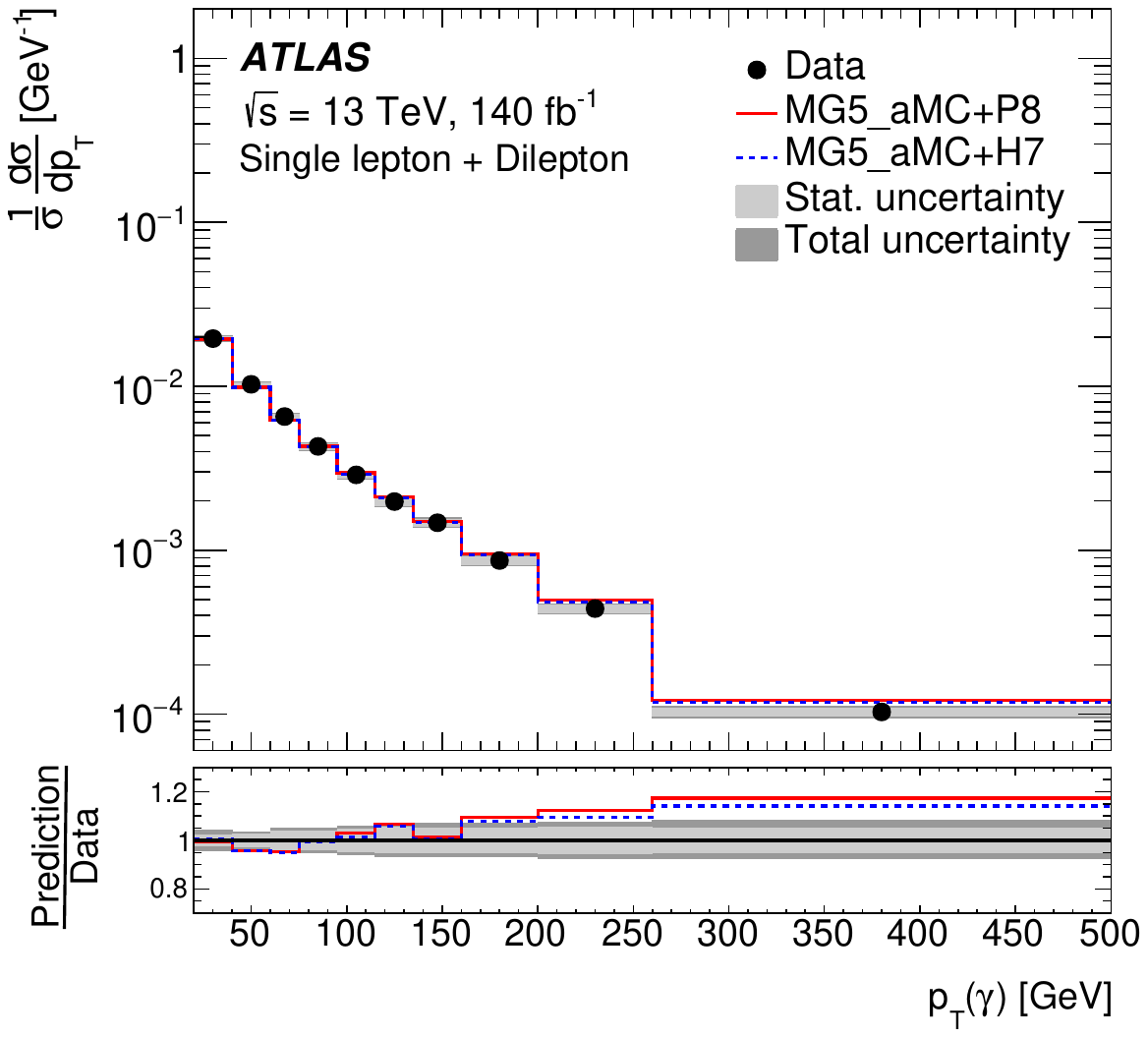}}\\
\subfloat[]{\includegraphics[width=0.40\textwidth]{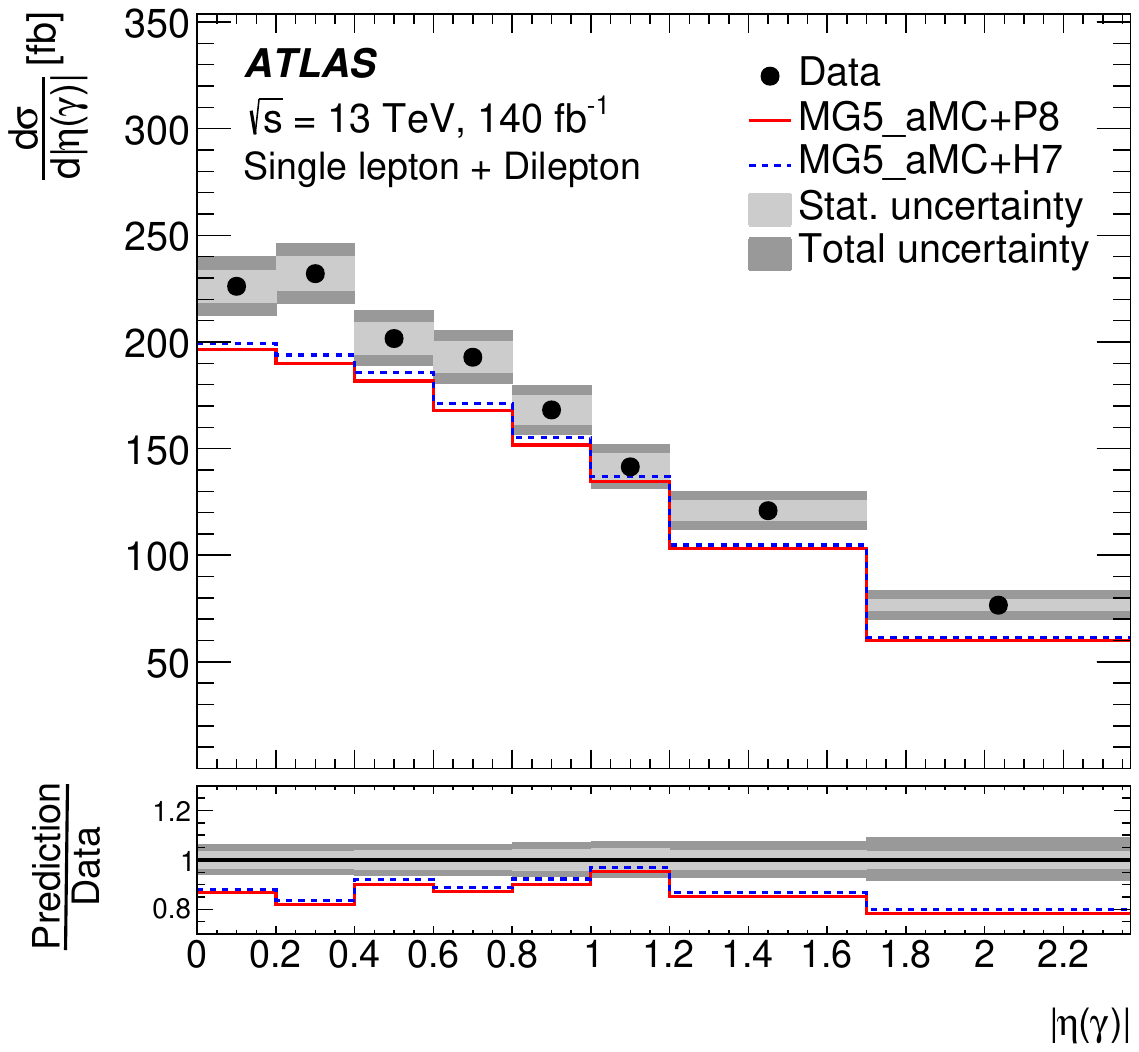}}%
\subfloat[]{\includegraphics[width=0.40\textwidth]{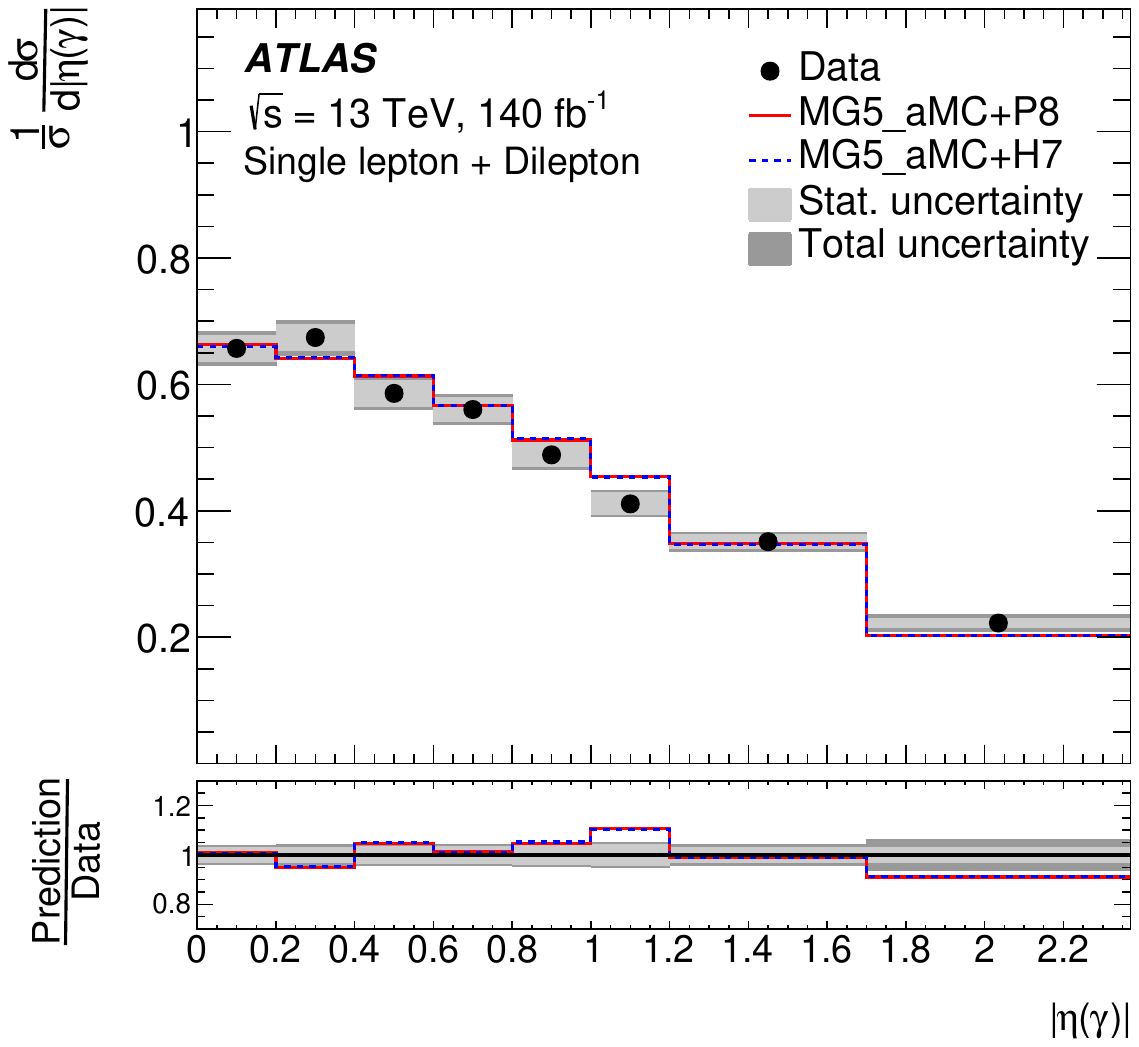}}\\
\caption{The (a, c) absolute and (b, d) normalised \tty production differential cross-sections measured in the combined fiducial phase space of the single-lepton and dilepton channels as a function of the photon (a, b) \pt and (c, d) $\eta$. Data are compared with the NLO \MGNLO simulation interfaced to \PYTHIA[8] and \HERWIG[7]. The lower panels show the ratios of the predictions to the data. The last bin of the photon \pt distributions includes the overflow events.}
\label{fig:tty_prod_SLDL}
\end{figure}

The breakdown of the categories of systematic uncertainties and the statistical one is illustrated in Figure~\ref{fig:grouped_unc_SLDL} for the absolute \tty production differential cross-sections measured in the combined fiducial phase space of the single-lepton and dilepton channels as a function of the photon \pt and $\eta$. The systematic uncertainties in the unfolded distributions are decomposed into signal modelling uncertainties, uncertainties related to the different background categories, experimental uncertainties, and statistical uncertainties. Among experimental uncertainties the largest effects come from the jet uncertainties followed by $b$-tagging and photon uncertainties.

\begin{figure}[ht]
\centering
\subfloat[]{\includegraphics[width=0.5\textwidth]{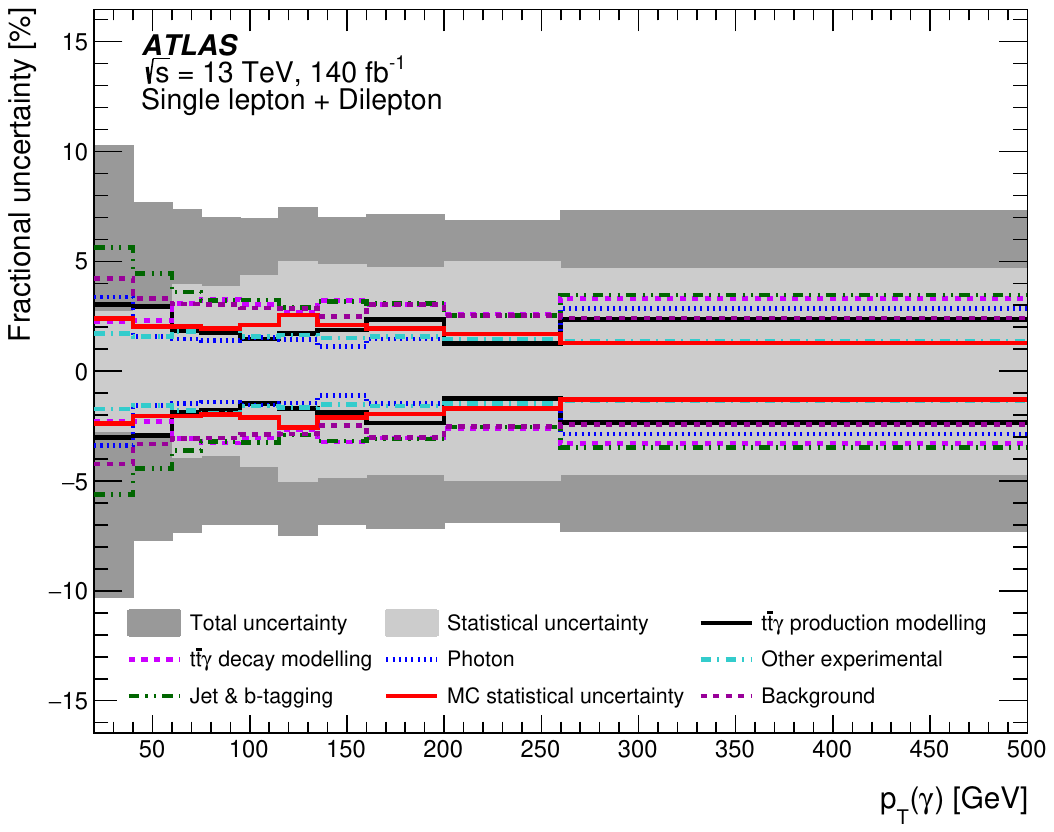}}%
\subfloat[]{\includegraphics[width=0.5\textwidth]{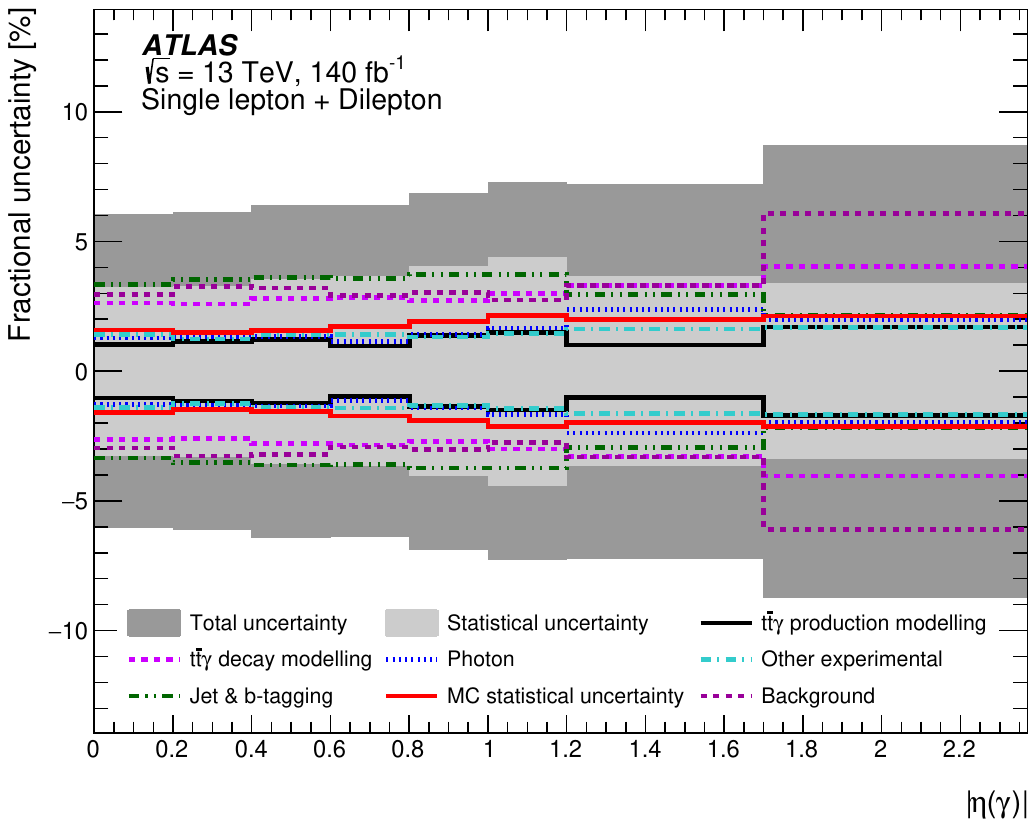}}
\caption{Contributions of the systematic uncertainties grouped in categories in each bin of the measurement of the absolute \tty production differential cross-sections measured in the combined fiducial phase space of the single-lepton and dilepton channels as a function of the photon (a) \pt  and (b) $\eta$. The `\tty decay modelling' category includes both the normalisation and modelling uncertainties and the `Background' category the uncertainties related to other prompt photon, fake photon and fake lepton backgrounds. The `Other experimental' category corresponds to the combined effect of lepton, \MET, luminosity and pile-up uncertainties. The relative uncertainties quoted are obtained by repeating the fit, fixing the set of nuisance parameters of the sources corresponding to each category to their post-fit values, and subtracting in quadrature the resulting uncertainty from the total uncertainty of the nominal fit. The total uncertainty is different from the sum in quadrature of the components due to correlations among nuisance parameters.}
\label{fig:grouped_unc_SLDL}
\end{figure}

The compatibility between the measured differential cross-sections and the predictions,
evaluated using a ratio of $\chi^2$ to the number of degrees of freedom (ndf) and the corresponding $p$-values, are summarised in Table~\ref{tab:chi2_ttyprod} for all measured variables, including those shown in Appendix~\ref{sec:appendix_diff_prod}. They are evaluated taking into account the statistical and systematic covariance matrices of the measurements and the number of degrees of freedom.

\begin{table}
\scriptsize
\centering
\caption{The $\chi^2$/ndf and $p$-values between the measured absolute and normalised cross-sections of \tty production and the NLO \MGNLO simulations interfaced to \PYTHIA[8] and \HERWIG[7]. They are obtained using the uncertainties of the measured distribution and their correlations.}
\label{tab:chi2_ttyprod}
\scalebox{0.9}{
\begin{tabular}{l | c c | c c | c c |  c c }
\toprule
& \multicolumn{4}{c}{Absolute cross-sections} & \multicolumn{4}{c}{Normalised cross-sections} \\
&  \multicolumn{2}{c}{MG5\_aMC@NLO+\Pythia[8]} & \multicolumn{2}{c}{MG5\_aMC@NLO+\Herwig[7]} &  \multicolumn{2}{c}{MG5\_aMC@NLO+\Pythia[8]} & \multicolumn{2}{c}{MG5\_aMC@NLO+\Herwig[7]}\\
Variables & $\chi^2$/ndf & $p$-value & $\chi^2$/ndf & $p$-value & $\chi^2$/ndf & $p$-value & $\chi^2$/ndf & $p$-value \\
\midrule
\multicolumn{9}{c}{Single-lepton and dilepton combined} \\
\midrule
$\pt(\gamma)$ &	 10.7/10 &	 0.38 &	 \phantom{0}9.3/10 &	 0.50 &	 11.6/9 &	 0.23 &	 \phantom{0}8.6/9 & 	 0.47 \\
$|\eta(\gamma)|$ &	 15.8/8\phantom{0} &	 0.04 &	 14.2/8\phantom{0} &	 0.08 &	 10.2/7 &	 0.18 &	 10.0/7 & 	 0.19 \\
\midrule
\multicolumn{9}{c}{Single-lepton channel} \\
\midrule
\pt($\gamma$) &	 11.7/10&	 0.31&	 10.4/10&	 0.40&	 43.9/9\phantom{0}&	 $<0.01\phantom{< \;}$ &	 32.1/9\phantom{0}& 	 $<0.01\phantom{< \;}$ \\

$|\eta(\gamma)|$ &	 11.8/8\phantom{0}&	 0.16&	 11.1/8\phantom{0}&	 0.20&	 \phantom{0}8.1/7\phantom{0}&	 0.33&	 \phantom{0}8.1/7\phantom{0}& 	 0.32\\

$\Delta R(\gamma, \ell)$ &	 10.5/7\phantom{0}&	 0.16&	 9.9/7&	 0.19&	 8.8/6&	 0.19&	 8.8/6& 	 0.19\\

$\Delta R(\gamma, b)_{\text{min}}$&	 12.5/5\phantom{0}&	 0.03&	 12.3/5\phantom{0}&	 0.03&	 7.6/4&	 0.11&	 9.0/4& 	 0.06\\

$\Delta R(\ell, j)_{\text{min}}$ &	 \phantom{0}6.3/5\phantom{0}&	 0.28&	 6.6/5&	 0.25&	 1.5/4&	 0.83&	 2.5/4& 	 0.65\\

$\pt(j_1)$ &	 12.6/5\phantom{0}&	 0.03&	 10.8/5\phantom{0}&	 0.06&	 8.2/4&	 0.08&	 9.7/4& 	 0.05\\
\midrule
\multicolumn{9}{c}{Dilepton channel} \\
\midrule
\pt($\gamma$) &	 \phantom{0}8.6/6\phantom{0} &	 0.20 &	 7.2/6 &	 0.31 &	 6.5/5 &	 0.26 &	 5.9/5 & 	 0.32 \\
$|\eta(\gamma)|$ &	 12.1/8\phantom{0} &	 0.15 &	 9.9/8 &	 0.27 &	 9.2/7 &	 0.24 &	 7.9/7 & 	 0.34 \\
$\Delta R(\gamma, \ell)_{\text{min}}$ &	 17.2/7\phantom{0} &	 0.02 &	 16.5/7\phantom{0} &	 0.02 &	 14.2/6\phantom{0} &	 0.03 &	 14.4/6\phantom{0} & 	 0.03 \\
$\Delta R(\gamma, b)_{\text{min}}$ &	 \phantom{0}7.8/5\phantom{0} &	 0.17 &	 5.0/5 &	 0.42 &	 1.4/4 &	 0.84 &	 0.8/4 & 	 0.93 \\
$\Delta R(\ell, j)_{\text{min}}$ &	 \phantom{0}9.3/5\phantom{0} &	 0.10 &	 6.4/5 &	 0.27 &	 5.4/4 &	 0.25 &	 3.7/4 & 	 0.45 \\
$\pt(j_1)$ &	 10.5/5\phantom{0} &	 0.06 &	 5.0/5 &	 0.42 &	 7.8/4 &	 0.10 &	 3.6/4 & 	 0.46 \\
\bottomrule
\end{tabular}
}
\end{table}

\FloatBarrier

\subsection{Differential distributions for the total \tty production and decay process}
\label{sec:total_tty_diff}
The measurements are also performed for the sum of the \tty production and \tty decay in the same phase space as the \tty production process alone at the stable particle level, similarly to previous measurements. The results are compared with the LO $2\to7$ MC simulations of the \tty process used in previous publications~\cite{TOPQ-2020-03, TOPQ-2017-14} described in Section~\ref{sec:modelling}.

The absolute and normalised cross-sections in the single-lepton channel are shown as functions of photon \pt, photon $|\eta|$ and leading jet \pt in Figure~\ref{fig:tty_diff_Ljets} and as functions of $\Delta R(\gamma, b)_{\text{min}}$, $\Delta R (\gamma, \ell)$ and $\Delta R (\ell,j)_{\text{min}}$ in Figure~\ref{fig:tty_diff_Ljets_DR}. Overall, the shapes of the $\Delta R$ and $|\eta|$ distributions are well described within the total uncertainties of the measurement, while the description of the \pt distributions by the LO  simulation is poor. Compared to the distributions measured for the \tty production process alone, the \pt of the photon is softer, while there are no significant differences between the shapes of the photon $\eta$. As expected, the $\Delta R$ distributions have more events at lower $\Delta R$ values when including the \tty decay events in the signal, where the photon can be emitted by a lepton. The corresponding distributions measured in the dilepton channel are shown in Appendix~\ref{sec:appendix_diff}.

\begin{figure}[ht]
\centering
\subfloat[]{\includegraphics[width=0.40\textwidth]{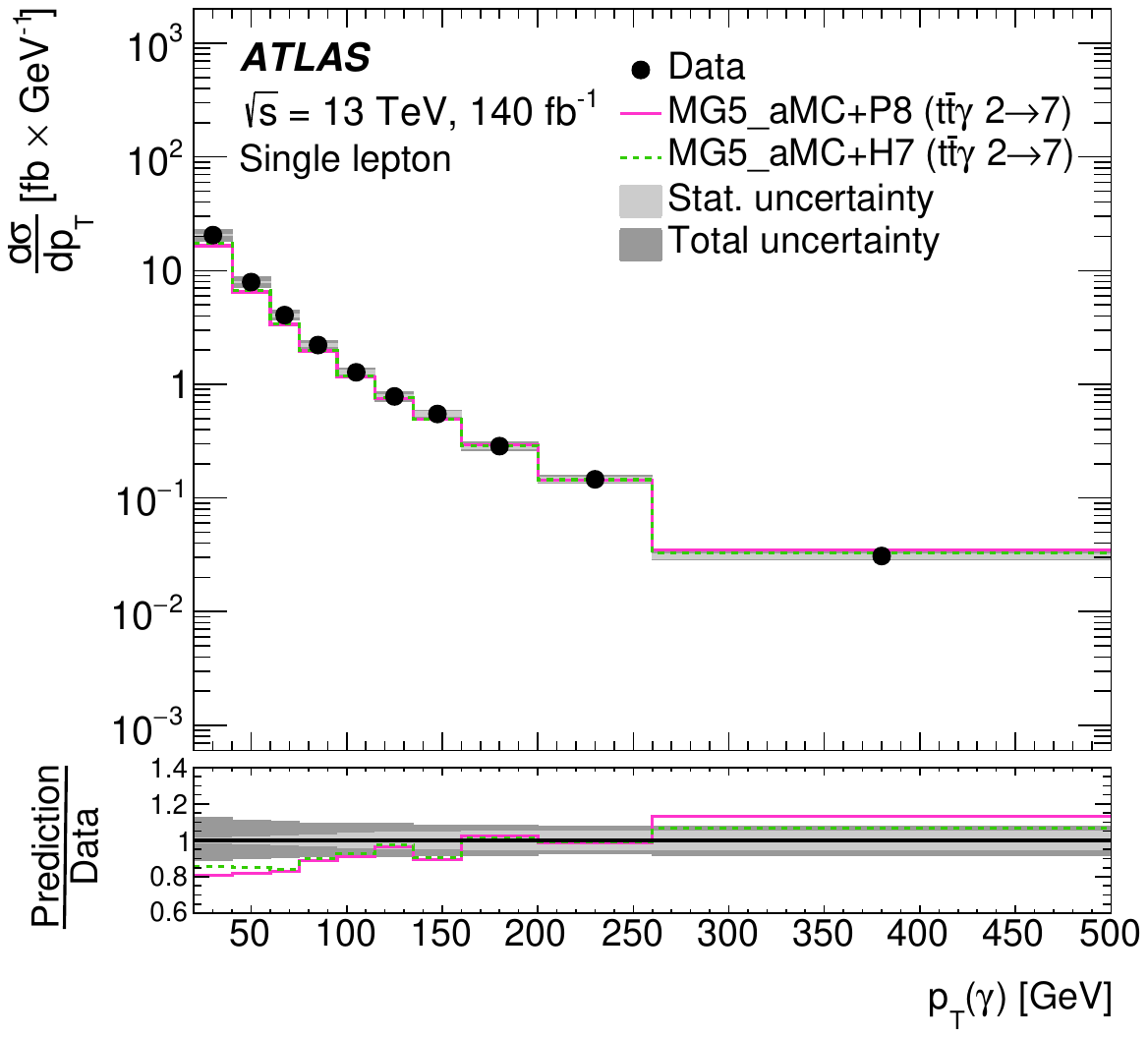}}%
\subfloat[]{\includegraphics[width=0.40\textwidth]{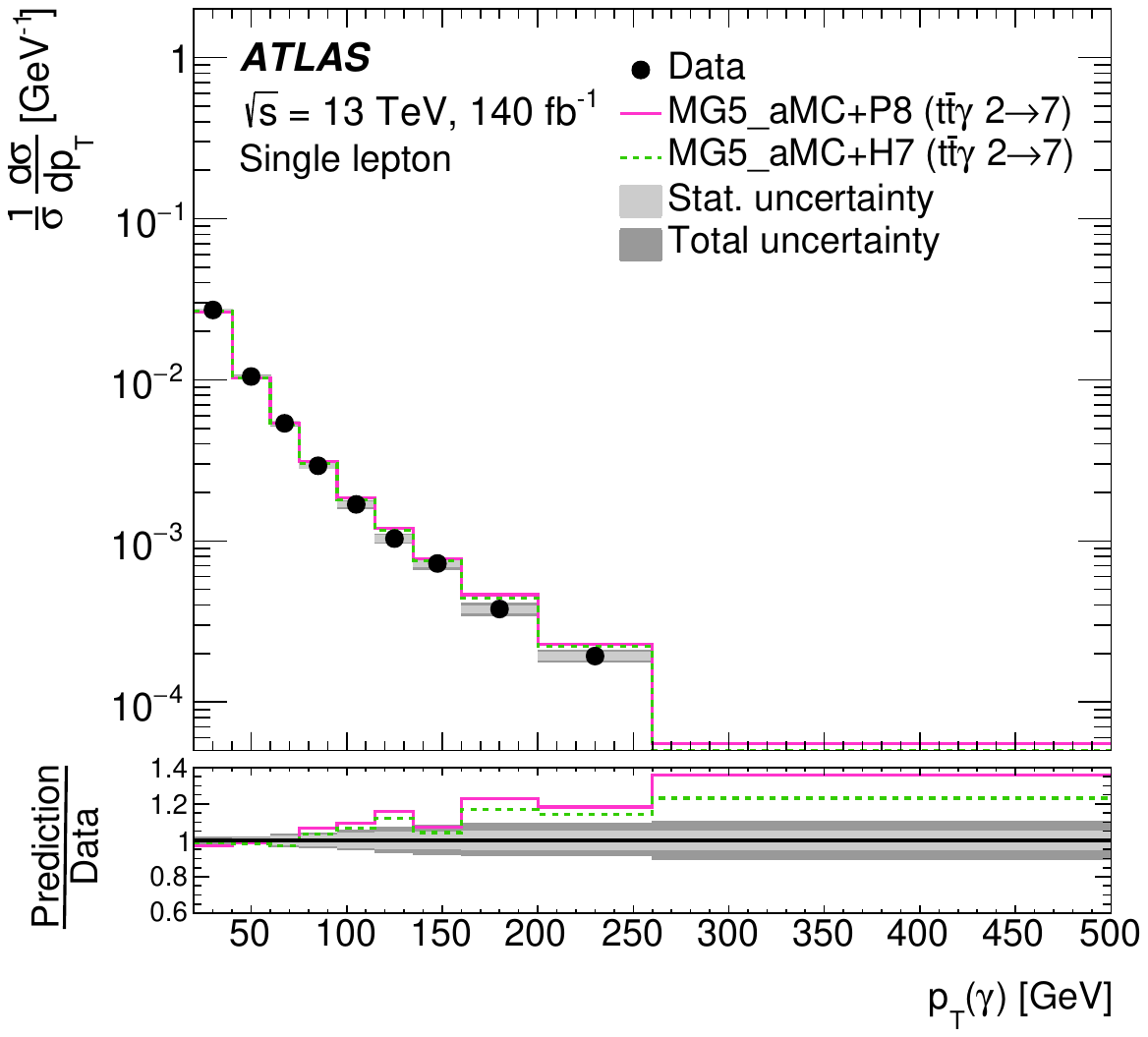}}\\
\subfloat[]{\includegraphics[width=0.40\textwidth]{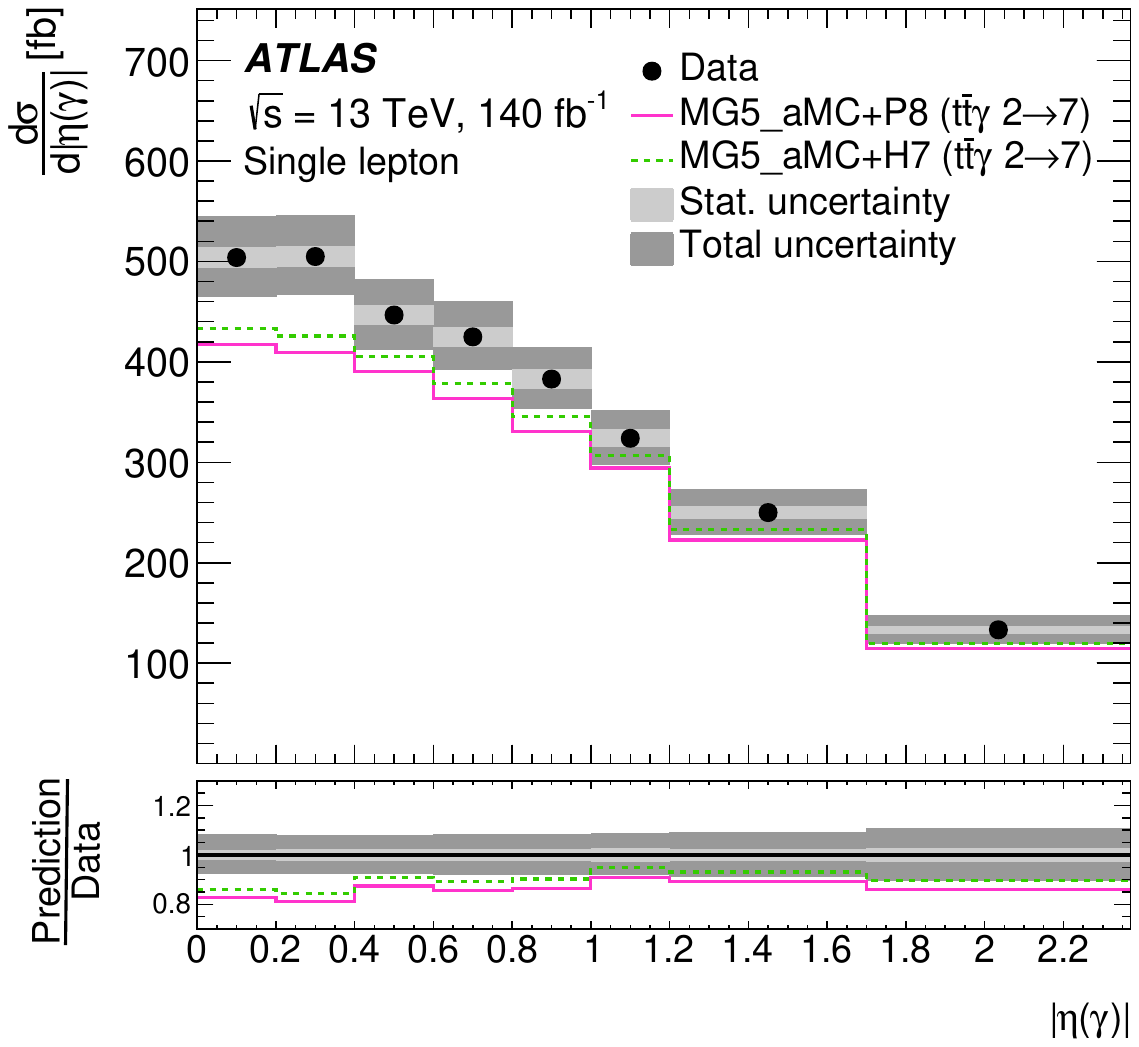}}%
\subfloat[]{\includegraphics[width=0.40\textwidth]{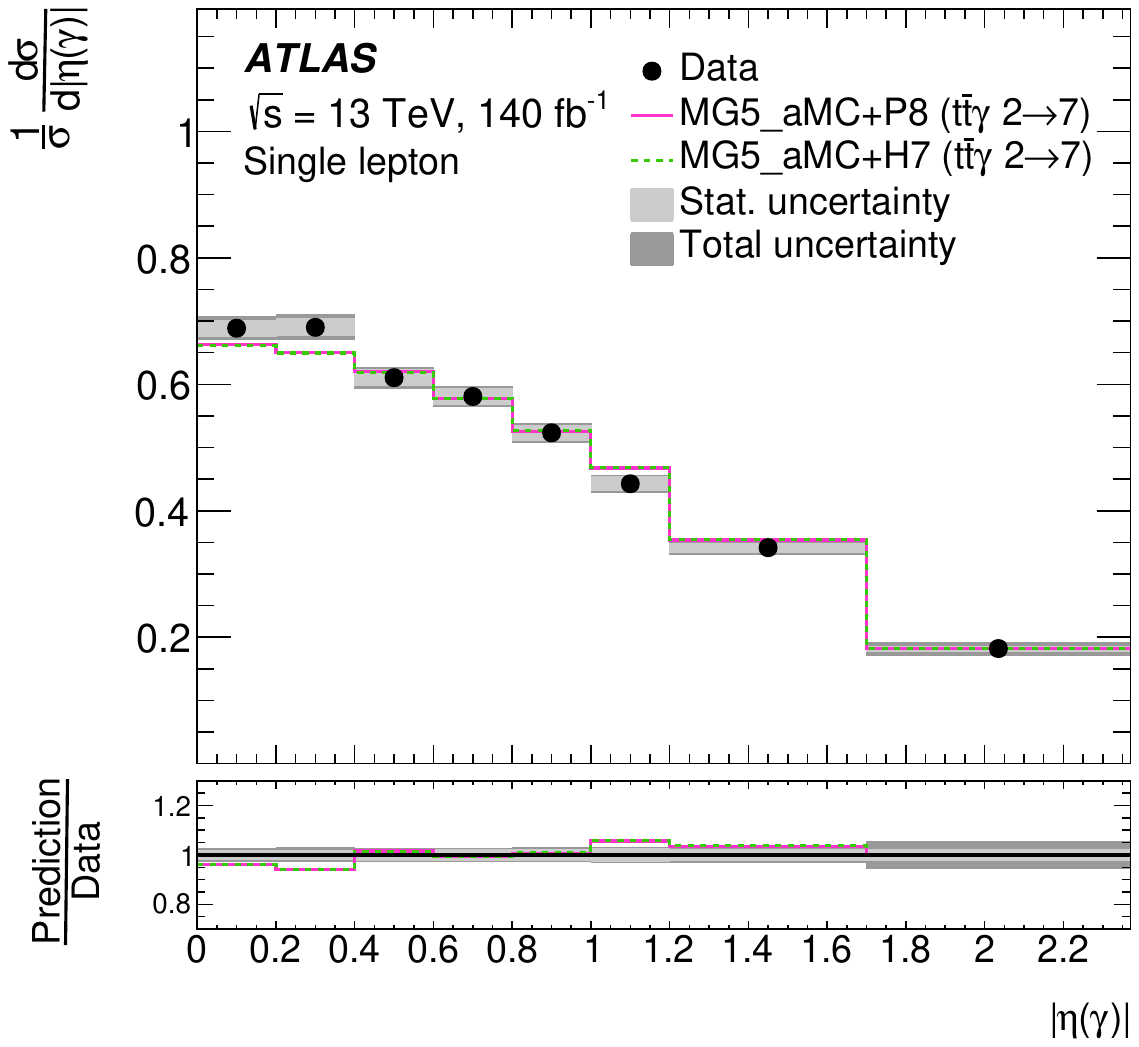}}\\
\subfloat[]{\includegraphics[width=0.40\textwidth]{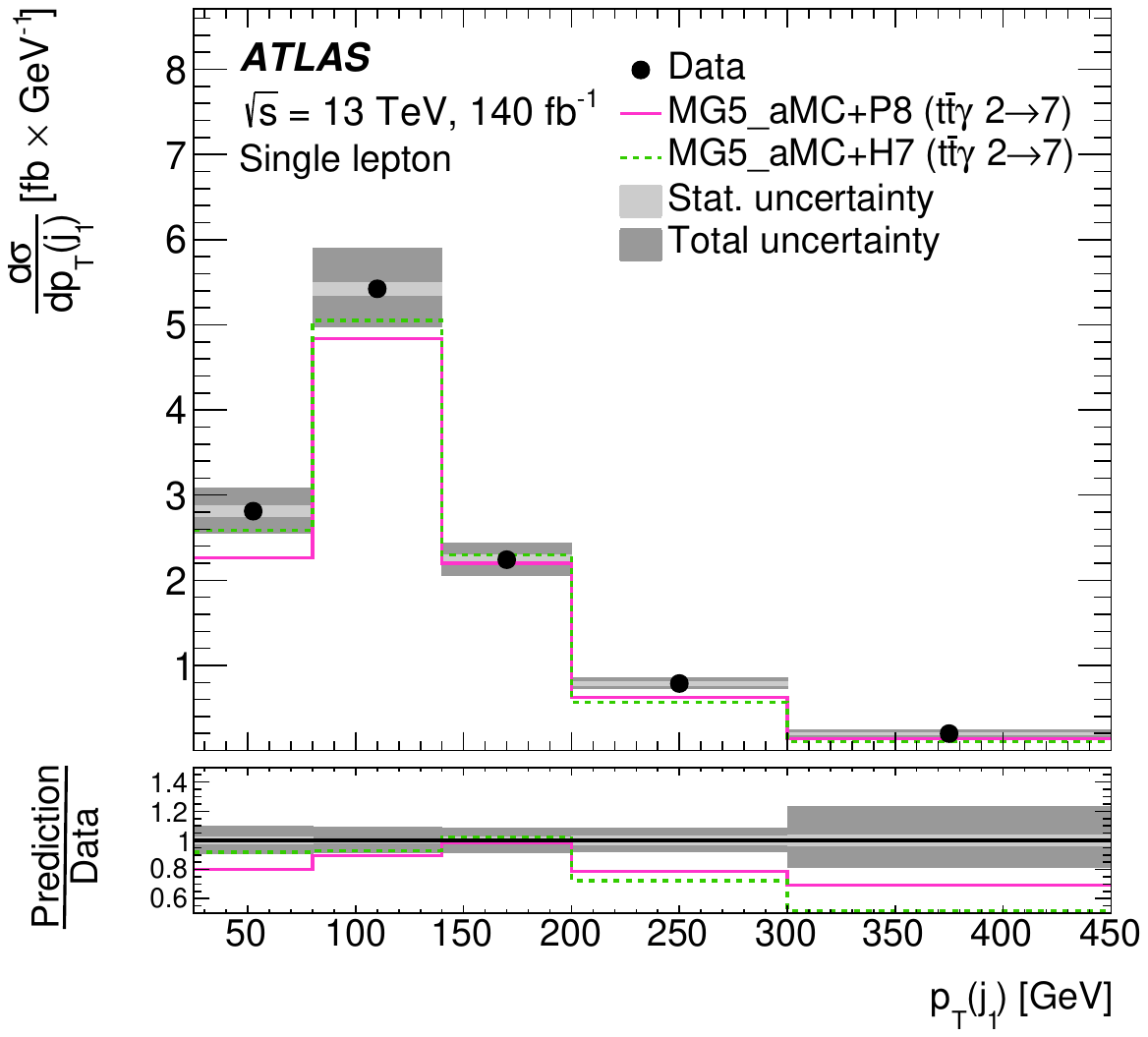}}%
\subfloat[]{\includegraphics[width=0.40\textwidth]{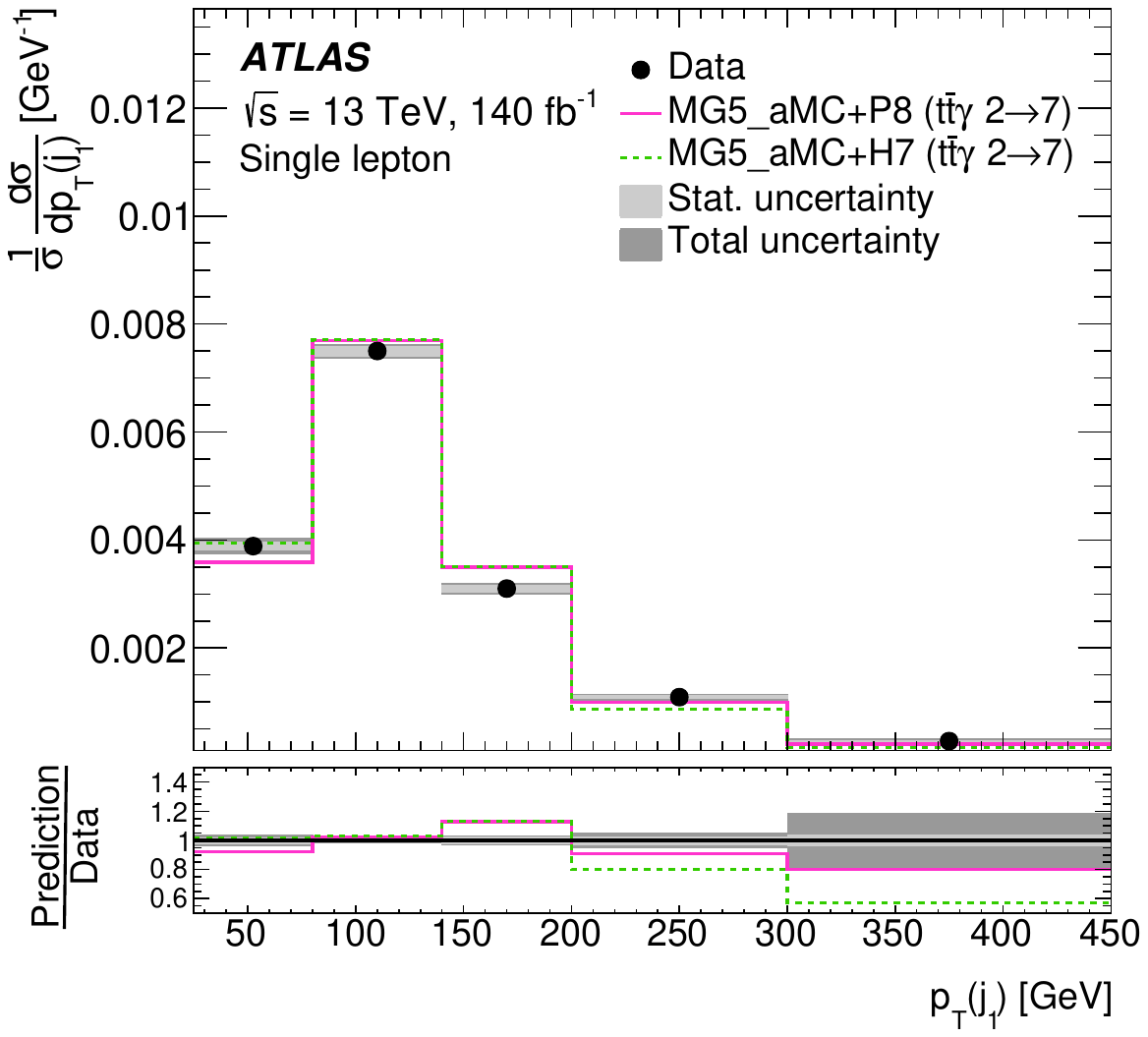}}
\caption{The (a, c, e) absolute and (b, d, f) normalised differential distributions of the total \tty production and decay cross-section measured in the fiducial phase space in the single-lepton channel as a function of the (a, b) photon \pt, (c, d) photon $|\eta|$ and (e, f) leading jet \pt. Data are compared with the \tty \MGNLO simulation at LO interfaced to \PYTHIA[8] and \HERWIG[7]. The lower panels show the ratios of the predictions to the data. The last bin of the distributions includes the overflow events. }
\label{fig:tty_diff_Ljets}
\end{figure}

\begin{figure}[ht]
\centering
\subfloat[]{\includegraphics[width=0.40\textwidth]{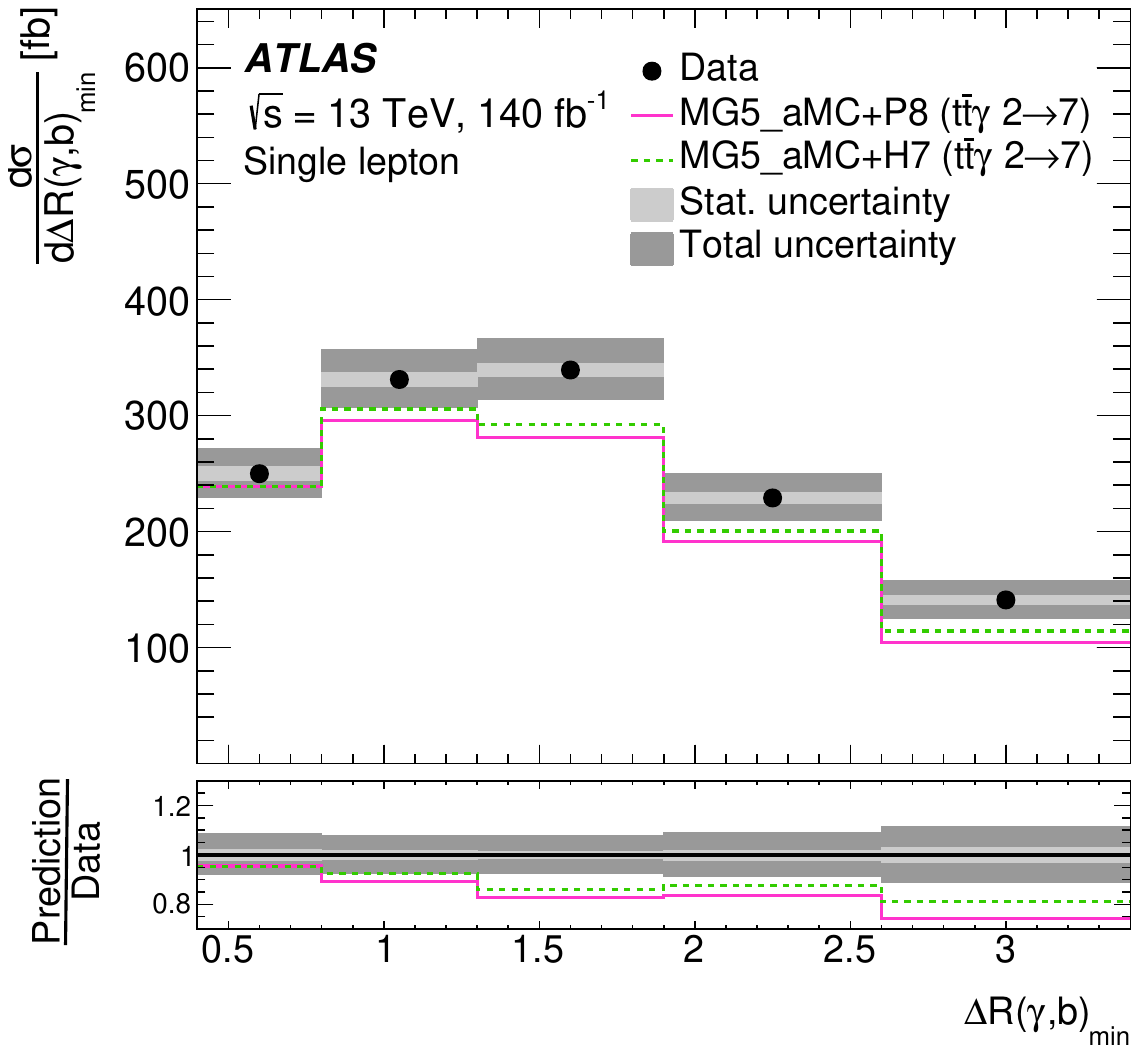}}%
\subfloat[]{\includegraphics[width=0.40\textwidth]{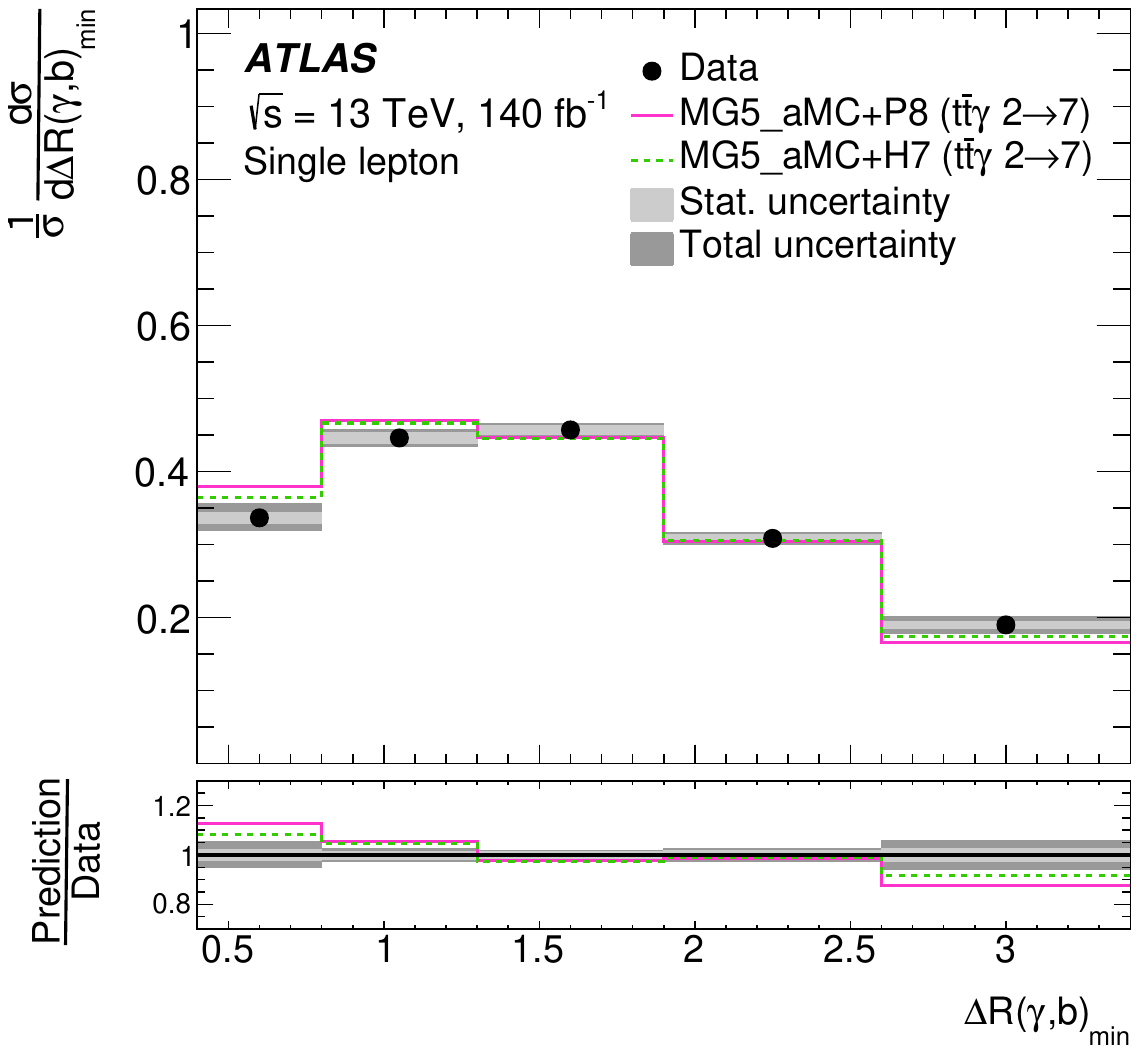}}\\
\subfloat[]{\includegraphics[width=0.40\textwidth]{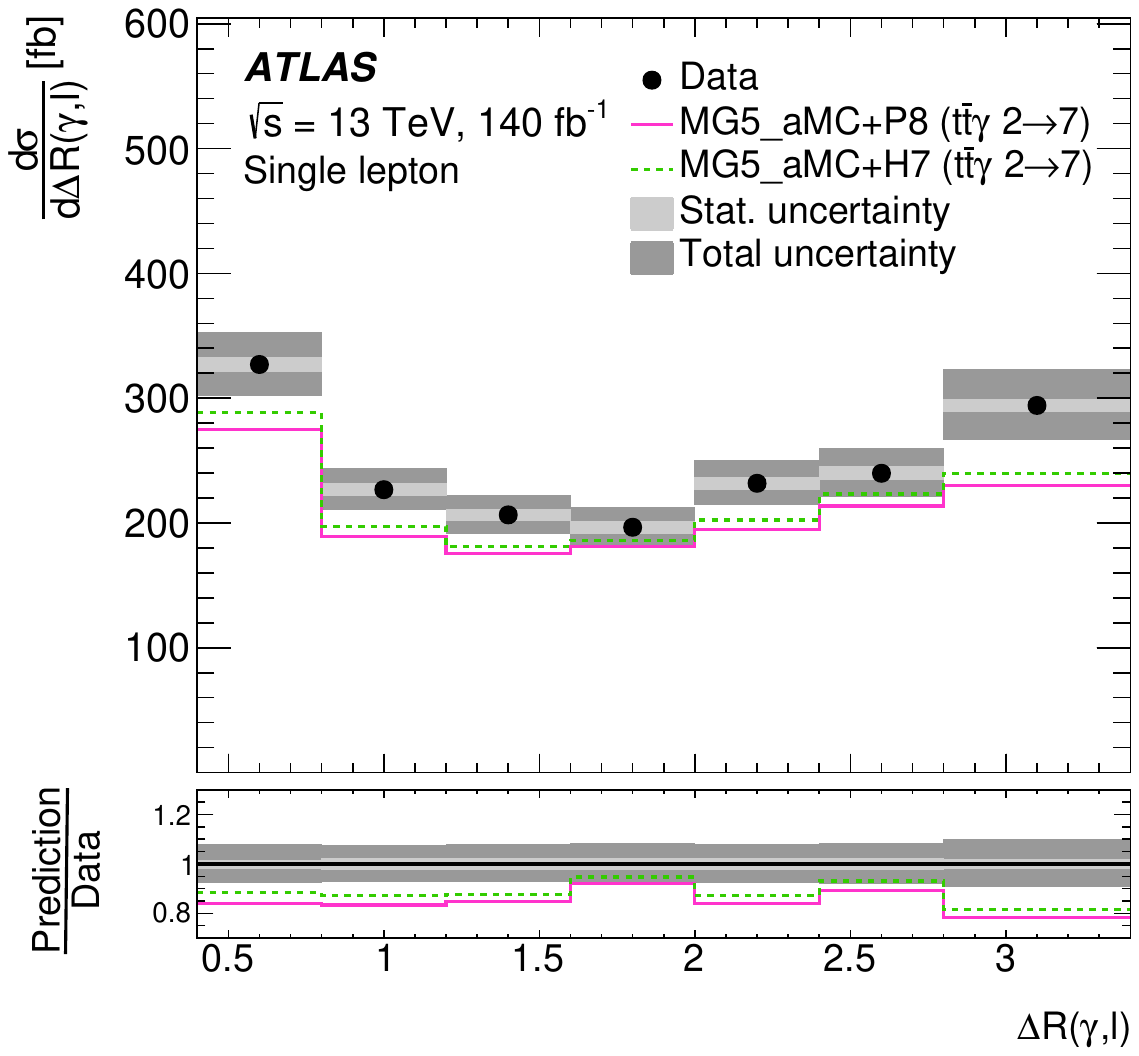}}%
\subfloat[]{\includegraphics[width=0.40\textwidth]{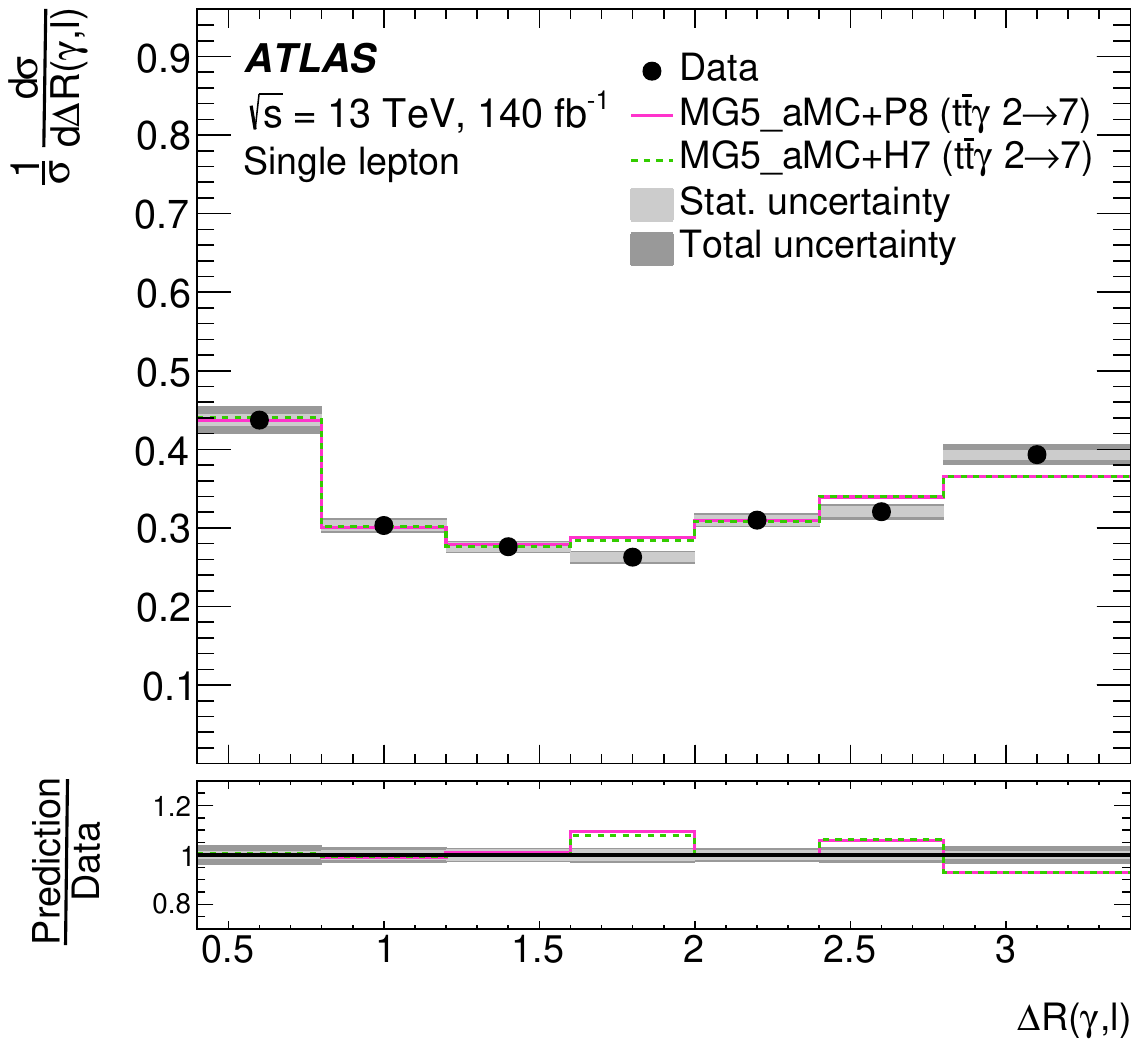}}\\
\subfloat[]{\includegraphics[width=0.40\textwidth]{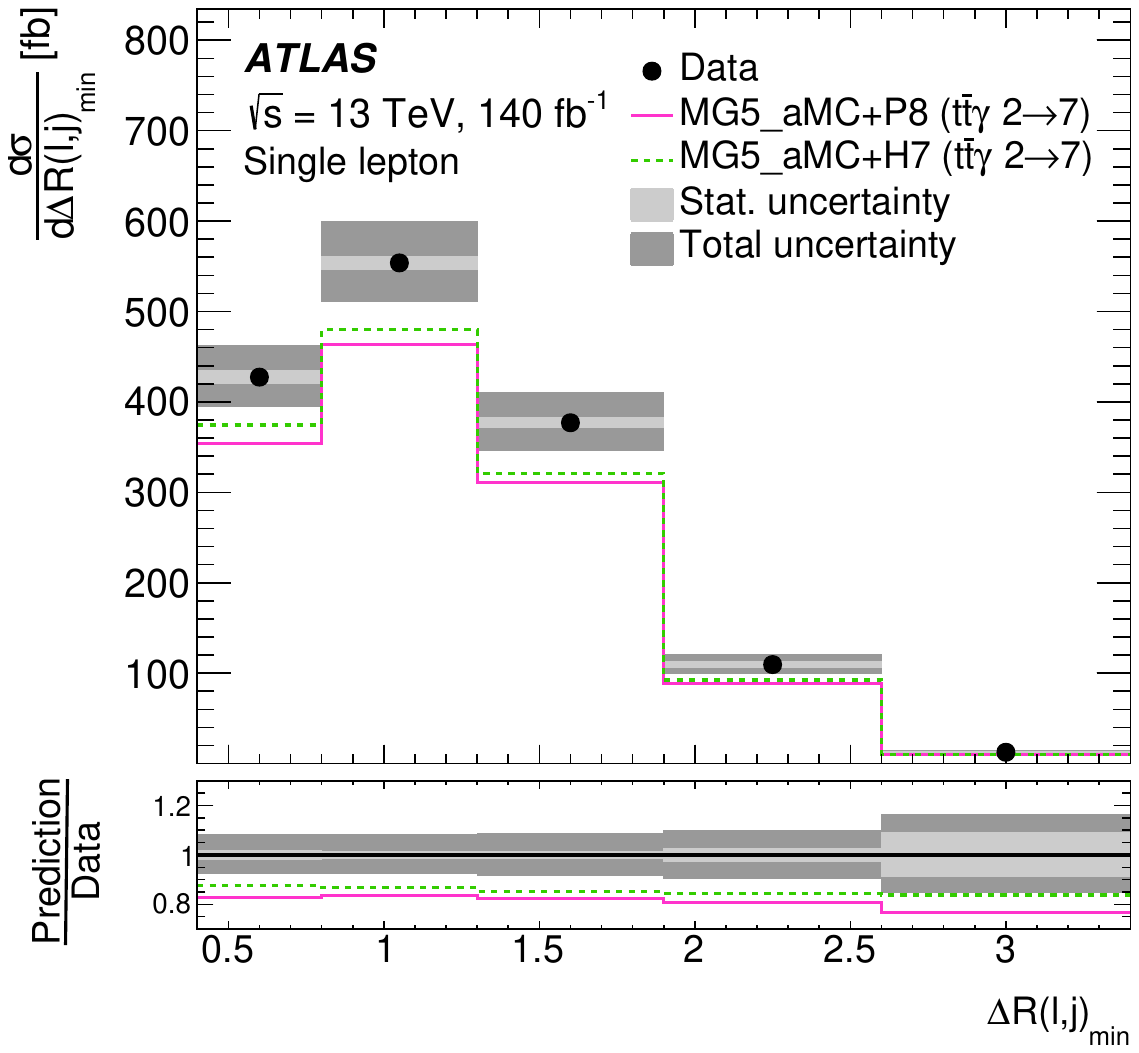}}%
\subfloat[]{\includegraphics[width=0.40\textwidth]{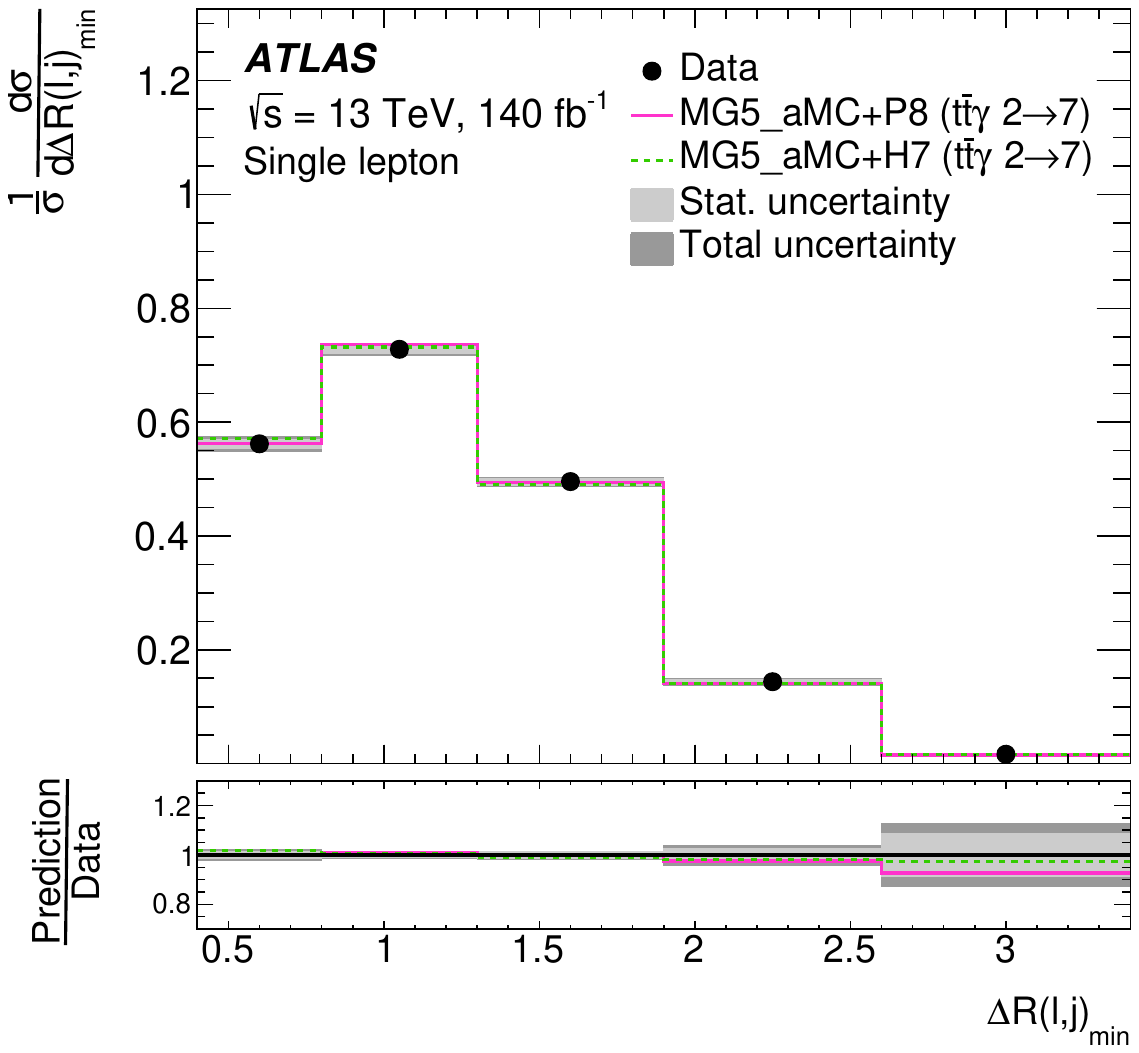}}\\
\caption{The (a, c, e) absolute and (b, d, f) normalised differential distributions of the total \tty production and decay cross-section measured in the fiducial phase space in the single-lepton channel as a function of (a, b) $\Delta R (\gamma, b)_{\text{min}}$, (c, d) $\Delta R (\gamma, \ell)$ and (e, f) $\Delta R (\ell, j)_{\text{min}}$. Data are compared with the \tty \MGNLO simulation at LO interfaced to \PYTHIA[8] and \HERWIG[7]. The lower panels show the ratios of the predictions to the data. The last bin of the distributions includes the overflow events.}
\label{fig:tty_diff_Ljets_DR}
\end{figure}

Following the previous publications by the ATLAS and CMS collaborations~\cite{TOPQ-2017-14, TOPQ-2020-03, CMS-TOP-21-004}, the cross-sections in the dilepton channel are measured as a function of additional variables. The absolute and normalised cross-sections as functions of $|\Delta \eta(\ell\ell)|$, $\Delta \phi (\ell\ell)$, and the \pt of the dilepton system are presented in Figure~\ref{fig:tty_diff_DL_dilepton}, while Figure~\ref{fig:tty_diff_DL_DR} shows the $\Delta R$ distributions between the photon and the leading and subleading lepton ordered by \pt. The qualitative description of the $\Dphill$ and $|\Delta \eta(\ell\ell)|$ distributions by the simulation is in agreement with the observations of the previous measurements performed in a different phase space at parton~\cite{TOPQ-2020-03} or particle level~\cite{TOPQ-2017-14}.

\begin{figure}[ht]
\centering
\subfloat[]{\includegraphics[width=0.40\textwidth]{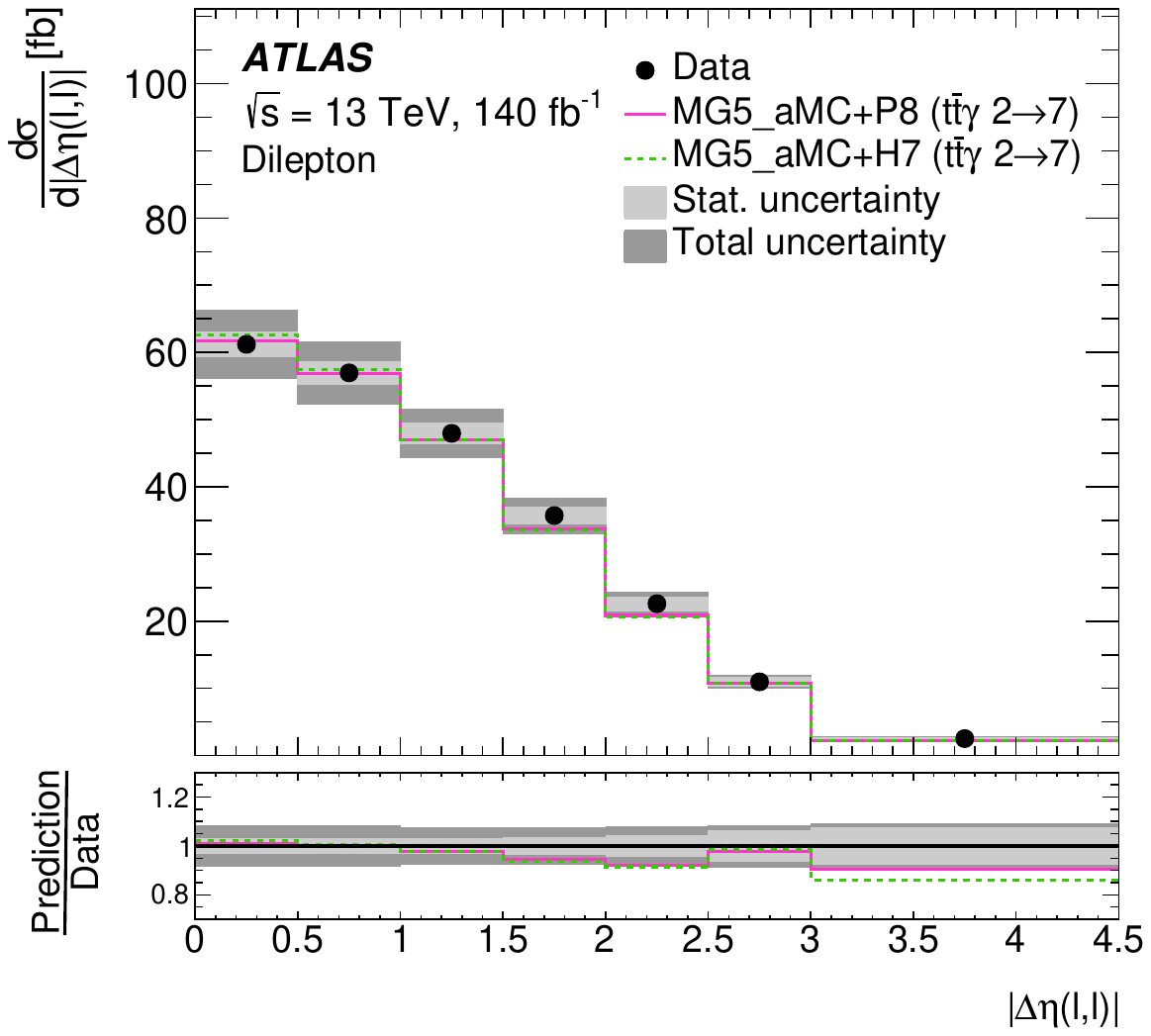}}%
\subfloat[]{\includegraphics[width=0.40\textwidth]{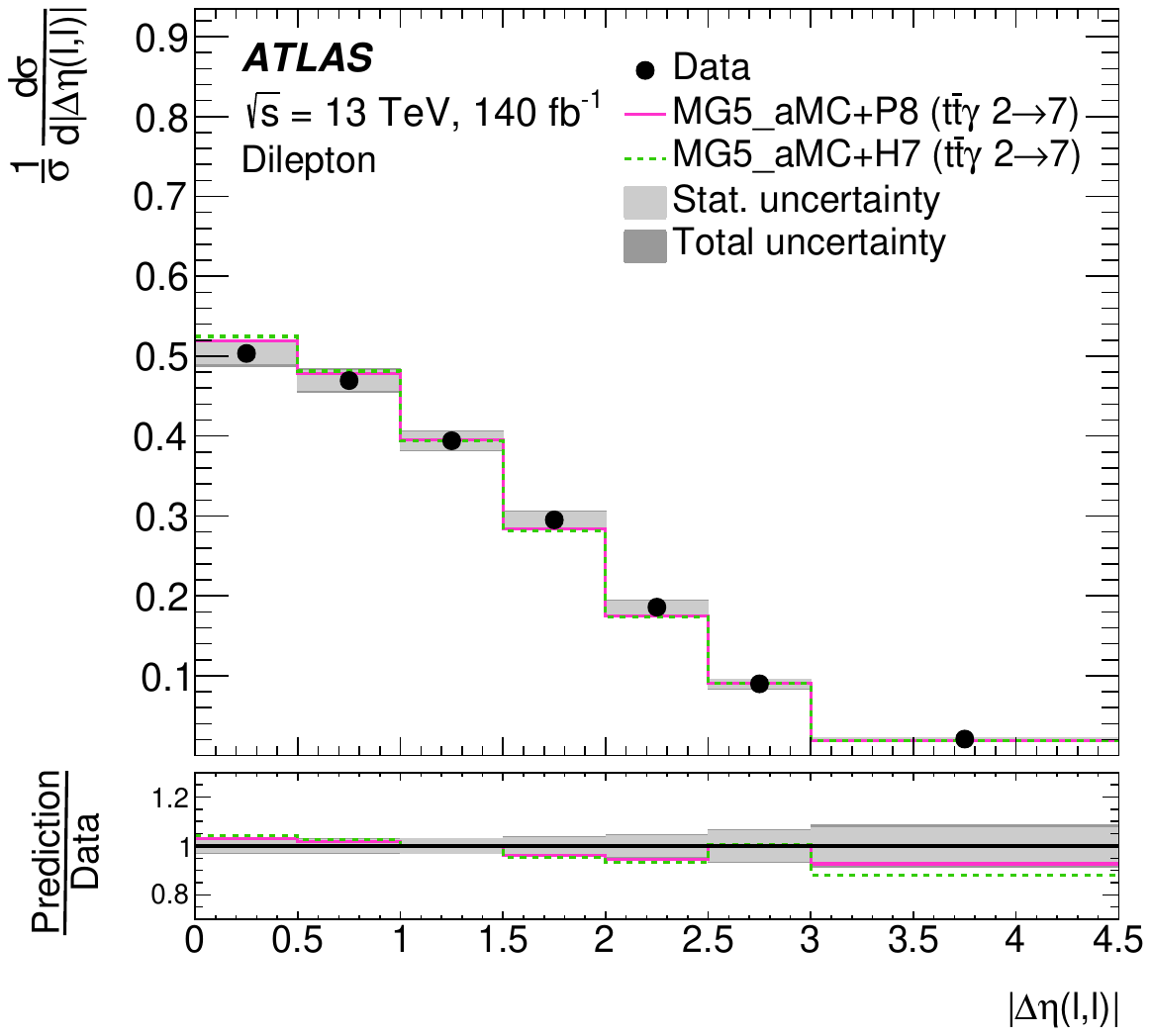}}\\
\subfloat[]{\includegraphics[width=0.40\textwidth]{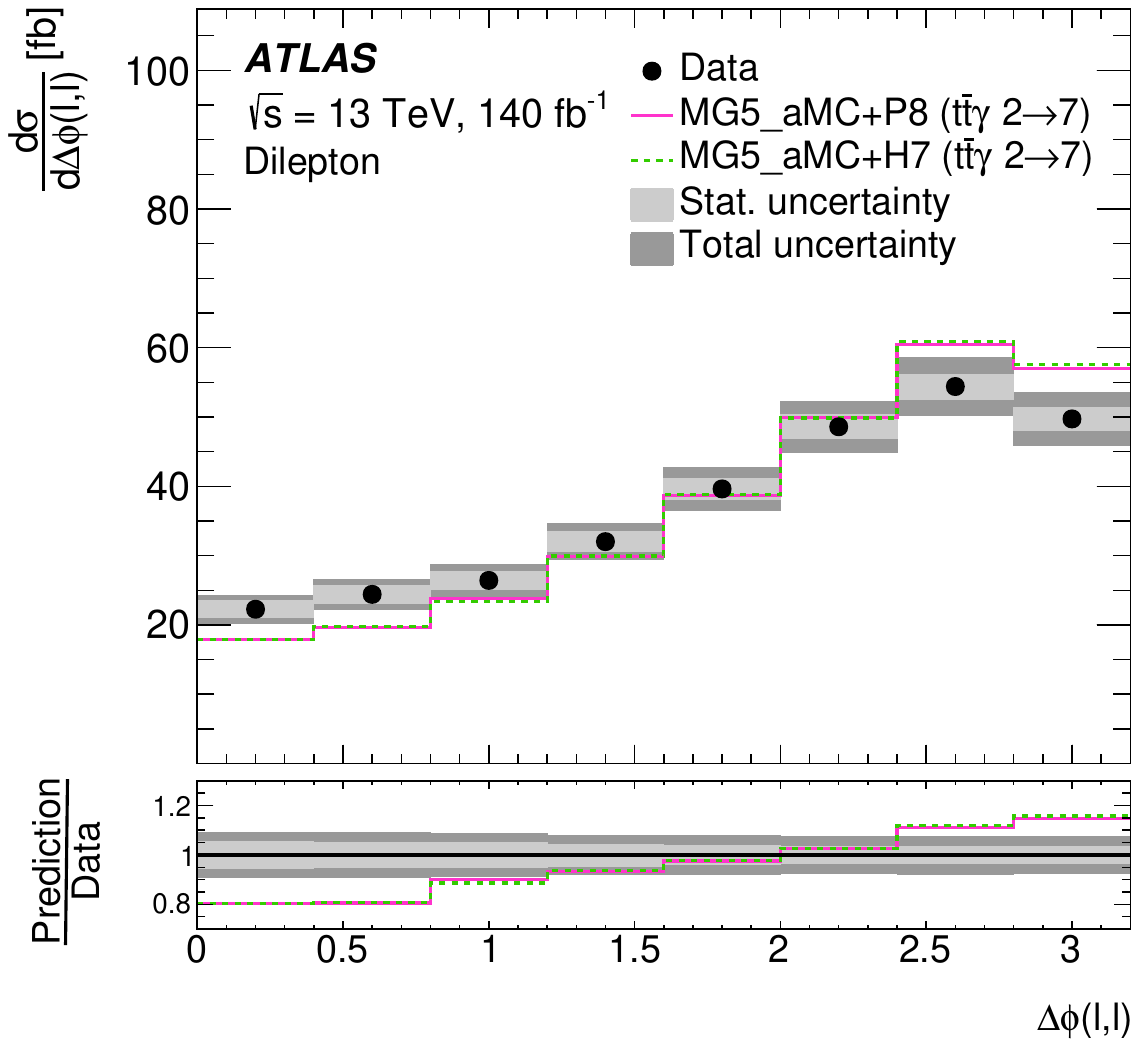}}%
\subfloat[]{\includegraphics[width=0.40\textwidth]{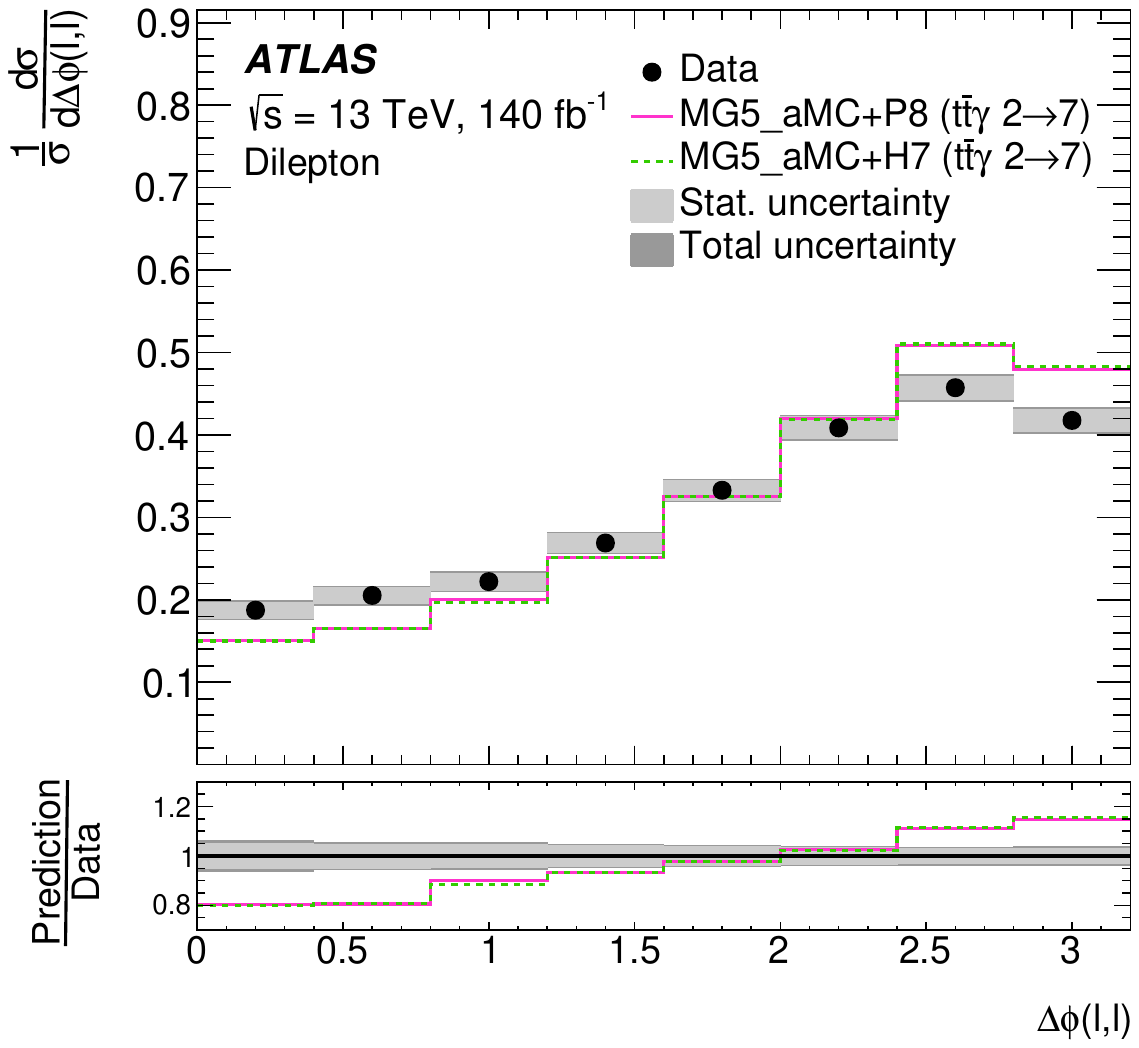}}\\
\subfloat[]{\includegraphics[width=0.40\textwidth]{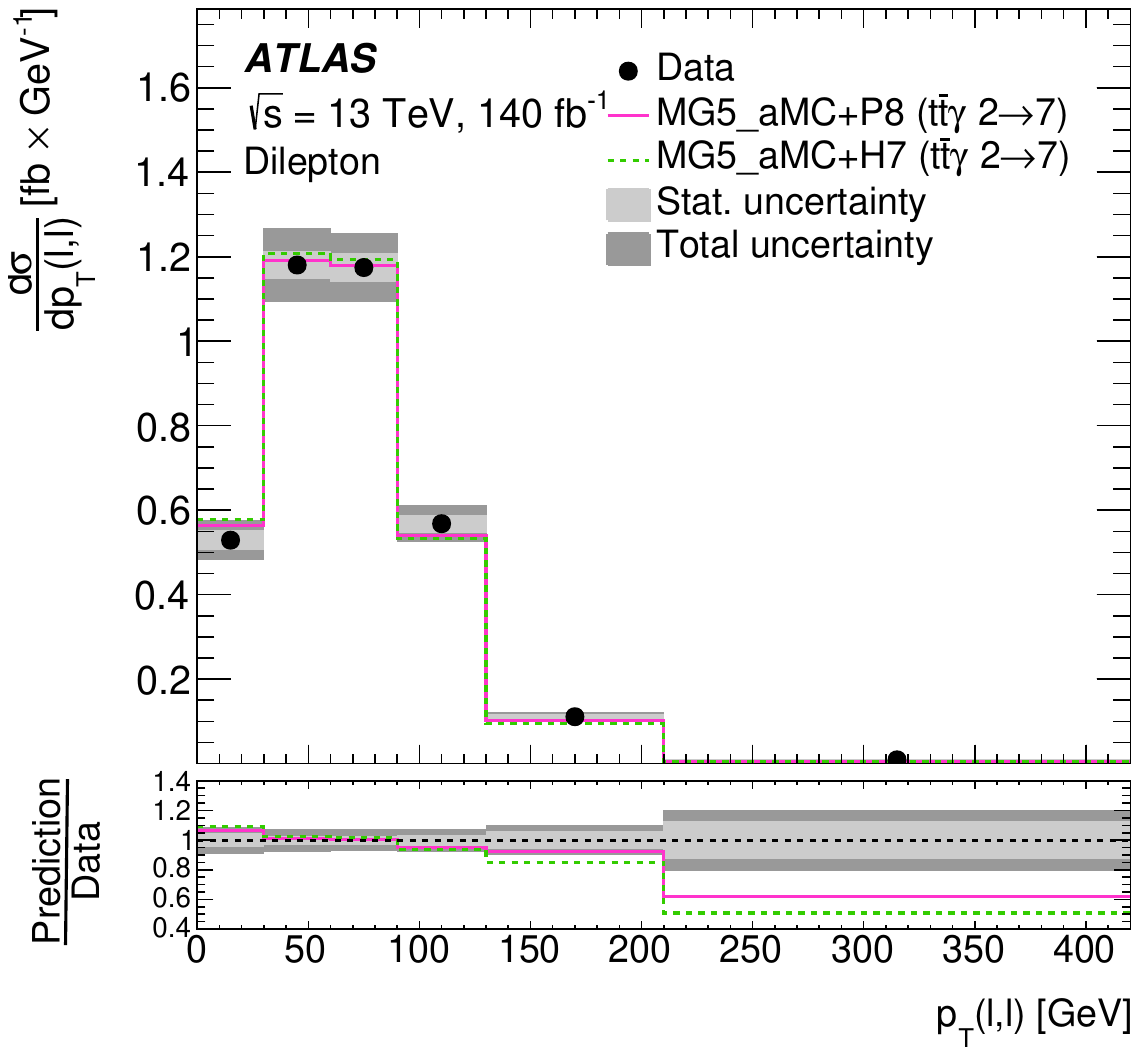}}%
\subfloat[]{\includegraphics[width=0.40\textwidth]{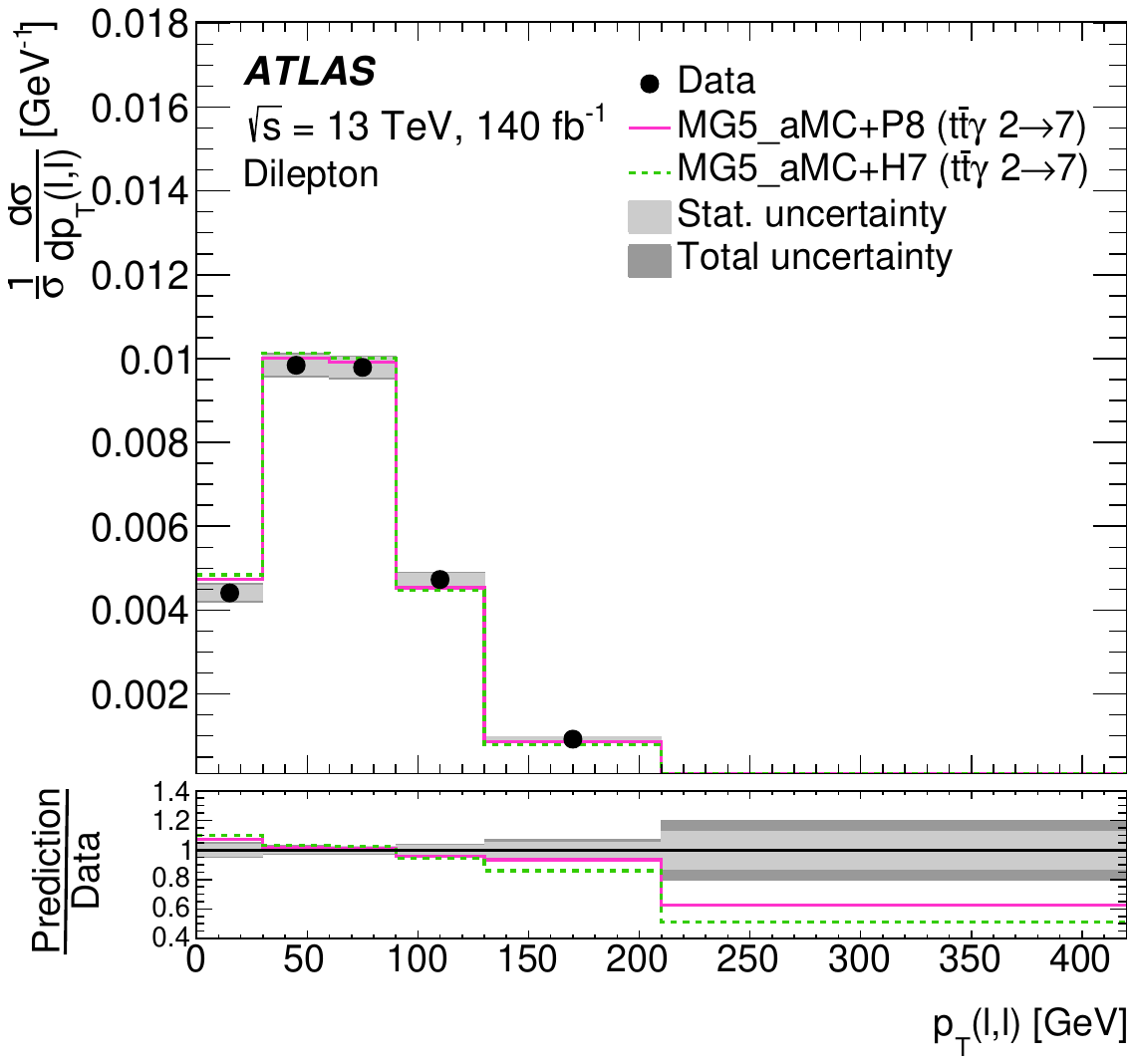}}\\
\caption{The (a, c, e) absolute and (b, d, f) normalised differential cross-sections of the of the total \tty production and decay process measured in the fiducial phase space in the dilepton channel as a function of (a, b) $|\Delta \eta(\ell\ell)|$, (c, d) $\Delta \phi (\ell\ell)$, and the (e, f) \pt of the dilepton system. Data are compared with the \tty \MGNLO simulation at LO interfaced to \PYTHIA[8] and \HERWIG[7]. The lower panels show the ratios of the predictions to the data. The last bin of the distributions includes the overflow events.}
\label{fig:tty_diff_DL_dilepton}
\end{figure}

\begin{figure}[ht]
\centering
\subfloat[]{\includegraphics[width=0.40\textwidth]{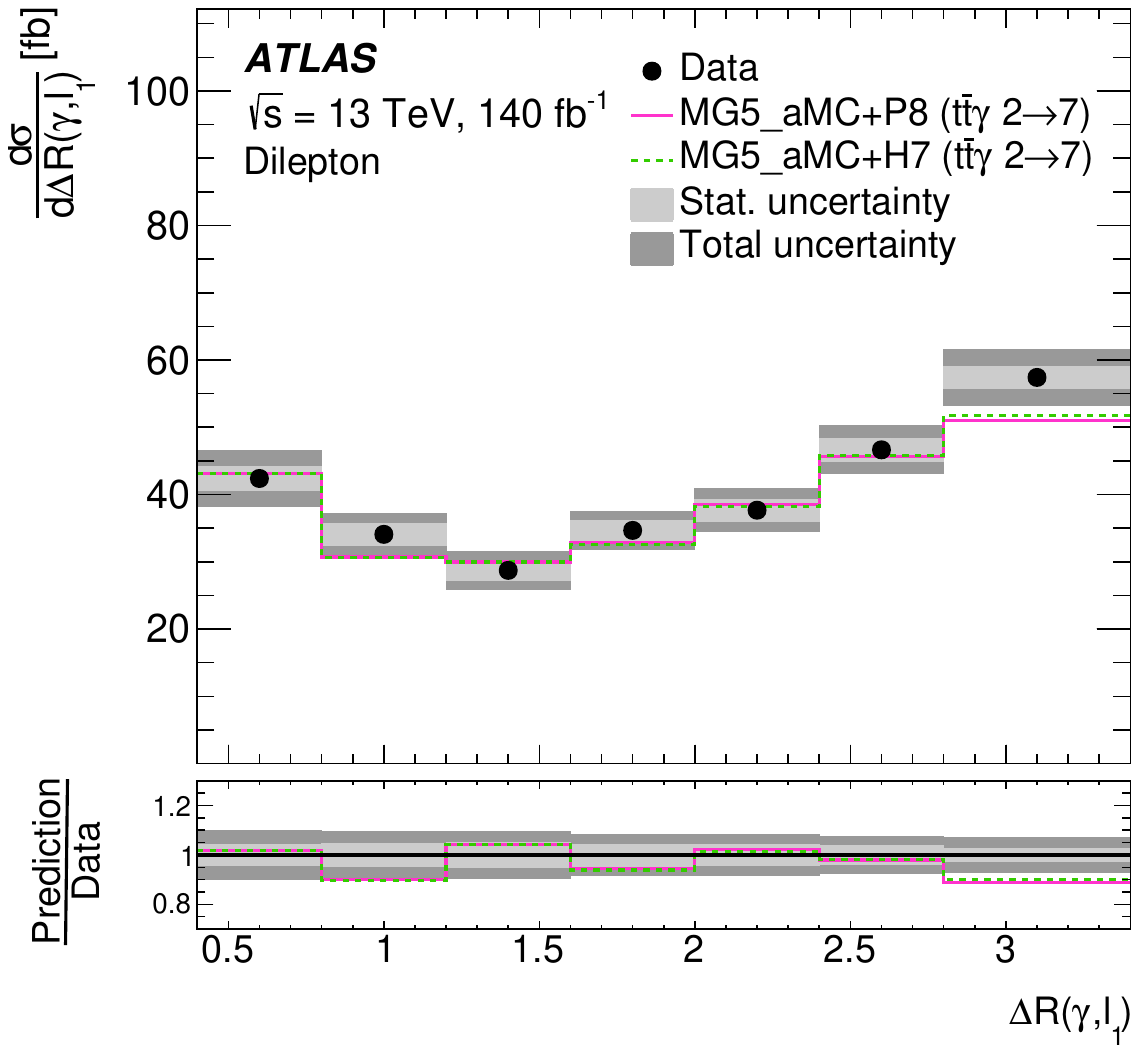}}%
\subfloat[]{\includegraphics[width=0.40\textwidth]{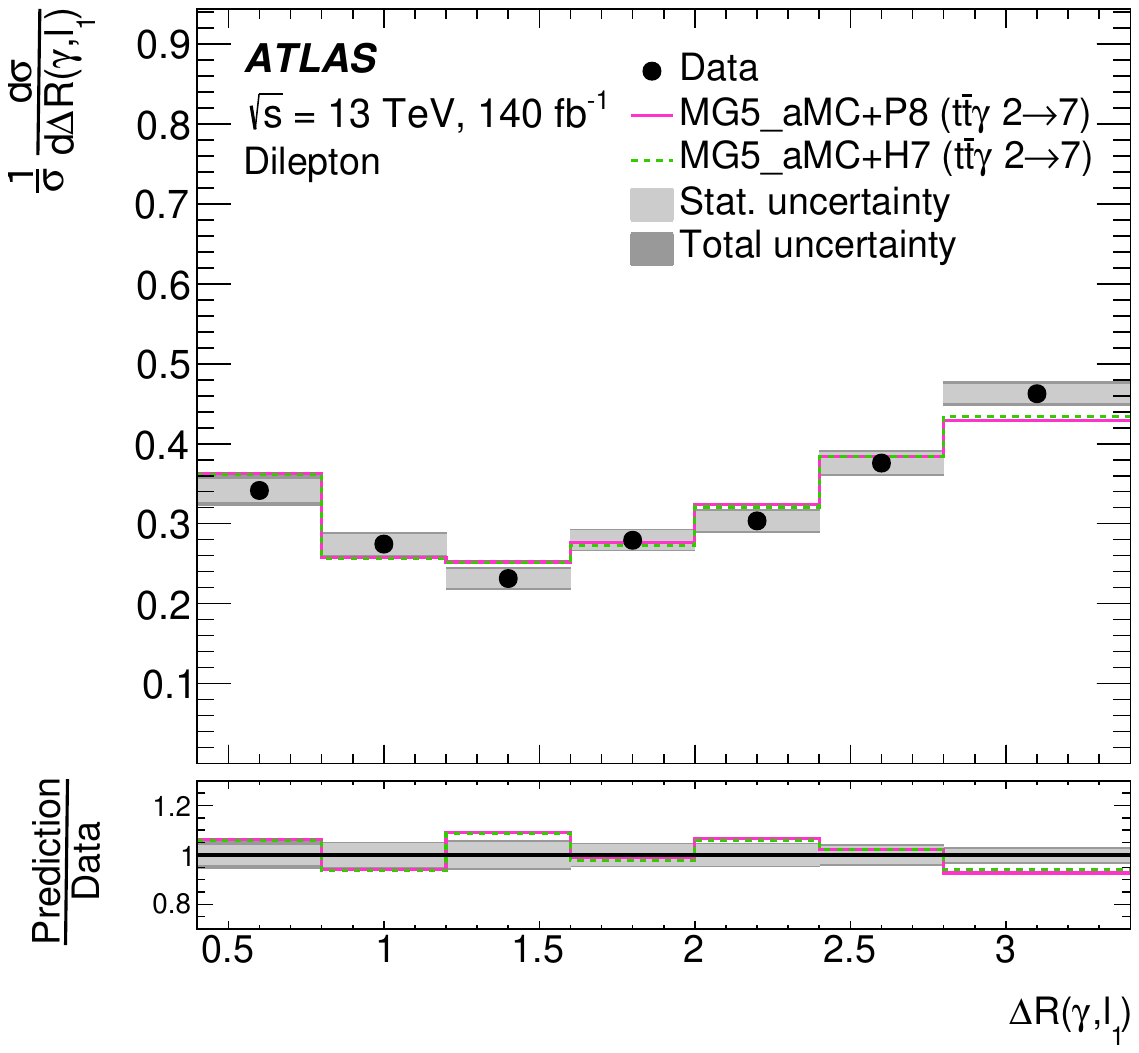}}\\
\subfloat[]{\includegraphics[width=0.40\textwidth]{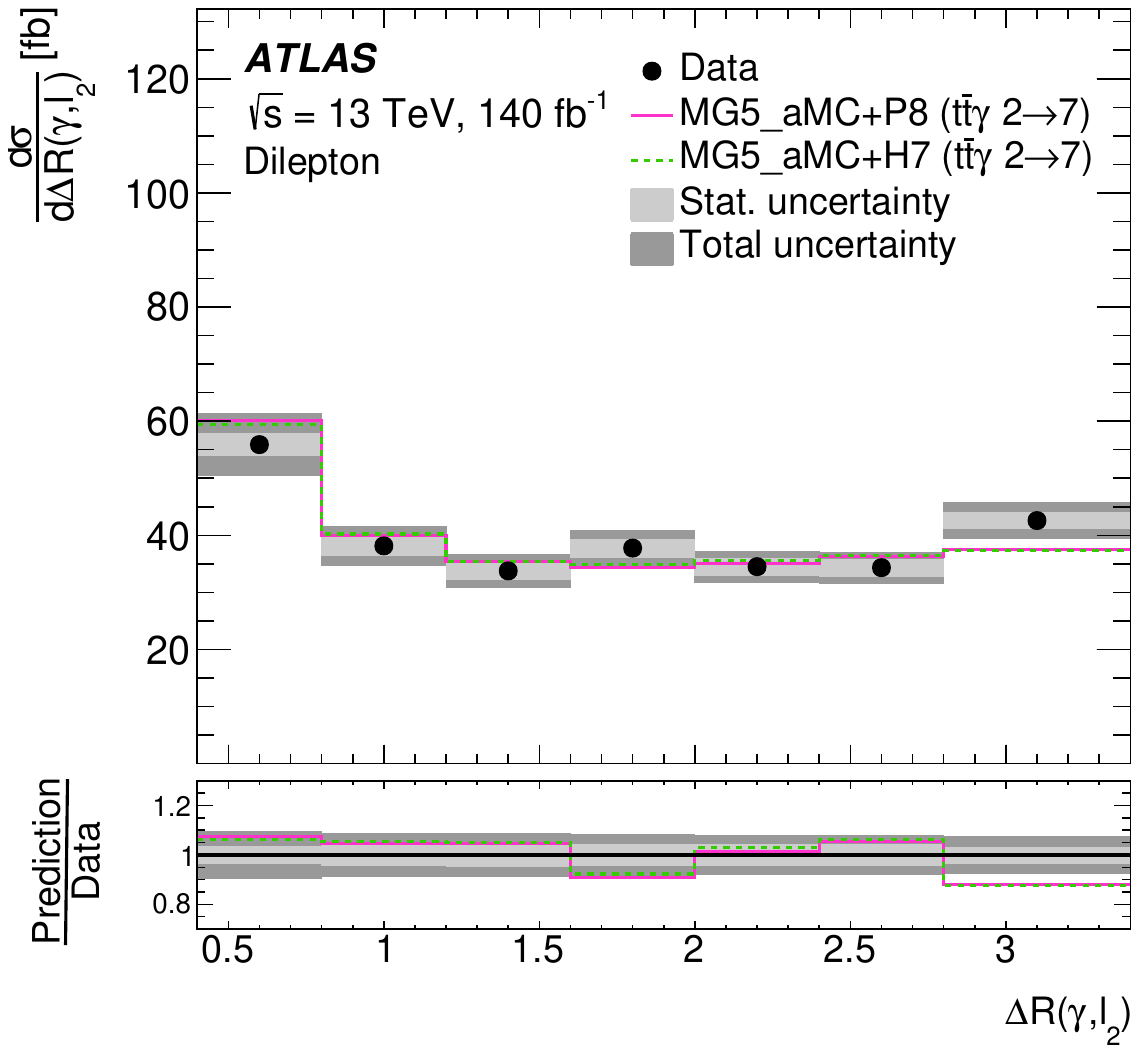}}%
\subfloat[]{\includegraphics[width=0.40\textwidth]{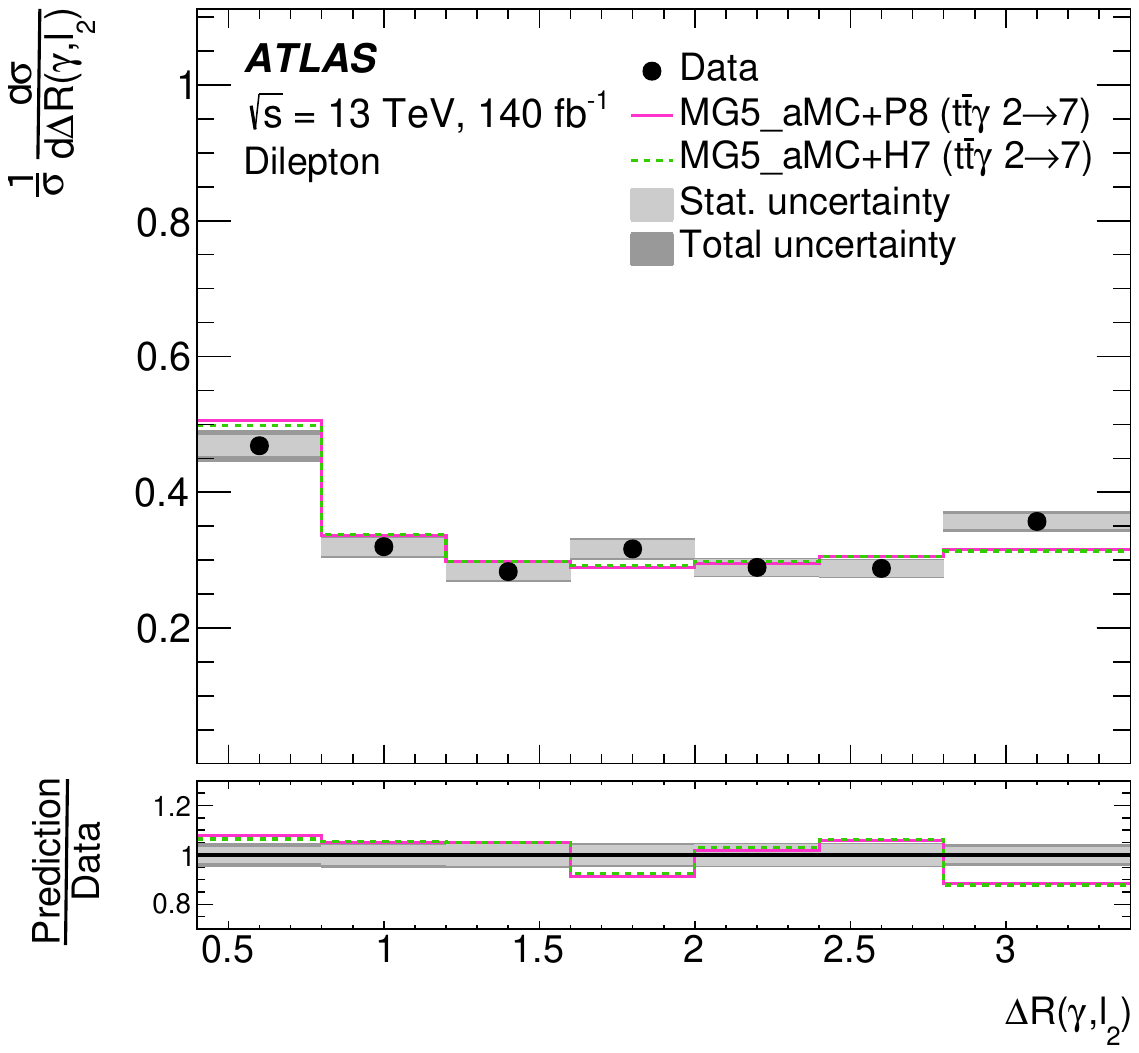}}
\caption{The (a, c) absolute and (b, d) normalised differential cross-sections of the of the total \tty production and decay process measured in the fiducial phase space in the dilepton channel as a function of (a, b) $\Delta R (\gamma, \ell_1)$ and (c, d) $\Delta R (\gamma, \ell_2)$. Data are compared with the \tty \MGNLO simulations at LO interfaced to \PYTHIA[8] and \HERWIG[7]. The lower panels show the ratios of the predictions to the data. The last bin of the distributions includes the overflow events.}
\label{fig:tty_diff_DL_DR}
\end{figure}

The compatibility between the measured differential cross-sections and the predictions is also quantified using $\chi^2$/ndf and $p$-values, which are summarised in Table~\ref{tab:chi2_tty} for all variables considered, including those in Appendix~\ref{sec:appendix_diff}. On average, $p$-values corresponding to the LO $2\to7$ simulation are lower than those for the \MGNLO simulation presented in Table~\ref{tab:chi2_ttyprod} pointing to a somewhat poorer description of the data provided by the LO $2\to7$ simulation.

\begin{table}
\scriptsize
\centering
\caption{The $\chi^2$/ndf and $p$-values between the measured absolute and normalised cross-sections of the total \tty production and decay process and the LO $2\rightarrow 7$ \MGNLO simulation interfaced to \PYTHIA[8] and \HERWIG[7]. They are obtained using the uncertainties of the measured distribution and their correlations.}
\scalebox{0.9}{
\begin{tabular}{l | c c | c c | c c |  c c }
\toprule
& \multicolumn{4}{c}{Absolute cross-sections} & \multicolumn{4}{c}{Normalised cross-sections} \\
&  \multicolumn{2}{c}{MG5\_aMC@NLO+\Pythia[8]} & \multicolumn{2}{c}{MG5\_aMC@NLO+\Herwig[7]} &  \multicolumn{2}{c}{MG5\_aMC@NLO+\Pythia[8]} & \multicolumn{2}{c}{MG5\_aMC@NLO+\Herwig[7]}\\
Variables & $\chi^2$/ndf & $p$-value & $\chi^2$/ndf & $p$-value & $\chi^2$/ndf & $p$-value & $\chi^2$/ndf & $p$-value \\
\midrule
\multicolumn{9}{c}{Single-lepton channel} \\
\midrule
$\pt(\gamma)$ &	 12.6/10&	 0.25&	 \phantom{0}8.5/10&	 0.58&	 21.2/9\phantom{0}&	 $0.01$ &	 12.1/9& 	 0.21 \\

$|\eta(\gamma)|$	& 13.5/8\phantom{0}&	 0.10&	 13.3/8\phantom{0}&	 0.10&	 12.0/7&	0.10&	 12.9/7& 	 0.08\\

$\Delta R(\gamma, \ell)$ &	 15.3/7\phantom{0}&	 0.03&	 14.0/7\phantom{0}&	 0.05&	 13.8/6&	 0.03&	 18.6& 	 $<0.01\phantom{< \;}$\\

$\Delta R(\gamma, b)_{\text{min}}$ &	 8.9/5&	 0.11&	 6.2/5&	 0.29&	 \phantom{0}9.3/4&	 0.05&	 \phantom{0}6.0/4& 	 0.20 \\

$\Delta R(\ell, j)_{\text{min}}$	& 4.9/5&	 0.43&	 3.1/5&	 0.68&	 \phantom{0}0.8/4&	 0.93&	 \phantom{0}0.8/4& 	 0.94 \\

$\pt(j_1)$	& 25.4/5\phantom{0}&	 $<0.01\phantom{< \;}$ &	 43.0/5\phantom{0}&	 $<0.01\phantom{< \;}$ &	 27.2/4&	 $<0.01\phantom{< \;}$ &	 45.0/4& 	 $<0.01\phantom{< \;}$ \\
\midrule
\multicolumn{9}{c}{Dilepton channel} \\
\midrule
\pt($\gamma$) &	 7.6/6&	 0.27&	 4.9/6&	 0.56&	 6.7/5&	 0.24&	 4.7/5& 	 0.45\\

$|\eta(\gamma)|$	& 5.2/8&	 0.73&	 6.0/8&	 0.64&	 5.4/7&	 0.61&	 6.3/7& 	 0.50\\

$\Delta R(\gamma, \ell)_{\text{min}}$ &	 23.6/7\phantom{0}&	 $<0.01\phantom{< \;}$ &	 22.8/7\phantom{0}&	 $<0.01\phantom{< \;}$ &	 20.1/6\phantom{0}&	 $<0.01\phantom{< \;}$ &	 19.6/6\phantom{0}& 	 $<0.01\phantom{< \;}$ \\

$\Delta R(\gamma, \ell_1)$ &      	 10.1/7\phantom{0}&	 0.18&	 8.8/7&	 0.27&	 9.8/6&	 0.13&	 8.5/6& 	 0.21\\

$\Delta R(\gamma, \ell_2)$ &      	 14.8/7\phantom{0}&	 0.04&	 15.1/7\phantom{0}&	 0.03&	 14.3/6\phantom{0}&	 0.03&	 14.7/6\phantom{0}& 	 0.02\\

\Detall &	 3.9/7&	 0.79&	 6.7/7&	 0.46&	 3.1/6&	 0.80&	 5.5/6& 	 0.48\\

\Dphill &	 35.4/8\phantom{0}&	 $<0.01\phantom{< \;}$ &	 37.8/8\phantom{0}&	 $<0.01\phantom{< \;}$ &	 35.3/7\phantom{0}&	 $<0.01\phantom{< \;}$ &	 37.5/7\phantom{0}& 	 $<0.01\phantom{< \;}$ \\

$\pt(\ell,\ell)$	& 6.7/6&	 0.35&	 12.9/6\phantom{0}&	 0.04&	 5.9/5&	 0.32&	 11.5/5\phantom{0}& 	 0.04\\

$\Delta R(\gamma, b)_{\text{min}}$ &	 1.8/5&	 0.87&	 3.7/5&	 0.59&	 1.8/4&	 0.77&	 3.7/4& 	 0.45\\

$\Delta R(\ell, j)_{\text{min}}$	& 6.1/5&	 0.30&	 9.2/5&	 0.10&	 10.0/4\phantom{0}&	 0.04&	 12.8/4\phantom{0}& 	 0.01\\

$\pt(j_1)$	& 10.8/5\phantom{0}&	 0.05&	 19.2/5\phantom{0}&	 $<0.01\phantom{< \;}$ &	 9.8/4 &	 0.04&	 17.6/4\phantom{0}& 	 $<0.01\phantom{< \;}$ \\
\bottomrule
\end{tabular}
\label{tab:chi2_tty}
}
\end{table}

\FloatBarrier
\section{EFT interpretation}
\label{sec:EFT}
Physics phenomena beyond the SM might only manifest themselves at an energy scale $\Lambda$ that is larger than the scales probed at the LHC\@. In that case, the new states are expected to be produced virtually. In the EFT approach, these virtual effects are parameterized by adding higher-dimensional operators to the SM Lagrangian. Assuming lepton number conservation, the lowest-order EFT operators to contribute have dimension six. Since the interaction strength of an operator of dimension $d$ is proportional to $\Lambda^{4-d}$, conventionally chosen to be 1~\TeV, higher-dimensional operators are suppressed. The \tty production process is expected to be sensitive to several EFT operators~\cite{Bylund:2016phk}, but it has the highest sensitivity to the dipole operators \ctB, \ctW, coupling to the weak hypercharge and isospin gauge bosons, respectively. The \ttZ production is also modified by the \ctB and \ctW operators.
The electric and magnetic dipole couplings can be expressed in terms of these dimension-6 EFT operators. In particular, the Lagrangian describing the $\ttbar X$ vertex (with $X=\gamma, Z$) can be written as:
\begin{equation}
\mathcal{L}_{t\bar{t}X}= e\bar{t}\left[
\gamma^\mu \left(C_{1,\text{V}}^X+\gamma_5C_{1,\text{A}}^X\right)
+\dfrac{i\sigma^{\mu\nu}q_\nu}{m_t}\left(C_{2,\text{V}}^X+\gamma_5 C_{2,\text{A}}^X\right)
\right] tX_\mu,
\end{equation}
where $m_\text{t}$ is the top quark mass. The couplings $C_{1,\text{V}}^X$ and $C_{1,\text{A}}^X$ are fixed in the SM by the quantum numbers of the top quark and the $\text{SU}(2)\times \text{U}(1)$ electroweak symmetry. The $C_{2,\text{V}}^X$ and $C_{2,\text{A}}^X$ couplings represent the electric and magnetic dipole moments of the top quark (and their weak equivalents), which are absent from the SM at tree-level and receive very small values from higher-order corrections. These anomalous dipole couplings can be expressed as function of the EFT operators as:

\begin{align}
C_{2,\text{V}}^Z &= \dfrac{v^2m_t}{\sqrt{2}c_ws_wm_Z\Lambda^2}\Re\left[C_{tZ}\right], \,\,\,\, C_{2,\text{A}}^Z = \dfrac{v^2m_t}{\sqrt{2}c_ws_wm_Z\Lambda^2}\Im\left[C_{tZ}\right],\\
C_{2,\text{V}}^\gamma &= \dfrac{\sqrt{2}vm_t}{e\Lambda^2}\Re\left[C_{t\gamma}\right],
\,\,\,\,\,\,\,\,\,\,\,\,\,\,\,\,\,\,\,\,\,\,\, C_{2,\text{A}}^\gamma = \dfrac{\sqrt{2}vm_t}{e\Lambda^2}\Im\left[C_{t\gamma}\right],\\
\end{align}
where $s_w$ ($c_w$) is the sine (cosine) of the Weinberg angle, $m_Z$ is the $Z$ boson mass, $v$ is the vacuum expectation value and \ctY and \ctZ are the EFT operators, which are linear combinations of the dipole operators \ctB, \ctW introduced in Ref.~\cite{Bylund:2016phk,Aguilar-Saavedra:2018ksv}:
\begin{align}
\ctZ &= c_w\cdot\ctW - s_w\cdot\ctB, \\
\ctY &= s_w\cdot\ctW + c_w\cdot\ctB.
\end{align}

Those relationships hold for both the real and imaginary parts of the coefficients.

Since the \ctB, \ctW operators most significantly modify the rate and the shape of the photon \pt distribution in \tty production, this variable is used to set limits on them. This is illustrated in Figure~\ref{fig:photonpt_CtX}, where the absolute cross-section as a function of \pt (see Figure~\ref{fig:tty_prod_SLDL} (a)) is compared with the SM prediction and predictions where one of the \ctB, \ctW EFT operators is different from zero.

\begin{figure}[!htbp]
\centering
\includegraphics[width=0.6\textwidth]{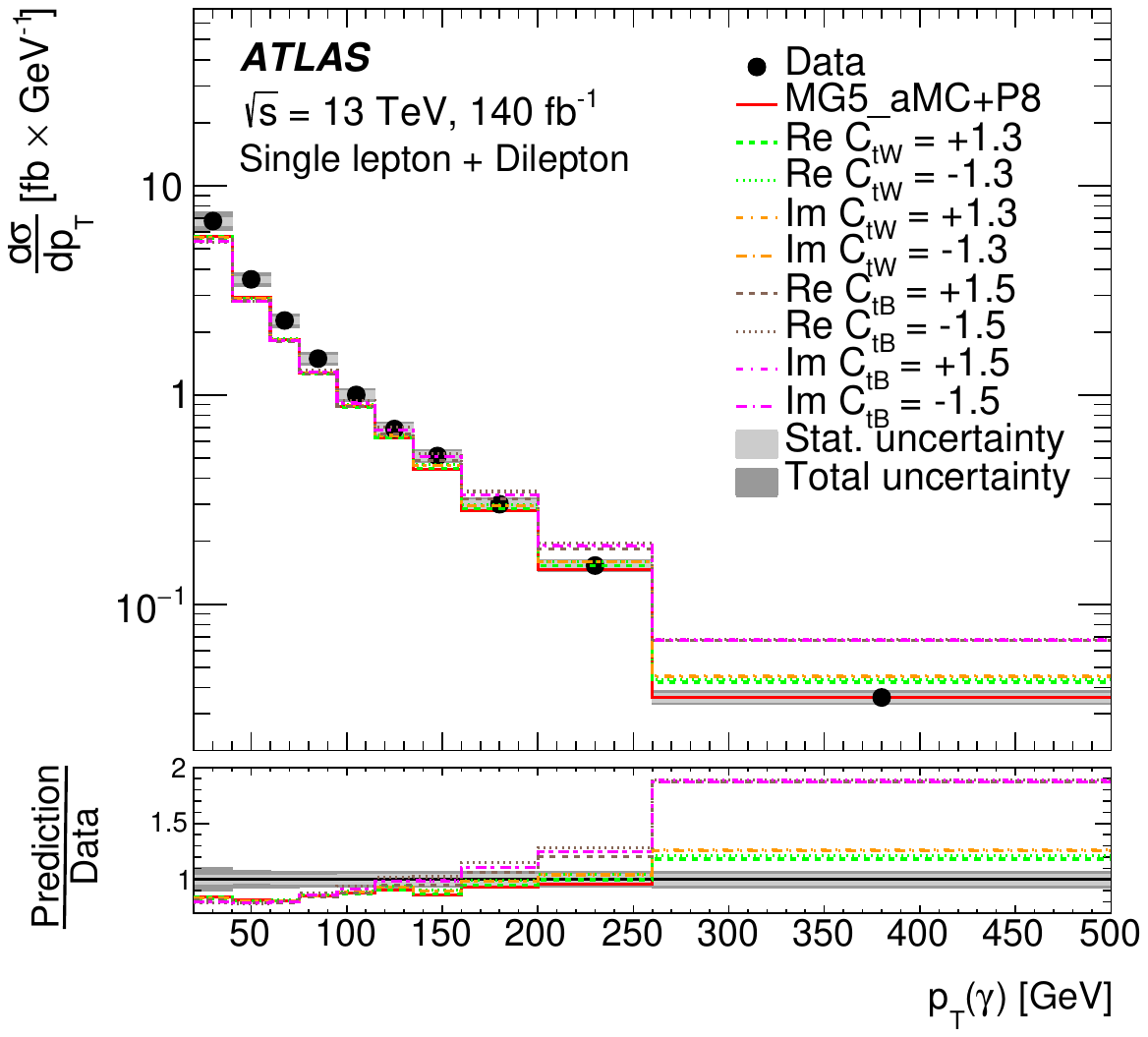}
\caption[]{Comparison of the photon \pt distribution from the combined measurement in the single-lepton and dilepton channels shown in Figure~\ref{fig:tty_prod_SLDL} with the SM prediction and predictions where one EFT parameter is different from zero. The values of the coefficients correspond to the largest values considered in the simulation of the samples. The lower panel shows the ratio of the prediction to the data.}
\label{fig:photonpt_CtX}
\end{figure}

The \tty production cross-section in each bin of the corresponding distributions can be parameterised as
\begin{equation}\nonumber
\sigma = \sigma_\text{SM} + \sum_i C_i \sigma_i + \sum_{i,j} C_i C_j \sigma_{ij}.
\end{equation}
The coupling parameters of the EFT operators considered are denoted by $C_i$, $\sigma_i$ corresponds to the cross-section of the interference of diagrams with one EFT vertex with diagrams from the SM, and the cross-section $\sigma_{ij}$ represents the interference of two diagrams with one EFT vertex each or the squares of the amplitudes with one effective vertex for $i = j$.
In the following, contributions with $i=j$ are referred to as quadratic terms, while $i\neq j$ are referred to as cross-terms.

The EFTfitter tool~\cite{EFTfitter2016} is used to obtain the best-fit values of the EFT parameters in a Bayesian statistical framework. For each bin of the \pt distribution, the SMEFT prediction of the SM term ($\sigma_\text{SM}$), which is simulated at LO in QCD, is scaled to the NLO cross-section of the nominal \tty production sample, thus taking into account both shape and normalisation effects. The real and imaginary parts of the \ctB and \ctW operators are considered separately.

The fit is performed with the full parameterisation of the \tty production cross-section including the linear and quadratic terms and the cross-terms introduced above, for all the operators simultaneously. The analysis shows very little sensitivity when only the linear terms%
, which represent the interference terms, are considered in the fit when compared with the full quadratic fit.
Uncertainties related to the modelling of the EFT samples are not considered in the fit. The results of the global fit (\textit{global mode} in the figures) to the four coefficients are shown in Figure~\ref{fig:results2D} as two-dimensional contours. The measured values are in good agreement with the predictions of the SM\@. As expected, a linear relationship between \ctW and \ctB is observed (left column).

\begin{figure}[!htbp]
\centering
\subfloat[]{\includegraphics[width=0.33\textwidth]{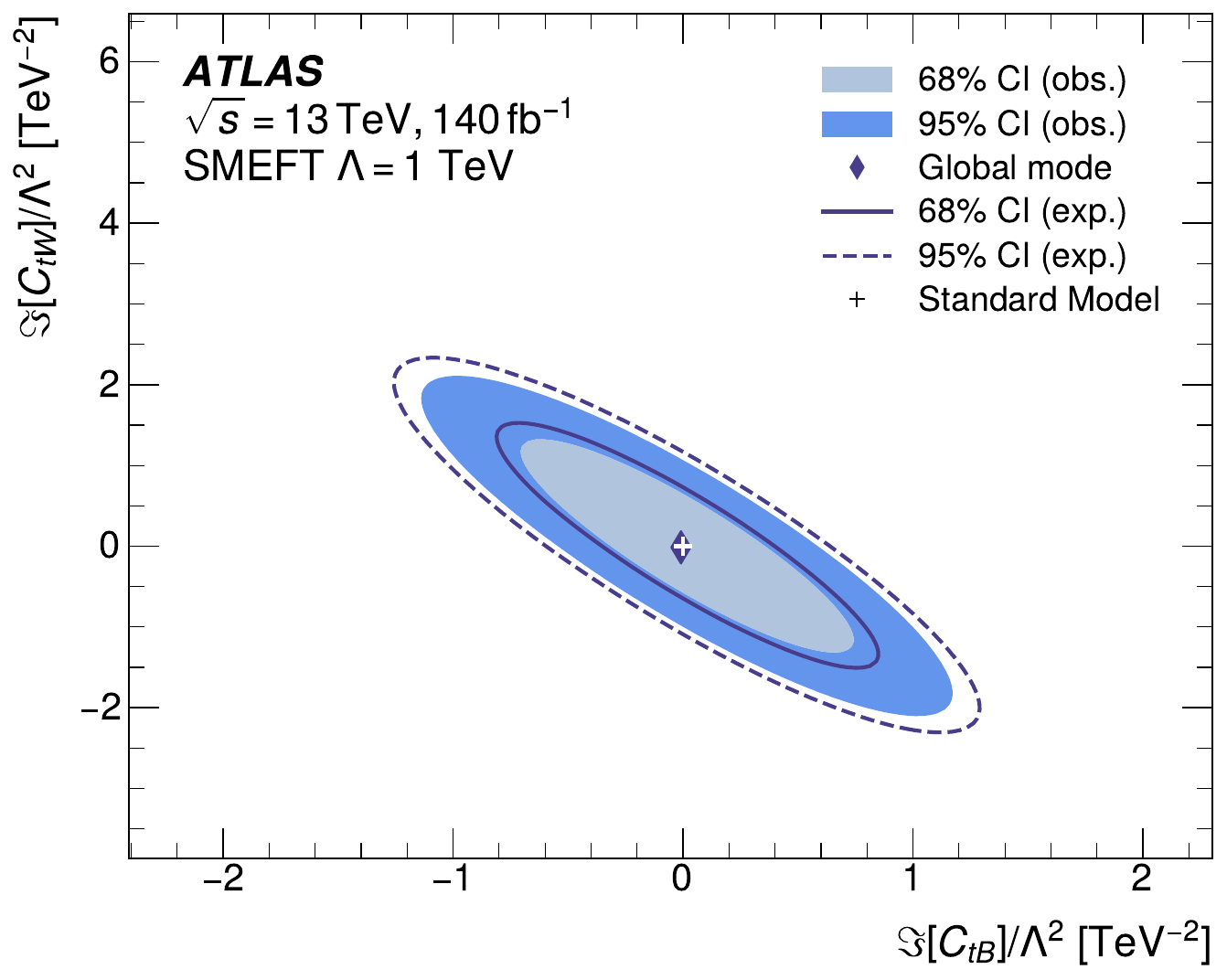}}%
\subfloat[]{\includegraphics[width=0.33\textwidth]{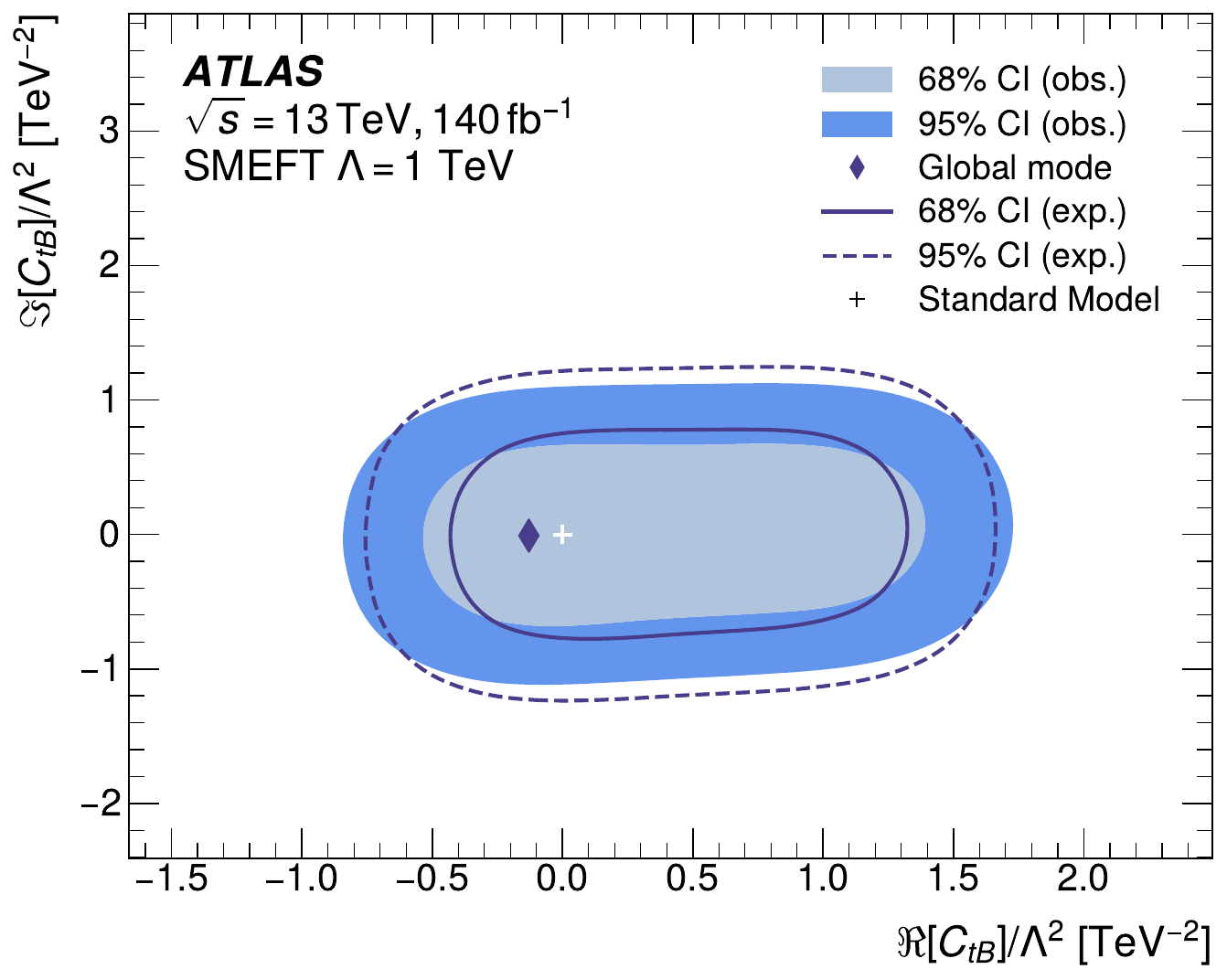}}%
\subfloat[]{\includegraphics[width=0.33\textwidth]{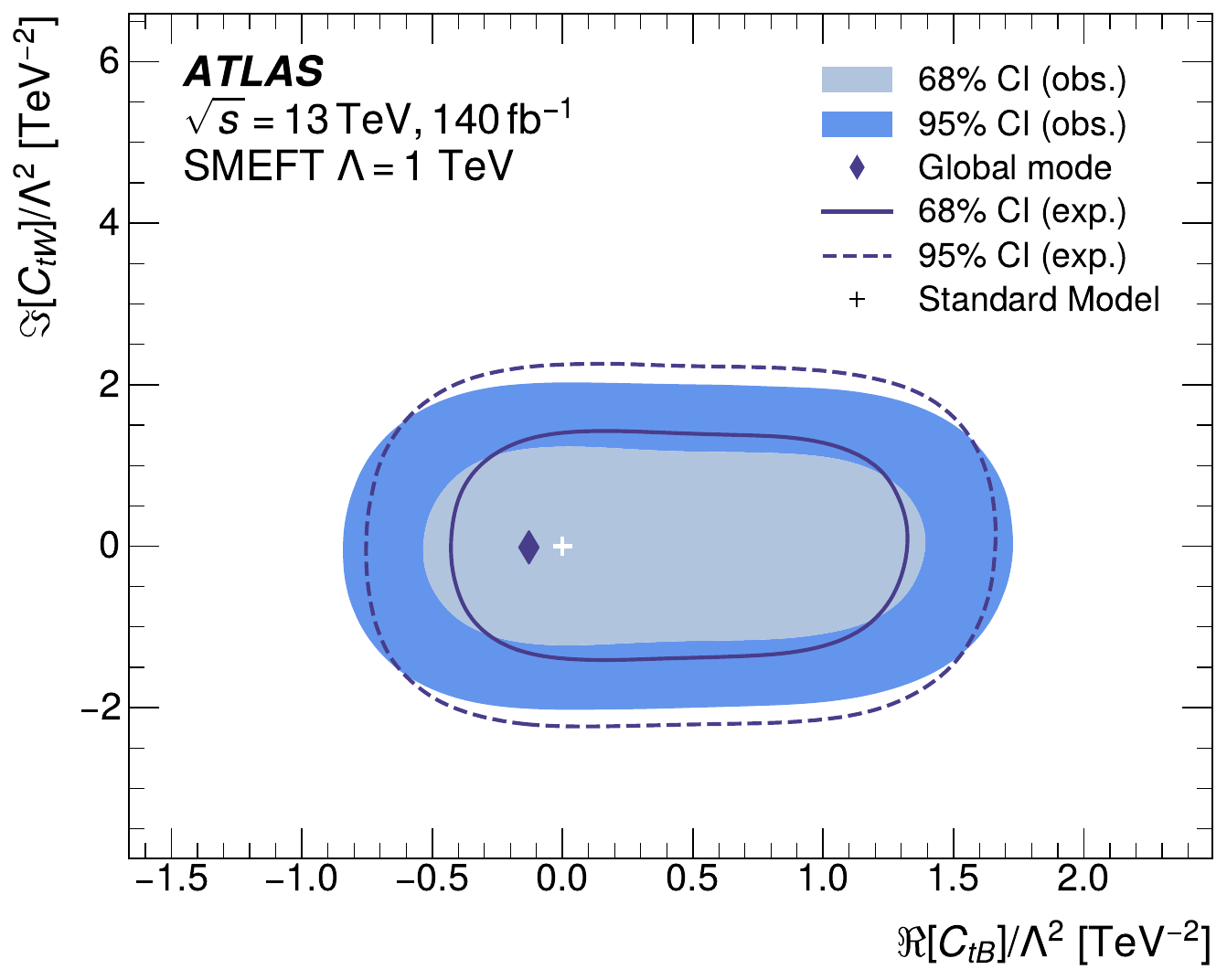}}\\
\subfloat[]{\includegraphics[width=0.33\textwidth]{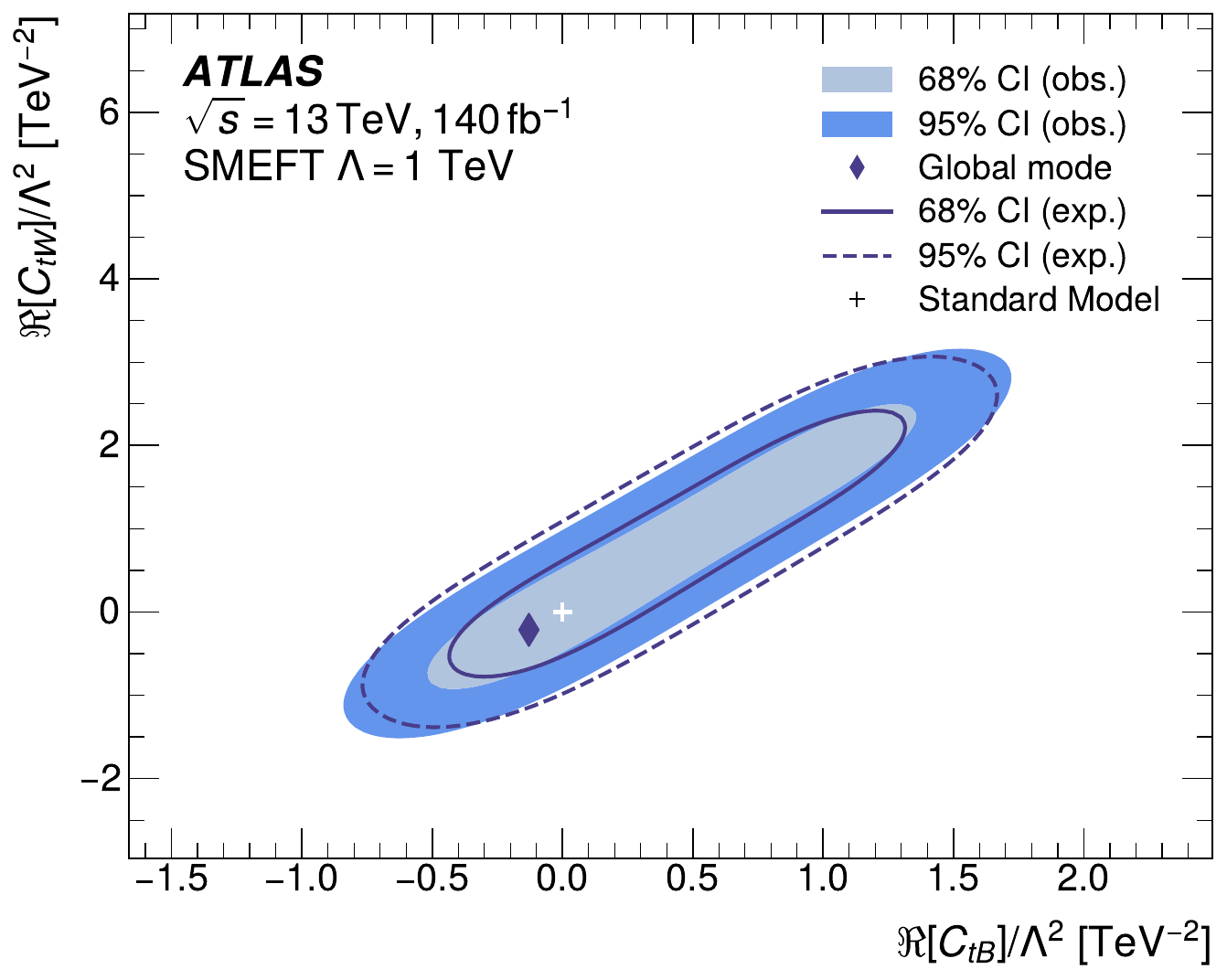}}%
\subfloat[]{\includegraphics[width=0.33\textwidth]{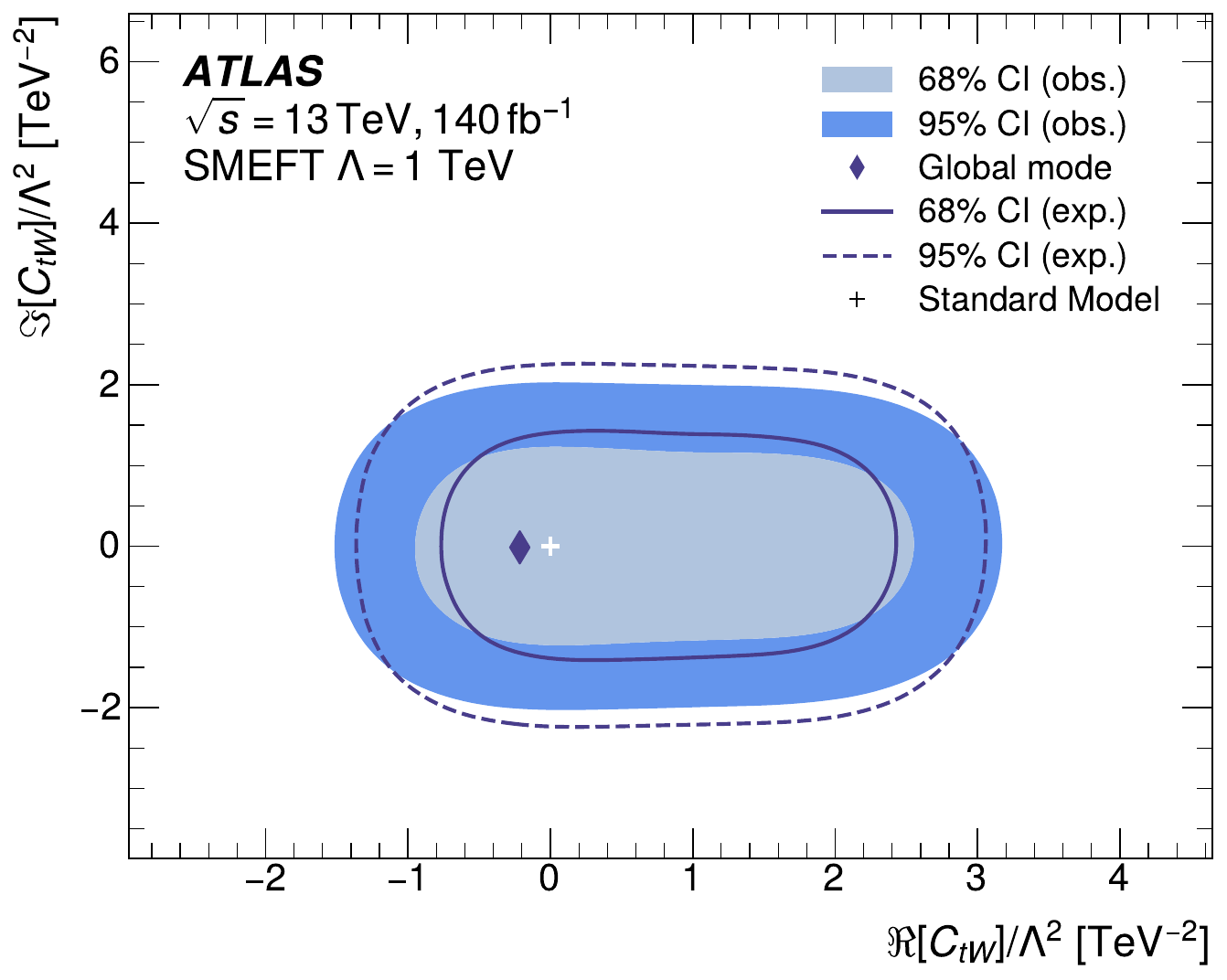}}%
\subfloat[]{\includegraphics[width=0.33\textwidth]{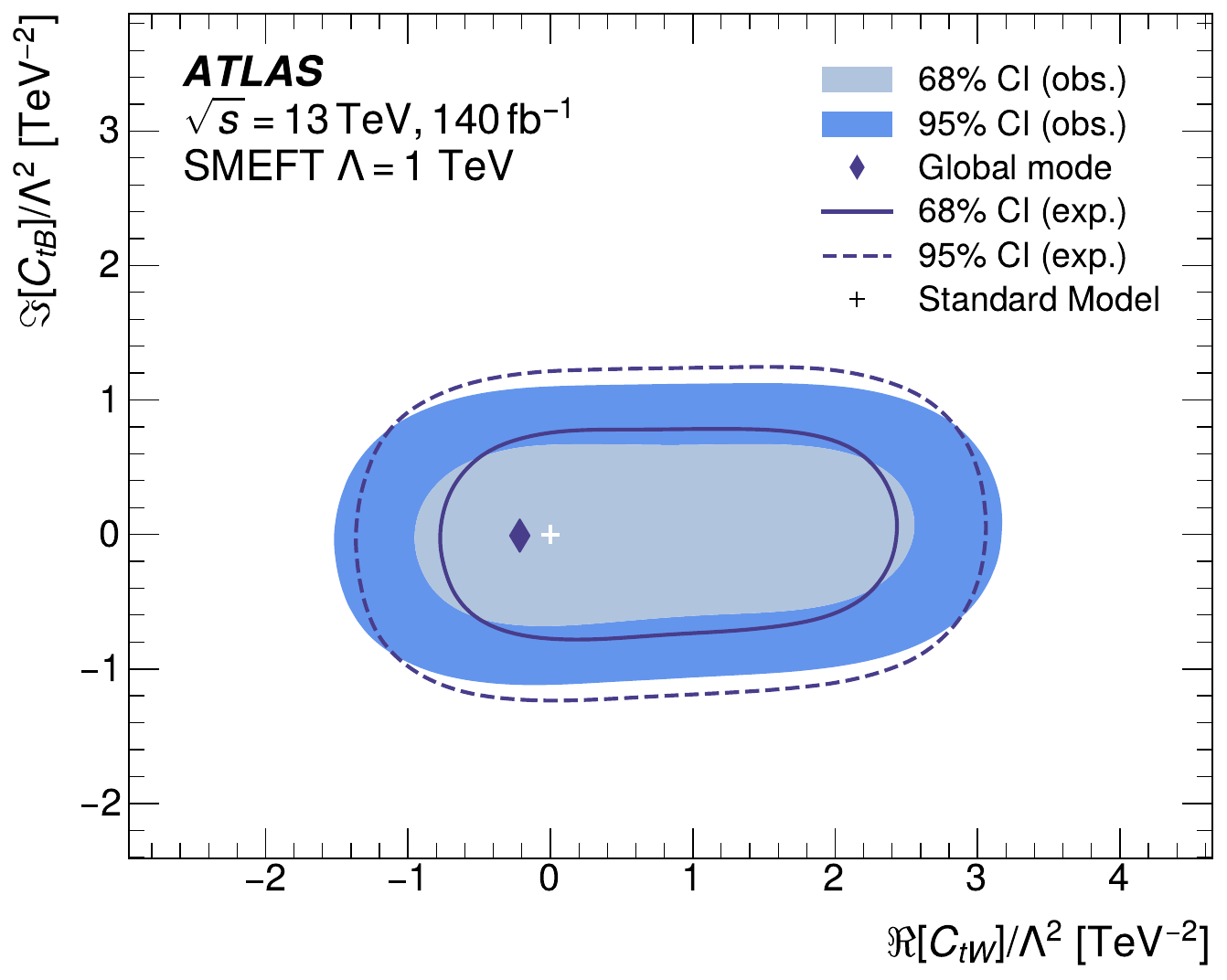}}
\caption[]{Two-dimensional marginalised posteriors for the \ctW and \ctB operators from the quadratic fit, indicating the $68\%$ and $95\%$ credible intervals, extracted using the photon \pt distribution from the combined measurement in the single-lepton and dilepton channels shown in Figure~\ref{fig:tty_prod_SLDL}. The pairs of operators shown are (a) $\Im{\left[C_{tB}\right]}$,$\Im{\left[C_{tW}\right]}$, (b) $\Re{\left[C_{tB}\right]}$,$\Im{\left[C_{tB}\right]}$, (c) $\Re{\left[C_{tB}\right]}$,$\Im{\left[C_{tW}\right]}$, (d) $\Re{\left[C_{tB}\right]}$,$\Re{\left[C_{tW}\right]}$, (e) $\Re{\left[C_{tW}\right]}$,$\Im{\left[C_{tW}\right]}$ and (f) $\Re{\left[C_{tW}\right]}$,$\Im{\left[C_{tB}\right]}$.}
\label{fig:results2D}
\end{figure}

Additionally, the fits where only one EFT parameter is probed at a time, while all others are set to zero, referred to as independent fits, are carried out using the quadratic parameterisations. The observed and expected $68\%$ and $95\%$ credible intervals for the \ctW and \ctB operators from both sets of fits, the marginalised quadratic fits and the independent quadratic fits, are given in Table~\ref{tab:EFTLimits_ttgamma}. The marginalised values are obtained by integrating the posterior probability distribution over the other coefficients. The best-fit values for the different coefficients are in agreement with the predictions of the SM despite a slightly larger measured \tty production cross-section than the NLO calculation.
The constraints on the \ctB coefficient are roughly a factor of two better than those on \ctW.

\begin{table}[!htb]
\footnotesize
\caption{Observed and expected $68\%$ and $95\%$ credible intervals on $C/\Lambda^{2}$~[TeV$^{-2}$] for the \ctW and \ctB operators obtained from the \tty measurement,
comparing the results obtained from the marginalised quadratic fit (`marg.') and the independent quadratic fits (`indep.').
Also shown are the best-fit values (global mode) for each operator.}
\label{tab:EFTLimits_ttgamma}
\def\arraystretch{1.7}
\begin{center}

\begin{tabular}{lllllll}
\toprule
\multicolumn{2}{l}{Wilson coefficient} & $68\%$ CI (exp.) & $95\%$ CI (exp.) & $68\%$ CI (obs.) & $95\%$ CI (obs.) & Best-fit \\
\midrule

\multirow{2}{*}{$\Re{\left[C_{tW}\right]}$}
& $\mathcal{O}(\Lambda^{-4})$ (marg.) & $\left[-0.35,1.9\right]$ & $\left[-1.0,2.7\right]$ & $\left[-0.55,1.9\right]$ & $\left[-1.2,2.8\right]$ & $-0.21$ \\
& $\mathcal{O}(\Lambda^{-4})$ (indep.) & $\left[-0.52,0.60\right]$ & $\left[-0.90,0.98\right]$ & $\left[-0.44,0.46\right]$ & $\left[-0.78,0.84\right]$ & $-0.01$ \\
\addlinespace[0.5em]

\multirow{2}{*}{$\Im{\left[C_{tW}\right]}$}
& $\mathcal{O}(\Lambda^{-4})$ (marg.) & $\left[-1.1,1.1\right]$ & $\left[-1.9,2.0\right]$ & $\left[-0.95,0.95\right]$ & $\left[-1.8,1.8\right]$ & $-0.01$ \\
& $\mathcal{O}(\Lambda^{-4})$ (indep.) & $\left[-0.58,0.56\right]$ & $\left[-0.96,0.94\right]$ & $\left[-0.48,0.44\right]$ & $\left[-0.84,0.80\right]$ & $-0.01$ \\
\addlinespace[0.5em]

\multirow{2}{*}{$\Re{\left[C_{tB}\right]}$}
& $\mathcal{O}(\Lambda^{-4})$ (marg.) & $\left[-0.20,1.0\right]$ & $\left[-0.58,1.5\right]$ & $\left[-0.30,1.0\right]$ & $\left[-0.66,1.5\right]$ & $-0.13$ \\
& $\mathcal{O}(\Lambda^{-4})$ (indep.) & $\left[-0.31,0.31\right]$ & $\left[-0.51,0.52\right]$ & $\left[-0.26,0.23\right]$ & $\left[-0.44,0.44\right]$ & $-0.03$ \\
\addlinespace[0.5em]

\multirow{2}{*}{$\Im{\left[C_{tB}\right]}$}
& $\mathcal{O}(\Lambda^{-4})$ (marg.) & $\left[-0.58,0.62\right]$ & $\left[-1.1,1.1\right]$ & $\left[-0.50,0.54\right]$ & $\left[-0.96,0.98\right]$ & $-0.01$ \\
& $\mathcal{O}(\Lambda^{-4})$ (indep.) & $\left[-0.32,0.31\right]$ & $\left[-0.53,0.51\right]$ & $\left[-0.26,0.24\right]$ & $\left[-0.45,0.44\right]$ & $-0.01$ \\
\addlinespace[0.5em]

\bottomrule
\end{tabular}

\end{center}
\end{table}

The \ttZ production is also modified by the \ctB and \ctW operators as mentioned above, providing complementary constraining power. Thus, the limits on EFT parameters are also set using a simultaneous measurement of the photon \pt and the $Z$ boson \pt using the input data and simulations from the \ttZ production measurement of Ref.~\cite{ATLAS:2023eld}. The $Z$ boson \pt distribution is measured in \ttZ events with two, three and four leptons in the final state and is corrected to particle level in a fiducial phase space. The event selection and reconstruction is designed to avoid any statistical overlap with the \tty measurement to ease the combination. Both measurements consider the same sources of the systematic uncertainties for all the common objects. Thus,
in the simultaneous profile-likelihood fit to the photon and $Z$ boson distributions, all experimental uncertainties are treated as correlated between the measurements. The modelling uncertainties of the signal and background processes are uncorrelated owing to the different MC simulations used in both measurements.

After the fit, there are up to 10\% correlations between the unfolded bins of the \ttZ and \tty distributions, arising from the correlation of the experimental uncertainties. The measured cross-sections and post-fit values of the systematic uncertainties are not significantly more constrained compared to the results of the individual \tty and \ttZ measurements (see Ref.~\cite{ATLAS:2023eld}).

The observed and expected $68\%$ and $95\%$ credible intervals for the \ctW and \ctB operators obtained from the combined measurement are summarised in Table~\ref{tab:eft-limits-nominal}. The quadratic limits on the real and imaginary parts of the coefficients are similar, with \ctB being slightly more constrained than \ctW. The combination with \ttZ slightly improves the limits, in particular the credible intervals for \ctW obtained with the independent quadratic fits are reduced by up to 20\%. For better visualisation, the results of the marginalised quadratic fits obtained using the photon \pt in \tty and in combination with \ttZ are shown in Figure~\ref{fig:eft-limits-nominal}.

\begin{table}[!htb]
\footnotesize
\caption{Observed and expected $68\%$ and $95\%$ credible intervals  on $C/\Lambda^{2}$~[TeV$^{-2}$] for the \ctW and \ctB operators obtained from a combined \ttZ and \tty measurements comparing the results obtained from the marginalised global quadratic fit (`marg.') and the independent quadratic fits (`indep.'). Also shown are the best-fit values (global mode) for each operator.}
\label{tab:eft-limits-nominal}
\def\arraystretch{1.7}
\begin{center}
\begin{tabular}{lllllll}
\toprule
\multicolumn{2}{l}{Wilson coefficient} & $68\%$ CI (exp.) & $95\%$ CI (exp.) & $68\%$ CI (obs.) & $95\%$ CI (obs.) & Best-fit \\
\midrule

\multirow{2}{*}{$\Re{\left[C_{tW}\right]}$} & $\mathcal{O}(\Lambda^{-4})$ (marg.) & $\left[-0.50,1.2\right]$ & $\left[-1.0,2.2\right]$ & $\left[-0.35,1.8\right]$ & $\left[-1.1,2.5\right]$ & $\hphantom{-}1.61$ \\
& $\mathcal{O}(\Lambda^{-4})$ (indep.) & $\left[-0.44,0.54\right]$ & $\left[-0.78,0.86\right]$ & $\left[-0.28,0.32\right]$ & $\left[-0.54,0.60\right]$ & $\hphantom{-}0.02$ \\
\addlinespace[0.5em]

\multirow{2}{*}{$\Im{\left[C_{tW}\right]}$} & $\mathcal{O}(\Lambda^{-4})$ (marg.) & $\left[-1.1,0.45\right]$ & $\left[-1.7,1.2\right]$ & $\left[-1.3,0.15\right]$ & $\left[-1.8,1.1\right]$ & $-0.89$ \\
& $\mathcal{O}(\Lambda^{-4})$ (indep.) & $\left[-0.48,0.50\right]$ & $\left[-0.82,0.82\right]$ & $\left[-0.32,0.30\right]$ & $\left[-0.58,0.58\right]$ & $-0.01$ \\
\addlinespace[0.5em]

\multirow{2}{*}{$\Re{\left[C_{tB}\right]}$} & $\mathcal{O}(\Lambda^{-4})$ (marg.) & $\left[-0.28,0.70\right]$ & $\left[-0.60,1.2\right]$ & $\left[-0.34,1.2\right]$ & $\left[-0.68,1.4\right]$ & $\hphantom{-}0.94$ \\
& $\mathcal{O}(\Lambda^{-4})$ (indep.) & $\left[-0.28,0.32\right]$ & $\left[-0.49,0.52\right]$ & $\left[-0.24,0.19\right]$ & $\left[-0.41,0.39\right]$ & $-0.03$ \\
\addlinespace[0.5em]

\multirow{2}{*}{$\Im{\left[C_{tB}\right]}$} & $\mathcal{O}(\Lambda^{-4})$ (marg.) & $\left[-0.34,0.62\right]$ & $\left[-0.78,1.0\right]$ & $\left[-0.18,0.72\right]$ & $\left[-0.64,1.0\right]$ & $\hphantom{-}0.46$ \\
& $\mathcal{O}(\Lambda^{-4})$ (indep.) & $\left[-0.31,0.30\right]$ & $\left[-0.51,0.51\right]$ & $\left[-0.23,0.21\right]$ & $\left[-0.40,0.40\right]$ & $-0.01$ \\
\addlinespace[0.5em]

\bottomrule
\end{tabular}

\end{center}
\end{table}

\begin{figure}[!htb]
\centering
\includegraphics[width=0.5\textwidth]{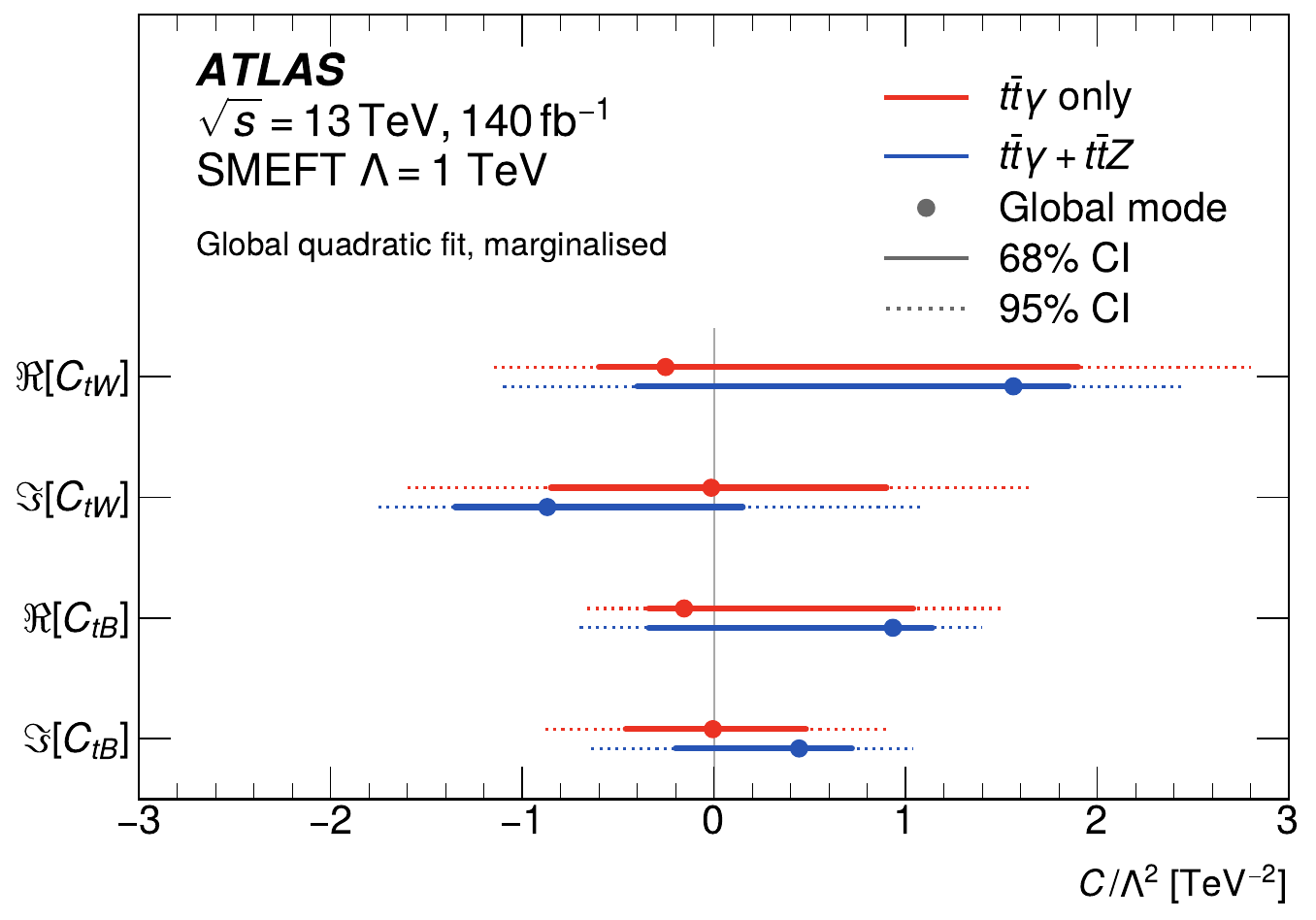}
\caption{Comparison of the $68\%$ and $95\%$ credible intervals obtained in the marginalised quadratic fits using the photon \pt in \tty production and using the simultaneous measurement of the \pt of the $Z$ boson and the \pt of the photon. Also shown are the best-fit values (global mode) for each operator. }
\label{fig:eft-limits-nominal}
\end{figure}

To gauge the importance of each individual measurement in the combination, and to better visualise the interplay between the different directions of EFT sensitivity, the quadratic fits are repeated in terms of the operators \ctZ and \ctY. The \ctZ coefficient is probed by the \ttZ measurement, while \tty is sensitive to \ctY. The comparison of the individual \ttZ and \tty exclusion contours and the combination are shown in Figure~\ref{fig:eft-2d-posteriors-rotated-quadratic-comparison} as a function of \ctZ and \ctW. The shape of the \ttZ exclusion contours is driven by the fact that the \ttZ measurement is mostly sensitive to $\ctZRe$. The imaginary parts of the operators, $\ctZIm$ and $\Im{\left[\ctY\right]}$, do not interfere with the SM (CP-odd dipole terms) and, therefore, only the quadratic terms with positive contributions affect the differential cross-section. The \ttZ measurement is not able to resolve the \ctZ and \ctY, a degenerate structure is observed in e.g., the marginalised posterior of \ctY (Figure~\ref{fig:eft-2d-posteriors-rotated-quadratic-comparison} (a)). It is resolved by the addition of the \tty measurement, as expected.
Both \ttZ and \tty lead to the same contour shape in the plane of \ctZRe versus \ctZIm (Figure~\ref{fig:eft-2d-posteriors-rotated-quadratic-comparison} (f)), while their combination exhibits a different structure, which highlights the importance of combining measurements with complementary directions of EFT sensitivity.

\begin{figure}[!htb]
\begin{center}
\subfloat[]{\includegraphics[width=0.32\textwidth]{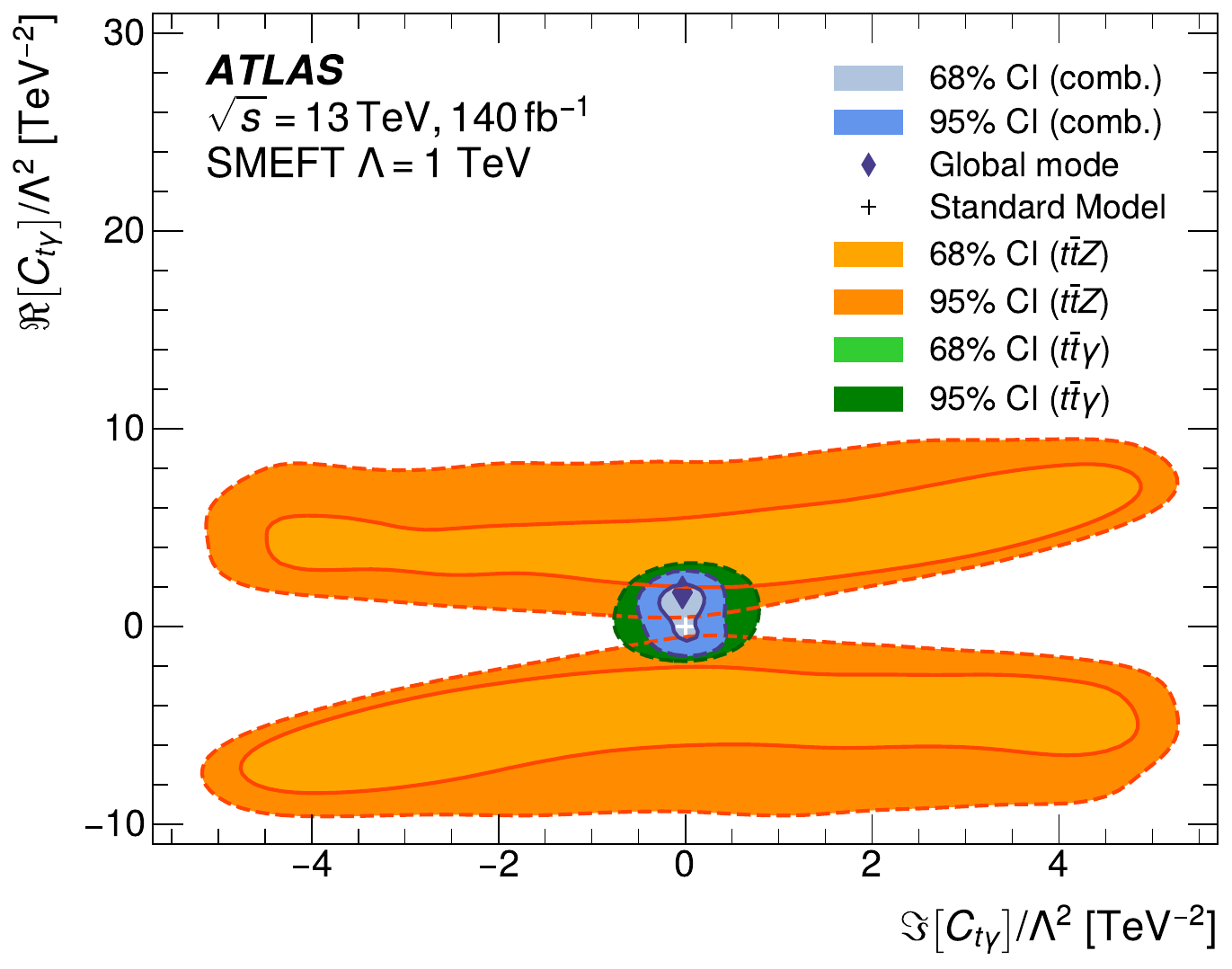}}
\subfloat[]{\includegraphics[width=0.32\textwidth]{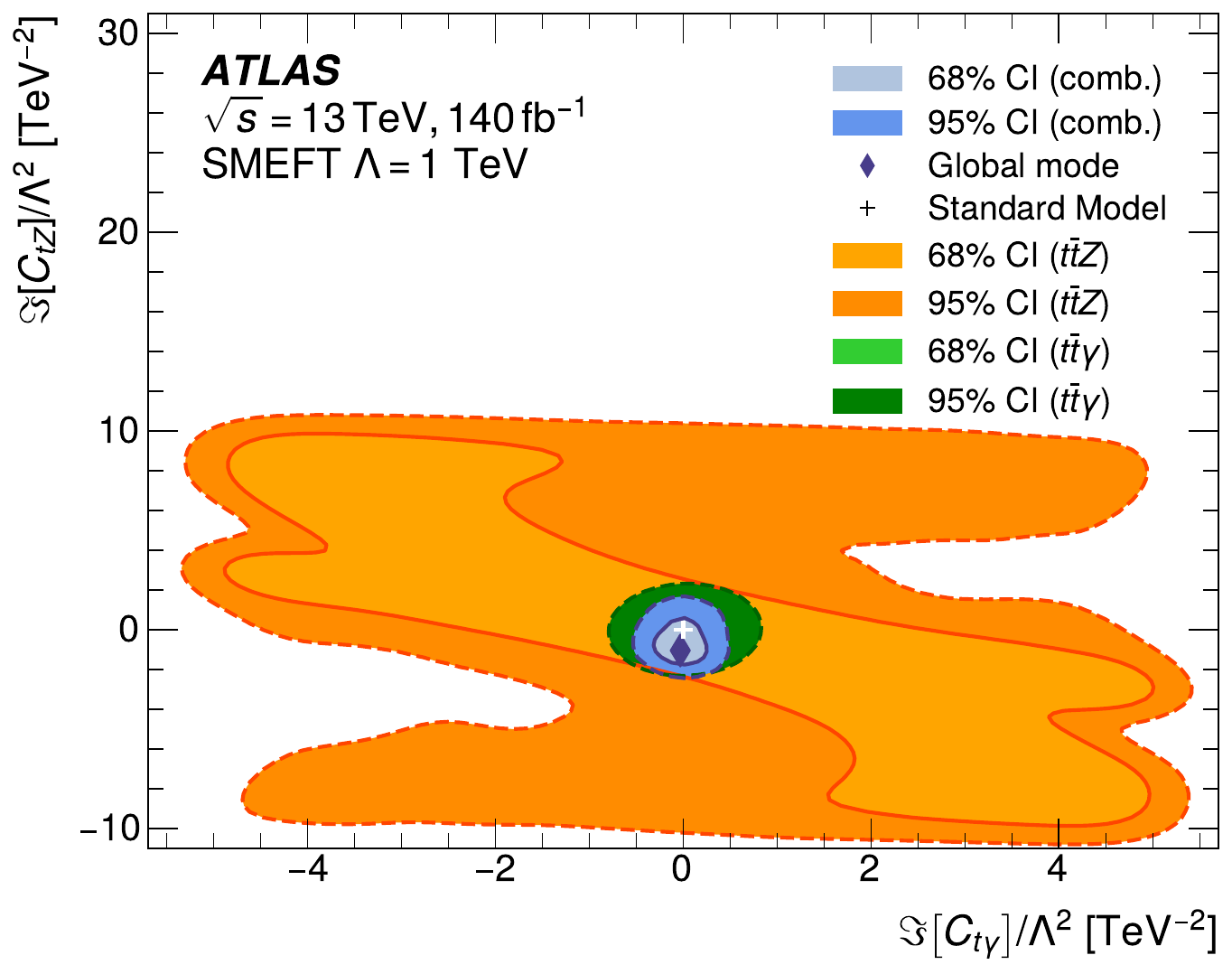}}
\subfloat[]{\includegraphics[width=0.32\textwidth]{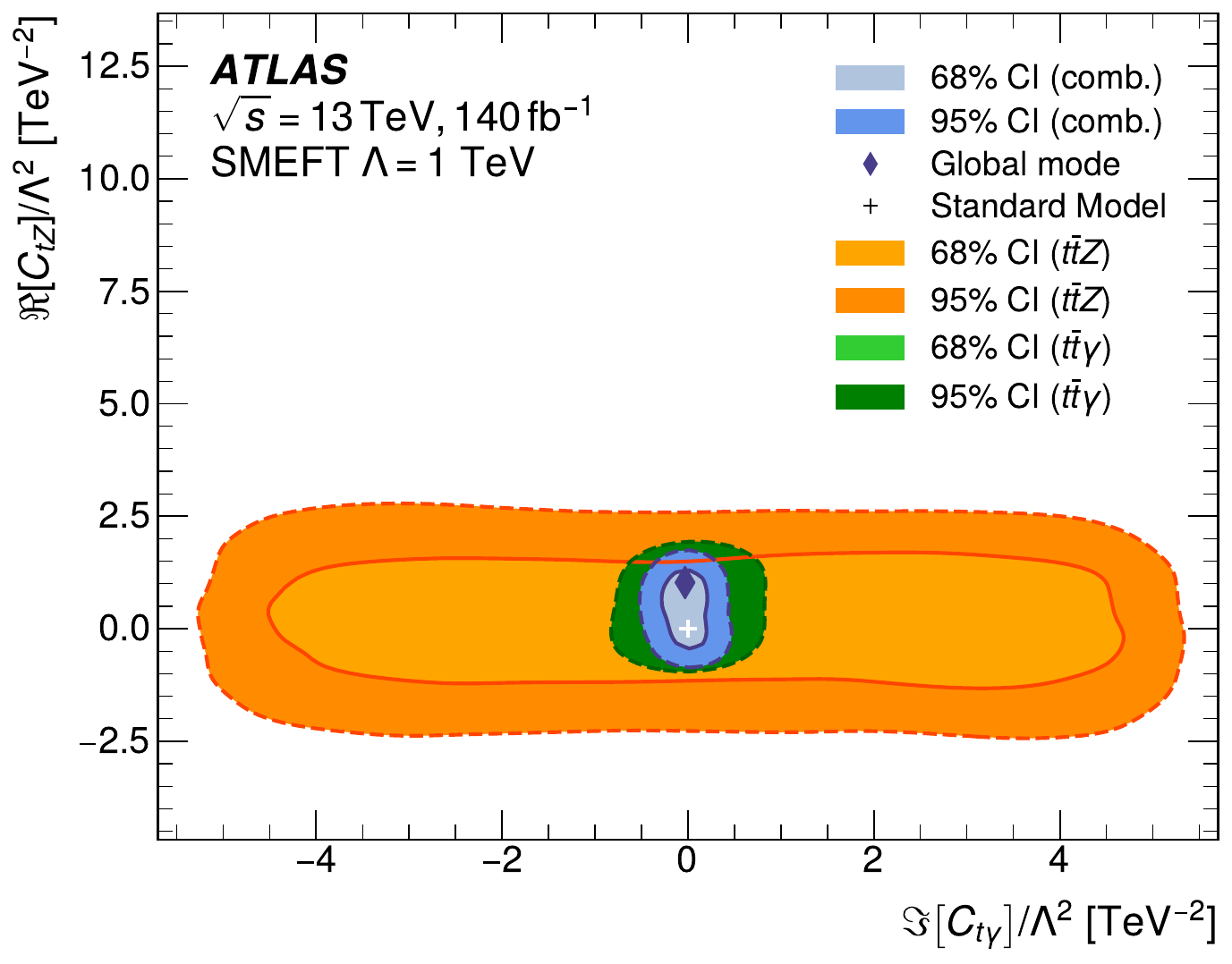}}\\
\subfloat[]{\includegraphics[width=0.32\textwidth]{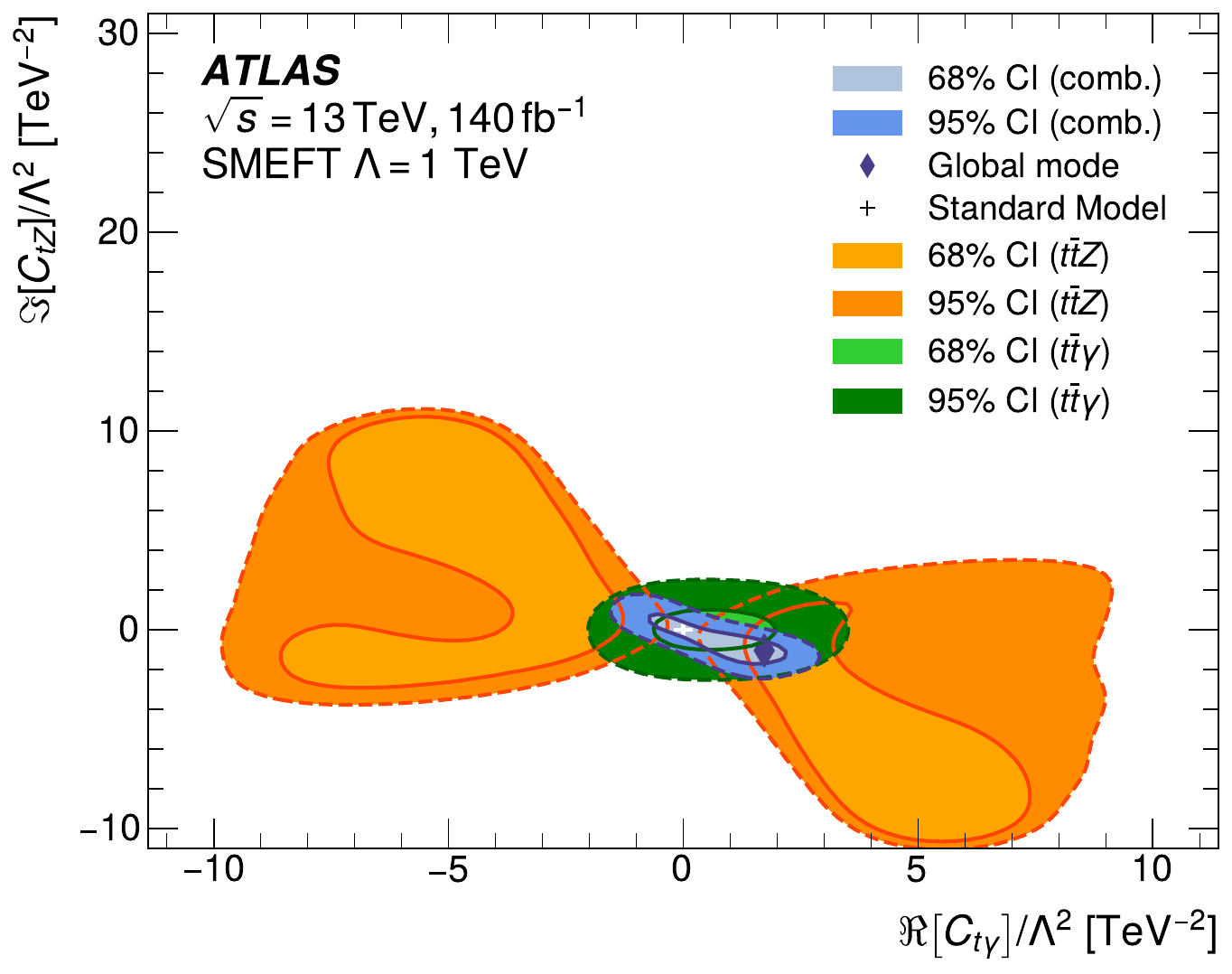}}
\subfloat[]{\includegraphics[width=0.32\textwidth]{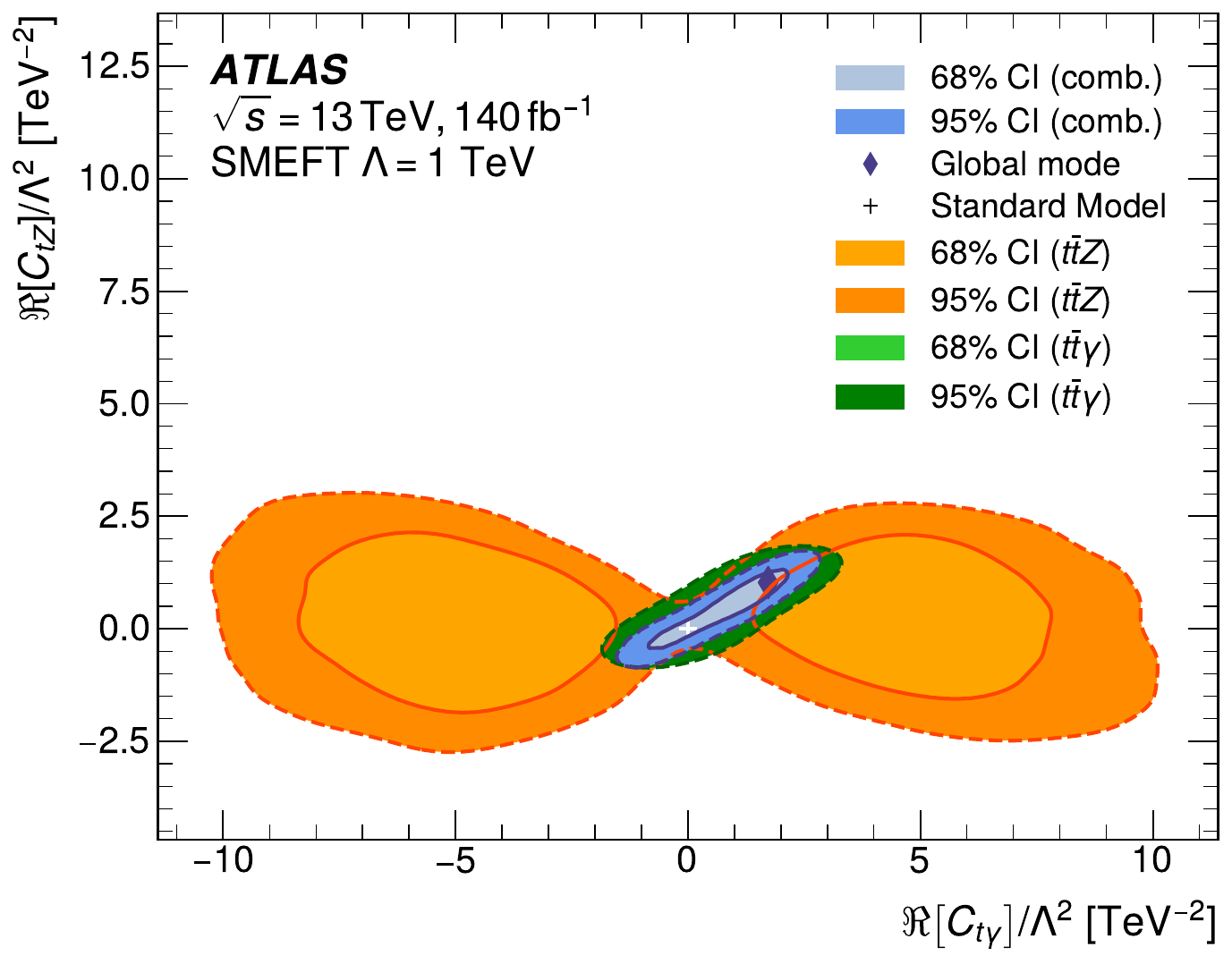}}
\subfloat[]{\includegraphics[width=0.32\textwidth]{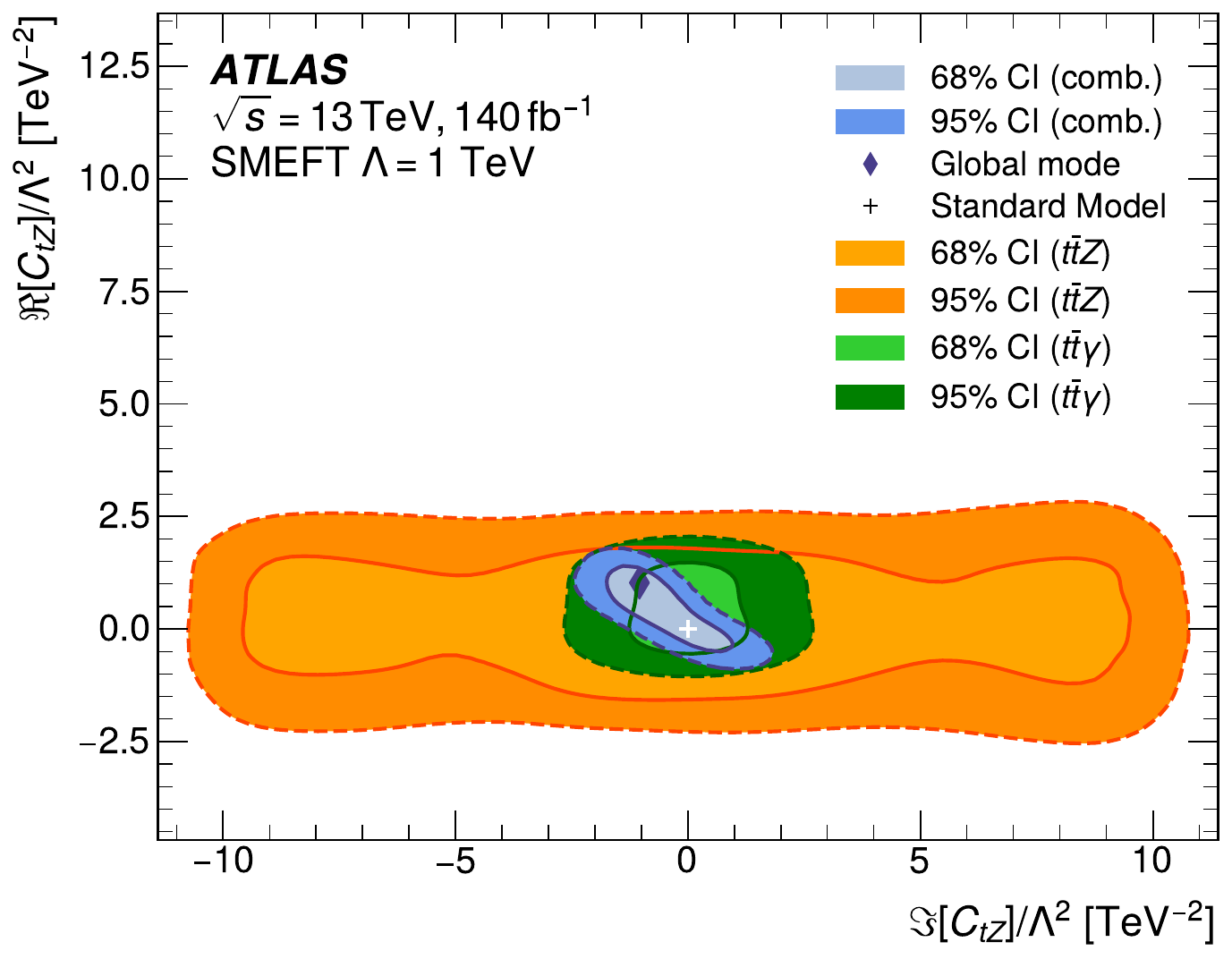}}
\end{center}
\caption{Two-dimensional marginalised posteriors for the \ctY and \ctZ operators from the quadratic fit, indicating the $68\%$ and $95\%$ credible intervals. The pairs of operators shown are (a) $\Im{\left[\ctY\right]}$,$\Re{\left[\ctY\right]}$, (b) $\Im{\left[\ctY\right]}$,$\Im{\left[\ctZ\right]}$, (c) $\Im{\left[\ctY\right]}$,$\ctZRe$, (d) $\Re{\left[\ctY\right]}$,$\ctZIm$, (e) $\Re{\left[\ctY\right]}$,$\ctZRe$ and (f) $\ctZIm$,$\ctZRe$. The individual contributions of the \ttZ and \tty measurements are displayed separately from the result of the full combination.
Also shown are the best-fit value (global mode) and the SM prediction.}
\label{fig:eft-2d-posteriors-rotated-quadratic-comparison}
\end{figure}

\FloatBarrier
\section{Conclusion}
\label{sec:conclusion}
Detailed measurements of the production cross-section of top quark pairs in association with a
photon are presented, using a data sample of proton--proton collisions collected with the ATLAS detector at the LHC at $\sqrt{s} = 13$~\TeV, corresponding to a total integrated luminosity of \intlumitot. The cross-section of the \tty production process, where the photon is radiated from one of the incoming quarks, or the top quark, is measured for the first time. The inclusive and differential measurements are performed in the single-lepton and dilepton decay channels in fiducial regions at particle level for photon $\pt > 20$~\gev.
The combined inclusive fiducial cross-section for the \tty production process
is $319 \pm 4 \, ({ \text{stat}}) ^{+15}_{-14}({\mathrm{syst}})\, \text{fb}$. The combined \tty cross-section for events with photon radiated at any stage of the process is $788 \pm 5\, ({ \text{stat}})\, ^{+38}_{-37} \,({\mathrm{syst}})\, \text{fb}$. The MC predictions are in agreement with the measurements within the total uncertainties, although they slightly underestimate the cross-sections.

The differential cross-sections, absolute and normalised, are measured for six variables in the single-lepton and up to eleven variables in the dilepton channel for the \tty topology with the photon radiated from an initial-state parton or one of the top quarks and for the \tty process regardless of the origin of the photon. They characterise the photon, lepton and jet  kinematics as well as the angular separation between the reconstructed objects of the \tty event. The shapes of the measured cross-sections are mostly well described by the MC predictions.

The differential cross-sections as a function of the photon \pt are interpreted in the framework of the SM effective field theory, setting limits on  parameters related to the electroweak dipole moments of the top quark. The EFT interpretation is also performed by using a simultaneous fit to the photon \pt and the $Z$ boson \pt spectra measured in the $t\bar{t}$ production in association with a $Z$ boson.

\section*{Acknowledgements}
%

%
%

%
%

We thank CERN for the very successful operation of the LHC and its injectors, as well as the support staff at
CERN and at our institutions worldwide without whom ATLAS could not be operated efficiently.

The crucial computing support from all WLCG partners is acknowledged gratefully, in particular from CERN, the ATLAS Tier-1 facilities at TRIUMF/SFU (Canada), NDGF (Denmark, Norway, Sweden), CC-IN2P3 (France), KIT/GridKA (Germany), INFN-CNAF (Italy), NL-T1 (Netherlands), PIC (Spain), RAL (UK) and BNL (USA), the Tier-2 facilities worldwide and large non-WLCG resource providers. Major contributors of computing resources are listed in Ref.~\cite{ATL-SOFT-PUB-2023-001}.

We gratefully acknowledge the support of ANPCyT, Argentina; YerPhI, Armenia; ARC, Australia; BMWFW and FWF, Austria; ANAS, Azerbaijan; CNPq and FAPESP, Brazil; NSERC, NRC and CFI, Canada; CERN; ANID, Chile; CAS, MOST and NSFC, China; Minciencias, Colombia; MEYS CR, Czech Republic; DNRF and DNSRC, Denmark; IN2P3-CNRS and CEA-DRF/IRFU, France; SRNSFG, Georgia; BMBF, HGF and MPG, Germany; GSRI, Greece; RGC and Hong Kong SAR, China; ISF and Benoziyo Center, Israel; INFN, Italy; MEXT and JSPS, Japan; CNRST, Morocco; NWO, Netherlands; RCN, Norway; MEiN, Poland; FCT, Portugal; MNE/IFA, Romania; MESTD, Serbia; MSSR, Slovakia; ARIS and MVZI, Slovenia; DSI/NRF, South Africa; MICIU/AEI, Spain; SRC and Wallenberg Foundation, Sweden; SERI, SNSF and Cantons of Bern and Geneva, Switzerland; NSTC, Taipei; TENMAK, T\"urkiye; STFC/UKRI, United Kingdom; DOE and NSF, United States of America.

Individual groups and members have received support from BCKDF, CANARIE, CRC and DRAC, Canada; CERN-CZ, PRIMUS 21/SCI/017 and UNCE SCI/013, Czech Republic; COST, ERC, ERDF, Horizon 2020, ICSC-NextGenerationEU and Marie Sk{\l}odowska-Curie Actions, European Union; Investissements d'Avenir Labex, Investissements d'Avenir Idex and ANR, France; DFG and AvH Foundation, Germany; Herakleitos, Thales and Aristeia programmes co-financed by EU-ESF and the Greek NSRF, Greece; BSF-NSF and MINERVA, Israel; Norwegian Financial Mechanism 2014-2021, Norway; NCN and NAWA, Poland; La Caixa Banking Foundation, CERCA Programme Generalitat de Catalunya and PROMETEO and GenT Programmes Generalitat Valenciana, Spain; G\"{o}ran Gustafssons Stiftelse, Sweden; The Royal Society and Leverhulme Trust, United Kingdom.

In addition, individual members wish to acknowledge support from CERN: European Organization for Nuclear Research (CERN PJAS); Chile: Agencia Nacional de Investigaci\'on y Desarrollo (FONDECYT 1190886, FONDECYT 1210400, FONDECYT 1230812, FONDECYT 1230987); China: Chinese Ministry of Science and Technology (MOST-2023YFA1605700), National Natural Science Foundation of China (NSFC - 12175119, NSFC 12275265, NSFC-12075060); Czech Republic: PRIMUS Research Programme (PRIMUS/21/SCI/017); EU: H2020 European Research Council (ERC - 101002463); European Union: European Research Council (ERC - 948254, ERC 101089007), Horizon 2020 Framework Programme (MUCCA - CHIST-ERA-19-XAI-00), European Union, Future Artificial Intelligence Research (FAIR-NextGenerationEU PE00000013), Italian Center for High Performance Computing, Big Data and Quantum Computing (ICSC, NextGenerationEU); France: Agence Nationale de la Recherche (ANR-20-CE31-0013, ANR-21-CE31-0013, ANR-21-CE31-0022, ANR-22-EDIR-0002), Investissements d'Avenir Labex (ANR-11-LABX-0012); Germany: Baden-Württemberg Stiftung (BW Stiftung-Postdoc Eliteprogramme), Deutsche Forschungsgemeinschaft (DFG - 469666862, DFG - CR 312/5-2); Italy: Istituto Nazionale di Fisica Nucleare (ICSC, NextGenerationEU), Ministero dell'Università e della Ricerca (PRIN - 20223N7F8K - PNRR M4.C2.1.1); Japan: Japan Society for the Promotion of Science (JSPS KAKENHI JP21H05085, JSPS KAKENHI JP22H01227, JSPS KAKENHI JP22H04944, JSPS KAKENHI JP22KK0227); Netherlands: Netherlands Organisation for Scientific Research (NWO Veni 2020 - VI.Veni.202.179); Norway: Research Council of Norway (RCN-314472); Poland: Polish National Agency for Academic Exchange (PPN/PPO/2020/1/00002/U/00001), Polish National Science Centre (NCN 2021/42/E/ST2/00350, NCN OPUS nr 2022/47/B/ST2/03059, NCN UMO-2019/34/E/ST2/00393, UMO-2020/37/B/ST2/01043, UMO-2021/40/C/ST2/00187, UMO-2022/47/O/ST2/00148); Slovenia: Slovenian Research Agency (ARIS grant J1-3010); Spain: BBVA Foundation (LEO22-1-603), Generalitat Valenciana (Artemisa, FEDER, IDIFEDER/2018/048), Ministry of Science and Innovation (MCIN \& NextGenEU PCI2022-135018-2, MICIN \& FEDER PID2021-125273NB, RYC2019-028510-I, RYC2020-030254-I, RYC2021-031273-I, RYC2022-038164-I), PROMETEO and GenT Programmes Generalitat Valenciana (CIDEGENT/2019/023, CIDEGENT/2019/027); Sweden: Swedish Research Council (VR 2018-00482, VR 2022-03845, VR 2022-04683, VR grant 2021-03651), Knut and Alice Wallenberg Foundation (KAW 2017.0100, KAW 2018.0157, KAW 2018.0458, KAW 2019.0447, KAW 2022.0358); Switzerland: Swiss National Science Foundation (SNSF - PCEFP2\_194658); United Kingdom: Leverhulme Trust (Leverhulme Trust RPG-2020-004); United States of America: U.S. Department of Energy (ECA DE-AC02-76SF00515), Neubauer Family Foundation.

%
%


%
\clearpage
\appendix
\part*{Appendix}
\addcontentsline{toc}{part}{Appendix}
\section{Variables used in the training of the neural networks}
\label{sec:appendix_NN}

\begin{table}[htbp]
\centering
\caption{%
List of the input variables used in the training of the NNs in the  (left) single-lepton and (right) dilepton channels. The variables are ordered by separation power between \tty production and \tty decay or all background processes, respectively.}
\scalebox{0.90}{
\begin{tabular}{l  l }
\hline
Single lepton & Dilepton \\
\midrule
Invariant mass of photon and lepton & $\Delta R$ between photon and closest lepton \tabularnewline[0.5em]
Photon \pt & Invariant mass of photon and closest lepton \tabularnewline[0.5em]
$\Delta R$ of photon and lepton & Photon \pt \tabularnewline[0.5em]
Invariant mass of photon and leading PCBT $b$-jet & Invariant mass of photon and closest $b$-jet \tabularnewline[0.5em]%
Sum of invariant masses of the reconstructed top quark \\
and antiquark (4 variables) & Photon energy \tabularnewline[0.5em]
Photon energy & Scalar sum of \pt of all jets \tabularnewline[0.5em]
Sum of squared differences between the top-quark pole mass & \pt and energy of the two jets with highest \pt \\
and reconstructed $t\bar{t}$ mass (4 variables) & (4 variables)\tabularnewline[0.5em]
Invariant mass of all jets, the lepton and the photon & $\Delta R$ of photon and closest $b$-jet \tabularnewline[0.5em]
$H_{\text{T}}$ & \MET \tabularnewline[0.5em]%
Reconstructed leptonic $W$ boson \pt & Number of jets\tabularnewline[0.5em]%
\pt and energy of four jets with highest \pt (8 variables) & Photon $\eta$ \tabularnewline[0.5em]%
$\Delta R$ between photon and closest b-jet & Number of $b$-jets\tabularnewline[0.5em]
\MET & Photon $\phi$\tabularnewline[0.5em]
Invariant mass of lepton and closest $b$-jet\tabularnewline[0.5em]
Number of jets\tabularnewline[0.5em]
Transverse mass of leptonic $W$ boson\tabularnewline[0.5em]
$\Delta R$ between lepton and closest $b$-jet\tabularnewline[0.5em]
Invariant mass of reconstructed $W$ bosons, shifted\\
by the $W$ boson (2 variables)\tabularnewline[0.5em]
Photon $\eta$\tabularnewline[0.5em]
PCBT distributions of the four jets with the highest scores \\(4 variables)\tabularnewline[0.5em]%
Photon conversion type\tabularnewline[0.5em]%
Number of $b$-jets\tabularnewline[0.5em]
Photon $\phi$\tabularnewline[0.5em]
\hline
\end{tabular}
}
\end{table}

\begin{figure}[!htb]
\centering
\subfloat[]{\includegraphics[width=0.45\textwidth]{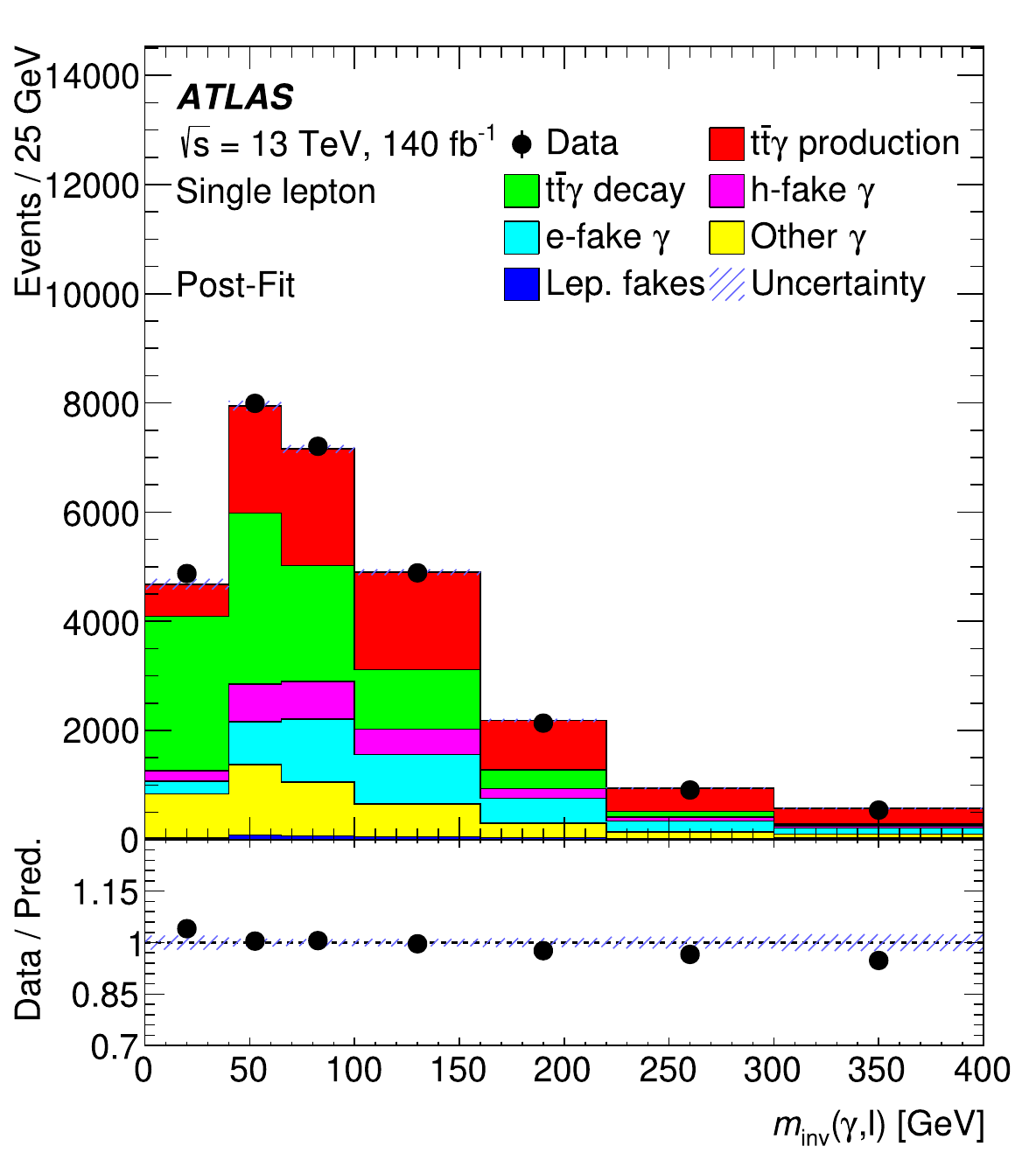}}%
\subfloat[]{\includegraphics[width=0.45\textwidth]{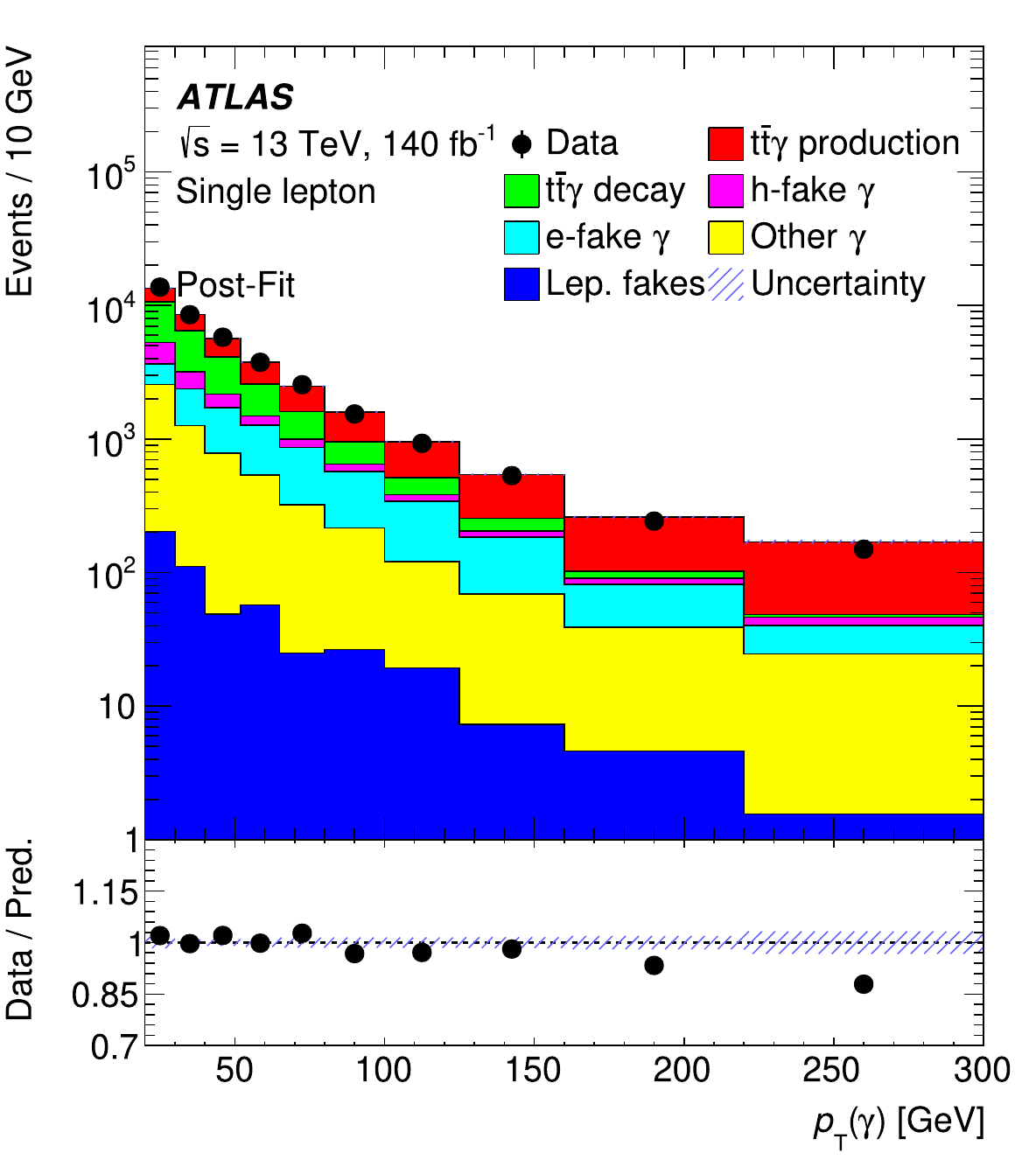}}\\
\subfloat[]{\includegraphics[width=0.45\textwidth]{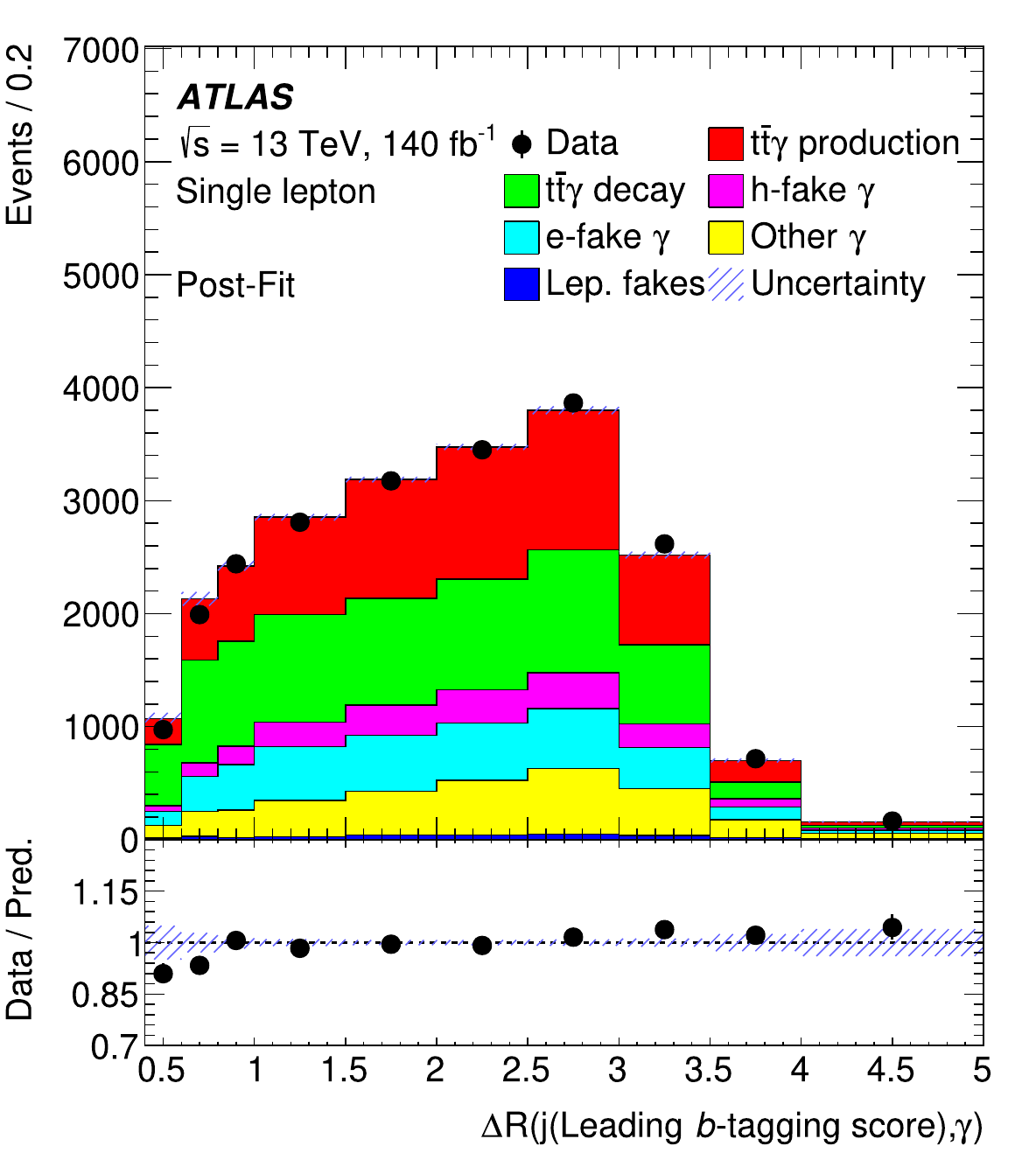}}%
\subfloat[]{\includegraphics[width=0.45\textwidth]{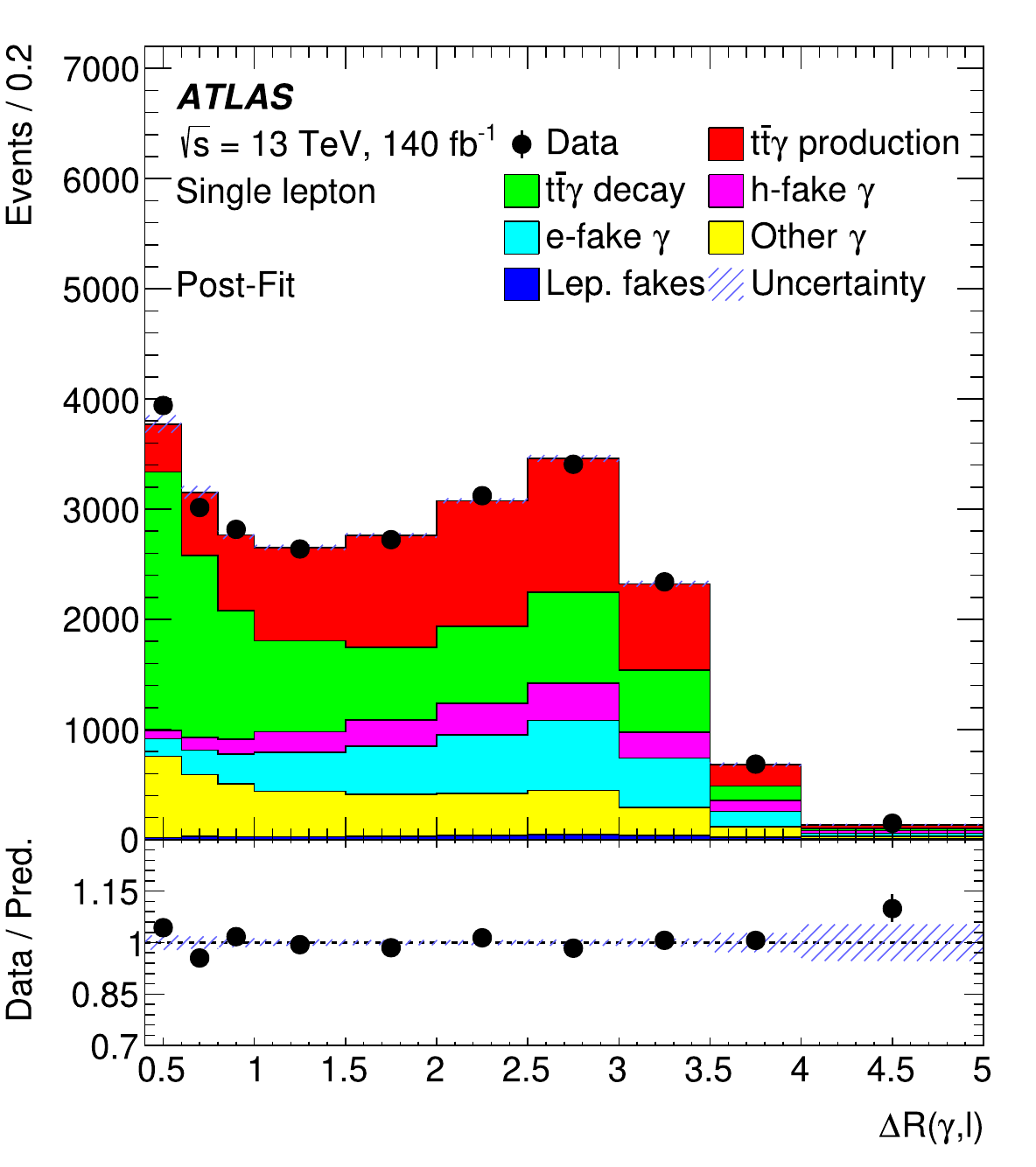}}\\
\caption{Distributions of the four input variables used in the training of the NN in the single-lepton channel with the largest separation power between \tty production and \tty decay after the fit to data for the measurement of the \tty production cross-section: (a) invariant mass of the photon and lepton system, (b) photon \pt, (c) $\Delta R$ between the photon and the jet with the highest $b$-tagging score and  (d) $\Delta R (\gamma, \ell)$. The uncertainty band represents the total post-fit uncertainty in the prediction.  The lower panels show the ratios of the data to the total post-fit predictions.}
\end{figure}

\begin{figure}[!htb]
\centering
\subfloat[]{\includegraphics[width=0.45\textwidth]{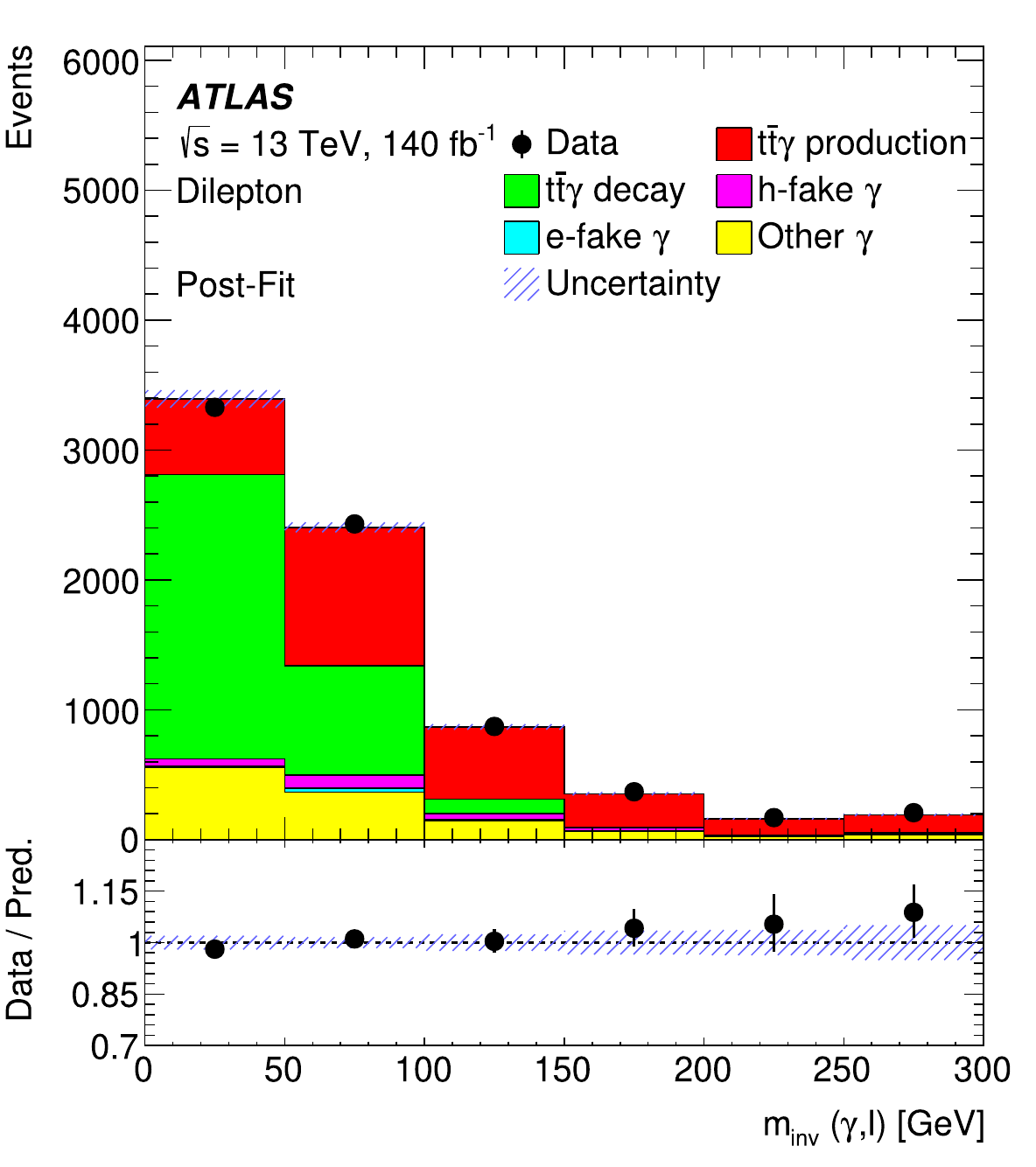}}%
\subfloat[]{\includegraphics[width=0.45\textwidth]{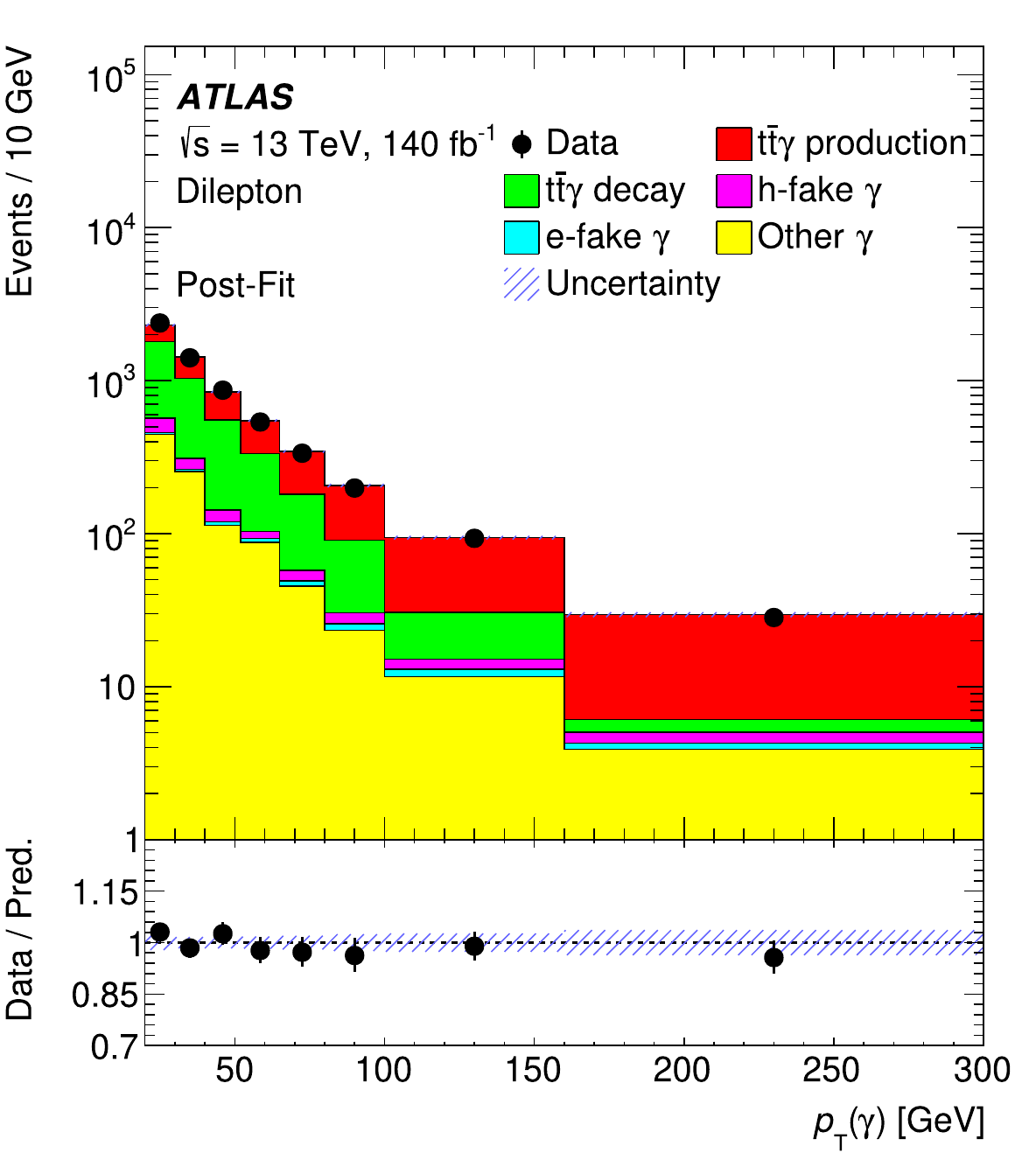}}\\
\subfloat[]{\includegraphics[width=0.45\textwidth]{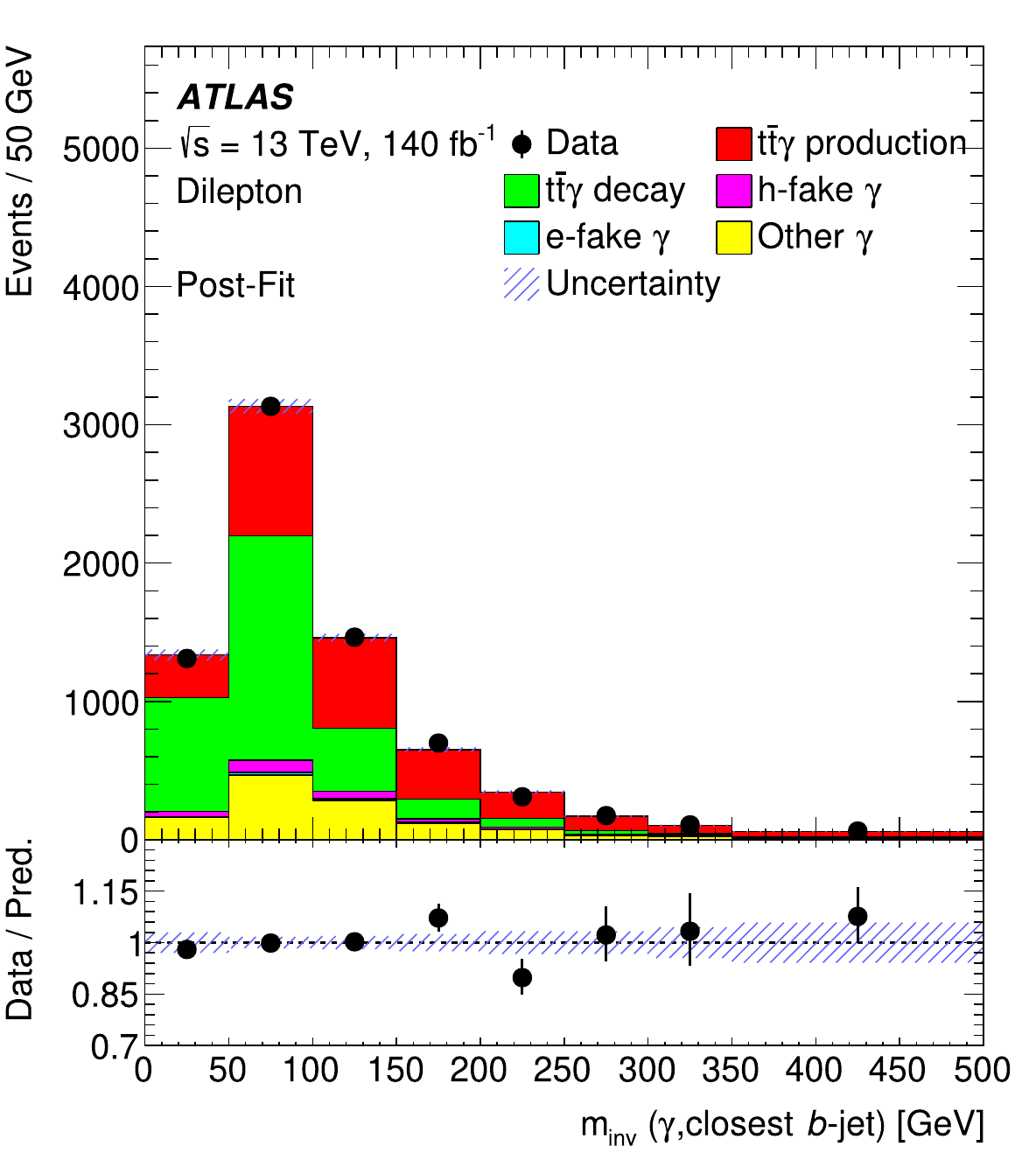}}%
\subfloat[]{\includegraphics[width=0.45\textwidth]{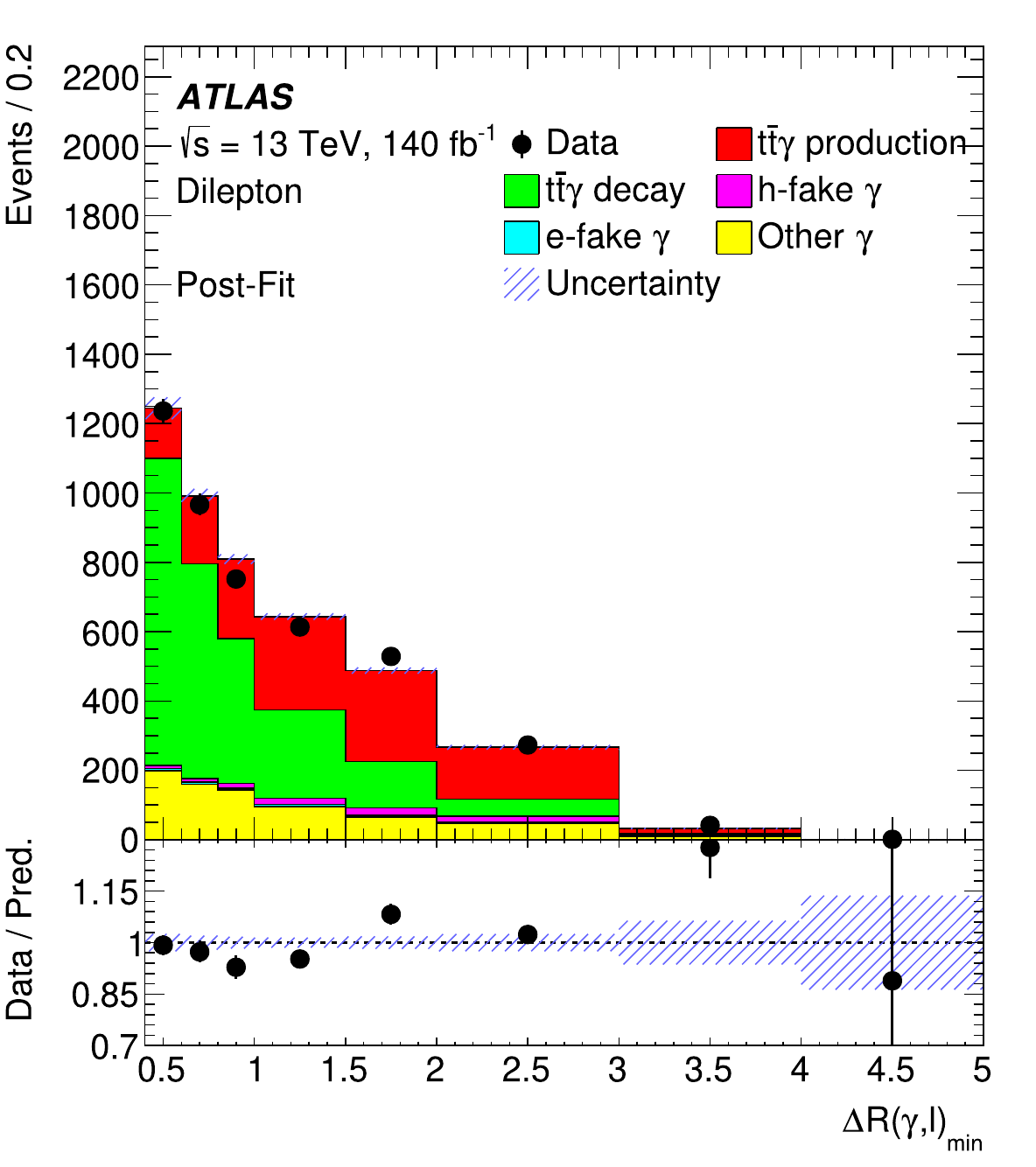}}\\  \caption{Distributions of the four input variables used in the training of the NN in the dilepton channel with the largest separation power between \tty production and the total background after the fit to data for the measurement of the \tty production cross-section: (a) invariant mass of the photon and the closest lepton system, (b) photon \pt, (c) invariant mass of the photon and the closest $b$-jet and (d) $\Delta R (\gamma, \ell)_{\text{min}}$. The uncertainty band represents the total post-fit uncertainty in the prediction.  The lower panels show the ratios of the data to the total post-fit predictions.}
\end{figure}
\FloatBarrier

\section{Differential cross-sections for the \tty  production process}
\label{sec:appendix_diff_prod}

The absolute and normalised cross-sections of the \tty production as a function of photon \pt and $|\eta|$ and leading jet \pt are shown in Figure~\ref{fig:ttyprod_diff_Ljets_app} (single-lepton channel) and Figure~\ref{fig:ttyprod_diff_DL_app} (dilepton channel). They are compared with the
\MGNLO simulation at NLO interfaced to \PYTHIA[8] and \HERWIG[7].
The agreement between the data and both predictions is similar. Both simulations
predict a harder photon \pt distribution, while they describe the shape of the photon $|\eta|$ and the \pt of a leading jet well.

\begin{figure}[ht]
\centering
\subfloat[]{\includegraphics[width=0.40\textwidth]{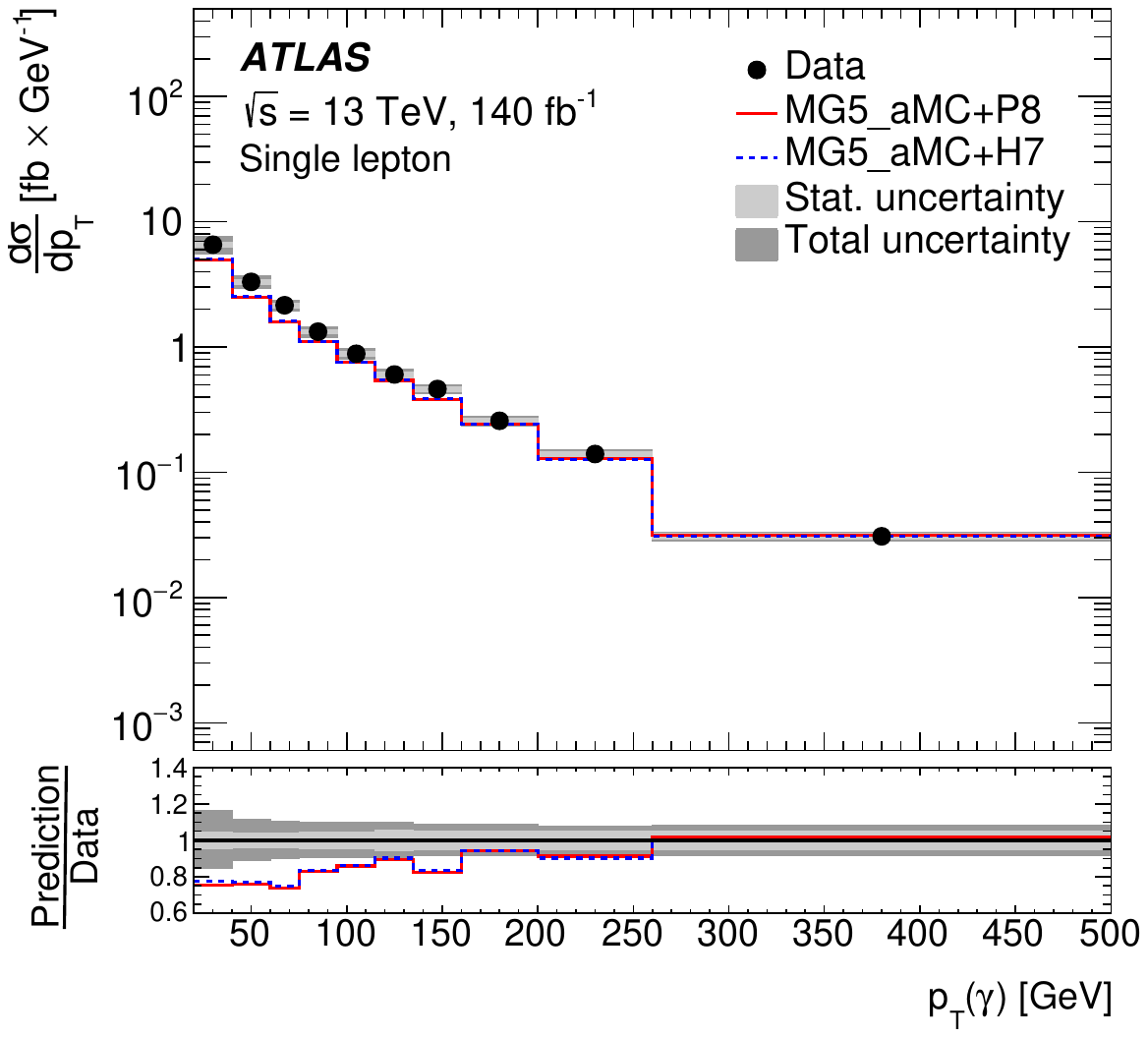}}%
\subfloat[]{\includegraphics[width=0.40\textwidth]{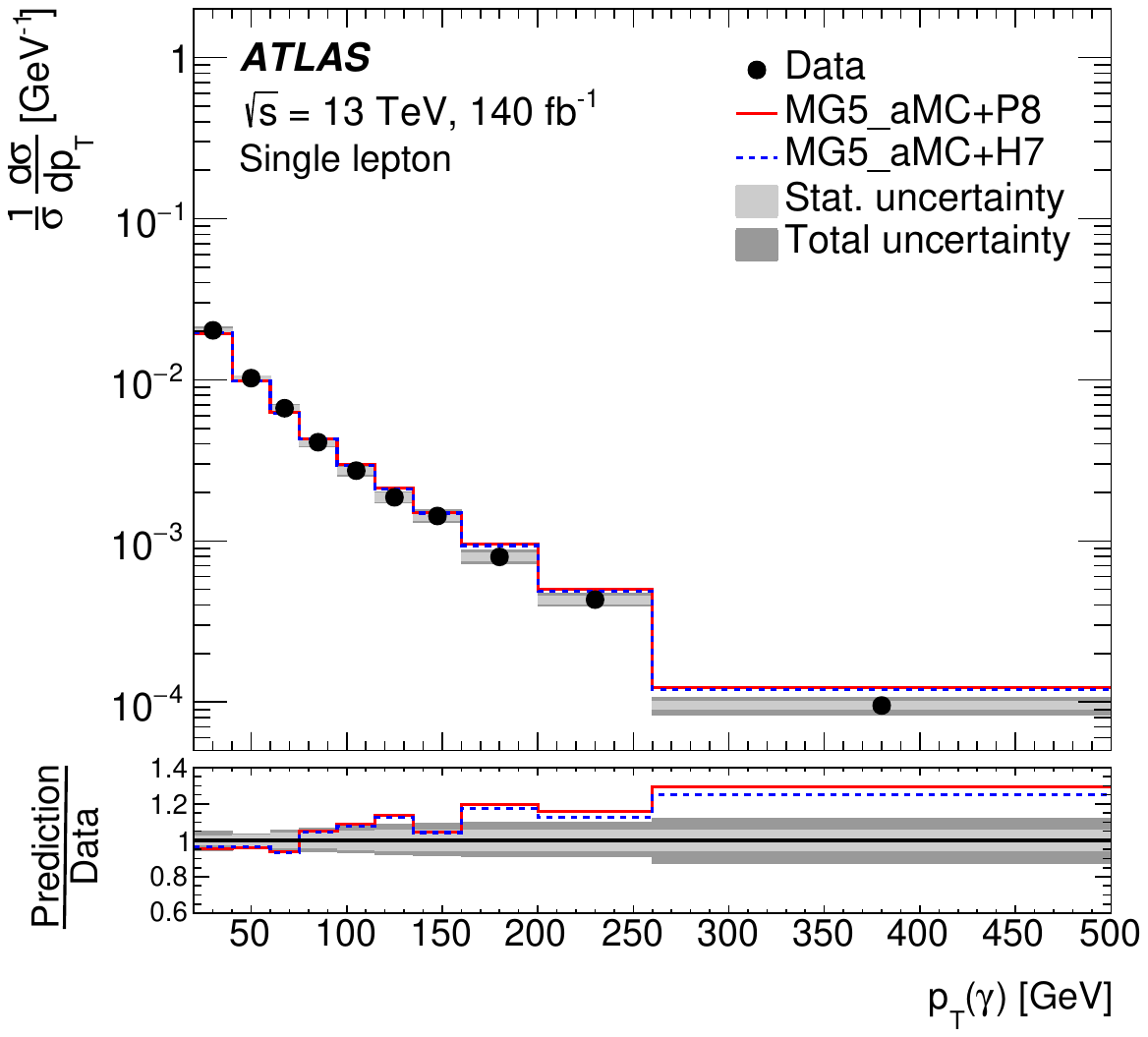}}\\
\subfloat[]{\includegraphics[width=0.40\textwidth]{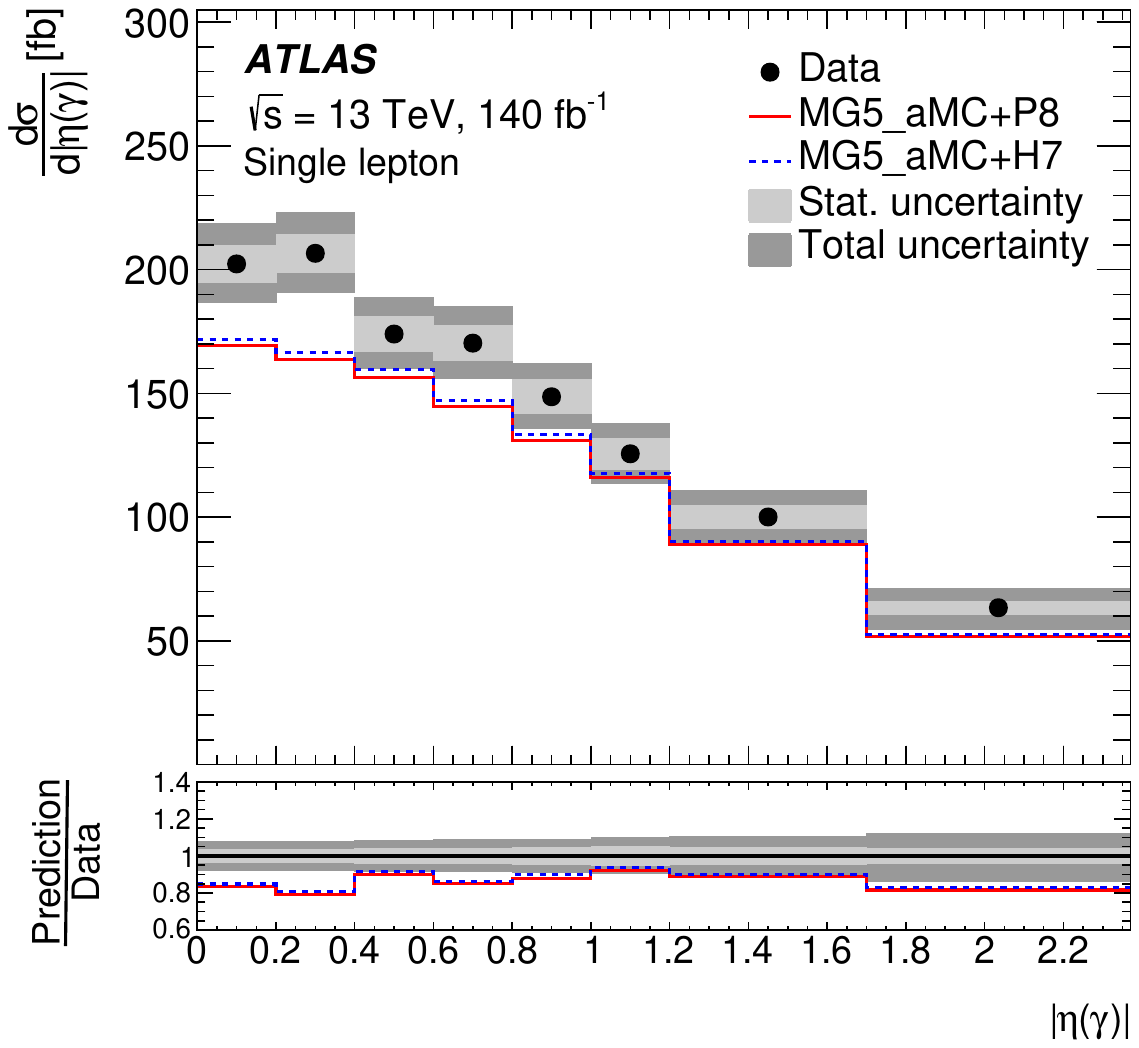}}%
\subfloat[]{\includegraphics[width=0.40\textwidth]{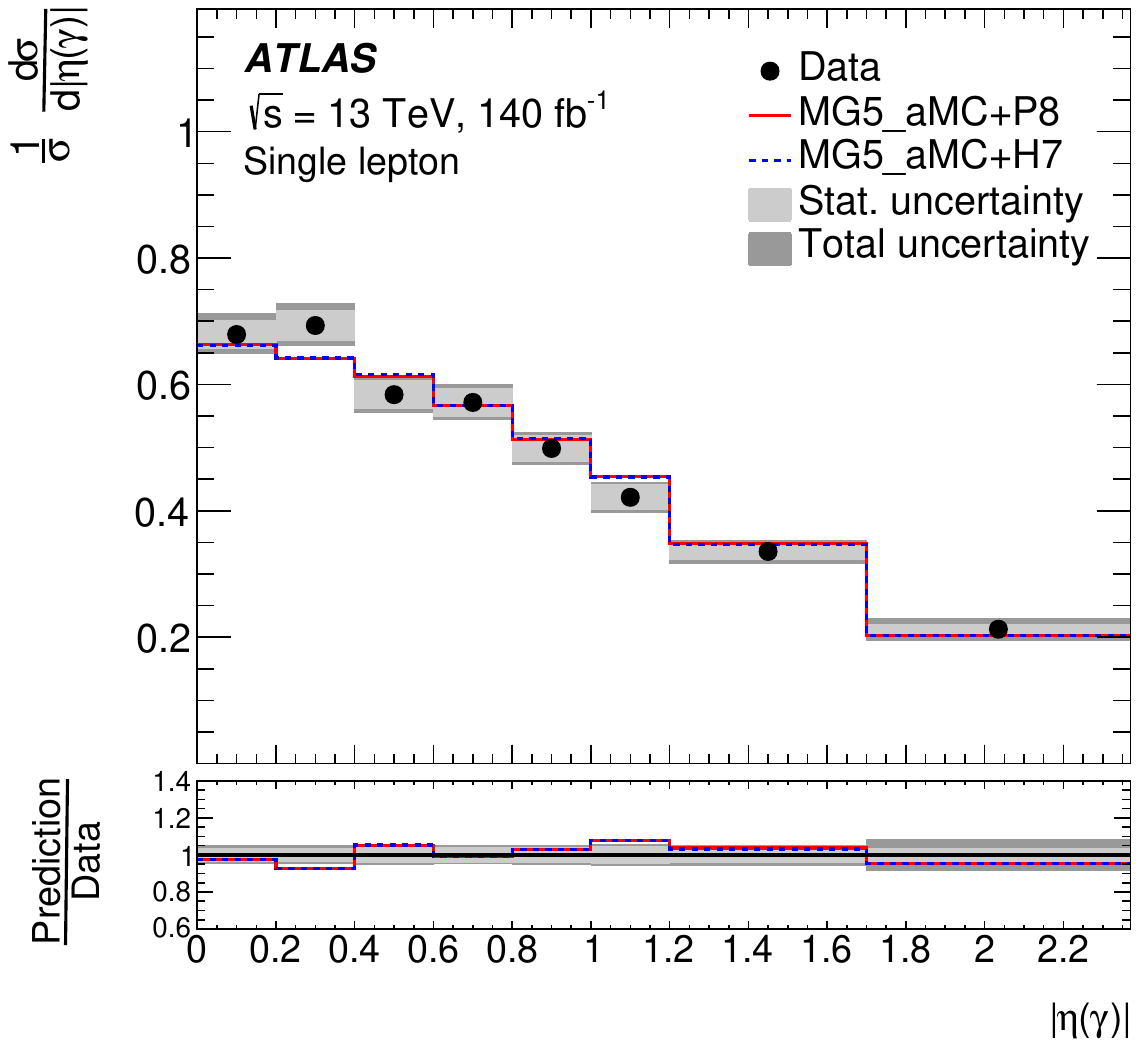}}\\
\subfloat[]{\includegraphics[width=0.40\textwidth]{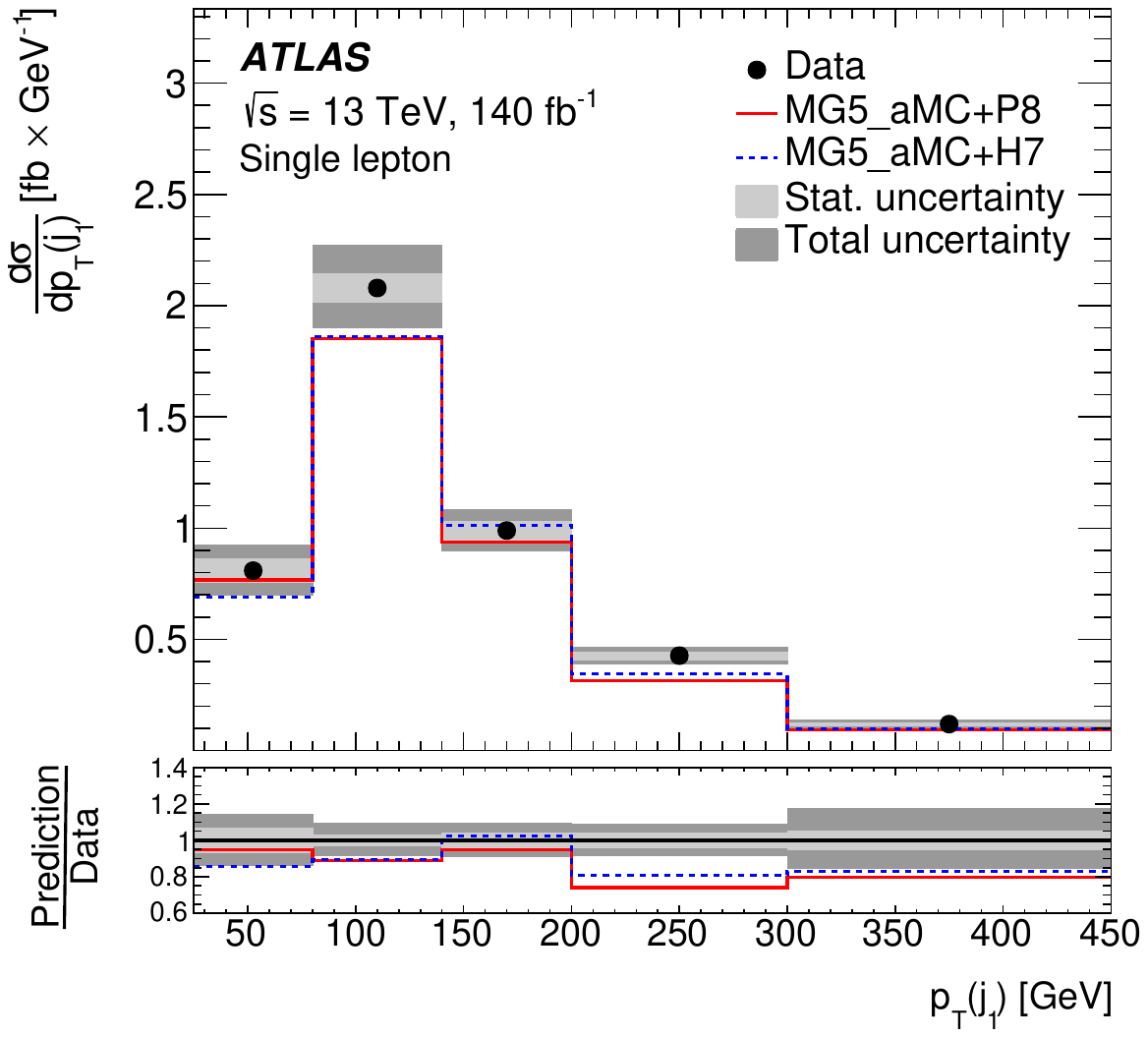}}%
\subfloat[]{\includegraphics[width=0.40\textwidth]{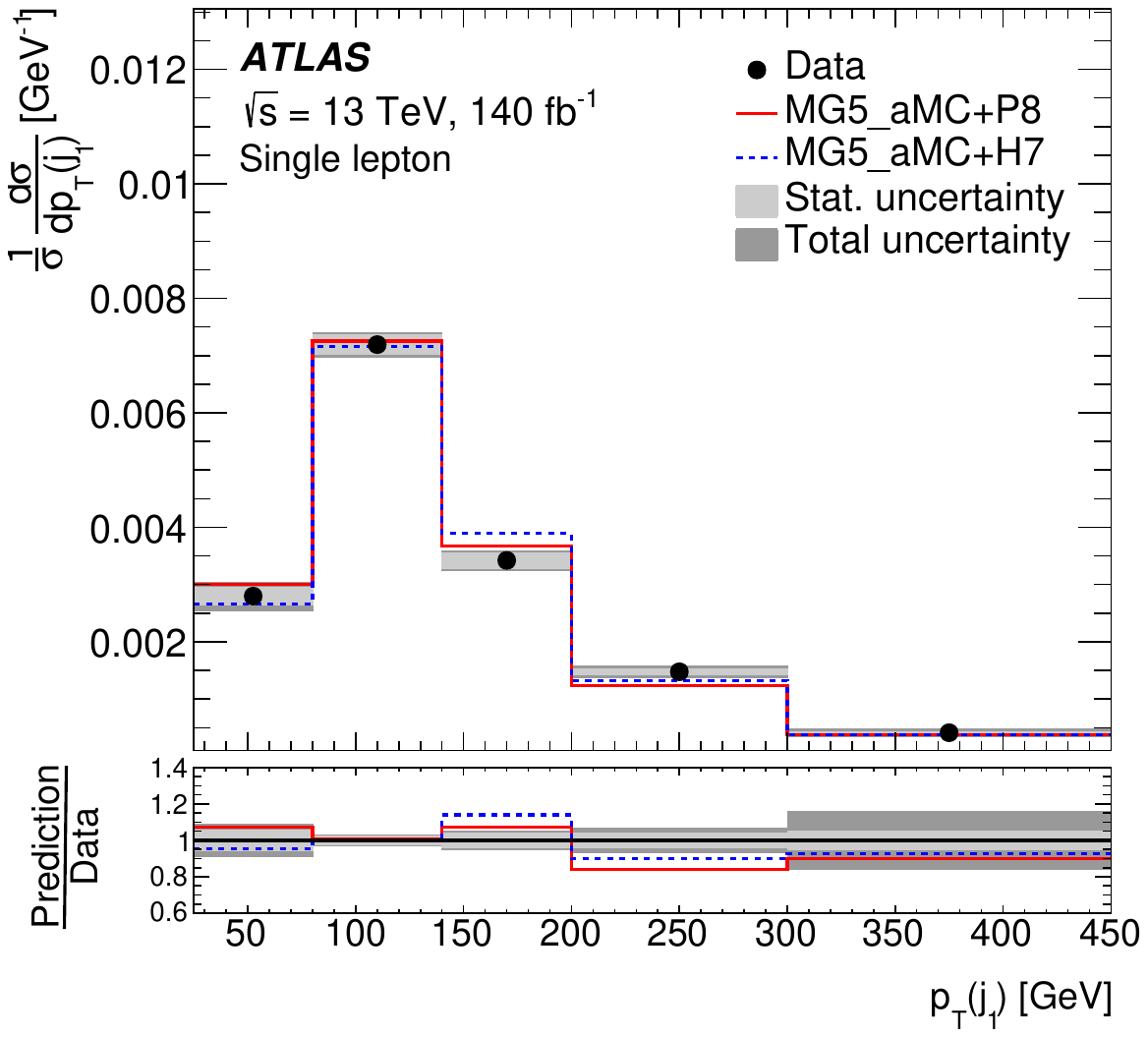}}\\
\caption{The (a, c, e) absolute and (b, d, f) normalised differential cross-section measured in the fiducial phase space in the single-lepton channel as a function of the (a, b) photon \pt, (c, d) photon $|\eta|$ and (e, f) \pt of the leading jet. Data are compared with the NLO \MGNLO simulation interfaced to \PYTHIA[8] and \HERWIG[7]. The lower panels show the ratios of the predictions to the data. The last bin of the distributions includes the overflow events.}
\label{fig:ttyprod_diff_Ljets_app}
\end{figure}

\begin{figure}[ht]
\centering
\subfloat[]{\includegraphics[width=0.40\textwidth]{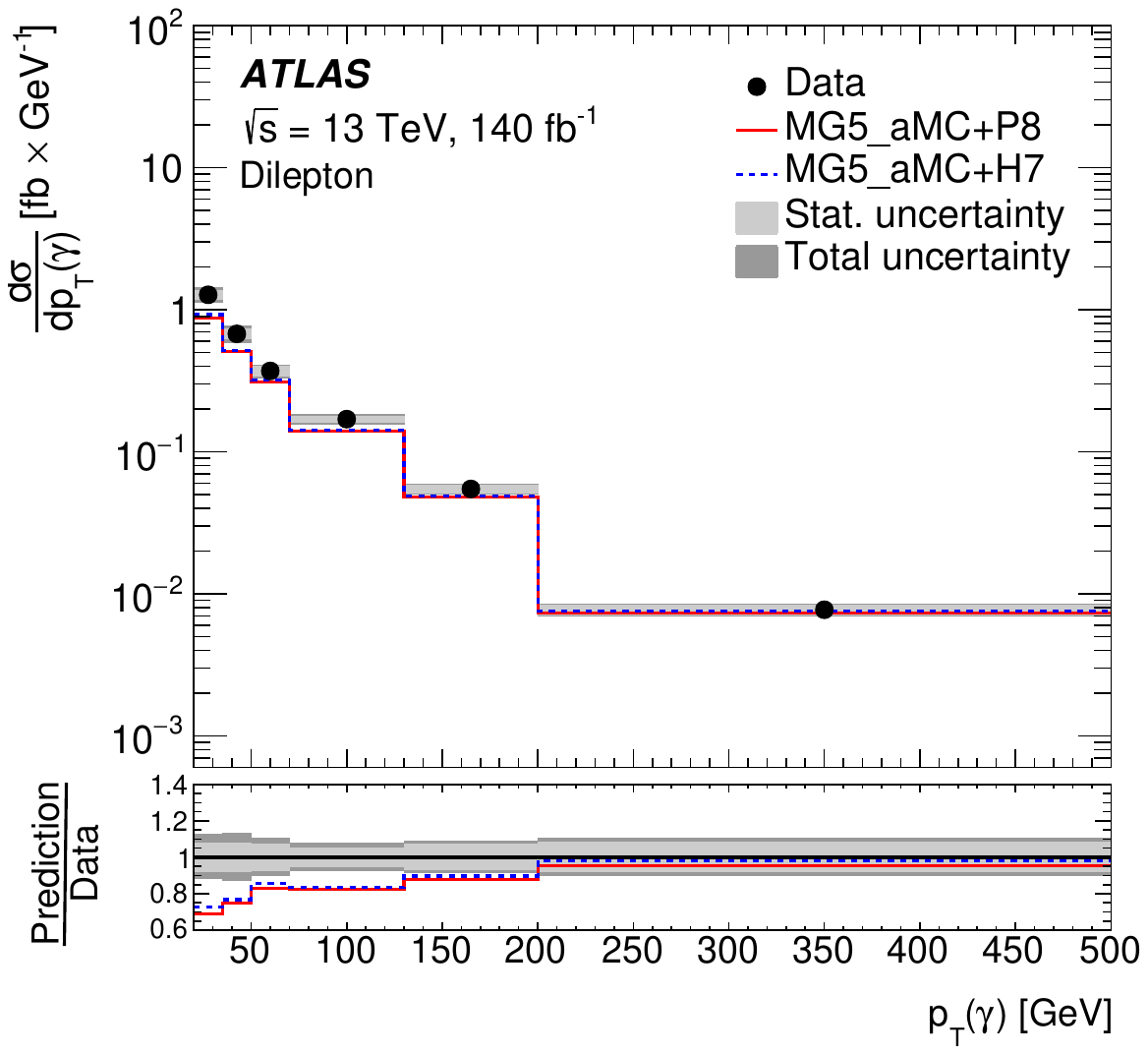}}%
\subfloat[]{\includegraphics[width=0.40\textwidth]{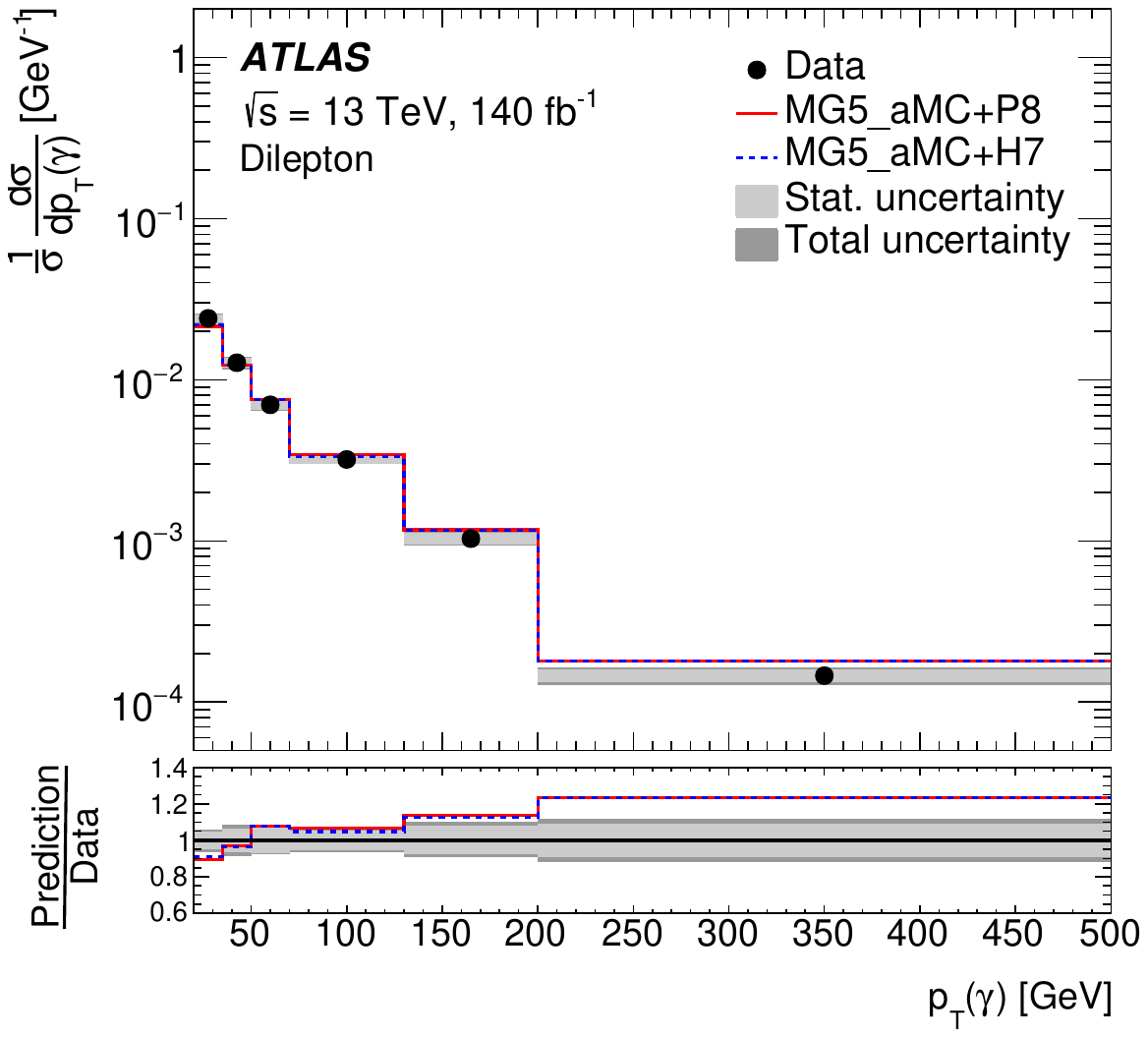}}\\
\subfloat[]{\includegraphics[width=0.40\textwidth]{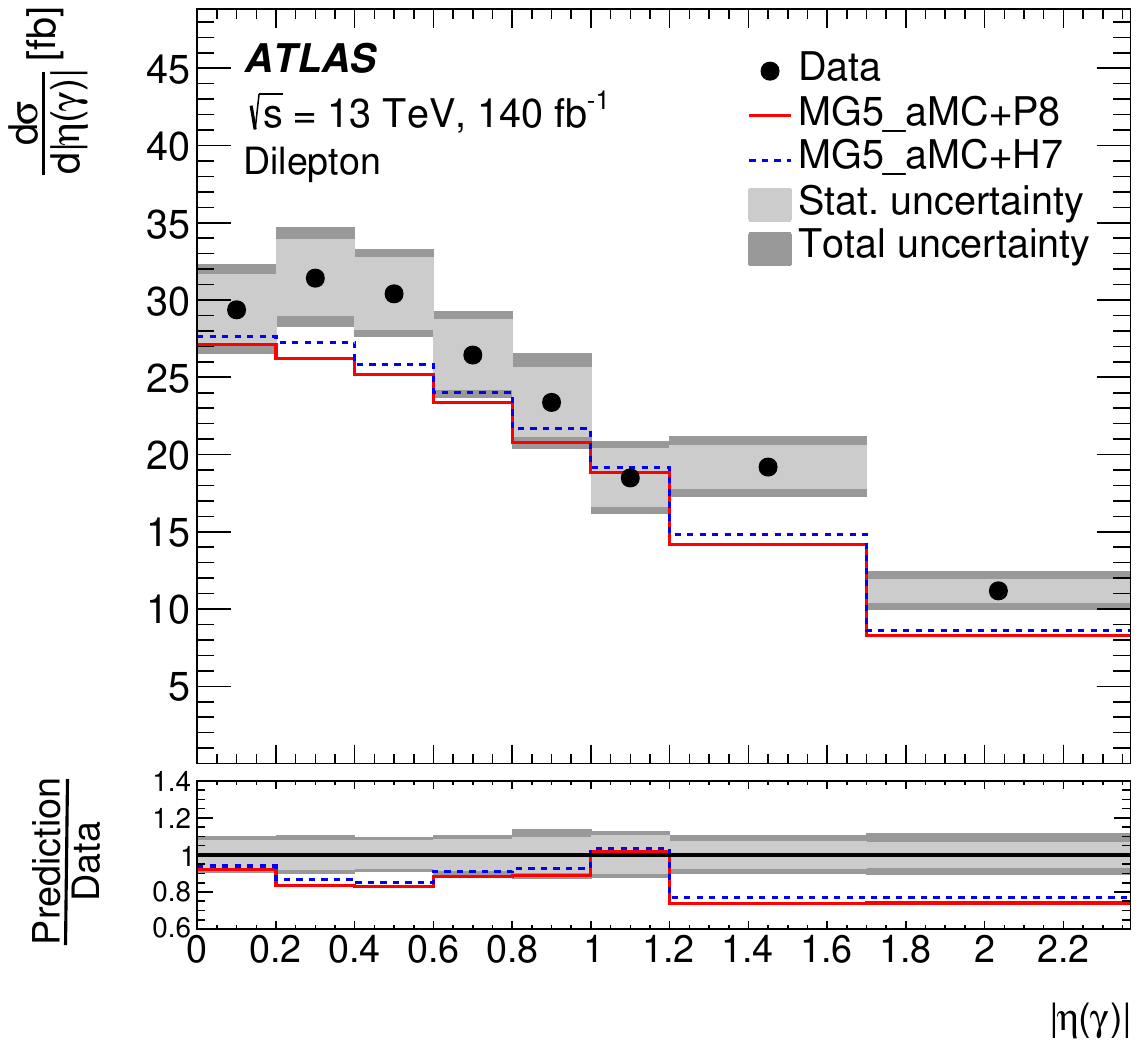}}%
\subfloat[]{\includegraphics[width=0.40\textwidth]{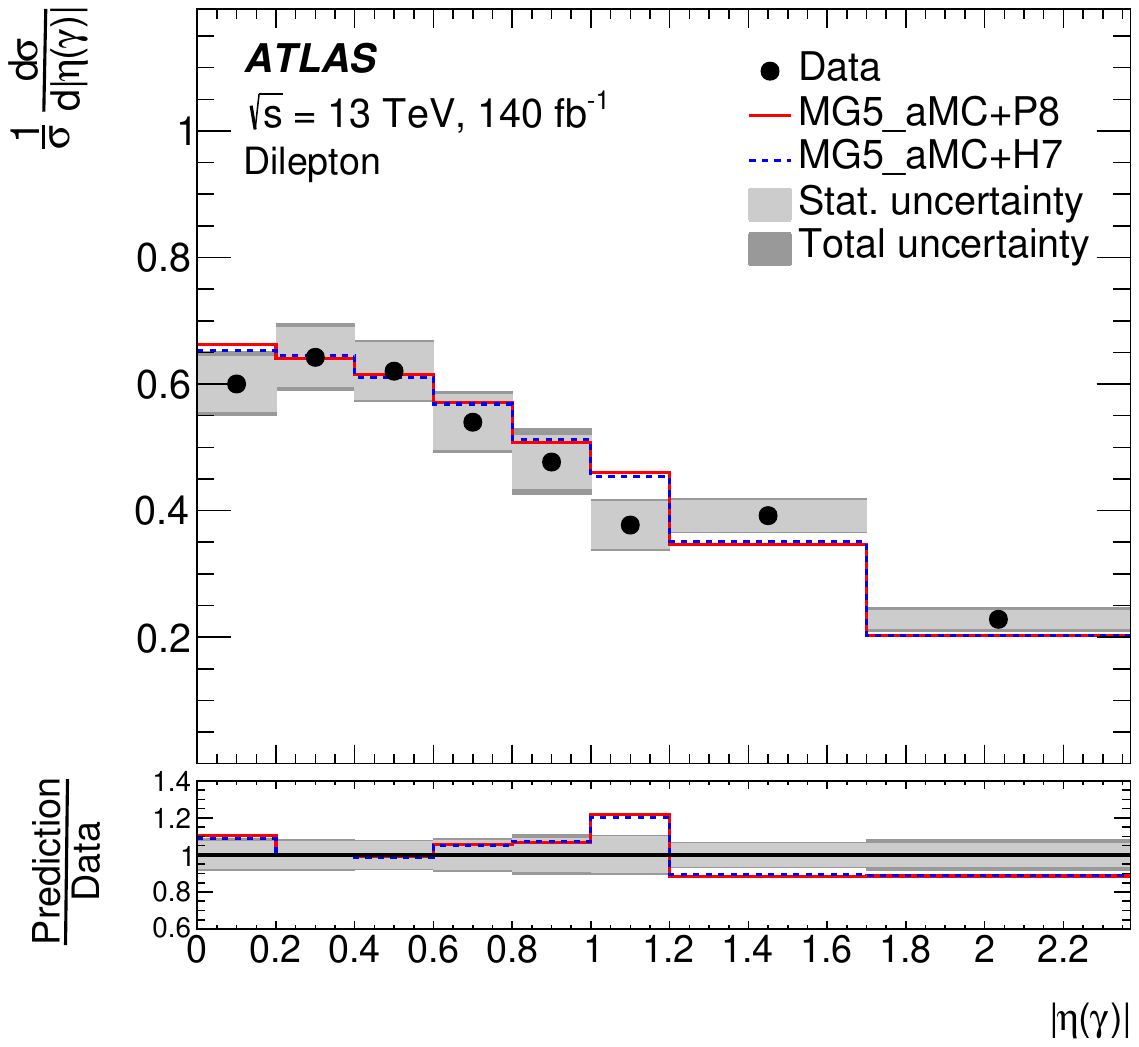}}\\
\subfloat[]{\includegraphics[width=0.40\textwidth]{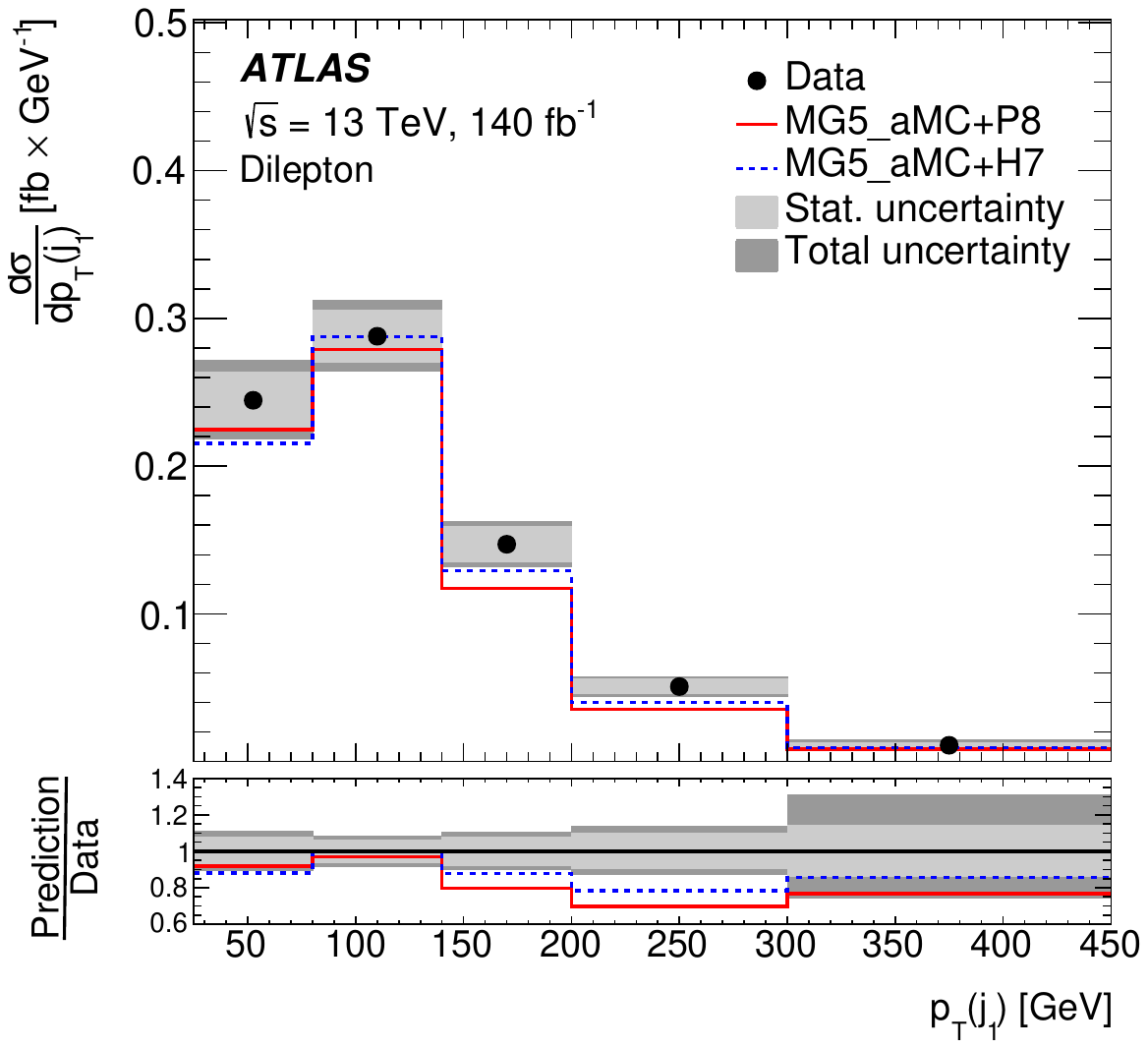}}%
\subfloat[]{\includegraphics[width=0.40\textwidth]{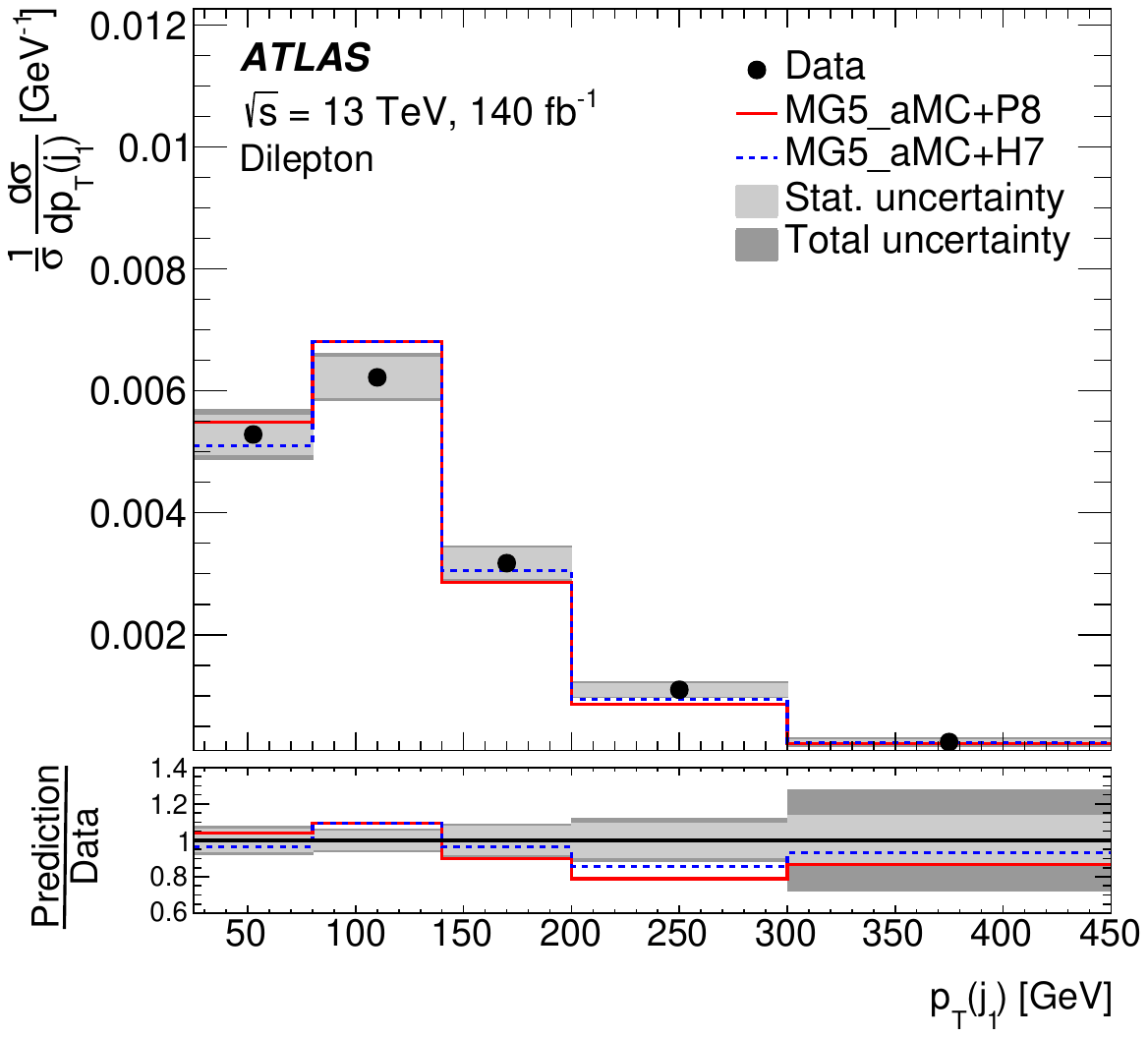}}
\caption{The (a, c, e) absolute and (b, d, f) normalised differential cross-section measured in the fiducial phase space in the dilepton channel as a function of the (a, b) photon \pt, (c, d) photon $|\eta|$ and (e, f) leading jet \pt. Data are compared with the NLO \MGNLO simulation interfaced to \PYTHIA[8] and \HERWIG[7]. The lower panels show the ratios of the predictions to the data. The last bin of the distributions includes the overflow events.}
\label{fig:ttyprod_diff_DL_app}
\end{figure}

\FloatBarrier

\section{Differential cross-sections for the combined \tty  production and decay process}
\label{sec:appendix_diff}
The absolute and normalised cross-sections for the sum of the \tty production and \tty decay measured in the same phase space as the \tty production process alone are shown as functions of photon \pt, photon $|\eta|$ and leading jet \pt in Figure~\ref{fig:tty_diff_DL_app} and as functions of $\Delta R (\gamma, b)_{\text{min}}$, $\Delta R (\gamma, \ell)$ and $\Delta R (\ell,j)_{\text{min}}$ in Figure~\ref{fig:tty_diff_DL2_app} for the dilepton channel.

\begin{figure}[ht]
\centering
\subfloat[]{\includegraphics[width=0.40\textwidth]{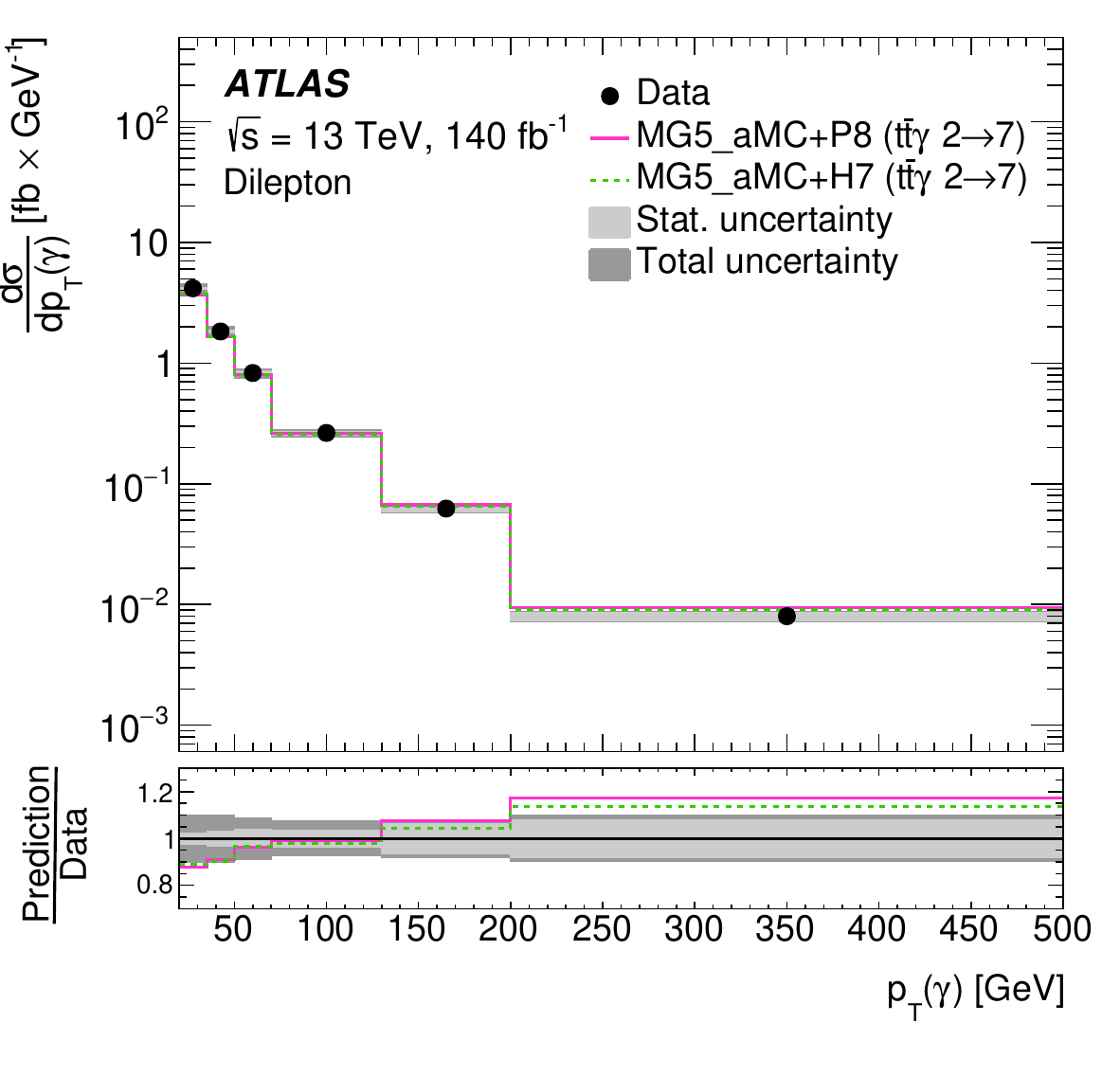}}%
\subfloat[]{\includegraphics[width=0.40\textwidth]{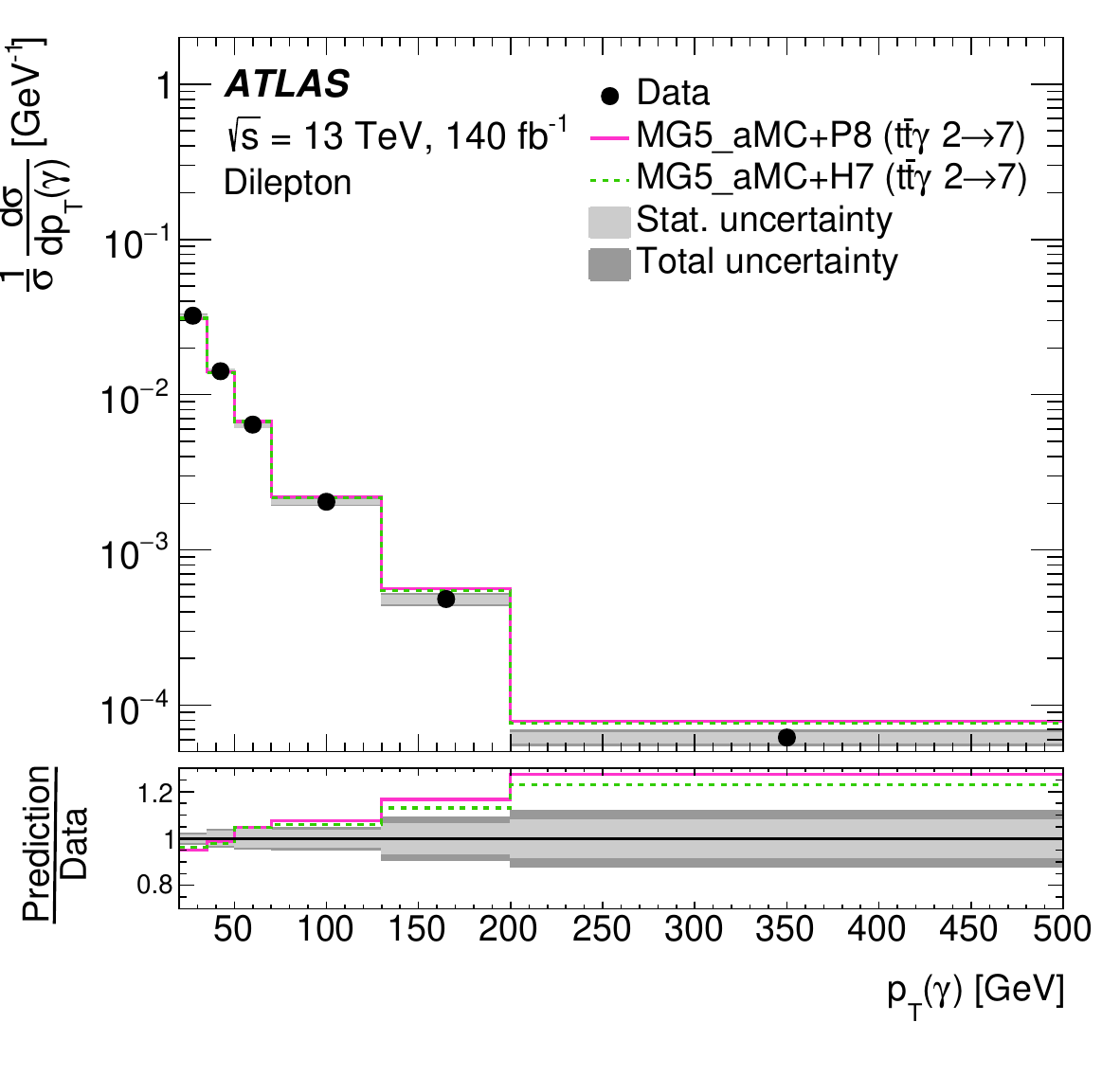}}\\
\subfloat[]{\includegraphics[width=0.40\textwidth]{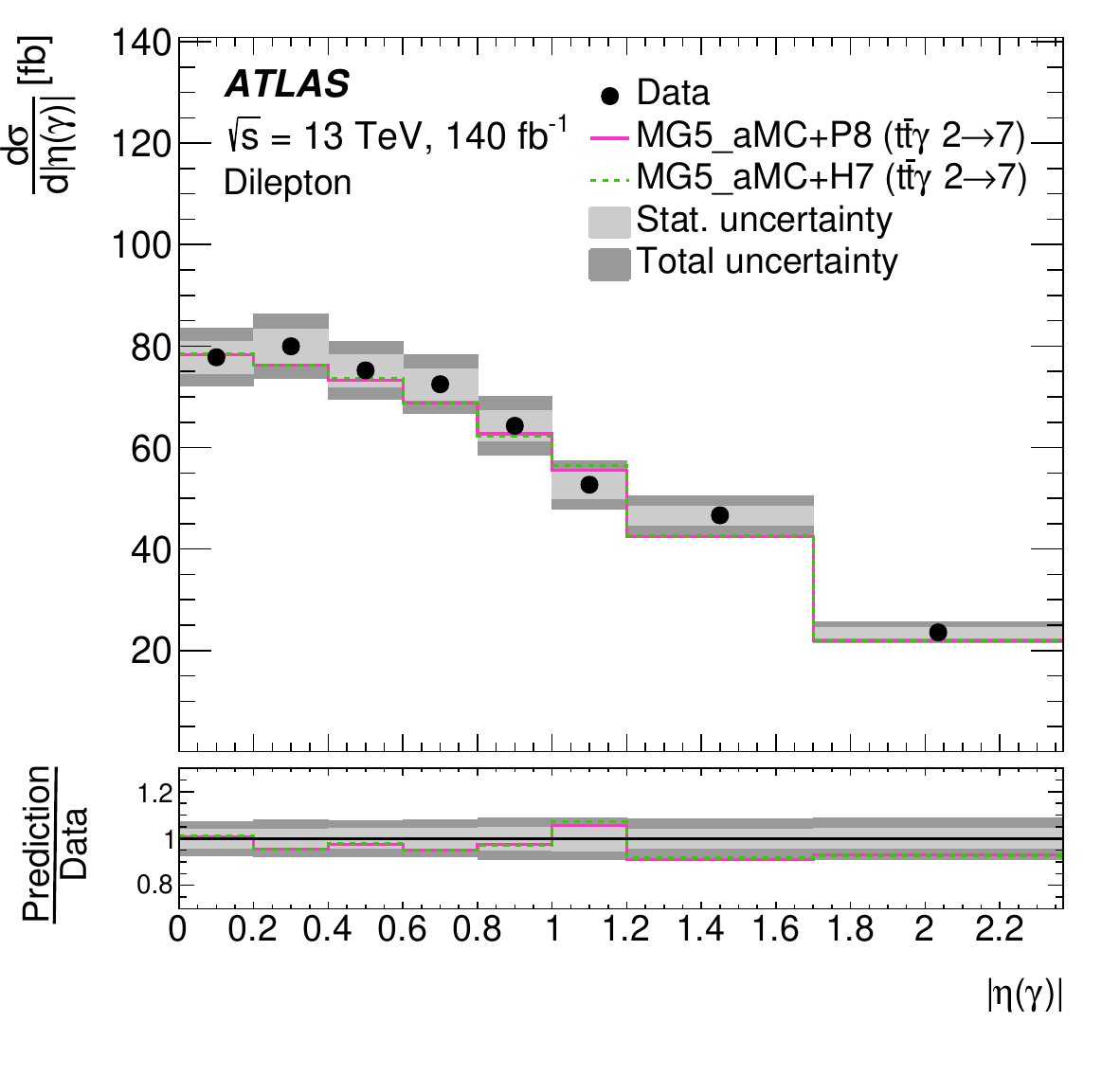}}%
\subfloat[]{\includegraphics[width=0.40\textwidth]{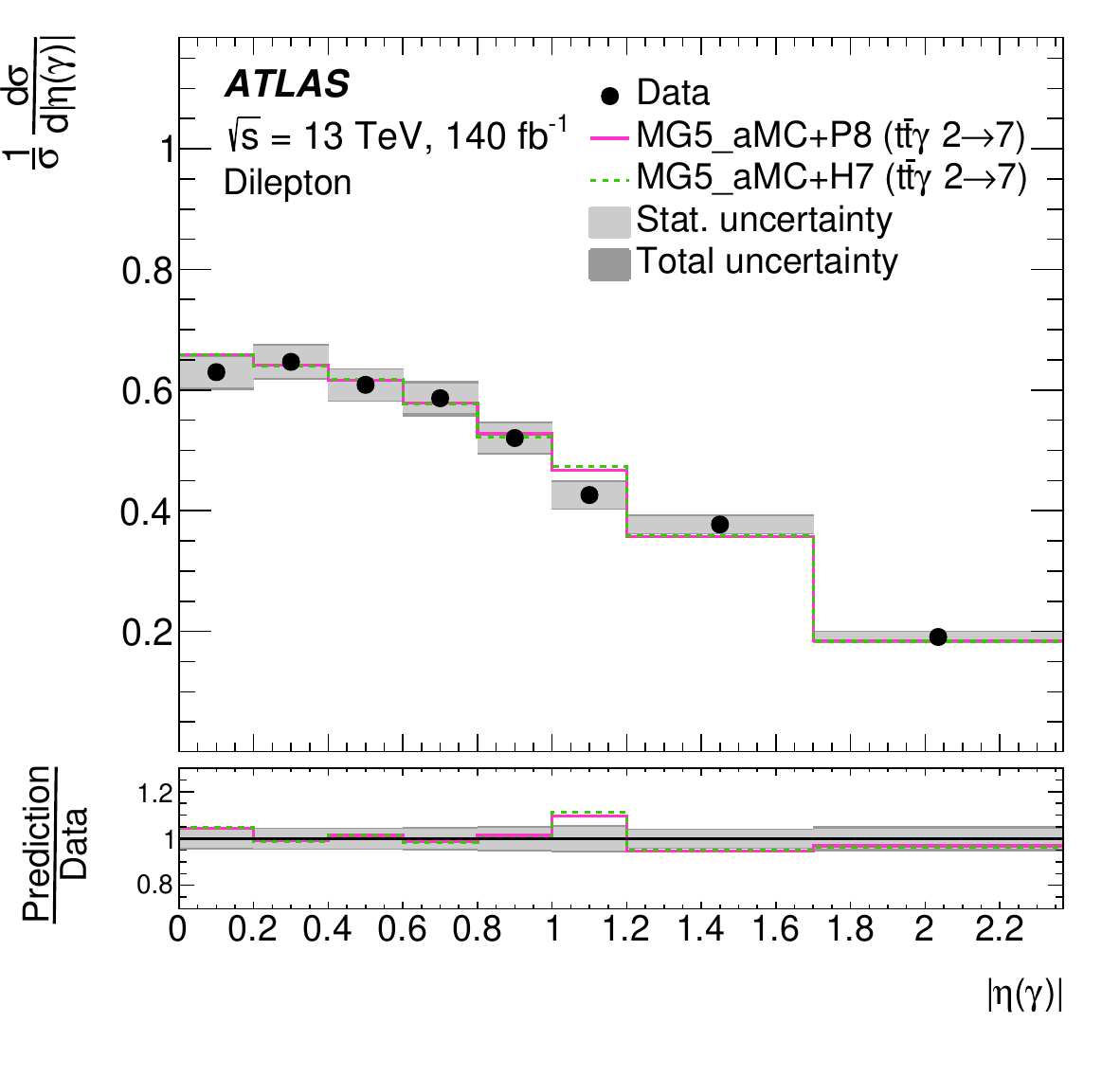}}\\
\subfloat[]{\includegraphics[width=0.40\textwidth]{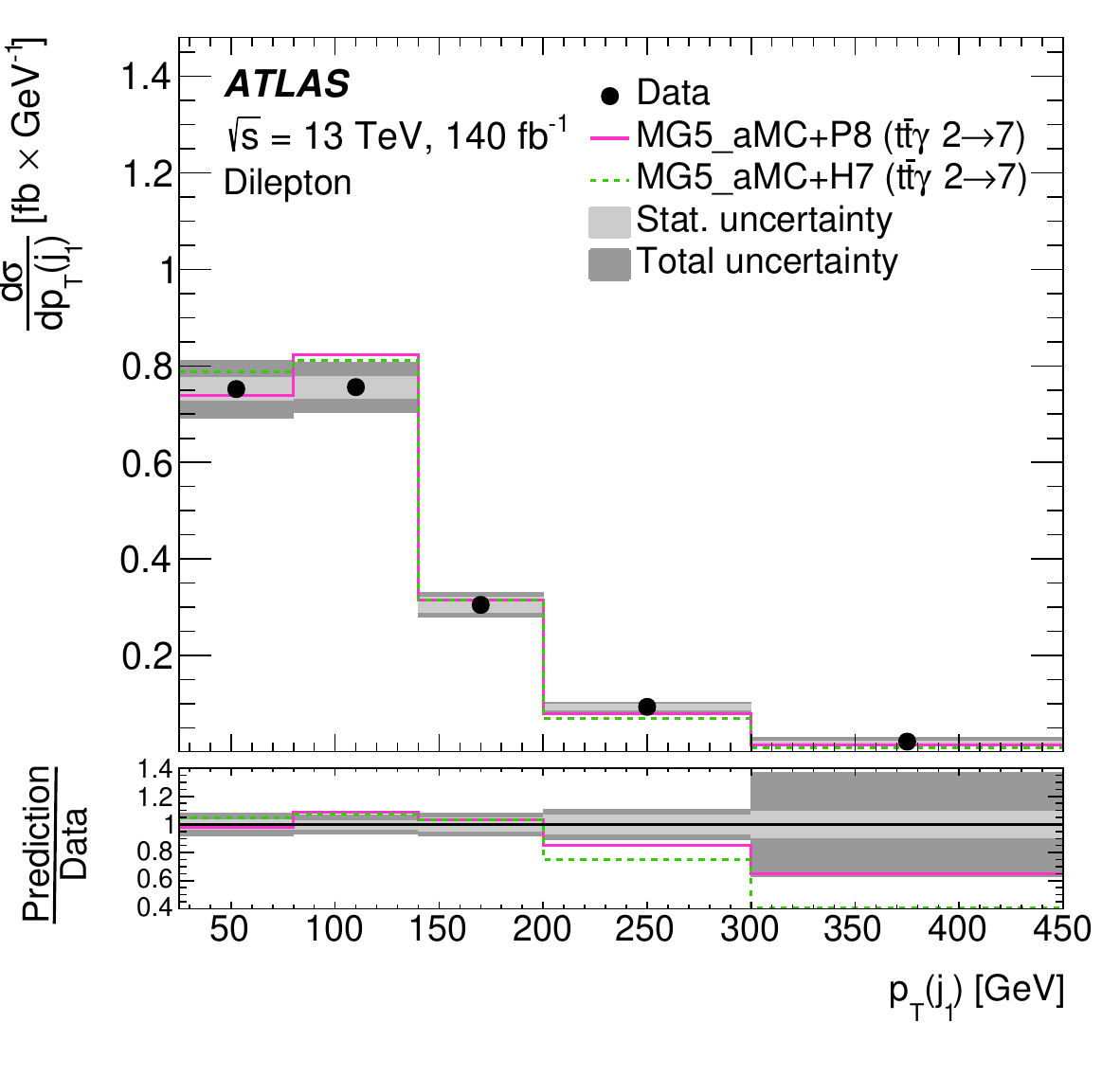}}%
\subfloat[]{\includegraphics[width=0.40\textwidth]{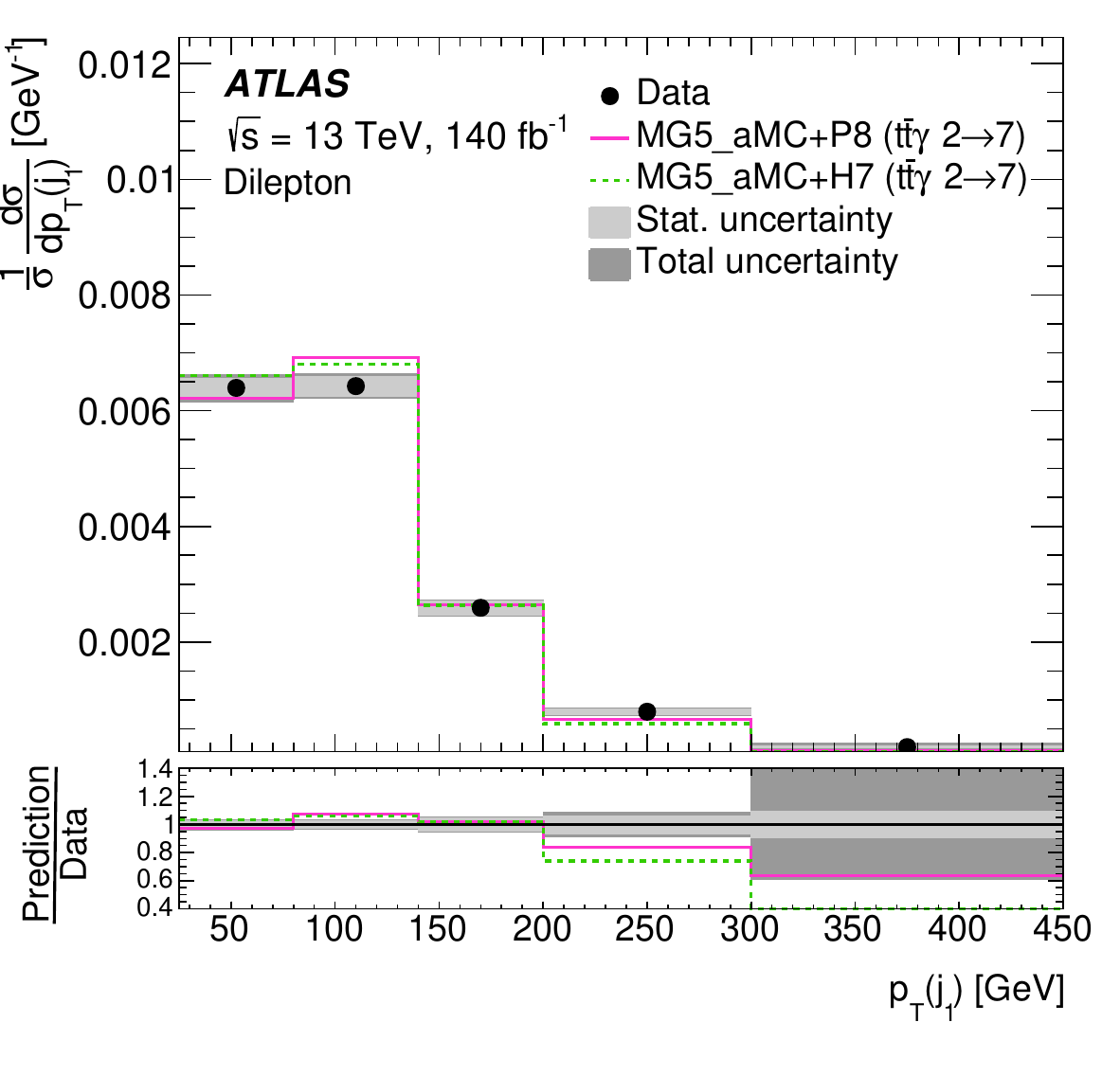}}
\caption{The (a, c, e) absolute and (b, d, f) normalised differential cross-sections of the total \tty production and decay measured in the fiducial phase space in the dilepton channel as a function of the (a, b) photon \pt, (c, d) photon $|\eta|$ and (e, f) leading jet \pt. Data are compared with the \tty \MGNLO simulation at LO (\tty $2\rightarrow 7$ process) interfaced to \PYTHIA[8] and \HERWIG[7]. The lower panels show the ratios of the predictions to the data. The last bin of the distributions includes the overflow events.}
\label{fig:tty_diff_DL_app}
\end{figure}

\begin{figure}[ht]
\centering
\subfloat[]{\includegraphics[width=0.40\textwidth]{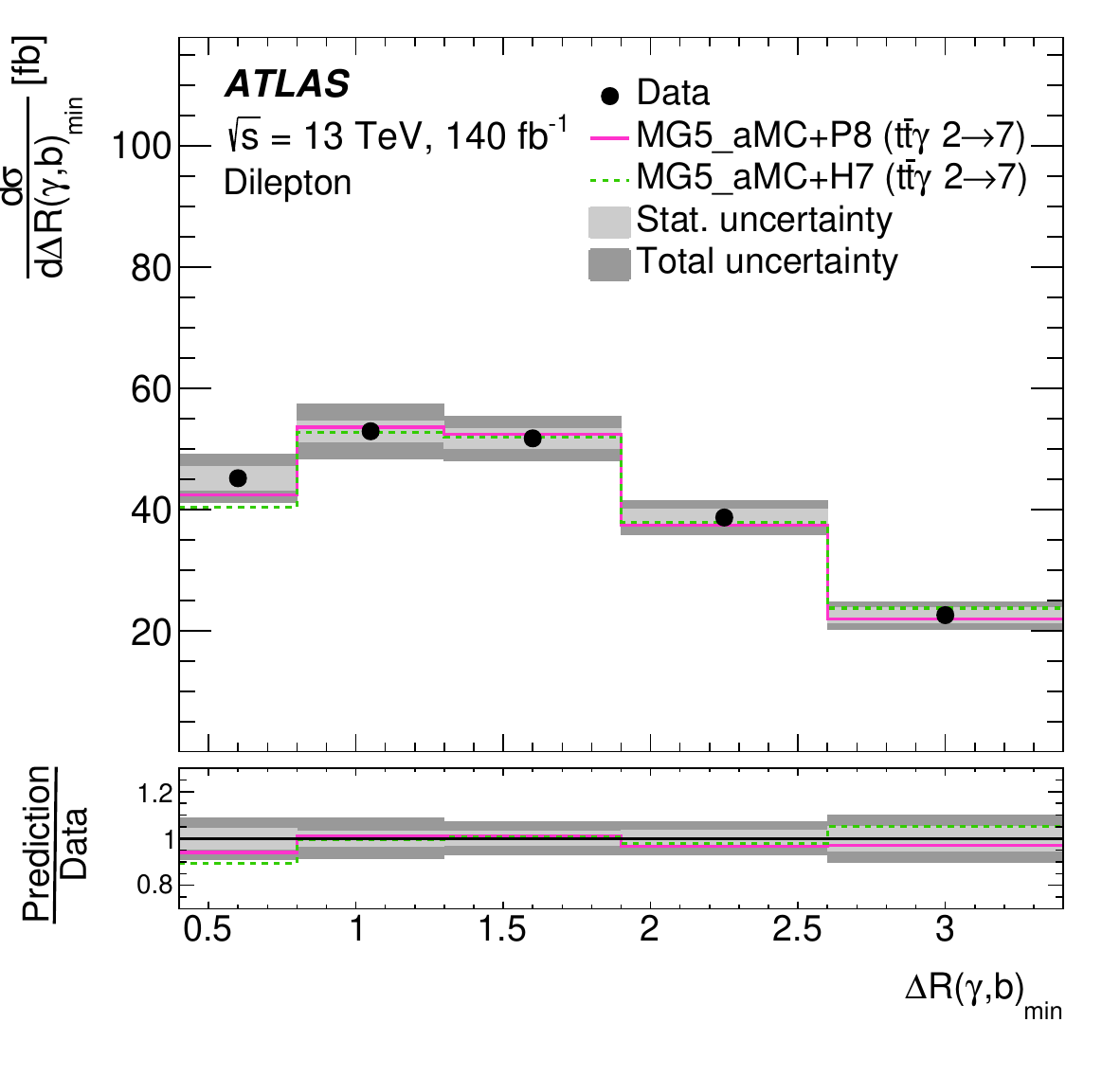}}%
\subfloat[]{\includegraphics[width=0.40\textwidth]{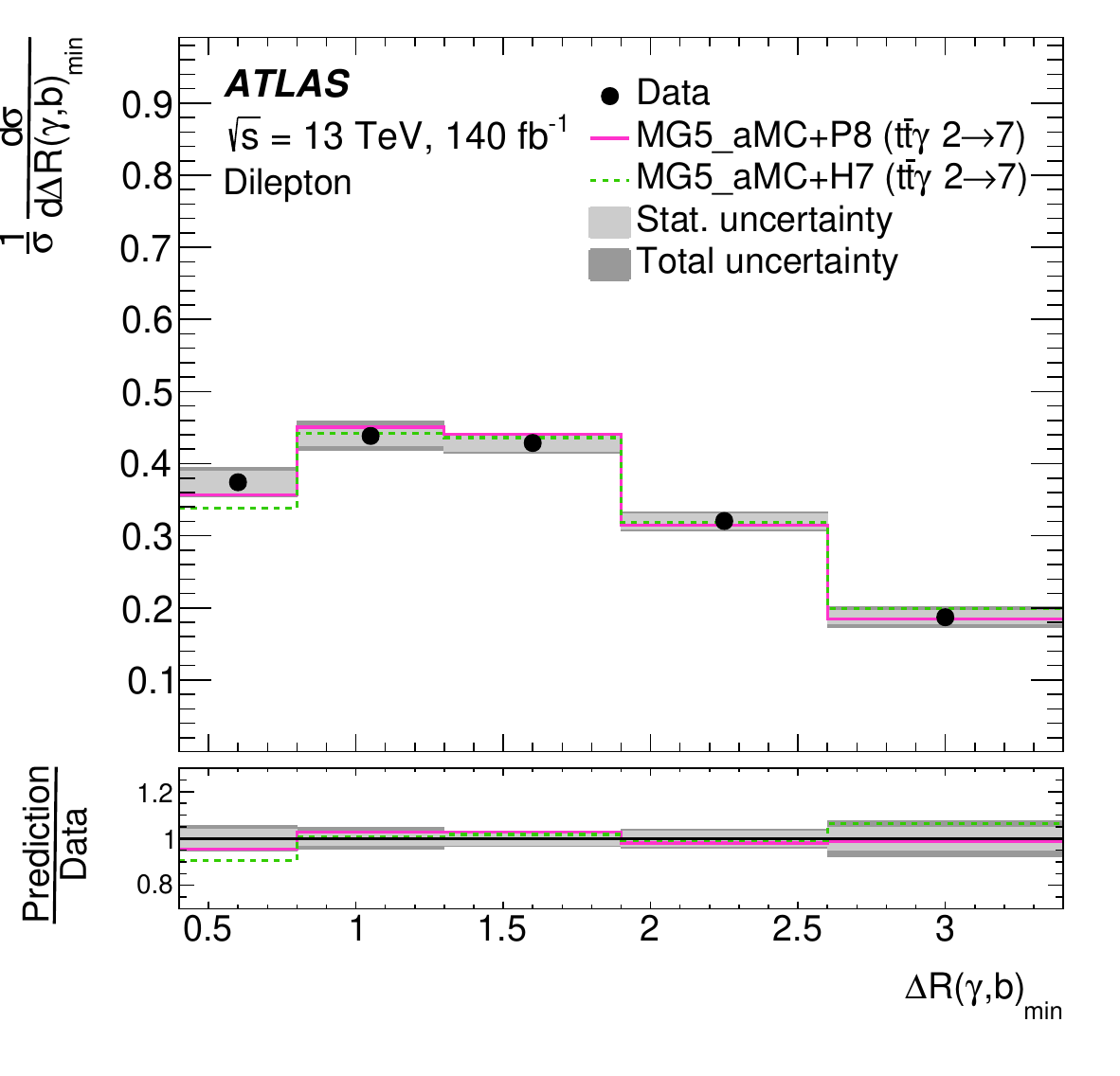}}\\
\subfloat[]{\includegraphics[width=0.40\textwidth]{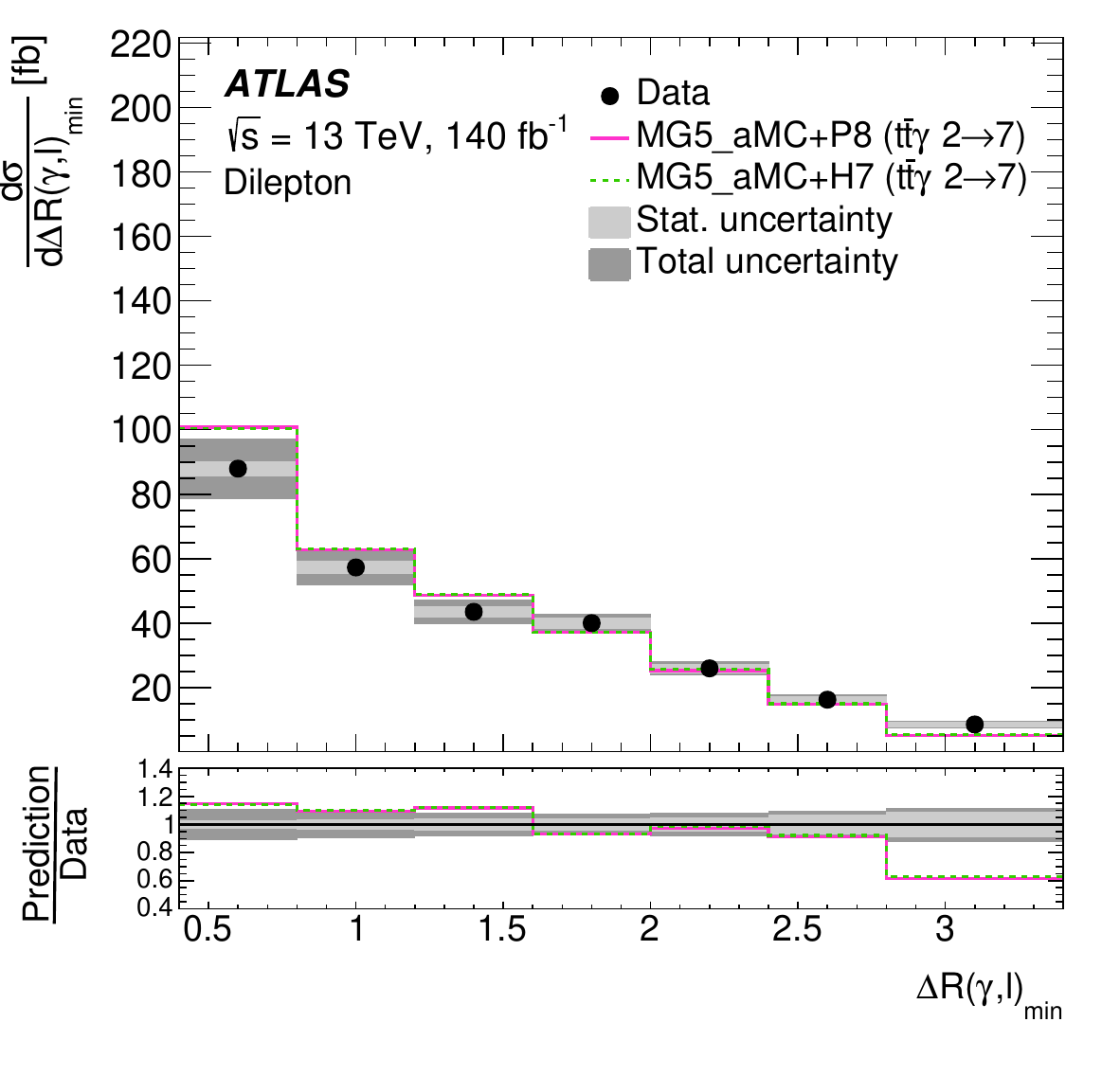}}%
\subfloat[]{\includegraphics[width=0.40\textwidth]{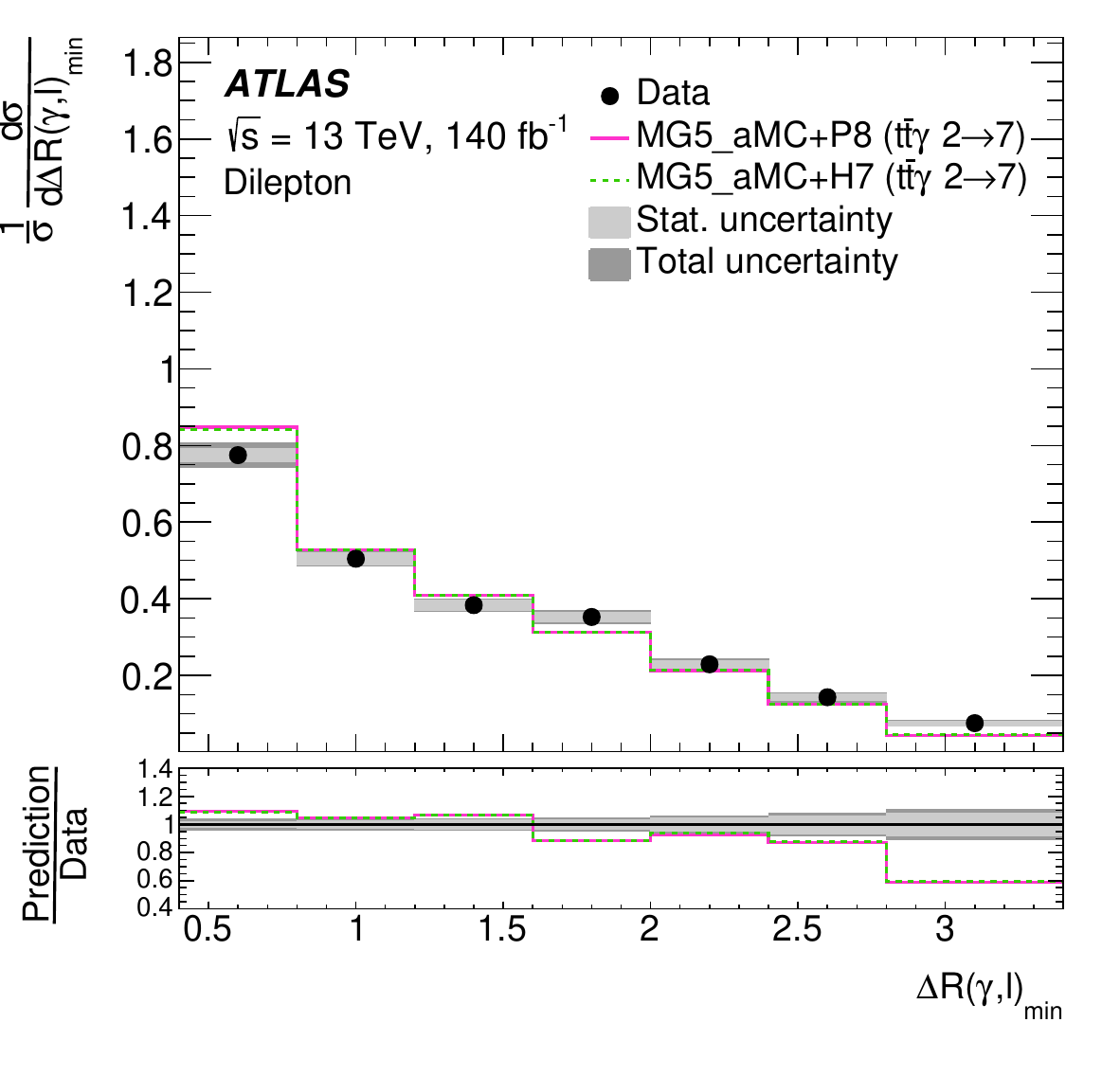}}\\
\subfloat[]{\includegraphics[width=0.40\textwidth]{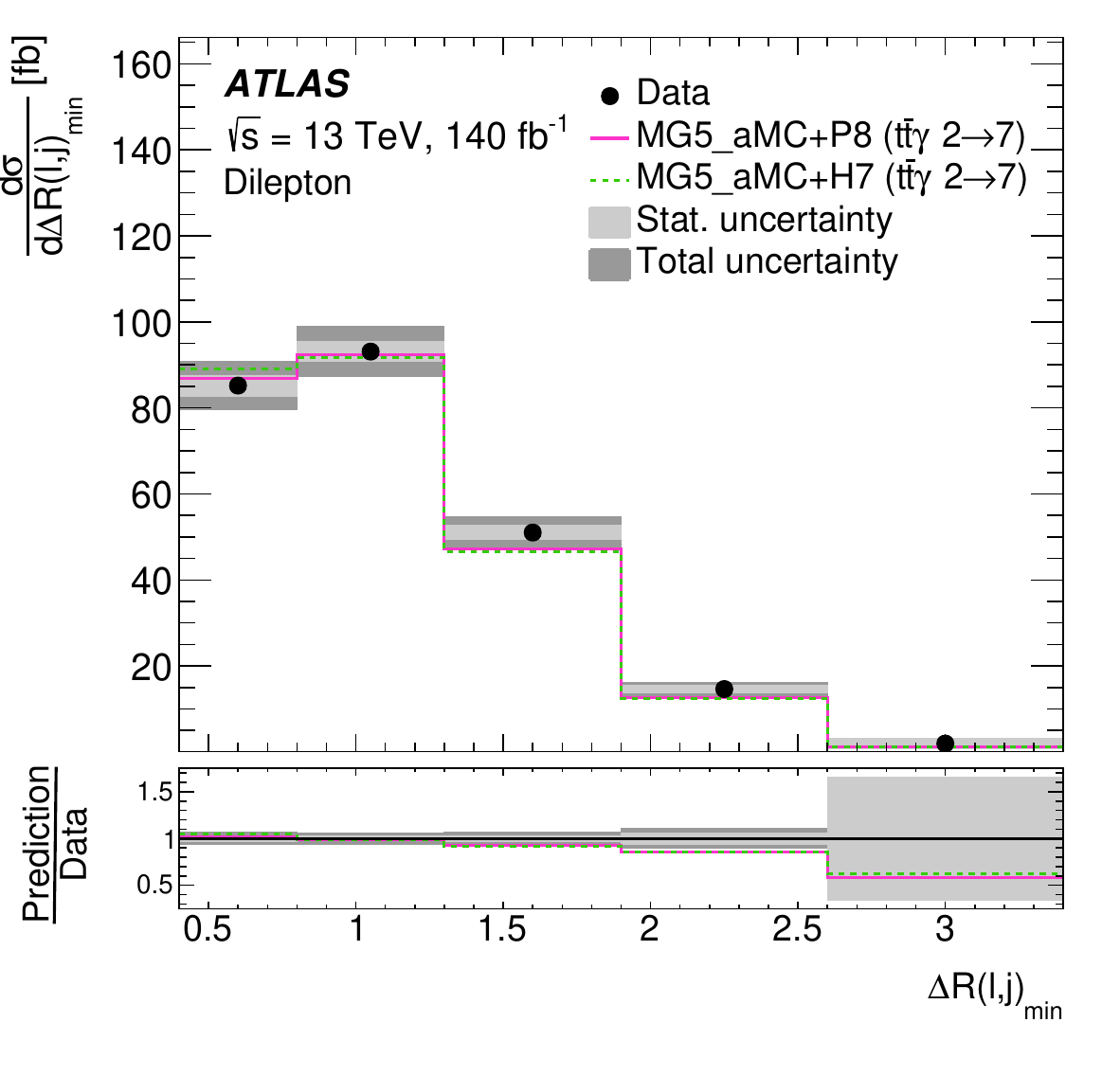}}%
\subfloat[]{\includegraphics[width=0.40\textwidth]{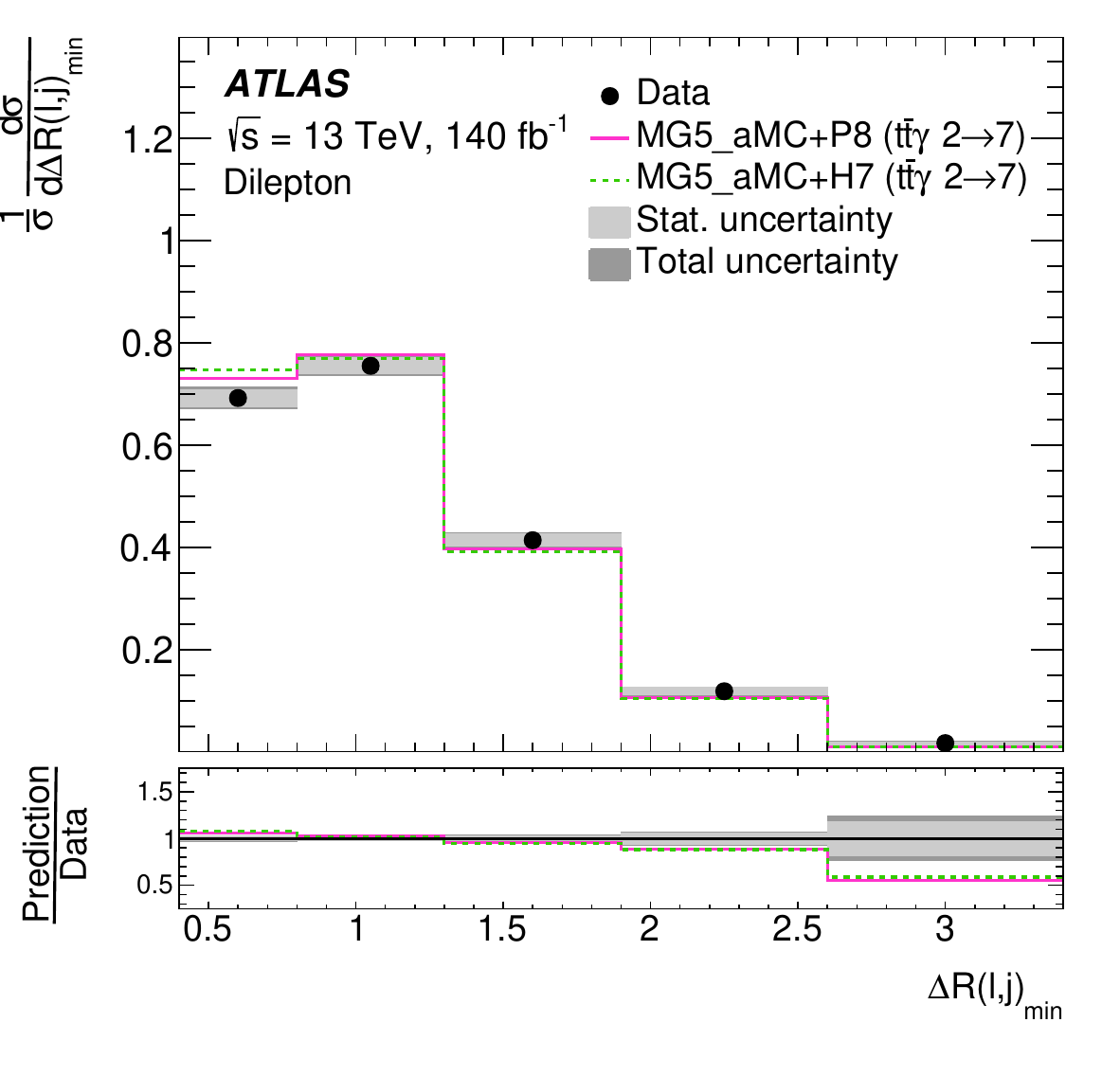}}\\
\caption{The (a, c, e) absolute and (b, d, f) normalised differential cross-sections of the total \tty production and decay measured in the fiducial phase space in the dilepton channel as a function of (a, b) $\Delta R(\gamma, b)_{\text{min}}$, (c, d) $\Delta R (\gamma, \ell)$ and (e, f) $\Delta R (\ell,j)_{\text{min}}$. Data are compared with the \tty \MGNLO simulation at LO (\tty $2\rightarrow 7$ process) interfaced to \PYTHIA[8] and \HERWIG[7]. The lower panels show the ratios of the predictions to the data. The last bin of the distributions includes the overflow events.}
\label{fig:tty_diff_DL2_app}
\end{figure}

\FloatBarrier
\printbibliography
\clearpage
 
\begin{flushleft}
\hypersetup{urlcolor=black}
{\Large The ATLAS Collaboration}

\bigskip

\AtlasOrcid[0000-0002-6665-4934]{G.~Aad}$^\textrm{\scriptsize 103}$,
\AtlasOrcid[0000-0001-7616-1554]{E.~Aakvaag}$^\textrm{\scriptsize 16}$,
\AtlasOrcid[0000-0002-5888-2734]{B.~Abbott}$^\textrm{\scriptsize 121}$,
\AtlasOrcid[0000-0002-1002-1652]{K.~Abeling}$^\textrm{\scriptsize 55}$,
\AtlasOrcid[0000-0001-5763-2760]{N.J.~Abicht}$^\textrm{\scriptsize 49}$,
\AtlasOrcid[0000-0002-8496-9294]{S.H.~Abidi}$^\textrm{\scriptsize 29}$,
\AtlasOrcid[0009-0003-6578-220X]{M.~Aboelela}$^\textrm{\scriptsize 44}$,
\AtlasOrcid[0000-0002-9987-2292]{A.~Aboulhorma}$^\textrm{\scriptsize 35e}$,
\AtlasOrcid[0000-0001-5329-6640]{H.~Abramowicz}$^\textrm{\scriptsize 152}$,
\AtlasOrcid[0000-0002-1599-2896]{H.~Abreu}$^\textrm{\scriptsize 151}$
% ...
\bigskip
\\

% ...
$^{16}$Department for Physics and Technology, University of Bergen, Bergen; Norway.\\
% ...
$^{29}$Physics Department, Brookhaven National Laboratory, Upton NY; United States of America.\\
% ...
$^{35}$$^{(a)}$Facult\'e des Sciences Ain Chock, Universit\'e Hassan II de Casablanca;$^{(b)}$Facult\'{e} des Sciences, Universit\'{e} Ibn-Tofail, K\'{e}nitra;$^{(c)}$Facult\'e des Sciences Semlalia, Universit\'e Cadi Ayyad, LPHEA-Marrakech;$^{(d)}$LPMR, Facult\'e des Sciences, Universit\'e Mohamed Premier, Oujda;$^{(e)}$Facult\'e des sciences, Universit\'e Mohammed V, Rabat;$^{(f)}$Institute of Applied Physics, Mohammed VI Polytechnic University, Ben Guerir; Morocco.\\
% ...
$^{44}$Physics Department, Southern Methodist University, Dallas TX; United States of America.\\
% ...
$^{49}$Fakult\"{a}t Physik , Technische Universit{\"a}t Dortmund, Dortmund; Germany.\\
% ...
$^{55}$II. Physikalisches Institut, Georg-August-Universit\"{a}t G\"ottingen, G\"ottingen; Germany.\\
% ...
$^{103}$CPPM, Aix-Marseille Universit\'e, CNRS/IN2P3, Marseille; France.\\
% ...
$^{121}$Homer L. Dodge Department of Physics and Astronomy, University of Oklahoma, Norman OK; United States of America.\\
% ...
$^{151}$Department of Physics, Technion, Israel Institute of Technology, Haifa; Israel.\\
$^{152}$Raymond and Beverly Sackler School of Physics and Astronomy, Tel Aviv University, Tel Aviv; Israel.\\
% ...
$^{a}$ Also Affiliated with an institute covered by a cooperation agreement with CERN.\\
$^{b}$ Also at An-Najah National University, Nablus; Palestine.\\
$^{c}$ Also at Borough of Manhattan Community College, City University of New York, New York NY; United States of America.\\
$^{d}$ Also at Center for High Energy Physics, Peking University; China.\\
$^{e}$ Also at Center for Interdisciplinary Research and Innovation (CIRI-AUTH), Thessaloniki; Greece.\\
$^{f}$ Also at Centro Studi e Ricerche Enrico Fermi; Italy.\\
% ...

\end{flushleft}


%

%
%

%

%
%
%

%

\end{document}